# THÈSE

PRÉSENTÉE À

## L'Université de Pau et des Pays de l'Adour

ÉCOLE DOCTORALE DES SCIENCES EXACTES ET DE LEURS APPLICATIONS

Par
**Mourad Kmimech**

POUR OBTENIR LE GRADE DE
DOCTEUR

SPÉCIALITÉ : INFORMATIQUE

# Vérification d'assemblages de composants logiciels : Application aux modèles de composants UML2.0 et Ugatze

Soutenue le : 17 Décembre 2010

Devant la commission d'examen composée de :

**Composition du jury**

| | |
|---|---|
| *Président :* | Bernard Coulette |
| *Rapporteurs :* | Bernard Coulette<br>Jean-Pierre Giraudin |
| *Examinateurs :* | Eric Cariou<br>Noureddine Belkhatir |
| *Directeurs de thèse :* | Mohamed Tahar Bhiri<br>Philippe Aniorté |

# Remerciements

Une thèse est un travail long, demandant beaucoup d'investissement personnel, et surtout parsème de doutes. C'est aussi, avant tout, une expérience humaine à part entière. Je tiens donc à remercier l'ensemble des personnes qui ont contribué, parfois sans le savoir, à l'achèvement de ce travail.

Je tiens à remercier tous les membres du jury. Un très grand merci à M. **Jean-Pierre Giraudin**, Professeur à l'Université Pierre Mendès France et M. **Bernard Coulette**, Professeur à l'Université de Toulouse II-Le Mirail d'avoir accepté de rapporter ce travail. J'ai apprécié la profondeur de la relecture et la pertinence de vos commentaires et de vos remarques qui ont permis d'améliorer la qualité de ce document. Merci à M. **Eric Cariou**, Maître de conférences à l'Université de Pau et des Pays de l'Adour et M. **Noureddine Belkhatir** Professeur à l'Université Pierre Mendès France d'avoir accepté d'examiner mon travail. Enfin, je remercie M. **Bernard Coulette** Professeur à l'Université de Toulouse II-Le Mirail de m'avoir accordé l'honneur d'être le président de mon jury.

Je tiens à remercier M. **Philippe Aniorté**, Professeur à l'Université de Pau et des Pays de l'Adour pour m'avoir accueilli dans son laboratoire et pour avoir accepté de diriger cette thèse. Je te remercie pour m'avoir fait confiance tout au long de ces années.

Je tiens à remercier M. **Mohamed Tahar Bhiri,** Maître assistant à l'Université de Sfax de m'avoir encadré durant ces années de mastère et de thèse. Je n'aurais pu rêver meilleur encadreur, toujours disponible, toujours motivé, toujours de bons conseils. J'ai énormément apprécié les années de travail sous ta direction. Merci pour ton soutien sans lequel je n'aurais jamais réussi à aller au bout, tes conseils toujours lumineux et ta patience face à mes passages à vide. Merci aussi pour le temps que tu m'as consacré au jour le jour pendant ces années de thèse et ton amour contagieux de la recherche. Merci enfin pour l'amitié que tu m'as offerte.

Je remercie tout particulièrement mes deux amis Mohamed et Walid. Merci à vous pour vos aides et vos encouragements tout particulièrement pendant les périodes de doute.

Je tiens à remercier M. Mabrouk Ben Nacer d'avoirs mis à ma disposition son bureau à la faculté des sciences de Sfax durant mes dernières années de thèse.

Je tiens à remercier Nesrine Ben Ammar et Dridi Sahar pour leurs aides lors des traductions.

De manière plus personnelle, un très grand merci à ma future femme Imen pour son soutien et sa patience pendant toutes ces années de thèse.

Ces années de travail n'auraient pu être réalisées sans un soutien extérieur et infini de ma famille. Mes parents d'abord, Ismail et Nabiha sans lesquels je n'aurai jamais pu en arriver là. Merci pour cela, merci aussi pour votre compréhension et vos sacrifices. Merci aussi à mes frères Hichem et Mohamed et ma sœur Houdé, ainsi qu'à leurs petites familles (Yacine, Amine, Ilef, Islem, Yomna et Oswa qui n'ai pas encore né et qui par coïncidence naîtra peut être le jour de ma soutenance). Je leur prie d'accepter aussi dans ces quelques lignes mes excuses pour l'éloignement toutes ces dernières années

Je remercie tous mes amis qui ont pu m'épauler ces dernières années, merci Khaled, Loussif, Zied, Mohamed Ali, Habib, Haffed, Hattab, Ezzedine, Makrem, Naoufel, Mabrouk, Nourredine, Tawfik, l'oncle Nacer, Maher, Ahmed, recevez ici toute mon amitié

# Sommaire

# Liste des figures et tables

# Introduction générale

## 1. Problématique

Le paradigme composant est apparu après le paradigme objet. Dans les architectures à composants, l'unité de décomposition est le composant. A l'instar de la notion d'objet, il n'existe pas de définition universelle de la notion de composant. Cependant trois aspects fondamentaux liés à la notion de composant sont largement admis [Szyperski, 2002]. Premièrement un composant décrit d'une façon explicite les services qu'il offre ainsi que les services qu'il requiert. Deuxièmement, un composant est une entité composable. Troisièmement, un composant est une entité capable d'être déployée sur une plate-forme d'exécution, indépendamment des autres composants.

L'approche par composants vise la réutilisation par assemblage aisé et cohérent des composants. Vis-à-vis de l'approche par objets, cette approche déplace la complexité d'un graphe de classes (hiérarchie de classes, redéfinition de méthodes et relation client) vers des points de connexion entre les composants en distinguant de façon nette deux types de composants : composant serveur et composant client. Afin de vérifier la cohérence – absence de contradiction- d'un assemblage de composants, une approche contractuelle basée sur des contrats d'assemblage établis entre les composants serveurs et les composants clients est préconisée. Celle-ci distingue quatre niveaux de contrats d'assemblage [Beugnard, 1999], [Beugnard, 2005] : contrats syntaxiques, contrats sémantiques, contrats de synchronisation et contrats de qualité de services (Propriétés Non-Fonctionnelles : PNF). Cette approche contractuelle inter-composants est perçue comme un prolongement à la conception par contrats (Design by Contracts) célèbre dans le monde OO et supportée par divers langages comme Eiffel [Meyer, 1992], OCL [OMG, 2005a] et JML [Leavens, 2000].

Le cadre général de cette thèse est la vérification de la cohérence d'un assemblage de composants en adoptant une approche contractuelle. Précisément, nous abordons deux instances de ce problème de vérification de la cohérence de l'assemblage de composants : l'assemblage de composants UML2.0 et l'assemblage de composants Ugatze[1].

## 2. Proposition

La démarche partagée par la plupart des travaux existants dans le domaine de la vérification de la cohérence d'assemblages de composants est l'emploi des techniques et des outils généraux tels que B [Abrial, 1996] et CSP [Hoare, 1985]. Pour y parvenir, de tels travaux [Lanoix, 2008a], [André, 2010], [Mouakher, 2008] proposent des traductions plus ou moins systématiques de modèles de composants source vers le formalisme cible. Ceci oblige l'architecte à manipuler des concepts liés au formalisme cible tels que machine abstraite, machine de raffinement, primitives de composition de machines, obligations de preuve, preuve interactive, événement, processus, composition de processus, non déterministe et relations de raffinement. En outre, le passage d'un modèle semi-formel comme UML2.0 vers des méthodes formelles générales comme B et CSP se heurte aux problèmes suivants :

- difficulté d'obtenir une spécification formelle conservant le plus possible la structure de la spécification semi-formelle,

[1] Le modèle Ugatze est issu de notre équipe de recherche.

- difficulté d'obtenir une spécification formelle suffisamment claire pour être facilement lisible et exploitable.

- difficulté d'animer des spécifications formelles afin d'obtenir des ''bons'' modèles

Dans cette thèse, nous préconisons une approche favorisant la continuité entre le modèle source et le modèle cible. Les deux modèles de composants sources retenus sont: UML2.0 et Ugatze. Ils sont considérés comme **des modèles semi-formels**.

Afin de vérifier la cohérence de l'assemblage de composants de ces deux modèles, cette thèse préconise leur traduction vers **des modèles de composants formels** comme Acme/Armani [Garlan, 2000] et Wright [Allen, 1997].

En ce qui concerne la vérification d'assemblages de composants UML2.0, nous visons trois types de contrats : syntaxiques ou encore structurels, de synchronisation et de qualité. Le modèle UML2.0 à vérifier est décrit en utilisant le diagramme de composants pour les aspects structuraux, une extension de PSM (Protocol State Machine) appelée PoSM (Port State Machine) [Samek, 2005] pour les aspects comportementaux et CQML [Aagedal, 2001] pour les aspects non-fonctionnels. La vérification des contrats syntaxiques et de qualité de services établis dans cette thèse pour le modèle de composants UML2.0 est confiée à l'évaluateur des prédicats supporté par la plateforme AcmeStudio [ABLE, 2009]. Tandis que la vérification des contrats de synchronisation est confiée au model-checker FDR [FDR2, 2003] en passant par notre traducteur de Wright vers CSP : Wr2fdr. En outre, afin d'ouvrir le modèle de composants UML2.0 sur **l'analyse dynamique** (vis-à-vis de l'analyse statique basée sur les contrats), nous avons conçu, réalisé et testé en utilisant une approche de type IDM (précisément la plateforme AMMA [AMMA, 2005] autour d'ATL [Jouault, 2006]) un outil de transformation de Wright vers Ada : Wright2Ada.

En ce qui concerne le modèle de composants Ugatze, nous visons essentiellement la vérification des propriétés structurelles. Nous avons formalisé en Acme les concepts structuraux venant d'Ugatze et nous avons établi des contrats d'assemblage décrits comme des propriétés invariantes en Armani.

## 3. Organisation de la thèse

Cette thèse comporte huit chapitres. Dans le chapitre 1, nous introduisons les notions de base : composant, contrat et classification des contrats à quatre niveaux (syntaxique, sémantique, synchronisation et qualité de services). Ensuite, nous étudions les moyens de description et de vérification de ces contrats. Enfin, nous proposons une approche de vérification de la cohérence d'assemblages de composants semi-formels UML2.0 et Ugatze. Pour des raisons de continuité, notre approche utilise des traductions des modèles de composants UML2.0 et Ugatze vers des modèles de composants formels Acme/Armani et Wright.

Dans le chapitre 2, nous présentons les modèles retenus : UML2.0, Ugatze, CQML, Acme/Armani et Wright.

Dans le chapitre 3, nous proposons deux démarches : *VerifComponentUML2.0* et *VerifComponentUgatze*. La démarche *VerifComponentUML2.0* permet de vérifier la cohérence d'un assemblage de composants UML2.0 vis-à-vis des contrats syntaxiques, de qualité de services et synchronisation. Quant à la démarche *VerifComponentUgatze*, elle offre un cadre permettant la vérification de l'assemblage de composants Ugatze vis-à-vis des contrats syntaxiques.

Dans le chapitre 4, nous proposons une traduction du modèle de composants UML2.0 en Acme/Armani afin de vérifier la cohérence d'assemblages de composants UML2.0 vis-à-vis des contrats syntaxiques et structurels [Kmimech, 2009a], [Kmimech, 2009d], [Kmimech, 2009e]. La vérification de ces contrats est confiée à l'évaluateur de prédicats supporté par la plateforme AcmeStudio [ABLE, 2009].

Le chapitre 5 a pour objectif de formaliser en Acme/Armani un assemblage de composants UML2.0 dotés des PNF décrites en CQML afin de vérifier sa cohérence : chaque PNF exigée doit avoir sa réciproque (PNF offerte) dans l'assemblage de composants traité.

Dans le chapitre 6, nous proposons une maintenance évolutive et corrective de l'outil [Wr2fdr, 2005]. En effet, suite à l'utilisation de l'outil Wr2fdr, nous avons remarqué que l'outil génère des erreurs liées aux propriétés 2 (absence d'interblocage sur les connecteurs) et 3 (absence d'interblocage sur les rôles). En plus, les propriétés 1 (cohérence des ports avec le Calcul) et 8 (compatibilité port/rôle) ne sont pas traitées par cette version de l'outil. Vu l'importance de cet outil, nous avons contacté les auteurs de Wright, expliqué les problèmes rencontrés et récupéré le source de cet outil afin de le corriger et de le compléter.

Dans le chapitre 7, nous proposons une approche IDM permettant de transformer une architecture logicielle décrite à l'aide de l'ADL formel Wright vers un programme concurrent Ada comportant plusieurs tâches exécutées en parallèle. Pour y parvenir, nous avons élaboré deux méta-modèles en Ecore : le méta-modèle de Wright et le méta-modèle partiel d'Ada. De plus, nous avons conçu, réalisé et testé un programme Wright2Ada permettant de transformer un modèle source Wright conforme à son méta-modèle Wright vers un modèle cible Ada conforme au méta-modèle partiel Ada.

Dans le chapitre 8, nous proposons une approche de traduction du modèle de composants semi-formel Ugatze vers le modèle de composants Acme/Armani. Ceci autorise la vérification des contrats syntaxiques et structurels d'un assemblage de composants Ugatze.

De plus, cette thèse comporte les annexes suivantes :

- L'annexe A décrit la sémantique statique de la partie structurelle d'Ada,
- L'annexe B formalise en OCL la sématique statique de la partie comportementale d'Ada,
- L'annexe C présente la programmation en ATL de la traduction des aspects comportementaux de Wright (décrits en CSP) en Ada,
- L'annexe D fournit en entier le programme Wright2Ada en ATL permettant de transformer de la spécification Wright vers du code Ada,
- L'annexe E fournit la grammaire de l'ADL Wright décrite en Xtext,
- L'annexe F fournit des fragments en Xpand corespondant aux instructions de la partie exécutive d'Ada,
- Enfin, l'annexe G fournit le template de génération de code Ada.

## 4. Publications

### Conférences Internationales

[1] Mourad Kmimech, Mohamed Tahar Bhiri, Philippe Aniorté, Abdelmajid Benhamadou. «*Formalization of Ugatze component model*». Second International Conference on Web and Information Technologies, Kerkena (Tunisia), June 2009.

[2] Mourad Kmimech, Mohamed Tahar Bhiri, Mohamed Graiet, Philippe Aniorté. «*Formalization in Acme of UML model components*». In International Conference on Computer Science and Information Systems (ATINER), Athens (Greece), July 2009.

[3] Mourad Kmimech, Mohamed Tahar Bhiri, Philippe Aniorté. «*Checking component assembly in Acme: an approach applied on UML2.0 components model*». In 4nd IEEE International Conference on Software Engineering Advances (ICSEA'2009). IEEE CS Press, Porto (Portugal), September 2009.

[4] Mohamed Graiet, Raoudha Maraoui Mourad Kmimech, Mohamed Tahar Bhiri, Walid Gaaloul. «*Towards an approach of formal verification of mediation protocol based on Web services*». The 12th International Conference on Information Integration and Web-based Applications & Services (iiWAS2010), Paris, Novembre 2010.

[5] Raoudha Maraoui, Mohamed Graiet, Mourad Kmimech, Mohamed Tahar Bhiri, Béchir El Ayeb. «*Formalization of Mediation Protocol for Web Services Composition with ACME/ARMANI ADL*». The Second International Conferences on Advanced Service Computing (SERVICE COMPUTATION 2010), Lisbon (Portugal), November 2010.

### Conférences nationales

[1] Mourad Kmimech, Mohamed Tahar Bhiri, Philippe Aniorté, Abdelmajid Ben Hamadou. «*Vers une formalisation du métamodèle de composants Ugatze*», Congrès INFORSID'O6, Workshop OCM-SI'06 (Objet Composant Modèle pour les Systèmes d'Information), Hammamet (Tunisie), Mai 2006.

[2] Mourad Kmimech. «*Contractualiser les composants Ugatze*», Congrès INFORSID'O7, Forum Jeunes Chercheurs, Perros-Guirec (France), 22 Mai 2007.

[3] Mourad Kmimech, Mohamed Tahar Bhiri, Philippe Aniorté, Abdelmajid Ben Hamadou. «*Une approche de contractualisation des composants Ugatze*», Congrès INFORSID'O7, Workshop OCM-SI'07 (Objet Composant Modèle pour les Systèmes d'Information), Perros-Guirec (France), Mai 2007.

[4] Mourad Kmimech, Mohamed Tahar Bhiri, Mohamed Graiet, Philippe Aniorté. «*Vérification d'assemblage de composants UML2.0 à l'aide d'Acme*». In workshop LMO/SafeModels, Nancy (France), Mars 2009.

[5] Mourad Kmimech, Mohamed Tahar Bhiri, Philippe Aniorté. «*Une approche de vérification d'assemblage de composants : application au modèle Ugatze*». In workshop Inforsid/ERTSI'09, Toulouse, Mai 2009.

[6] Mounira Belmabrouk, Mourad Kmimech, Mohamed Tahar Bhiri. «*Modélisation par objets des ROBDDs*». MajecStic 2010, Marseille, Octobre 2010.

# **Chapitre 1 :** Les architectures à composants

## 1.1 Introduction

Ce chapitre comporte cinq sections. La première section introduit les aspects fondamentaux da la notion de composant. La deuxième section présente la notion de contrat entre un composant serveur et un composant client en se basant sur la conception par contrat (Design by Contract) [Meyer, 1992]. La troisième section présente une classification de contrats à quatre niveaux : contrats syntaxiques, contrats sémantiques, contrats de synchronisation et contrats de qualité de services. Une telle classification est considérée comme un prolongement à la conception par contrat. La quatrième section aborde la vérification statique et dynamique d'assemblages de composants. Enfin, la cinquième section propose notre approche de vérification d'assemblages de composants décrits par des modèles de composants semi-formels comme UML2.0 et Ugatze.

## 1.2 Notion de composant

Le paradigme composant est apparu après le paradigme objet. Dans les architectures à composants, l'entité composable est le composant. À l'instar de la notion d'objet, il n'existe pas de définition universelle de la notion de composant. Cependant Szyperski a donné une définition, largement admise, de la notion de composant : "A software component is a unit of composition with contractually specified interfaces and explicit context dependencies only. A software component can be deployed independently and is subject to composition by third parties'' [Szyperski, 2002]. Cette définition exhibe trois aspects fondamentaux d'un composant logiciel. Premièrement un composant décrit d'une façon explicite les services qu'il offre ainsi que les services qu'il requiert. Les services exigés par un composant représentent ses dépendances vis-à-vis de son environnement -les autres composants-. Ceci est considéré comme un plus comparé à la notion d'objet. Les services offerts et/ou exigés par un composant sont regroupés sous forme d'interfaces contractualisées (cf. section 1.3).

Deuxièmement, un composant est une entité composable. Cela signifie qu'une application à base de composants est perçue comme un assemblage de composants. Un tel assemblage doit être cohérent c.à.d. respectant plusieurs types de contrats (cf. section 1.4).

Troisièmement, un composant est une entité capable d'être déployée sur une plate-forme d'exécution, indépendamment des autres composants. Ceci constitue une autre différence vis-à-vis de l'approche par objets. En effet, une application orientée objet est souvent issue d'un environnement de développement homogène.

## 1.3 Notion de contrat

Afin de formaliser les relations conceptuelles fortes (client et héritage) entre les classes, Bertrand Meyer a introduit le paradigme de la conception par contrat (Design by Contract) [Meyer, 1992], [Meyer, 1997]. En effet, son langage Eiffel supporte d'une façon native la conception par contrat. En Eiffel, les prédicats **require** (précondition), **ensure** (postcondition) et **invariant**[2] (invariant) permettent de décrire un contrat dit contrat client entre l'objet client d'une méthode (ou routine en Eiffel) et l'objet serveur (ou fournisseur) qui implante cette méthode. La table 1.1 inspirée de [Meyer, 1997] explicite les droits et obligations de la relation entre le client et le serveur.

| | **Client** | **Serveur** |
|---|---|---|
| **Obligation** | satisfaire la precondition | satisfaire la postcondition |
| **Droit** | le résultat de l'exécution de la méthode est correct | l'état initial de la méthode est correct |

**Table 1.1** : Contrat : Droit est obligation des deux parties

En outre, Eiffel formalise la relation entre classe ascendante et descendante via le contrat d'héritage : possibilité d'affaiblissement de précondition (**require else**), de renforcement de postcondition (**ensure then**) et de renforcement d'invariant. Notons au passage que la conception par contrat est une application pratique des travaux de Hoare liés à la spécification pré/post des programmes [Hoare, 1969].

Le développement de la notion de composant logiciel a offert une opportunité d'appliquer cette vision contractuelle afin de vérifier la cohérence d'un assemblage de composants. Dans la suite, nous allons décrire les différents types de contrats souhaités afin d'analyser un assemblage de composants logiciels.

## 1.4 Classification des contrats

Les travaux décrits dans [Beugnard, 1999], [Beugnard, 2005] proposent une classification des contrats selon 4 niveaux. Cette classification est considérée comme un prolongement des propositions de Bertrand Meyer sur la conception par contrat.

Les niveaux de contrats pour les composants sont illustrés par la Figure 1.1 inspirée de [Beugnard, 1999]. Chaque niveau englobe les obligations des niveaux inférieurs.

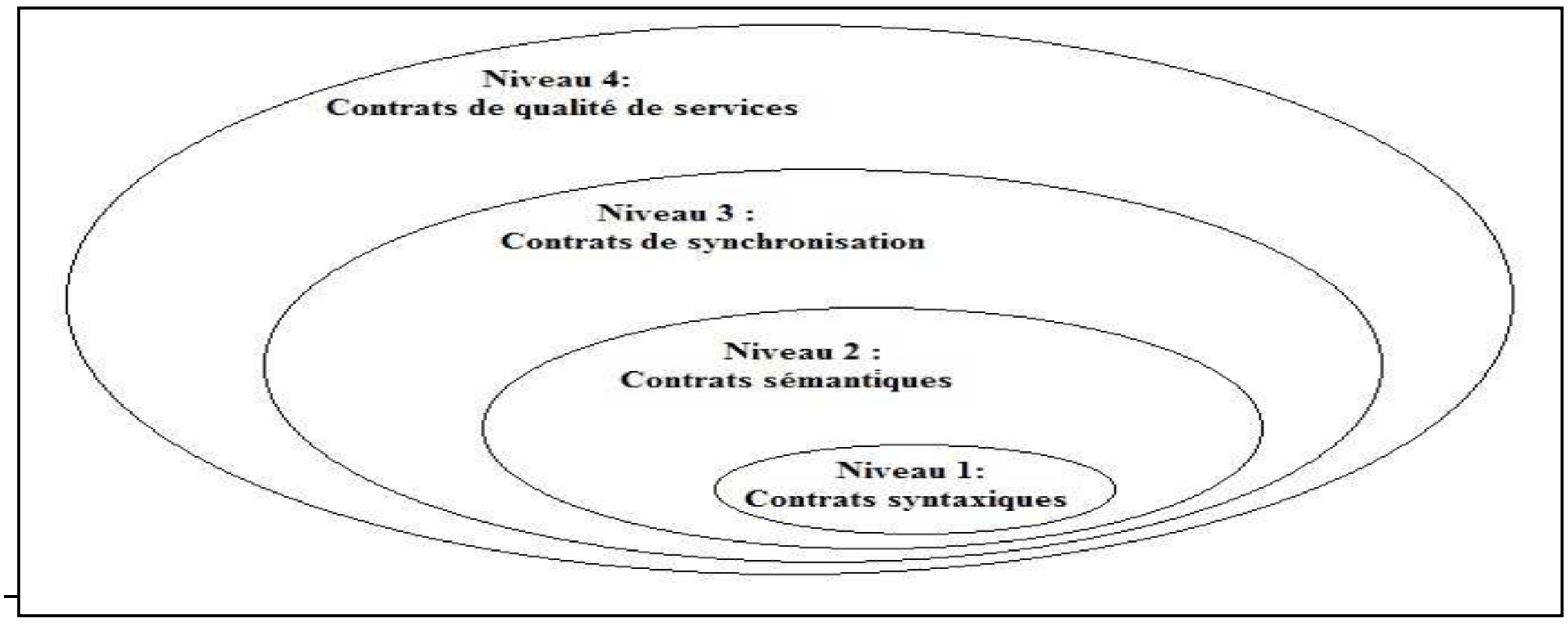

[2] Un invariant est une propriété commune à toutes les méthodes offertes par une classe. Il doit être satisfait durant la vie d'un objet. Il doit être établi par toutes les méthodes de création fournies par la classe.

**Figure 1.1 :** Niveaux de contrats pour les composants

### 1.4.1 Les contrats syntaxiques

Les contrats syntaxiques permettent de vérifier la conformité entre les signatures des opérations des interfaces. La signature d'une opération peut comporter les éléments suivants :

- nature de l'opération : opération de construction, consultation ou modification,
- paramètres formels : pour chaque paramètre, trois informations à prendre en considération à savoir son type, sa position et sa nature logique (in, out et in/out),
- exceptions levées.

Les incohérences détectées sont liées principalement à l'incompatibilité de types (type mismatch) en tenant compte des possibilités de typage offertes par le langage de description d'architectures utilisé. De même, nous pouvons étendre ce niveau en incluant les contrats structurels. De tels contrats expriment des contraintes liées aux règles de composition structurelle qui varient en fonction du modèle de composants traité. Par exemple, dans un assemblage UML2.0, un connecteur d'assemblage doit établir un lien entre une interface offerte et une interface requise de deux composants différents.

#### 1.4.2 Les contrats sémantiques

La sémantique d'une opération offerte/requise figurant au sein d'une interface offerte/requise peut être décrite en utilisant la conception par contrat : pré-condition, post-condition et invariant. Une telle sémantique peut être exprimée en utilisant un langage de contraintes de type OCL [Warmer, 2003].

#### 1.4.3 Les contrats de synchronisation

Les contrats de synchronisation s'intéressent à l'enchaînement des opérations acceptées et/ou demandées [Samek, 2005]. Ces contrats peuvent être décrits en utilisant des formalismes à base d'algèbres de processus, IOLTS (Input Output Labeled Transition Systems) et PSM (Protocol State Machine).

#### 1.4.4 Les contrats de qualité de services

Les contrats de qualité de services permettent de décrire les propriétés non fonctionnelles souhaitées ou offertes par une opération, une interface ou un composant. Sachant qu'une propriété non fonctionnelle (PNF) d'une entité logicielle est une contrainte liée à l'implémentation et la présentation de ses fonctionnalités [Taylor, 2009]. Parmi les PNF, nous citons : performance, sûreté, disponibilité, fiabilité, complexité, réutilisabilité, extensibilité, etc. Plusieurs formalismes de description des PNF sont proposés tels que : CQML [Aagedal, 2001], un profil UML pour CQML [Aagedal, 2002], un profil UML pour la qualité de services [OMG, 2008].

## 1.5 Vérification statique et dynamique

Après avoir assemblé ses composants, l'architecte a besoin de vérifier si les composants qui interagissent entre eux respectent les divers contrats applicatifs : contrats syntaxiques, sémantiques, de synchronisation et QdS (Qualité de Services). Mais un assemblage de composants (ou architecture) peut être large et complexe. Ceci exclut une analyse manuelle afin d'identifier des violations des contrats applicatifs. Donc, un recours aux outils d'analyse statique et dynamique d'assemblages de composant s'impose. Plusieurs modèles de composants notamment formels tels que Wright [Allen, 1996], [Allen, 1997], Darwin [Magee, 1995] et SafArchie [Barais, 2005] offrent des outils statiques. En outre certains modèles de composants comme AADL [SAE, 2004] supportent des outils de génération de code permettant d'ouvrir ces modèles sur l'analyse dynamique : tests unitaires, d'intégration, systèmes et d'acceptation. En effet, les analyses dynamiques permettent entre-autres de résoudre le problème des interactions partiellement compatibles[3] non traité par la plupart des outils d'analyse statique.

Dans la suite, nous allons étudier les différentes approches permettant de vérifier les divers contrats (syntaxiques, sémantiques, synchronisation et QdS).

### 1.5.1 Vérification des contrats syntaxiques

Les contrats syntaxiques englobant les propriétés liées à la compatibilité des signatures des opérations (offertes et requises) et les propriétés structurelles limitant les connexions entre les composants. Ces contrats dépendent des possibilités de typage (types prédéfinis, constructeurs de types simples, constructeurs de types composés, redéfinition de méthode et surcharge) et des règles de composition du modèle de composants traité. Les modèles de composants qui supportent des langages de contraintes peuvent spécifier les contrats syntaxiques notamment les propriétés structurelles comme des propriétés invariantes. Celles-ci sont vérifiées à l'aide d'un évaluateur de prédicats logiques. Parmi ces modèles, nous citons UML2.0 [OMG, 2005b], Fractal [Bruneton, 2004] et Acme [Garlan, 2000] dotés respectivement d'un langage de contraintes OCL [OMG, 2005a], CCLJ [Collet, 2005] et Armani [Monroe, 2001]. Mais contrairement à CCLJ et en particulier à Armani, OCL n'est pas dédié à exprimer des contraintes sur des modèles à composants. Il est plutôt conçu pour spécifier des contraintes sur des modèles orientés objets.

### 1.5.2 Vérification des contrats sémantiques

Les contrats sémantiques liés aux opérations offertes et requises des composants connectés peuvent être décrits en utilisant un langage de contraintes de type OCL. La vérification de ces contrats est souvent confiée à des prouveurs de théorèmes. Par exemple, le travail décrit dans [Messabihi, 2010] préconise l'utilisation de la méthode formelle B afin de vérifier les propriétés sémantiques d'un assemblage de composants Kmelia. Les travaux présentés dans [Lanoix, 2008a], [Lanoix, 2008b], [Mouakher, 2008] proposent une approche systématique de développement formel par composants basée sur des schémas d'assemblages UML et B. L'architecture du système est modélisée à l'aide de différents diagrammes UML2.0 (diagrammes de structures composites, diagrammes de classes et diagrammes de séquences). Le comportement autorisé ou attendu des interfaces est décrit à l'aide de modèles B. L'interopérabilité entre composants est vérifiée pour chaque connexion entre interfaces fournie et requise de l'architecture en utilisant le raffinement B. En effet, la correction d'une connexion peut s'exprimer en termes de raffinement : l'interface fournie doit raffiner l'interface requise.

---

[3] L'analyse statique ne déclenche aucune erreur mais ne peut pas être terminée car au moins une des propriétés dépend de données dont les valeurs ne peuvent être connues que lors de l'exécution du logiciel [Waignier, 2010].

### 1.5.3 Vérification des contrats de synchronisation

Les contrats de synchronisation peuvent être spécifiés par des formalismes à base d'algèbres de processus, systèmes de transitions étiquetées et Protocol State Machines augmentés. La vérification de ces contrats est souvent confiée à des models-checkers [Zhang, 2010].

Les travaux présentés dans [Barros, 2005a], [Barros, 2005b] relatifs à l'ADL Fractal proposent un cadre formel permettant de garantir non seulement que l'assemblage des composants Fractive – une implémentation du modèle des composants Fractal –est sûr quand il est déployé, mais aussi en présence de changements dynamiques et reconfigurations ultérieures. Les aspects comportementaux des composants sont spécifiés à l'aide de systèmes de transitions étiquetées (LTS). La sémantique d'un composant non primitif est obtenue comme le produit des LTS de sous-composants avec le contrôleur. Des propriétés propres au modèle hiérarchique et distribué Fractive sont définies et vérifiables grâce à la boîte à outils CADP [Garavel, 2007].

Le travail présenté dans [Kramer, 2003] propose une algèbre de processus appelée FSP (Finite State Processes) permettant de décrire les aspects comportementaux des composants Darwin. Un outil d'analyse LTSA (Labelled Transition System Analyser) permet d'animer et de vérifier certaines propriétés relatives aux descriptions architecturales en Darwin. L'outil LTSA effectue une recherche exhaustive de l'espace d'états relatif à une description architecturale en Darwin pour détecter des états d'interblocage (état n'ayant pas de successeurs) et d'ERROR. En effet, le langage FSP possède la faculté de décrire des comportements incorrects lors de la spécification des composants Darwin.

[Plasil, 2002] définit un modèle formel permettant de décrire les aspects comportementaux des composants logiciels. Le modèle proposé prend la forme des expressions régulières. Une validation de ce modèle a été faite sur l'ADL SOFA (SOFA CDL) [Plasil, 1998]. Ainsi, des propriétés liées au raffinement de la spécification et à la conformité d'une implémentation à sa spécification peuvent être vérifiées sur des descriptions architecturales SOFA.

Les travaux présentés dans [Vergnaud, 2006], [Vergnaud, 2005] permettent d'ouvrir AADL (Architecture Analysis & Design Langage) [SAE, 2004] sur les réseaux de Petri afin d'étudier certaines propriétés structurelles telles que : l'assemblage des composants n'engendre pas d'interblocage, les données utilisées dans les sous-programmes sont définies de façon déterministe.

Kmelia [André, 2006] est un modèle à composants basé sur les services. Ces derniers sont des entités de première classe. Ceci permet de rapprocher Kmelia des architectures orientées services [ERL, 2005]. Dans Kmelia, un service possède une signature, des assertions (précondition et postcondition) et une description du comportement en utilisant un système de transitions étiquetées étendu (eLTS). Pour la vérification des propriétés liées à l'interopérabilité dynamique des composants, Kmelia offre des ouvertures sur les langages adaptés et outillés supportant la notion de processus comme MEC et Lotos/CADP.

Dans l'ADL Rapide [Luckham, 1995], le comportement d'un composant est spécifié via des patterns d'événements décrivant la relation entre des données reçues et envoyées. Rapide permet la simulation des descriptions architecturales et offre des outils pour l'analyse des résultats de ces simulations afin de détecter des erreurs potentielles.

L'ADL Wright [Allen, 1997], [Garlan, 2003] offre quatre concepts architecturaux : composant, connecteur, configuration et style. Un composant Wright (respectivement un connecteur) peut être doté d'une ou plusieurs interfaces appelées ports (respectivement rôles). Le comportement d'un composant Wright (respectivement d'un connecteur) est décrit localement à travers les ports (respectivement les rôles) et globalement à travers un calcul (respectivement glu) en utilisant une algèbre de processus de type CSP [Hoare, 2004]. Ainsi, moyennant la traduction d'une configuration Wright (architecture Wright) en CSP, des outils de vérification CSP comme FDR [FDR2, 2003] peuvent être utilisés pour analyser des architectures Wright.

Le travail décrit dans [Mencl, 2003] propose un nouveau concept appelé PoSM : Port State Machine. Celui-ci est considéré comme une extension au PSM supporté par le modèle de composants UML2.0. Le concept PoSM permet de spécifier l'enchaînement des événements initialisés (opérations appelées) et observés (opérations reçues) d'un port attaché à un composant UML2.0. Egalement ce travail propose un outil de vérification appliqué aux PoSM.

### 1.5.4 Vérification des contrats QdS

Rares sont les modèles de composants qui offrent des mécanismes permettant de décrire les PNF et les contrats de QdS.

Le modèle de composants AADL [SAE, 2004] introduit la notion de propriété. A chaque composant, nous pouvons associer des propriétés et leur donner des valeurs. Le plugin OSATE [SAE, 2008] permet l'analyse de propriétés spécifiques telles que : niveaux de sécurité et niveau de sûreté.

Le modèle de composants Acme offre des facilités permettant la description des PNF en utilisant notamment le concept *Property*. Également, Armani couplé à Acme permet de spécifier les contrats de QdS d'un assemblage de composants Acme. Hormis l'évaluateur des prédicats Armani faisant partie intégrante de la plate-forme AcmeStudio, Acme ne supporte pas d'outils spécifiques d'analyse des PNF.

Le langage CQML [Aagedal, 2001] est un langage permettant d'exprimer des PNF des différents modèles à composants. Dans ce travail, nous utilisons CQML afin d'associer des PNF aux composants UML2.0.

### 1.5.5 Bilan

La table 1.2 récapitule les possibilités des modèles de composants examinés vis-à-vis de la description et vérification des contrats applicatifs: syntaxiques, structurels, sémantiques, de synchronisation et de qualité de services.

| | **Modèles de composants** | **Outils de vérification** | **Commentaires** |
|---|---|---|---|
| **Contrats syntaxiques** | UML2.0/OCL<br>Fractal/CCLJ,<br>Acme/Armani | Évaluateur des prédicats logiques | OCL est plutôt adapté au monde OO |
| **Contrats sémantiques** | Kmelia→B<br>UML2.0→B | Prouveur interactif de l'atelier B | Gap entre le modèle source et cible |
| **Contrats de synchronisation** | Fractal/LTS | CADP | Le PSM ne peut décrire qu'un seul sens de communication |
| | Darwin/FSP | LTSA | |
| | AADL/réseaux de | | |
| | Petri | | Wright propose des contrats standards |
| | Kmelia/eLTS | MEC, Lotos | |
| | Wright/CSP | FDR | |
| | UML2.0/PSM | | |
| **Contrats de QdS** | AADL/Propriété | Plugin OSATE | AADL vise des PNF spécifiques : sécurité et sûreté |
| | Acme/Armani | Évaluateur des prédicats logiques | Le concept Property d'Acme |

**Table 1.2 :** Description et vérification des contrats applicatifs

Le modèle de composants semi-formel UML2.0 ne permet pas de décrire tous les aspects d'une application à base de composants. Il a besoin d'autres formalismes plus ou moins intégrables dans UML2.0 comme PoSM et CQML afin de spécifier les aspects comportementaux et non fonctionnels. Aucun modèle de composants ne couvre les divers contrats applicatifs. Par exemple, Wright vise la description et vérification des contrats de synchronisation. Tandis que le modèle Acme/Armani vise la description et la vérification des contrats syntaxiques, structurels et de qualité de services.

## 1.6 Approche proposée

Cette thèse a pour objectif d'apporter une contribution à la vérification de la cohérence d'assemblages de composants semi-formels comme UML2.0 et Ugatze. La démarche partagée par la plupart des travaux existants [Lanoix, 2008a], [Lanoix, 2008b], [Mouakher, 2008], [Messabihi, 2010] dans le domaine de vérification de la cohérence d'assemblages de composants est l'emploi des techniques et des outils généraux tels que B et CSP. Pour, y parvenir, de tels travaux proposent des traductions plus ou moins systématiques de modèles de composants source vers le formalisme cible. Ceci oblige l'architecte à manipuler des concepts liés au formalisme cible tels que machine abstraite, machine de raffinement, non-déterminisme, primitives de composition de machines, événement, processus et composition de processus. Dans cette thèse, nous préconisons une approche favorisant la continuité entre le modèle source et le modèle cible. Ainsi, cette thèse préconise la traduction de **deux modèles de composants semi-formels** (UML2.0 et Ugatze) vers **des modèles de composants formels Acme/Armani et Wright**.

Le modèle de composants Acme/Armani est choisi pour son pouvoir à spécifier et vérifier les contrats syntaxiques et de QdS. Quant au modèle de composants Wright, il est retenu pour son aptitude à spécifier et vérifier les contrats de synchronisation. En outre, notre approche favorise l'analyse dynamique d'assemblages de composants UML2.0 via notre outil IDM de transformation de Wright vers Ada.

## 1.7 Conclusion

Après avoir introduit les notions de composant, de contrat et de classification de contrats et étudié les moyens de description et de vérification des contrats, nous avons proposé une approche de vérification de la cohérence d'assemblages de composants décrits par des modèles de composants semi-formels comme UML2.0 et Ugatze. Pour des raisons de continuité, notre approche préconise la traduction de modèles de composants semi-formels (UML2.0 et Ugatze) vers des modèles de composants formels (Acme/Armani et Wright).

Dans le chapitre suivant, nous allons étudier les modèles de composants retenus : UML2.0, Ugatze, Acme/Armani et Wright. Nous allons étudier également les aspects fondamentaux du langage CQML en tant que langage de description des PNF dans différents modèles à composants en l'occurrence dans UML2.0.

# Chapitre 2 : Les modèles de composants retenus

## 2.1 Introduction

Dans ce chapitre, nous présentons les principaux langages et formalismes utilisés dans cette thèse.

Ce chapitre comporte cinq sections. La section 2.2 est consacrée à l'étude du modèle des composants préconisé par UML2.0. La section 2.3 présente le modèle de composants Ugatze. La section 2.4 présente le langage de description d'architectures Acme/Armani. La section 2.5 présente le langage de description d'architectures Wright. Enfin, la section 2.6 présente des travaux relatifs à la spécification des propriétés liées à la qualité de services en se focalisant sur CQML. Ce langage est défendu comme étant le plus approprié pour la description des propriétés non fonctionnelles des composants logiciels. De plus, il est intégrable en UML2.0.

## 2.2 Le Modèle de composants UML2.0

UML [Rumbaugh, 2005], [OMG, 2005b] est un langage de modélisation graphique, semi-formel normalisé, défini par l'OMG. UML est considéré comme un successeur des langages de modélisation trouvés dans les méthodes : Booch [Booch, 1993], Object-Oriented Software Engineering (OOSE) [Jacobson, 1992] et Object Modeling Technique (OMT) [Rumbaugh, 1997].

A l'heure actuelle UML2.0 [OMG, 2005b] propose un modèle de composants permettant de définir les spécifications des composants, ainsi que l'architecture des systèmes à développer. La version UML2.0 permet de structurer les aspects d'un système avec treize diagrammes officiels appartenant à différents niveaux d'abstraction. Ces derniers sont classifiés comme l'indique la Figure 2.1.

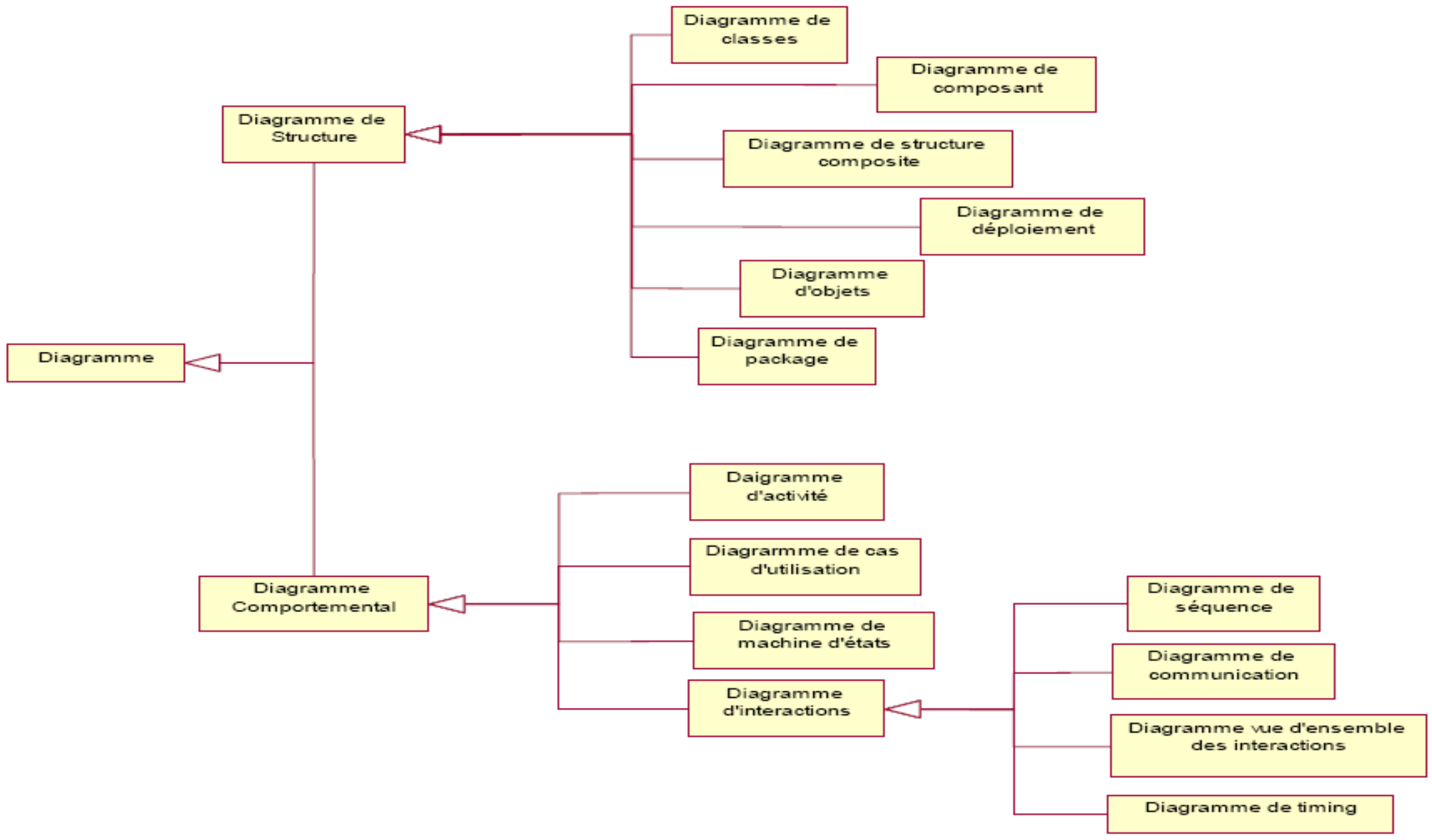

**Figure 2.1 :** Classification des diagrammes UML2.0

UML2.0 a introduit de nouvelles constructions qui le rendent plus adapté au développement à base de composants et à la spécification des descriptions architecturales.

## 2.2.1 Concepts structuraux

### 2.2.1.1 Composant UML2.0

La notion de composant a été un ajout majeur introduit dans le standard UML2.0 [OMG, 2005b]. Le composant permet en effet d'isoler des parties de logiciel en unités autonomes, avec des points de connexion bien définis. Les notions d'interface, de port ainsi que de connecteur sont des notions nouvellement introduites offrant des mécanismes utiles pour les architectures orientées composant. A travers ces mécanismes, UML2.0 apporte la modélisation par composants. Ces mécanismes d'assemblage de composants permettent de réaliser le vieux rêve du "légo logiciel", c'est-à-dire d'assembler des composants qui ne se connaissent pas obligatoirement, pour former un système englobant [Desfray, 2008].

Le document de la spécification d'UML [OMG, 2005a] définit le composant comme une unité modulaire, réutilisable, qui interagit avec son environnement par l'intermédiaire de points d'interactions appelés ports. Les ports sont typés par les interfaces : celles-ci contiennent un ensemble d'opérations et de contraintes. Les ports (et par conséquent les interfaces) peuvent être fournis ou requis. Enfin, le composant est voué à être déployé un certain nombre de fois, dans un environnement a priori non déterminé lors de la conception.

- **Exemple de composant :**

Le composant nommé Gestionnaire Produit (cf. Figure 2.2) est composé du composant « *Stock* » et du composant « *Distribution* ». Il offre les opérations offertes par les deux composants et requiert les opérations requises par les deux composants. Ce composant gère le produit à distribuer et s'occupe de la distribution physique [Cheesman, 2001].

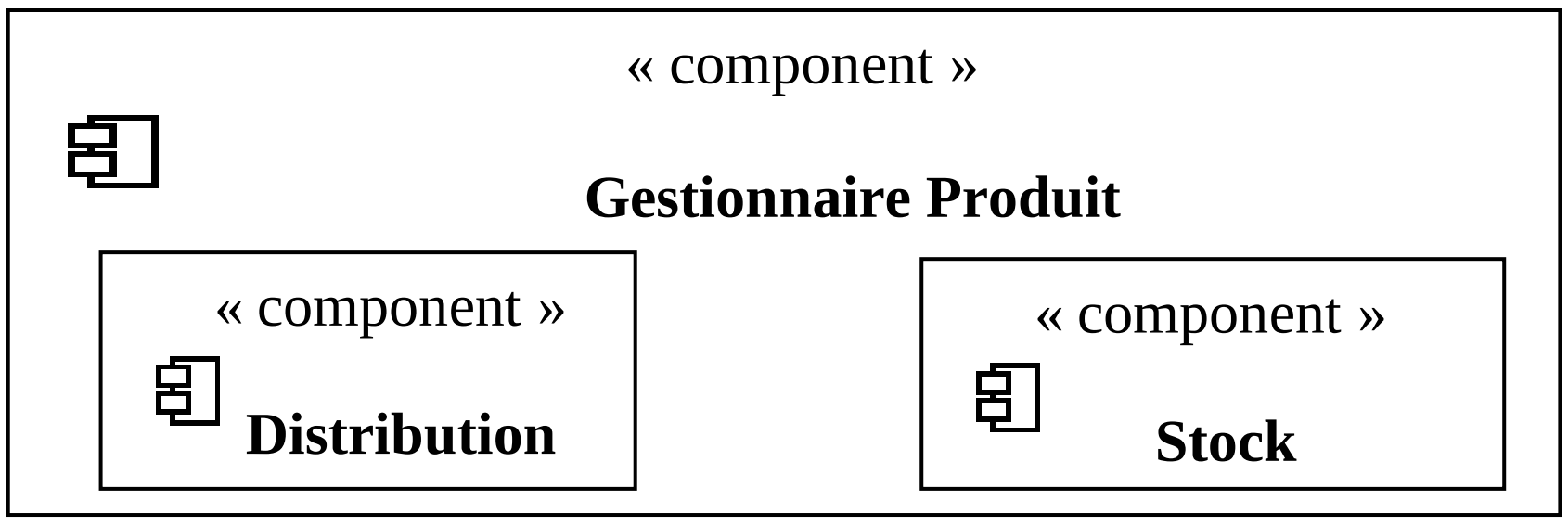

**Figure 2.2 :** Décomposition du composant Gestionnaire Produit

- **Structure composite :**

Il existe deux types de modélisation de composants dans UML2.0 : le composant atomique (ou basique) et le composant composite. Chaque modélisation définit une vue selon la spécification de la version d'UML2.0 [OMG, 2005b] : une vue de type boîte noire (vue externe) et une vue de type boîte blanche (vue interne).

- **La vue de type boîte noire** : cette vue considère qu'un composant est une entité d'encapsulation qui est uniquement caractérisée par ses interfaces requises et/ou fournies. Cette vue permet ainsi de montrer les propriétés publiques d'un composant (cf. Figure 2.3). Notons également que d'autres diagrammes UML (par exemple, diagramme de séquence, d'activités) peuvent être utilisés pour détailler le comportement du composant. En plus, une machine à états peut décrire le mode d'utilisation du port et/ou de l'interface ou le comportement du composant lui-même (cf. section 2.2.2.1).
- **La vue de type boîte blanche** : cette vue définit la structure interne du composant. Le composant est constitué de sous-composants appelés parties (*parts*) et peut contenir des connecteurs internes pour connecter ces sous-composants. Cette vue permet de montrer les propriétés privées d'un composant car la partie interne d'un composant est cachée et n'est accessible qu'au travers de ses interfaces (cf. Figure 2.4). Cette vue interne permet de montrer les différentes relations entre les différents éléments et les connecteurs qui les relient. Le lien entre les deux vues est réalisé par délégation des traitements à des connecteurs sur les ports qui sont connectés à des parties internes. La Figure 2.4 donne une représentation schématique d'un composant avec ses deux vues interne et externe.

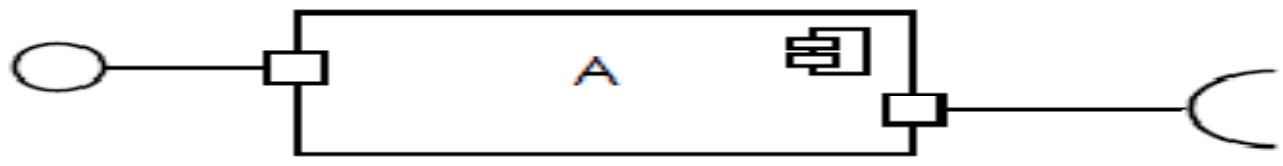

**Figure 2.3 :** Vue externe du composant en UML2.0

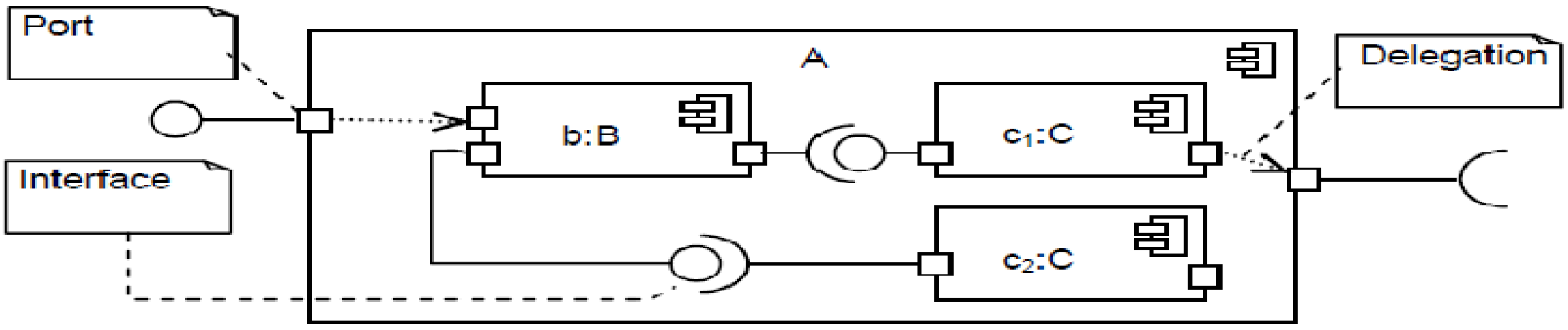

**Figure 2.4 :** Vue externe/interne du composant en UML2.0

## 2.2.1.2 Notion de port

Un port permet d'isoler un composant logiciel de son environnement en fournissant un point d'interaction adressable. Le port permet ainsi de circonscrire en un point précis les échanges entre le composant et son environnement extérieur ; il rassemble une ou plusieurs interfaces pour offrir ou requérir un élément de service [Accord, 2002]. On note aussi que le port est optionnel et permet d'accéder au connecteur comme s'il s'agissait d'un composant, notamment pour le configurer (par exemple pour affiner des paramètres de qualité de service).

Le comportement d'un port est issu de la composition des comportements de ses interfaces. Le comportement interne du composant ne doit être ni visible, ni accessible autrement que par ses ports. Les ports installés sur des composants ou classes peuvent être fournis ou requis. Une instance de composant peut avoir un (cf. Figure 2.5) ou plusieurs ports (cf. Figure 2.6).

- **Exemple 1 :**

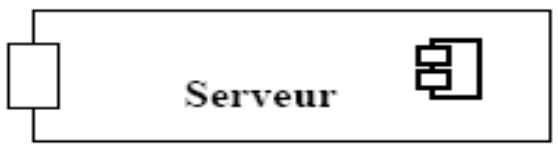

**Figure 2.5** Port sans interfaces

- **Exemple 2 :**

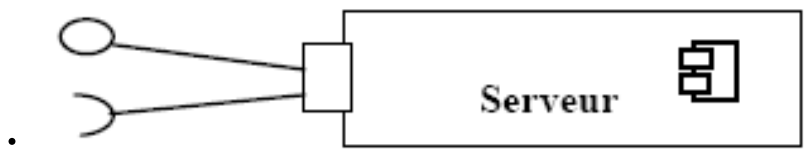

**Figure 2.6 :** Port avec interfaces

### 2.2.1.3 Notion d'interface

L'interface de composant en UML regroupe un ensemble non vide d'opérations qui spécifient des services offerts et requis par celui-ci sous la forme de signatures de méthodes. L'interface est un moyen d'expression des liens du composant ainsi que ses contraintes avec l'extérieur.

- **Interface Requise :** c'est une interface que le composant requiert de la part d'autres composants pour réaliser ses services fournis. Cette interface doit être connectée aux interfaces fournies des autres composants de l'environnement. Une telle interface peut être utilisée par le composant ou ses parties, ou bien être le type d'un de ses ports offerts.
- **Interface Fournie** *:* c'est une interface offerte qui englobe l'ensemble des services fournis par le composant à son environnement. Cette interface peut être implémentée soit directement par le composant ou bien par l'une de ses parties, ou bien être le type d'un de ses ports offerts.

Il y a trois façons pour représenter un composant doté d'une interface requise et d'une interface offerte. Les Figures 2.7, 2.8 et 2.9 illustrent les trois représentations appliquées sur un composant Planificateur qui offre une interface « *ActualiserPlans* » et exige une interface « *FaireReservation* » [Audibert, 2009].

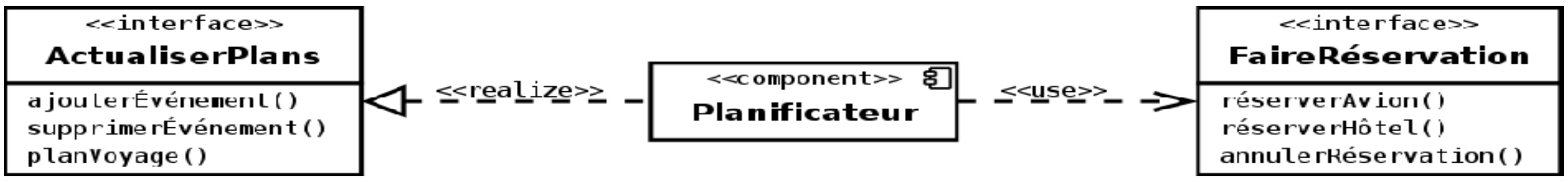

**Figure 2.7 :** Représentation 1

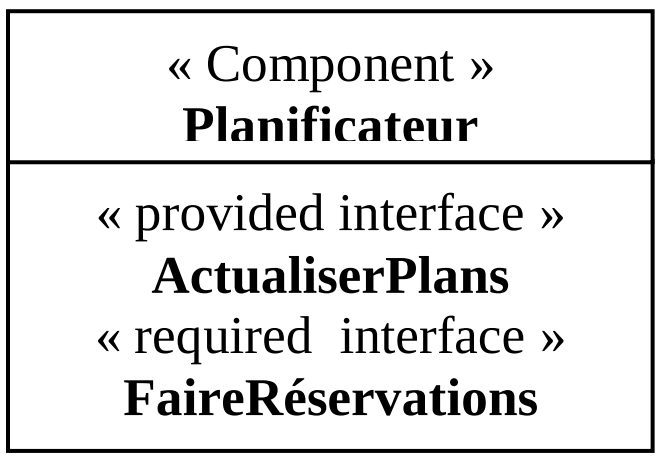

**Figure 2.8:** Représentation 2

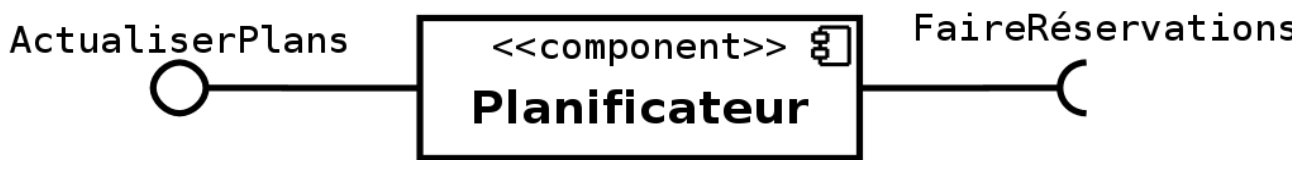

**Figure 2.9 :** Représentation 3

### 2.2.1.4 Notion de connecteur

Les interactions entre les composants sont décrites par des connecteurs. La connexion entre les ports requis et les ports fournis se fait au moyen de connecteurs. Le connecteur permet également de vérifier l'intégrité de la communication, c'est-à-dire de vérifier que les composants peuvent être connectés. Ainsi, il permet la réutilisation et l'adaptation des interfaces de composants déjà existants que l'on cherche à relier.

Deux types de connecteurs existent : le connecteur de délégation et le connecteur d'assemblage [Accord, 2002]. La distinction entre ces deux types relève de la nature des interfaces mises en connexion.

#### 2.2.1.4.1 Connecteur de Délégation

Le connecteur de délégation est un connecteur qui relie le contrat externe d'un composant (spécifié par ses ports) à la réalisation de ce comportement par les parties internes du composant. Il permet de lier un port du composant composite vers un port d'un composant situé à l'intérieur du composant composite : relier par exemple un port requis à un autre port requis (cf. Figure 2.10).

Un connecteur de délégation doit uniquement être défini entre les interfaces utilisées ou des ports de même type, c'est-à-dire entre deux ports ou interfaces fournis par le composant ou entre deux ports ou interfaces requis par le composant.

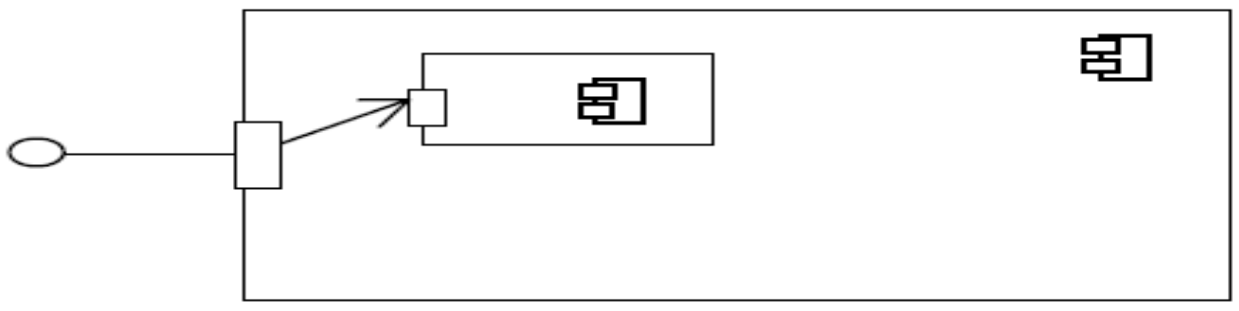

**Figure 2.10 :** Connecteur de délégation entre un port externe et un port interne

#### 2.2.1.4.2 Connecteur d'assemblage

Un connecteur d'assemblage est un connecteur entre deux composants qui définit qu'un composant fournit le service qu'un autre composant requiert. Un connecteur d'assemblage doit uniquement être défini à partir d'une interface requise ou d'un port vers une interface fournie ou un port (cf. Figure 2.11 et 2.12).

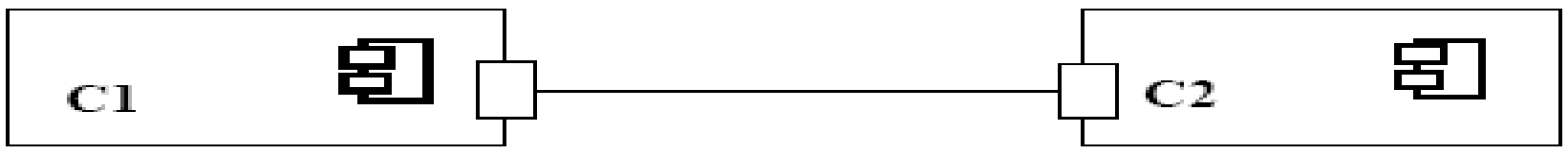

**Figure 2.11 :** Connecteur d’assemblage entre deux ports

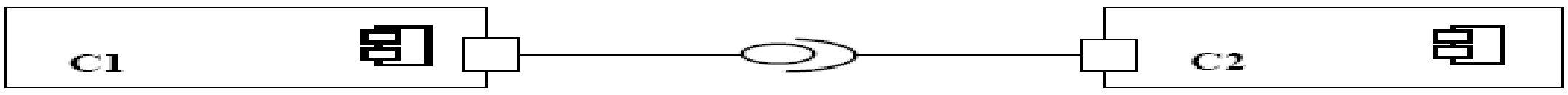

**Figure 2.12** Connecteur d’assemblage entre deux interfaces

## 2.2.2 Concepts comportementaux

### 2.2.2.1 Les machines à états finis

Rappelons qu’un composant fonctionnel est modélisé par un composant UML2.0. Les points d’interaction d’un composant fonctionnel avec son environnement sont modélisés par des ports UML2.0. Chaque port est typé par des interfaces offertes ou requises. Le comportement de chaque port est modélisé par un protocole «machine à états» d’UML2.0. En effet, UML2.0 [OMG, 2005b] introduit des machines d’états de protocole (PSM) pour décrire les séquences d’appels valides d’une instance. Les PSM sont une spécialisation des machines d'état UML, sans action, ni activité. Les transitions sont spécifiées en termes de pré/post conditions et d’invariants d’un état donné. En règle générale les PSM fournissent une description du comportement des composants et peuvent être combinés à un processus de raffinement pour générer des implémentations. En effet, [Lanoix, 2006] propose une approche pour le développement pas-à-pas des machines à états de protocole en utilisant des opérateurs qui préservent des propriétés comportementales. [Lanoix, 2006] introduit deux spécialisations de la relation de conformité du protocole proposé dans UML2.0 inspirées des travaux sur les méthodes formelles comme l'amélioration de spécifications et le raffinement de spécifications. Cependant les Protocoles de Transitions ont une restriction liée à l'activité auquel ils sont associés. En effet, Les PSM ne peuvent décrire qu’un seul sens de la communication. Cela signifie également qu’un PSM ne peut pas décrire les relations de communication entre les interfaces requises et fournies.

### 2.2.2.2 Les Ports States Machines

Le travail décrit dans [Mencl, 2003] propose un nouveau concept appelé PoSM : Port State Machine. Celui-ci est considéré comme une extension au PSM (Protocol State Machine) supporté par le modèle de composants UML2.0. Le concept PoSM permet de spécifier l’enchaînement des événements initialisés (opérations appelées) et observés (opérations reçues) d’un port attaché à un composant UML2.0. La Figure 2.13 représente la syntaxe abstraite d’un PoSM.

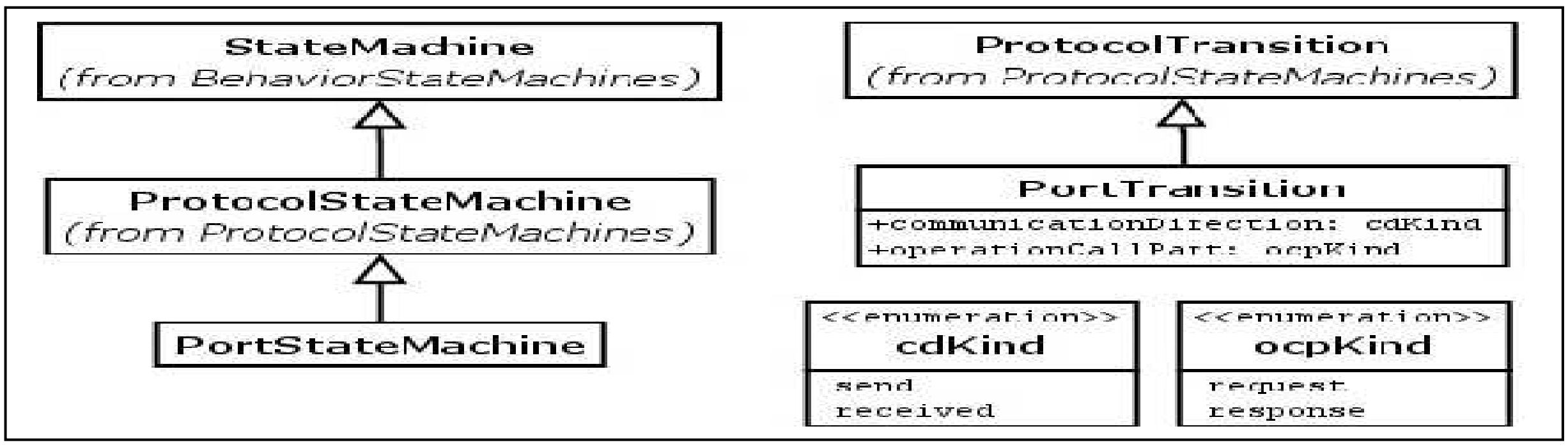

**Figure 2.13 :** Syntaxe abstraite d'un PoSM

### 2.2.3 Évaluation

Il est clair qu'UML2.0 favorise le développement basé sur le paradigme composant. UML2.0 propose de nombreux mécanismes au niveau du langage pour enrichir la définition de l'interface d'un composant. Si le langage OCL est maintenant convenablement adopté et intégré dans le modèle de composant, la définition d'information relative au comportement à tous les niveaux du composant (composant, port, interface) et le manque de cohérence entre ces informations gènent pour le moment l'analyse d'un assemblage de composants UML2.0. En effet, OCL n'est pas dédié à exprimer des contraintes sur des modèles à composants. Il est conçu plutôt pour spécifier des contraintes sur des modèles orientés objets.

Signalons également que des règles de cohérence liées à la bonne façon de constituer un assemblage de composants UML2.0 sont décrites et formalisées en OCL 2.0 [OMG, 2005b] au niveau du méta-modèle UML2.0. Mais ces règles sont loin d'être complètes. En effet, le travail décrit dans [Malgouyres, 2005] propose des nouvelles règles de cohérence liées à la bonne utilisation du diagramme de structures composites UML2.0. En outre, la vérification de ces règles se heurte au manque d'ateliers avec support OCL 2.0. En effet, les outils de preuve de contraintes OCL sont peu nombreux [Brucker, 2008], et nécessitent la traduction d'OCL dans des logiques mathématiques. Il est donc opportun de pouvoir exprimer ces contraintes invariantes dans un autre formalisme.

Les PoSM (Ports States Machines) peuvent être utilisés avec profit pour spécifier aussi bien les comportements partiels des ports que le comportement global des composants UML2.0.

## 2.3 Le modèle de composants Ugatze

Au cours de ces dernières années, l'équipe de recherche dirigé par Philippe Aniorté a développé et expérimenté (dans le cadre du projet européen ASIMIL [ASIMIL, 2002] un (méta)modèle de composants dénommé Ugatze, adapté à la réutilisation de composants logiciels autonomes, hétérogènes et distribués [Aniorté, 2004], [Seyler, 2004]. Les composants logiciels visés par le modèle Ugatze ne sont pas forcément conçus pour être réutilisés a priori : c'est la réutilisation a posteriori.

Le modèle de composants Ugatze est défini précisément via un méta-modèle, ce qui permet de manipuler des modélisations Ugatze via des outils dans le cadre d'un processus logiciel de type « Ingénierie Dirigée par les Modèles » (IDM ou MDE pour Model-Driven Engineering). Ces aspects n'étant pas développés dans cette thèse, nous renvoyons le lecteur intéressé à [Aniorté, 2004]. Ce (méta)modèle repose sur deux notions essentielles :

- L'interface du composant : c'est le résultat de la (re)spécification du composant, activité propre à la réutilisation [Cauvet, 1999]. La représentation de tous les composants à réutiliser est « unifiée » (au sens des besoins exprimés pour l'interopérabilité) via le méta-modèle,
- L'interaction entre composants : elle permet de gérer l'interopérabilité entre composants.

### 2.3.1 Les composants Ugatze

Le (méta)modèle Ugatze permet de construire des représentations de composants à réutiliser indépendantes des plates-formes d'origine des applications. La spécification des composants réutilisables repose sur le concept « d'interface » constituée de « points d'interaction ». Une syntaxe abstraite est généralement définie dans les termes d'un méta (méta)modèle : le Modèle MOF. L'interface du composant dans Ugatze (cf. Figure 2.14) est basée sur le principe d'abstraction, sur le découplage entre composants, sur la séparation entre donnée et contrôle et sur la multiplicité des modes d'interaction. Le découplage indique que les dépendances entre composants sont définies de manière externe à ceux-ci par l'intermédiaire de points d'interaction : points d'information et points de contrôle. Ugatze offre trois modes d'interaction principaux : les modes synchrone (OperationPoint : services fournis ou requis), asynchrone (SignalPoint : données reçues ou transmises) et le mode flux continu (StreamingPoint, dont seul le contenu apporte une sémantique). Ces trois modes sont portés par les points d'interaction de donnée et de contrôle.

### 2.3.2 Les points d'interaction

Un composant Ugatze est doté de plusieurs points d'interaction permettant à un composant d'interagir avec son environnement. Le (méta)modèle Ugatze propose essentiellement deux ensembles de points d'interaction : les points d'information et les points de contrôle.

#### 2.3.2.1 Les points d'information

Le (méta)modèle Ugatze propose les points d'information suivants (cf. Figure 2.15) :

- – les points d'entrée des données : Input Information Point (*IIP*), ce sont des points d'interaction sur lesquels les composants récupèrent le flux de données unidirectionnel,
- – Les points de sortie de données : Output Information Point (*OIP*), jouent un rôle symétrique en transférant le flot unidirectionnel de données vers l'extérieur du composant,
- – Les points d'opérations offertes : Provided Information Operation Point (*PIOP*). Ils reçoivent et répondent à des requêtes,
- – Les points d'opérations requises : Used Information Operation Point (*UIOP*). Ils déterminent les services requis pour l'exécution d'un composant Ugatze.

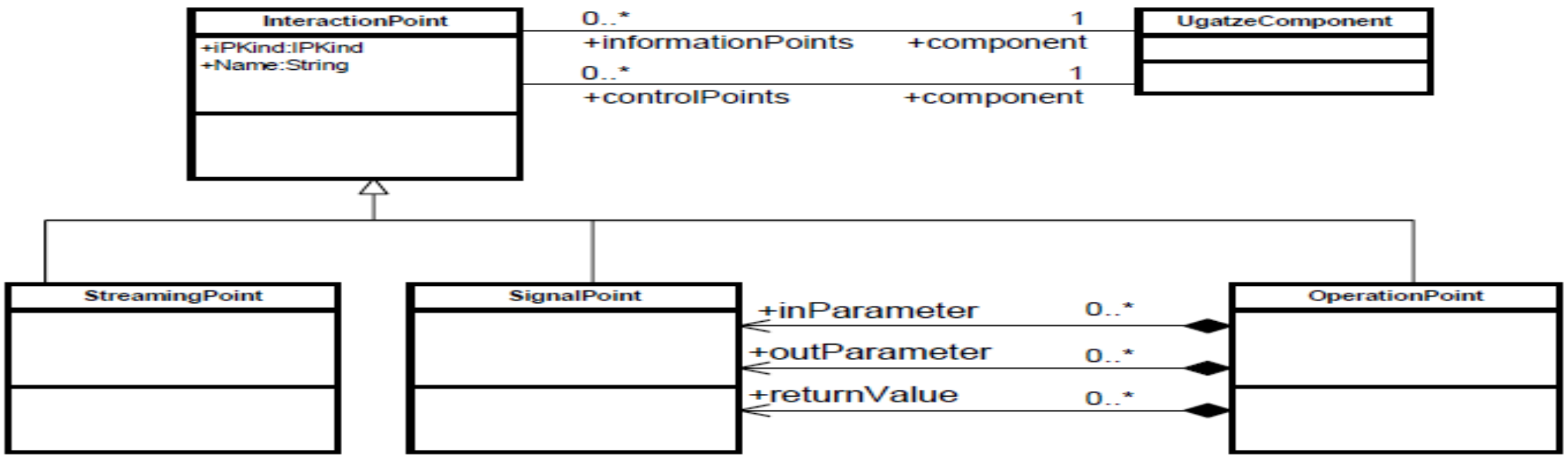

**Figure 2.14 :** Représentation UML de l'interface des composants Ugatze

#### 2.3.2.2 Les points de contrôle

Ugatze propose un ensemble de points de contrôle (cf. Figure 2.15) qui permettent à un composant de synchroniser un autre composant ou d'être synchronisé, à l'aide d'un message ou d'un événement. Ces points d'interaction sont appelés :

- *SEP* (Signal Emission Point) permettant au composant d'envoyer des signaux ou des messages (asynchrones),
- *SRP* (Signal Reception Point) permettant de recevoir ces signaux.

Ugatze inclut aussi la possibilité d'accéder dans la partie contrôle des composants à ces opérations de cycle de vie (création, recherche, activation voire mobilité…), et ce par l'intermédiaire des points d'opération :

- *PCOP* (Provided Controle Operation Point) pour les opérations de contrôle offertes,
- *UCOP* (Used Controle Operation Point) pour les opérations de contrôle requises.

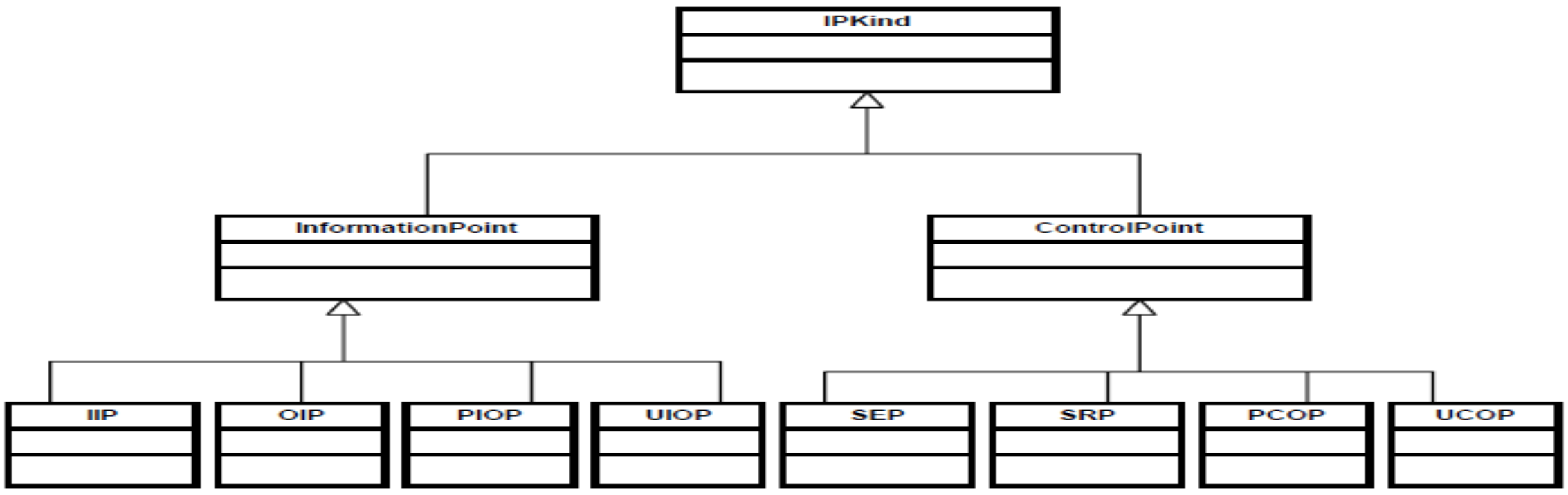

**Figure 2.15 :** Hiérarchie des points d'interaction

### 2.3.3 Les interactions Ugatze

Les interactions Ugatze permettent une intégration conceptuelle de composants autonomes. Une interaction Ugatze connecte un certain nombre (au minimum 2) de points d'interaction installés sur des composants Ugatze. Trois types d'interaction sont proposés par Ugatze : interaction directe, interaction ad hoc et interaction prédéfinie.

#### 2.3.3.1 Les interactions directes

Une interaction directe désigne une connexion directe entre deux (et seulement deux) points d'interactions. À l'instar des points d'interaction, les interactions directes se déclinent en deux catégories : interactions d'information et interactions de contrôle :

- L'interaction directe de donnée (directDataInteraction) concerne uniquement un *IOP* et un *IIP*, qui véhiculent le même type de données,
- L'interaction directe de contrôle (directControlInteraction) connecte un *SEP* et un *SRP*, avec des contraintes de typage sur l'information de contrôle véhiculée,
- L'interaction d'opération de contrôle (OperationControlInteraction) connecte un *PCOP* (Provided Control Operation Point) et un *UCOP* (Used Control Operation Point), avec également des contraintes de typage sur les points connectés,
- L'interaction d'opération d'information (OperationDataInteraction) connecte un *PIOP* (Provided Information Operation Point) et un *UIOP* (Used Information Operation Point), avec des contraintes de typage sur les points connectés.

#### 2.3.3.2 Les interactions ad hoc

Une interaction ad hoc permet une interopérabilité fonctionnelle entre les composants hétérogènes issus des divers environnements de production. Ces interactions ad hoc concernent aussi bien la partie donnée que contrôle d'un composant Ugatze. Une interaction ad hoc peut toucher à plusieurs (>=2) points d'interaction, elle encapsule souvent un traitement plus ou moins compliqué. Les interactions ad hoc se déclinent en trois catégories : fabrication d'information, interactions de contrôle complexe et interactions « mixtes ».

#### 2.3.3.3 Les interactions prédéfinies

Le (méta)modèle Ugatze propose des interactions prédéfinies telles que : le mécanisme multicast, la boîte aux lettres et la ressource partageable :

- **Le mécanisme multicast :** Parmi les propriétés des points d’interaction que propose Ugatze, il convient de remarquer que les points d’interaction sont “unicast”, c’est-à-dire que ceux-ci ne se “soucient” pas du nombre de composants récepteurs. Le mécanisme multicast permet à un composant d’envoyer la même donnée à plusieurs récepteurs, connectés au “multicast” par l’intermédiaire d’une interaction. Conceptuellement, ce mécanisme prend la forme d’une interaction à façon qui possède un *IIP* et plusieurs *OIP*, concernant le même type de données.
- **La boîte aux lettres** : Ce mécanisme de boîte aux lettres est utilisé lorsqu’une information (flux de données ou message) n’est pas transférée directement d’un composant à l’autre mais déposé dans la boîte en attendant qu’un autre composant la réclame. Concernant le dépôt dans cette boîte aux lettres, une interaction directe est utilisée pour connecter le composant émetteur et la boîte aux lettres. Quant au retrait, il est matérialisé par une interaction entre la sortie de la boîte aux lettres et un point d’entrée (*IIP*) du composant concerné. Au niveau des principes sous-jacents, ce dernier dispositif est typique des environnements où il y a exécution concurrente.
- **La ressource partageable :** La ressource partageable est un exemple typique de mécanisme de “contrôle”. Dans l’hypothèse où plusieurs composants utilisent la même ressource, chacun des composants impliqués doit tour à tour accéder à la ressource et la libérer lorsque son exécution est terminée. Les composants possèdent chacun un point d’accès à la ressource et un point de libération de la ressource. Deux points d’émission de signal *(SEP*) permettent aux composants de réserver puis de libérer la ressource, ces points sont appelés dans la pratique Resource Allocation Point *(RAP*) et Resource Release Point *(RRP*).

### 2.3.4 Représentation graphique et règles

Le (méta)modèle Ugatze contient une syntaxe abstraite, une syntaxe concrète graphique, ainsi que des règles de vérification. La syntaxe abstraite (ou (méta)modèle) est généralement définie dans les termes d'un méta (méta)modèle : le Modèle MOF.

#### 2.3.4.1 la syntaxe concrète

Une représentation graphique est associée au (méta)modèle Ugatze appelé : syntaxe concrète. La nécessité d'une telle notation n'est plus à démontrer, notamment dans le domaine des DSL (Domain Specific Language), définissant des langages spécifiques. Cette représentation est en effet indissociable de l'exploitation des modèles.

Nous présentons par la suite la représentation graphique d’un composant Ugatze et deux illustrations d’interaction issues du projet ASIMIL.

##### 2.3.4.1.1 Représentation graphique du point de vue interface

Comme exemple de composant issu du projet ASIMIL [ASIMIL, 2002], nous citons le composant MAS (Multi Agent System). Ce composant effectue des diagnostics sur le comportement de l’apprenant en cours d’apprentissage de pilotage ou de maintenance. Ces spécifications imposent à ce composant d’échanger de nombreuses informations hétérogènes avec son environnement. Un point d’entrée d’information (*IIP1* sur la Figure 2.16) récupère un couple d’informations constitué par une donnée concernant l’action effectuée par l’utilisateur et une autre concernant l’action requise par celui-ci. Le point d’interaction *IIP2*, donne à chaque instant les paramètres de l’avion en vol permettant de raisonner sur le comportement global de l’avion.

Grâce à un point d'opération requise (*FlightParameters* sur la Figure 2.16), le composant MAS peut effectuer des requêtes. Un point de sortie d'information appelé *IIOPDiag* permet d'envoyer aux autres composants le diagnostic du composant MAS. Ce diagnostic est représenté dans un message textuel. Dans le même temps, un flot d'information lui permet d'envoyer en continu ses diagnostics à un composant d'interface, et ce par l'intermédiaire d'un point de sortie d'information de type «flot de données» (StreamingPoint), *OIPStream* sur la Figure 2.16. Le Système Multi-Agents est doté de deux points de contrôle : un point d'allocation de ressource (*RAP*) et un point de libération de ressource (*RRP*) qui sont tous deux des points d'émission de signal (*SEP*). Il est également doté d'un point de réception de signal (*SRP*) qui permet à un autre composant de démarrer et stopper son exécution.

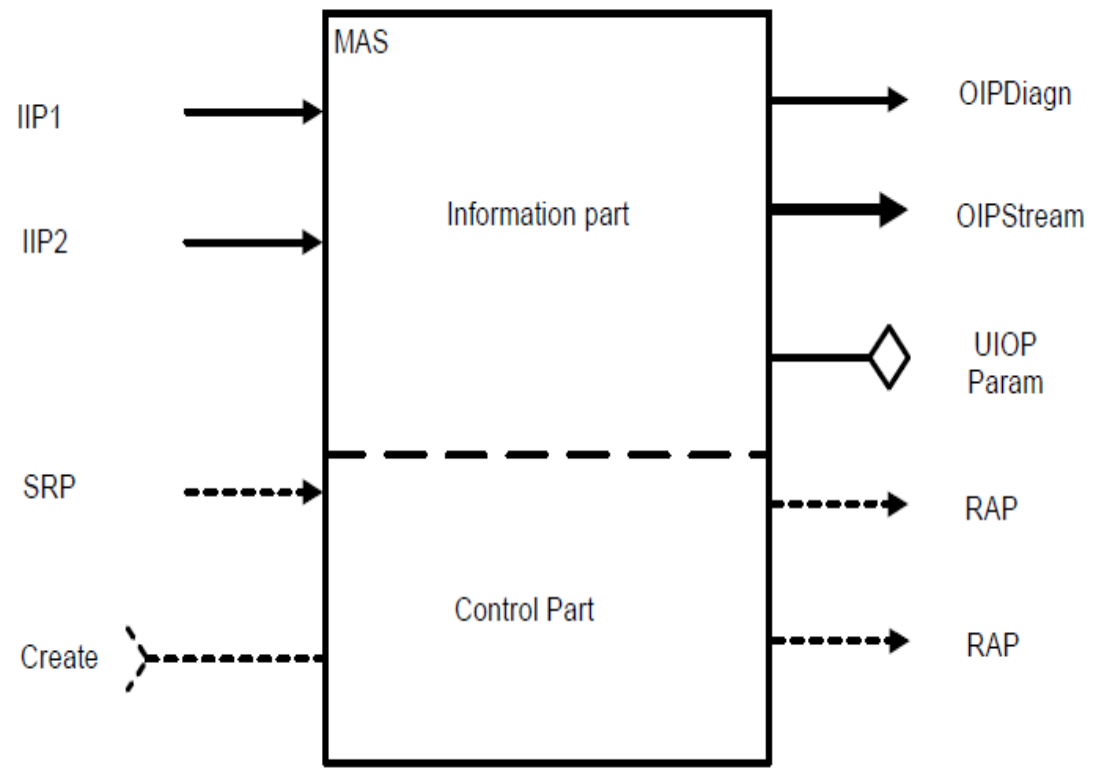

**Figure 2.16 :** Représentation du composant MAS avec la syntaxe concrète Ugatze

### 2.3.4.1.2 Représentation graphique du point de vue Interaction

Les points d'interaction, dont nous venons de décrire la représentation graphique, sont les éléments de base de la représentation graphique de l'interaction. En ce qui concerne le principe général de l'interaction, l'origine est un point d'interaction de sortie d'un composant, sa destination est un point d'interaction d'entrée d'un autre composant. La sémantique associée à ce graphique est la suivante : deux composants sont censés coopérer par le biais de l'interaction. Ce principe général est décliné de différentes manières en fonction du type de l'interaction. Par la suite, nous illustrons ces concepts par deux représentations graphique d'interaction.

La Figure 2.17 illustre un transfert direct d'information avec la syntaxe « *UgatzeGraphic* » à travers une interaction directe de données, reliant un *OIP* à un *IIP*. L'exemple est issu du projet ASIMIL qui illustre un simulateur de vol (*FSIMU*) qui envoie en continu des paramètres de vol que le système multi Agent (*MAS*) exploite. Notons que la représentation graphique d'une interaction directe de synchronisation est similaire à celle d'une interaction directe d'information.

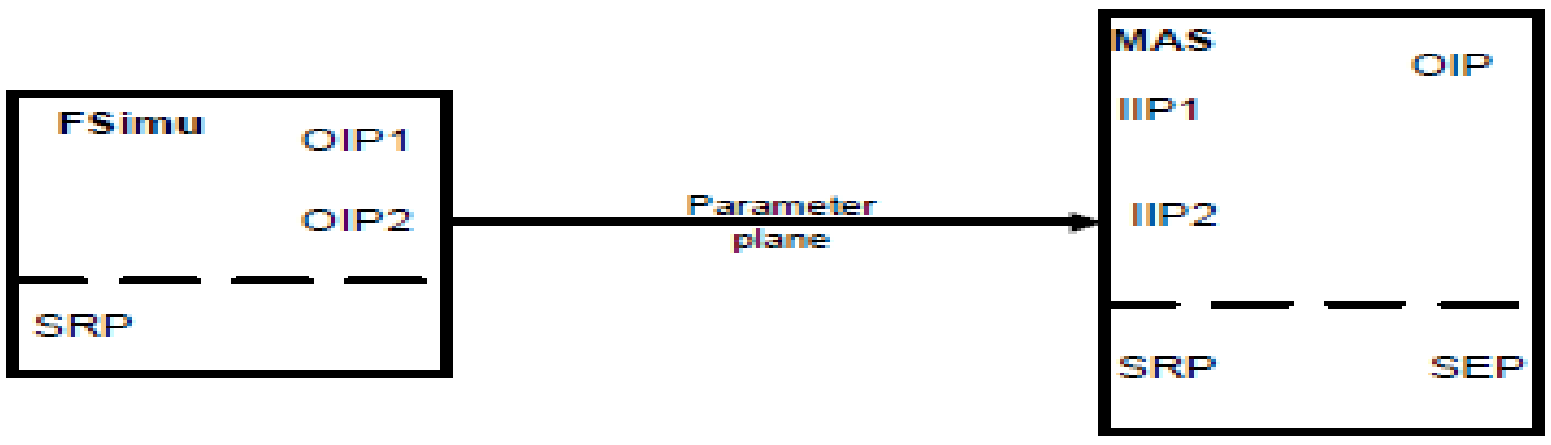

**Figure 2.17 :** Représentation graphique de l'interaction directe de données

La Figure 2.18 illustre une interaction d'opération de données avec la syntaxe « *UgatzeGraphic* ». L'interaction d'opération est un type d'interaction fréquemment rencontrée dans les modèles de composants "par contrat" et les langages de description d'architectures (ADL). De manière graphique, elle est symbolisée par une connexion entre un *UIOP* et un *PIOP*.

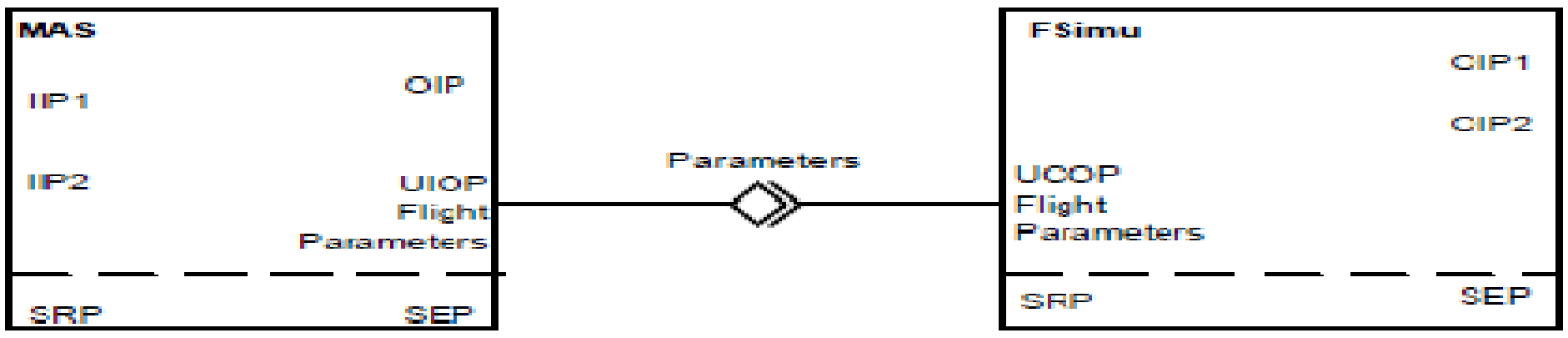

**Figure 2.18 :** Représentation graphique de l'interaction d'opération de données

### 2.3.4.2 Les règles de bonne utilisation

La syntaxe concrète de Ugatze est complétée par des règles. De telles règles ont pour objet de spécifier un ensemble de contraintes relatives aux différents éléments constitutifs du (méta)modèle, interface (points d'interaction) et interactions. Elles sont destinées à aider à sa bonne utilisation. Ces règles correspondent typiquement aux règles de bonne construction (« wellformedness rules ») que l'on retrouve dans la définition de méta-modèles. Elles sont décrites en langage naturel et à l'aide du langage OCL. Elles permettent de vérifier les modèles instanciés à partir du (méta)modèle.

### 2.3.5 Exemple d'assemblage de composants

Une architecture logicielle d'une application distribuée décrite en Ugatze est considérée comme un graphe biparti appelé graphe d'interactions comportant deux types de sommets : composant et interaction. Les composants et les interactions sont reliés par des arcs. Le graphe d'interactions décrit les aspects statiques d'une application Ugatze. Vu sous cet angle, il est similaire au diagramme de classes UML1.x ou au diagramme de composants UML2.0 [OMG, 2003]. La Figure 2.19 illustre un exemple de graphe d'interconnexion issu du projet ASIMIL [ASIMIL, 2002]. Le graphe d'interconnexion modélise l'intégration des quatre composants (MAS, Fsimu, TM, PFC) et la mise en œuvre des interactions directes, à façon, ou les mécanismes prédéfinis : multicast, boîte aux lettres (MailBox) et partage de ressource (sharedWindow), frabication d'information (BuildingInformation).

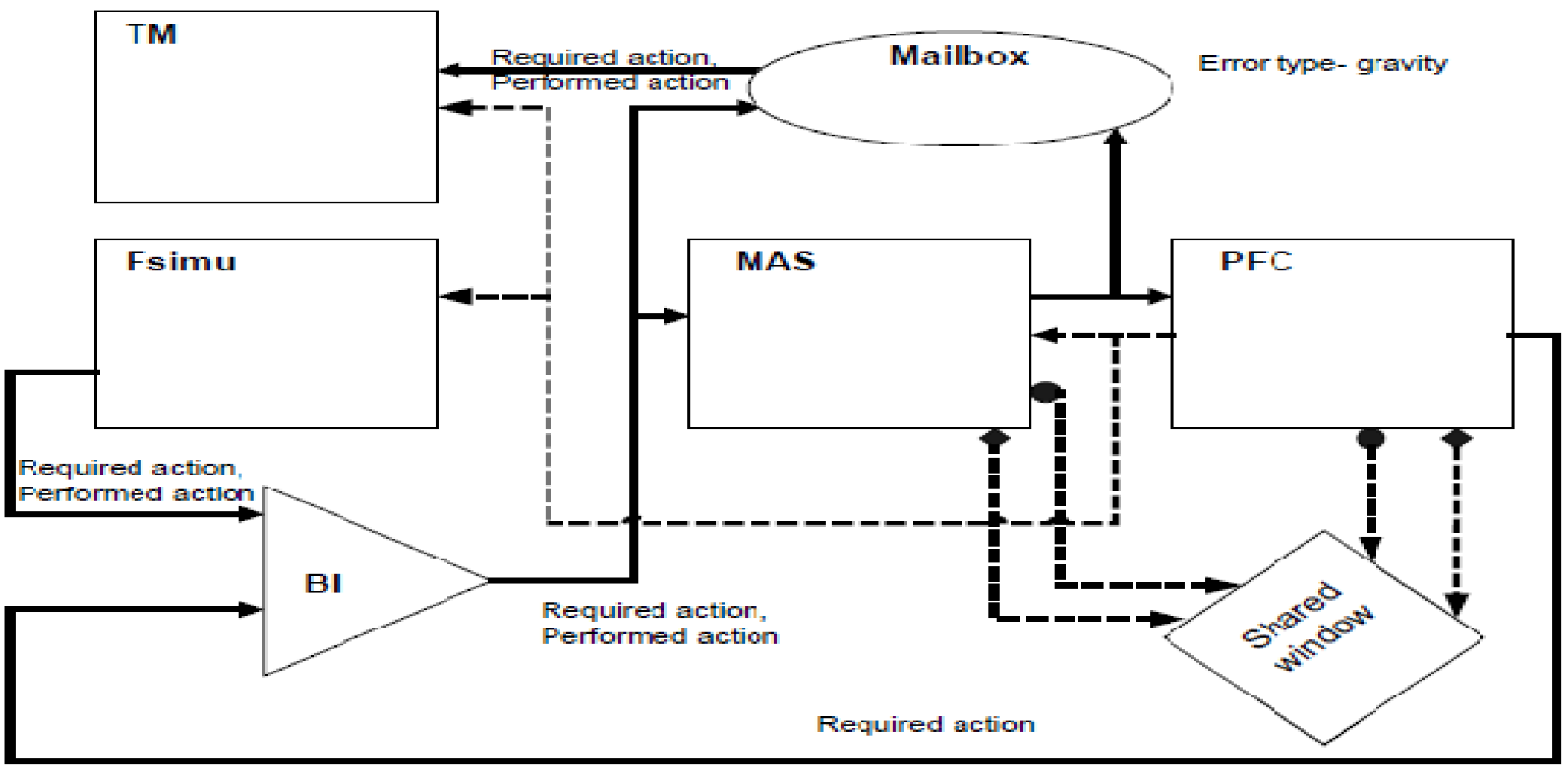

**Figure 2.19 :** Exemple de graphe d'interconnexion issu du Projet ASIMIL

On peut remarquer que tous les points d'interaction des interfaces des composants ne sont pas nécessairement utilisés. C'est par exemple le cas du point d'émission de signal du composant MAS. Ceci est une illustration des propriétés de variabilité du modèle. Chaque composant est réutilisable et adaptable pour différentes applications, en fonction de l'utilisation des points d'interaction des composants disponibles dans l'infrastructure de réutilisation. Pour plus de précision sur le modèle Ugatze nous renvoyons le lecteur aux références [Seyler, 2004], [Aniorté, 2004].

### 2.3.6 Evaluation

Actuellement, le modèle de composants Ugatze ne permet de décrire que les aspects syntaxiques et structurels d'un assemblage de composants Ugatze. Un enrichissement de ce modèle afin d'intégrer les aspects comportementaux, sémantiques et non fonctionnels s'impose. En outre, pour des raisons de vérification formelle, une ouverture d'Ugatze sur les modèles de composants formels comme Acme/Armani et Wright s'impose également.

## 2.4 Le langage de description d'architectures Acme

Acme [Garlan, 2000] est un langage de description d'architectures établi par la communauté scientifique dans le domaine des architectures logicielles. Il a pour buts principaux de fournir un langage pivot qui prend en compte les caractéristiques communes de l'ensemble des ADL, qui soit compatible avec leurs terminologies et qui propose un langage permettant d'intégrer facilement de nouveaux ADL. En effet, la plupart des langages fournissent des notions similaires comme le composant ou le connecteur [Accord, 2002]. Acme apparaît alors plus comme un langage fédérateur de ce qui existe que comme un langage réellement novateur. Il couvre la spécification des composants, des connecteurs, des configurations et éventuellement des styles architecturaux. Les auteurs d'Acme ont imaginé un langage sur lequel pourraient être converties toutes sortes de description d'architectures écrites suivant d'autres ADL, et depuis lequel une description pourrait être convertie suivant un ADL particulier pour bénéficier de ses outils supports. Bien sûr, pour opérer, le langage doit s'accompagner de la panoplie des convertisseurs adéquats.

### 2.4.1 Concepts structuraux

Acme fournit une ontologie architecturale consistant en sept éléments de conception : composant, connecteur, port, rôle, système, propriété et style d'architectures.

#### 2.4.1.1 Le concept composant

Un composant Acme représente l'unité de traitement ou de donnée d'une application. Par exemple, un client ou un serveur est un composant (cf. Figure 2.20). Il est spécifié par une interface composée de plusieurs types de ports. Chaque type de port identifie un point d'interaction entre le composant et son environnement. Il peut s'agir, par exemple, d'une simple signature d'une méthode ou d'un ensemble de procédures qui doivent être invoquées dans un ordre défini.

#### 2.4.1.2 Le concept Connecteur

Le connecteur représente l'interaction entre composants. Il s'agit d'un médiateur de communication qui coordonne les connexions entre composants. Les pipes, les appels de procédures, etc. sont des exemples de connecteurs simples. Cependant, des interactions plus complexes peuvent être représentées, comme une requête entre une base de données et une application. Un connecteur est spécifié par une interface composée d'un ensemble de rôles. Chaque rôle décrit les participants de l'interaction. La plupart des interactions sont binaires, ils possèdent deux rôles. Par exemple, le connecteur « RPC » de la Figure 2.21 est spécifié par deux rôles « *caller* » et « *callee* ».

#### 2.4.1.3 Le concept Système

Le système représente la configuration d'une application, c'est-à-dire l'assemblage structurel entre les composants et les connecteurs (cf. Figure 2.21). La structure d'un système est indiquée par un ensemble de composants, un ensemble de connecteurs, et un ensemble d'attachements. Un attachement lie un port d'un composant à un rôle de connecteur. L'exemple 1 de la Figure 2.20 décrit la représentation graphique de l'architecture client-serveur en Acme [Accord, 2002] et la Figure 2.21 correspond à la description de cette architecture en Acme. Ce système regroupe la définition de deux composants, *Client* et *Server*, ainsi que d'un connecteur *RPC*. Les ports de chaque composant sont précisés *send-request* pour le composant *Client* et *request-receive* pour le composant *Server*. Le connecteur *RPC* définit un ensemble de rôles et par quels composants ces rôles sont joués au travers de la clause attachement.

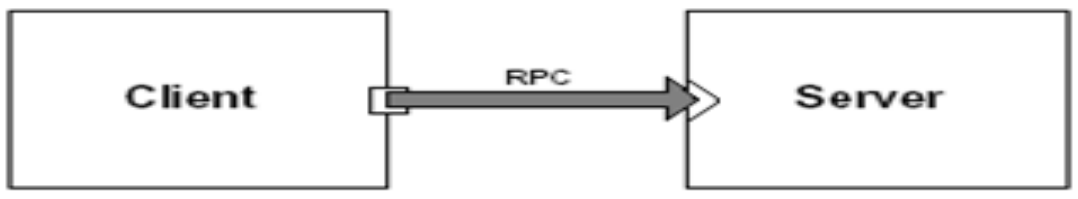

**Figure 2.20 :** Représentation graphique de l'architecture Client/Serveur en Acme

Acme permet la composition hiérarchique. En effet, un composant ou un connecteur peut avoir plusieurs descriptions de niveau plus bas. Ces descriptions sont appelées représentations, leurs rôle est d'établir correspondance entre l'interface externe d'un composant ou connecteur et ses représentations. Dans un cas simple de composant représentant un sous-système, une rep-map fait l'association entre les ports du composant et les ports des composants appartenant à ce sous-système.

#### 2.4.1.4 Style

Un style permet de définir une famille d'architectures qui sont reliées par des propriétés structurelles et sémantiques [Garlan, 1995]. L'objectif de ces styles est d'une part de fixer des règles et des contraintes topologiques, d'autre part de permettre à terme la réutilisation de code dans de nouveaux problèmes. En fait un tel style est voué à être le répertoire de base d'un architecte logiciel. Les buts visés par la notion de style sont :

– Aider à la compréhension : clarification de l'organisation du système si des structures conventionnelles sont utilisées,
– Guider le choix d'une solution « standard » adaptée à ses besoins,
– Faciliter la réutilisation,
– Permettre des analyses spécialisées.

```
System simple_cs = {
    Component Client = {
        Port send-request = {
        }
    }
    Component Server = {
        Port receive-request = {
        }
    }
    Connector rpc = {
        Role caller = {
        }
        Role callee = {
        }
    }
    Attachment Client.send-request to RPC.caller;
    Attachment Server.receive-request to RPC.callee;
}
```

**Figure 2.21 :** Description en Acme d'architectures client-serveur

Acme intègre parfaitement la notion de style telle que nous l'avons décrite précédemment. Un style de spécification architecturale (cf. 2.4.4) avec Acme et le langage de contraintes Armani qui lui est associé (cf. section 2.4.2) se compose :

– d'un ensemble de type de définitions,
– d'un ensemble de règles de conception (invariants et heuristiques),
– d'un ensemble d'analyses de conception,
– d'un ensemble minimal de sructures,
– de styles architecturaux qui prennent en compte à la fois la description du système par les éléments (composants, connecteurs,…) pour décrire par exemple un style client-serveur mais aussi les règles de conception (invariants, contraintes).

Les styles sont instanciés par des systèmes. Ces derniers :

– doivent respecter les invariants exprimés au niveau du style,
– héritent d'une structure minimale si elle est définie dans le style,
– doivent contenir des éléments instanciant des types définis au niveau du style,
– peuvent instancier plusieurs styles (seule la structure minimale d'un seul style est explicitement héritée dans ce cas).

### 2.4.1.5 Système d'annotation de propriétés

En Acme, l'intégration d'informations porteuses de sémantiques particulières : Description réalisée par un système d'annotations de propriétés (*Properties*). Une propriété possède un nom et peut avoir un type et une valeur à la déclaration. Le concept *property* d'Acme utilisable au niveau type et instance permet d'attacher des propriétés non fonctionnelles aux éléments architecturaux. Il est important de noter que ce modèle d'annotation de propriétés ne possède pas de sémantique, mais supporte tout type de sémantique. Une propriété peut même trouver sa source dans un autre ADL, dans ce cas, le type de cette propriété indiquant le langage d'origine et la valeur de la propriété est exprimée sous la forme d'une chaîne de caractères conformément à la syntaxe de l'ADL concerné. La Figure 2.22 illustre un exemple simple d'une description Acme dans sa forme textuelle[4] [Garlan, 1997]. Cette description définit les concepts évoqués précédemment, auxquels sont rattachées des annotations; elle représente l'architecture client-serveur décrite précédemment, intégrant des propriétés. Dans cet exemple, les propriétés sont utilisées de plusieurs manières. Elles font référence à des descriptions de style *Client-Serveur* définies dans d'autres systèmes (Aesop et Unicon). Elles font référence à des implémentations (par exemple : "CODE-LIB/client.c") et à un comportement défini en Wright [Allen, 1997] pour le connecteur *rpc*. Enfin, elles apportent des informations supplémentaires à l'architecture comme le type de communication : synchrone ou non.

```
System simple_cs = {
       Component client = {
              Port send-request;
              Properties {Aesop-style : style-id = client-server;
              UniCon-style : style-id = cs;
              source-code : external = "CODE-LIB/client.c" }
       }
       Component server = {
              Port receive-request;
              Properties { idempotence : boolean = true;
              max-concurrent-clients : integer = 1;
              source-code : external = "CODE-LIB/server.c" }
       }
       Connector RPC = {
              Roles {caller, callee}
              Properties { synchronous : boolean = true;
              max-roles : integer = 2;
              protocol : Wright = "..." }
       }
       Attachments {
       client.send-request to RPC.caller ;
       server.receive-request to RPC.callee
        }
}
```

**Figure 2.22 :** Description d'architectures client-serveur en Acme/Armani

Notons également que le concept propriété d'Acme utilisable au niveau type et instance permet d'attacher des propriétés non fonctionnelles aux éléments architecturaux.

## 2.4.2 Le langage Armani

Acme ne supporte pas la spécification de contraintes structurelles. Cependant, une extension nommée Acme/Armani [Garlan, 2000] autorise la spécification de contraintes structurelles. Armani [Garlan, 2000], [Monroe, 2001] est un langage de prédicats puissant. Ce langage couplé à Acme est basé sur des prédicats de logique du premier ordre. Il permet de décrire des propriétés architecturales sous forme d'invariant ou d'heuristique attachées à divers éléments architecturaux. Armani est similaire à OCL [Monroe, 2001] mais il fournit en plus un ensemble d'opérations de manipulation spécifiques aux architectures logicielles. Ces opérations facilitent la définition de contraintes qui portent sur des concepts comme les interactions entre composants, la conformité de types ou les relations d'héritage.

Armani est basé sur quatre concepts de base : les fonctions prédéfinies, les opérateurs, les quantificateurs et les fonctions de conception.

[4] Acme permet également une description graphique.

#### 2.4.2.1 Fonctions prédéfinies

Armani supporte un ensemble de fonctions prédéfinies. On peut classifier ces fonctions en quatre groupes :

– **Les fonctions de type** : ces fonctions concernent les types d'éléments. Par exemple, la fonction "*DeclaresType (E :Element, T :Type) :boolean*" permet de vérifier si l'élément *E* est de type *T* ou non,

– **Les fonctions de graphe** : ces fonctions concernent la connectivité d'éléments. Par exemple, la fonction "*Attached (E1 : Element, E2 :Element) :boolean*" permet de vérifier si l'élément *E1* est directement attaché à l'élément *E2* ou non,

– **Les fonctions de propriété** : ces fonctions concernent les propriétés. Par exemple, la fonction "*HasValue (P :Property) :boolean*" permet de vérifier si la propriété *P* possède une valeur ou non,

– **Les fonctions d'ensemble** : ce sont les fonctions appliquées sur les ensembles telles que **:** *union, intersection, contains, size, sum, select, collect***,** etc.

#### 2.4.2.2 Opérateurs

Similairement à OCL [Warmer, 2003], Armani supporte les opérateurs de comparaison (==, >, <,…), les opérateurs arithmétiques (+, -, /, mod, ...), les opérateurs logiques (and, or, ->, …) et l'opérateur de qualification (•). En plus, il supporte d'autres opérateurs spécifiques au domaine des architectures logicielle tels que; *Components, Connectors, Ports, AttachedPorts*, *AttachedRoles*, etc. Chaque opérateur spécifique est appliqué sur une catégorie d'éléments. Par exemple, l'opérateur "*Ports*" est appliqué sur un élément de catégorie "*Component*" et permet de retourner l'ensemble de tous les ports de ce composant.

#### 2.4.2.3 Quantificateurs

Armani supporte deux quantificateurs (*Forall* et *Exists*) qui permettent la quantification d'un prédicat. La Figure 2.23 illustre un exemple de manipulation de prédicats quantifiés. La première contrainte utilise le quantificateur *Forall* afin de vérifier que la propriété *secure* rattachée à tous les composants du système est définie à « true ». La deuxième contrainte utilise le quantificateur *Exists* afin de vérifier qu'il existe des connecteurs (parmi l'ensemble des connecteurs du système) déclaré de type *EventSystemType*.

**Forall** comp : CompType **in** sys**.**Components **|** comp.secure **= true;**

**Exists** conn : Connector **in** sys**.**Connectors **| declareType** (conn, EventSystemType)**;**

**Figure 2.23 :** Exemples d'expressions quantifiées en Armani

#### 2.4.2.4 Fonctions de conception

Armani offre la possibilité à l'architecte de définir des fonctions appropriées à son application. Ces fonctions sont appelées des fonctions de conception. Une fonction de conception est une simple fonction qui peut être utilisée dans la définition des contraintes architecturales. Armani offre deux constructions permettant la définition des fonctions de conception. Le premier **(***analysis*), permet la définition de fonctions de conception tout en utilisant les concepts Acme/Armani. Le second (*external analysis*), permet d'ouvrir Acme/Armani sur d'autres langages souvent impératifs comme Java.

### 2.4.3 L'outil AcmeStudio

Acme est supporté par un environnement de développement appelé AcmeStudio [ABLE, 2009]. Cet environnement implémente un outil d'analyse et de vérification des contraintes Armani : évaluateur des contraintes Armani. L'outil de vérification est directement intégré dans l'environnement AcmeStudio. Ce qui permet à l'outil de fournir des messages d'erreurs, de haut niveau d'abstraction, en relation avec la conception de l'architecture. De cette manière, l'architecte logiciel peut directement comprendre l'erreur et modifier sa conception.

Acme s'accompagne également d'une librairie d'API écrites en Java à l'intention des développeurs d'outils. Cette librairie contient les classes et les interfaces qui implémentent les entités d'Acme, et les moyens de les manipuler comme les *parsers* (et *unparser*), ou des utilitaires de traduction.

### 2.4.4 Exemple

La Figure 2.24 illustre la définition d'un style Pipe-Filter (*PipeFilterFam*). Ce style ou cette famille définit deux types de composants (*FilterT* et *UnixFilterT*), un type de connecteur (*PipeT*) et une contrainte qui vérifie que tout connecteur doit être de type *PipeT*.

```
Family PipeFilterFam = {
Component Type FilterT = {
             Ports { stdin; stdout; };
             Property throughput : int;
      };
Component Type UnixFilterT extends FilterT with {
      Port stderr;
      Property implementationFile : String;
      };
Connector Type PipeT = {
      Roles { source; sink; };
      Property bufferSize : int;
      };
Property Type StringMsgFormatT = Record [ size:int; msg:String; ];
Invariant Forall c in self.Connectors @ HasType(c, PipeT);
}
```

**Figure 2.24 :** Définition d'un style Pipe-Filter en Acme

La Figure 2.25 correspond à un système « *simplePF* » qui instancie le style « *PipeFilterFam* ». Le système « *simplePF* » est défini par des éléments instanciant des types définis au niveau du style « *PipeFilterFam* » et doit respecter les invariants exprimés au niveau du style. Le système « *simplePF* » démontre les différents intérêts de la notion de type. D'une part, la notion de type permet de définir plusieurs instances de type « *Filtre* » et ces instances héritent d'une structure minimale. D'autre part, le type « *Filtre* » peut être étendu en y ajoutant des propriétés. Par exemple, le composant « *showTracks*» définit une propriété « *implementationFile* ».

### 2.4.5 Évaluation

Acme est plus un médiateur entre ADL qu'un ADL. Il permet à n'importe quel outil qui supporte les sept concepts de base d'interagir avec d'autres outils. En fait, Acme fournit des bases simples qui forment un point de départ pour le développement de nouveaux ADL.

Par ailleurs, Acme est associé à un langage de prédicats assez puissant appelé Armani avec des fonctions appropriées au domaine de l'architecture logicielle. En effet, Armani est similaire à OCL [Monroe, 2001] mais il fournit en plus un ensemble d'opérations de manipulation spécifiques aux architectures logicielles. Ces opérations facilitent la définition de contraintes qui portent sur des concepts comme les interactions entre composants, la conformité de types ou les relations d'héritage.

```
System simplePF : PipeFilterFam = {
      Component smooth : FilterT = new FilterT
      Component detectErrors : FilterT;
      Component showTracks : UnixFilterT = new UnixFilterT extended with {
Property implementationFile : String = "IMPL_HOME/showTracks.c";
};
// Declare the system's connectors
      Connector firstPipe : PipeT;
      Connector secondPipe : PipeT;
// Define the system's topology
Attachments {
      smooth.stdout to firstPipe.source;
      detectErrors.stdin to firstPipe.sink;
      detectErrors.stdout to secondPipe.source;
      showTracks.stdin to secondPipe.sink;
      }
}
```

**Figure 2.25 :** Définition d'un système héritant du style Pipe-Filter en Acme

De plus, Acme présente un atout en offrant un langage et une boîte à outils servant de base à l'élaboration de nouveaux outils de construction. C'est un langage qui offre un double avantage aux constructeurs : il leur fournit une base solide est extensible permettant de réutiliser ou de stocker des éléments définis antérieurement comme par exemple les gabarits de conception ou les styles d'architecture.

En outre, la vocation d'Acme/Armani et la formalisation et la vérification d'architecture. Son but est de raisonner sur un niveau plus abstrait que sur l'implémentation. Cette abstraction permet au concepteur de mieux appréhender la complexité de l'architecture et de vérifier la cohérence de ces différents constituants. La définition d'un système revient donc à la définition des différents composants et connecteurs utilisés dans le système ainsi que la topologie de leurs interconnexions. Il faut toutefois aussi noter que le concept propriété d'Acme utilisable au niveau type et instance permet d'attacher des propriétés non fonctionnelles aux éléments architecturaux.

## 2.5 Le langage de description d'architectures Wright

Cette présentation est inspirée de la thèse de M. Graiet [Graiet, 2007].

### 2.5.1 Les concepts structuraux

Wright supporte des concepts architecturaux bien définis : composant, connecteur, configuration et style.

Un composant est une unité abstraite et indépendante. Il possède un type. Il comporte deux parties: la partie interface composée d'un ensemble de ports (*Ports*) qui fournissent les points d'interactions entre le composant et son environnement, et la partie calcul (*Computation*) qui décrit le comportement réel du composant. Les ports sont basés sur un ensemble d'événements émis et reçus. À chaque port est associé une description formelle par le langage CSP (Communicating Sequential Processes) [Hoare, 1985] spécifiant son comportement par rapport à l'environnement. Quant à la partie calcul, elle consiste à décrire ce que le composant fait du point de vue comportemental.

Un connecteur représente une interaction explicite et abstraite entre une collection de composants. Il possède un type. Il comporte deux parties : une interface constituée de points d'interactions appelés rôles (*Role*) et une partie qui représente la spécification d'assemblages (*Glue*). Le rôle indique comment se comporte un composant qui participe à l'interaction. La glu spécifie les règles d'assemblage entre un ensemble de composants pour former une interaction.

Les styles de Wright permettent de factoriser des caractéristiques communes (types de composants, types de connecteurs, propriétés sémantiques) à un ensemble des configurations.

### 2.5.2 Les concepts comportementaux

La formulation du comportement des composants et des connecteurs de façon informelle ne permet pas de prouver des propriétés non triviales sur l'architecture d'un système. Ainsi, pour spécifier le comportement et la coordination des composants et des connecteurs, Wright utilise une notation formelle basée sur le modèle de processus CSP. CSP est un modèle mathématique qui a pour but de formaliser la conception et le comportement de systèmes qui interagissent avec leur environnement de manière permanente. Il est basé sur de solides fondements mathématiques qui permettent une analyse rigoureuse.

Par la suite, nous présentons uniquement les notions essentielles de CSP utilisées dans Wright.

#### 2.5.2.1 Les événements

Dans le modèle CSP, tout est représenté par des événements. Un événement correspond à un moment ou une action qui présente un intérêt. CSP ne fait pas la distinction entre les événements initialisés et observés. Mais, CSP pour Wright le fait : Un événement initialisé s'écrit sous la forme ē ou _e. Un événement observé est noté e. Avec e représente le nom de l'événement. De plus, les événements peuvent transmettre des données : e?x et e!x, représentent respectivement les données d'entrée et de sortie. CSP définit un événement particulier noté √, qui indique la terminaison de l'exécution avec succès.

#### 2.5.2.2 Les processus

Pour définir un comportement, il faut pouvoir combiner les événements. Un processus correspond à la modélisation du comportement d'un objet par une combinaison d'événements et d'autre processus simples. Les principaux opérateurs fournis par CSP sont :

**-L'opérateur préfixe noté** ->: Le séquencement ou le préfixage est la façon la plus simple de combiner des événements. Un processus qui s'engage dans un événement e, puis se comporte comme le processus P, est noté « e->P ».

**-La récursion** : Par la possibilité de nommer un processus, il devient possible de décrire les comportements répétitifs très facilement. Nous décrivons par exemple le processus qui ne s'engage que dans l'événement e et qui ne s'arrête jamais par : P=e->P.

**-L'opérateur de choix externe ou déterministe noté** □ : Si nous avons le processus e-> P□u->Q et que l'environnement s'engage dans l'événement u, alors le processus s'engagera dans cet événement et se comportera comme le processus Q.

**-L'opérateur de choix interne ou non déterministe noté** Π: A l'inverse du choix déterministe, c'est le processus qui choisit de façon non déterministe le comportement à choisir parmi plusieurs. Cette fois le processus ē->PΠū->Q va choisir entre initialiser l'événement e et continuer comme P ou initialiser u et continuer comme Q. Il décide lui-même de ce choix sans se préoccuper de l'environnement.

**-L'alphabet** : l'alphabet fait référence à un processus et est noté αP, pour le processus P. L'alphabet d'un processus est l'ensemble des événements sur lequel le processus a une influence.

**Remarque :** Le symbole § désigne le processus de terminaison avec succès, ce qui veut dire que le processus s'est engagé dans un événement succès √ et s'est arrêté. Formellement, §= √→STOP (En CSP il est généralement noté «SKIP»).

## 2.5.2.3 Sémantique de Wright

L'ADLWright étant entièrement basé sur CSP au niveau de la spécification des interactions, sa sémantique l'est aussi. Il faut donc comprendre en premier lieu la sémantique de CSP.

### 2.5.2.3.1 Modélisation mathématique des processus CSP

Un processus CSP est un triplet (A, F, D) où A représente l'alphabet du processus, F représente ses échecs et D représente ses divergences.

**- Alphabet :** l'alphabet d'un processus représente l'ensemble des événements sur lequel le processus a une influence. De ce fait, si un événement n'est pas dans l'alphabet d'un processus, alors ce processus ne le connaît pas, ne le traite pas et donc l'ignore.

**- Echecs :** les échecs d'un processus sont une paire de traces et de refus. Une trace est une séquence d'événements permise par le processus. L'ensemble des traces possibles d'un processus P est noté traces(P). Un refus correspond à un ensemble d'événements proposés pour lequel le processus refuse de s'engager. Cet ensemble d'événements refusés par un processus P est noté refus(P). Cette notion de refus permet une distinction formelle entre processus déterministes et non déterministes. En effet, un processus déterministe ne peut jamais refuser un événement qu'il peut entamer alors qu'un processus non déterministe le peut.

**- Divergences :** une divergence d'un processus est définie comme une de ses traces quelconque après laquelle il y a un comportement chaotique. Ce comportement chaotique est représenté par le processus CHAOS :

$$CHAOS_A = STOP \prod (\forall x : A \bullet x \rightarrow CHAOS_A)$$

Ce processus peut se comporter comme n'importe quel autre. C'est le processus le plus non déterministe, le plus imprévisible, le plus incontrôlable de tous. La divergence est donc utilisée pour représenter des situations catastrophiques ou des programmes complètement imprédictibles (comme des boucles infinies).

#### 2.5.2.3.2 Les modèles sémantiques

Les trois principaux modèles sémantiques [Roscoe, 1997a], [Roscoe, 1997b] sont les traces, les échecs stables et les échecs-divergences.

Le modèle des traces associe à chaque processus les séquences finies d'événements admises par ce processus. Ce modèle permet donc de représenter les comportements possibles de processus sous forme de traces. Les traces du processus P sont dénotées par traces(P).

Le modèle des échecs stables associe à chaque processus P les couples de la forme (t, E), où t est une trace finie admise par P et E est l'ensemble des événements que le processus ne peut pas exécuter après avoir exécuté les événements de t. l'ensemble de ces couples est noté refus(P). Ce modèle permet de caractériser les blocages de P. En effet, si E est égal à l'ensemble des événements exécutables par P, alors P se retrouve bloqué.

Enfin, le modèle des échecs-divergences associe à chaque processus P l'ensemble de ses échecs stables et l'ensemble de ses divergences. Un processus P n'est divergent que s'il se trouve dans un état dans lequel les seuls événements possibles sont les événements internes. Cet état est dit divergent. L'ensemble des divergences de P noté divergences(P), est l'ensemble des traces t telles que le processus se retrouve dans un état divergent après avoir exécuté t. Si le processus est déterministe, alors divergences(P) est vide.

#### 2.5.2.3.3 Le raffinement CSP

Le raffinement consiste à calculer et à comparer les modèles sémantiques de deux processus. Le raffinement dépend donc du modèle considéré. Par exemple, dans le cas du modèle des échecs-divergences, si P et Q sont deux processus, alors Q raffine P, noté

$P \sqsubseteq_{FD} Q$ si : $\text{refus}(Q) \subseteq \text{refus}(P) \wedge \text{divergences}(Q) \subseteq \text{divergences}(P)$

Dans cet exemple, il n'est pas utile de comparer traces(P) et traces(Q), car par définition :

$\text{refus}(Q) \subseteq \text{refus}(P) \Rightarrow \text{traces}(Q) \subseteq \text{traces}(P)$

Intuitivement, Q est égal ou meilleur que P dans le sens où il a moins de risque d'échec et divergence. Q est plus prévisible et plus contrôlable que P, car si Q peut faire quelque chose d'indésirable ou refuser quelque chose, P peut le faire aussi. Concrètement, cela signifie qu'un observateur ne peut pas distinguer si un processus a été substitué à un autre.

##### 2.5.2.3.3.1 Utilisation de la sémantique de CSP dans Wright

La sémantique de Wright est entièrement basée sur CSP. En effet, chaque partie d'une architecture logicielle Wright est modélisée par un processus CSP. Pour analyser l'interaction de ces processus, il faut les combiner par l'opérateur || de CSP. Mais un problème émerge : la sémantique de CSP utilise des noms d'événements globaux. En effet, pour décrire que deux processus CSP interagissent, il suffit qu'ils partagent un nom d'événement identique. Tandis que les noms d'événements en Wright agissent comme des noms locaux propres aux composants et que l'interaction a lieu par les connecteurs.

Afin de réaliser cette correspondance d'événements locaux de Wright en événements globaux de CSP, Wright identifie et résout systématiquement les deux problèmes suivants :

1. Au niveau des instances : il peut exister plusieurs instances d'un même type. Ainsi des interactions indésirées peuvent avoir lieu en introduisant de multiples copies d'un même processus. Pour résoudre ce premier problème, il suffit de préfixer chaque nom d'événement par le nom de son instance. Ainsi un événement a :
    - 3 niveaux : N.P.e (nom du composant, nom du port, nom de l'événement), si le Calcul (Computation) utilise un événement du port P,

- 2 niveaux : N.e (nom du composant, nom de l’événement), si le Calcul (Computation) utilise un événement interne (non associé à un port).

2. Au niveau des liens : comme les noms d’événements utilisés dans les composants et connecteurs sont locaux, pour assurer une synchronisation entre un composant et un connecteur (indiqué par un lien) il faut pouvoir changer les noms d’événements afin de respecter le fait qu’en CSP deux processus ne communiquent que s’ils partagent un même événement. Pour résoudre ce deuxième problème, il suffit de renommer des événements du connecteur par les noms des événements des composants correspondants : si nous avons un nom d’événement pour un connecteur Conn.Role.e et que le rôle de ce connecteur est lié à un port d’un composant Comp.Port, alors nous voulons que le nom de l’événement du connecteur Conn.Role.e soit renommé Comp.Port.e.

### 2.5.3 Vérification d’architectures logicielles Wright

### 2.5.3.1 Description informelle des propriétés Wright

Nous allons maintenant expliciter informellement l’ensemble des propriétés[5] intégrées dans Wright sur la cohérence et la complétude des différents élements d’une architecture (composant, connecteur et configuration).

#### 2.5.3.1.1 Cohérence

La cohérence consiste à vérifier que tous les éléments décrivant l’architecture logicielle (les composants, les connecteurs et la configuration) sont cohérents.

#### 2.5.3.1.1.1 Cohérence d’un composant

Nous avons vu dans la description d’un composant qu’un composant est constitué par deux parties la partie interface décrivant les ports et la partie calcul, ainsi, nous vérifions la cohérence d’un composant en nous assurant que le calcul obéit aux règles d’interaction définies par les ports.

Le premier aspect décrit par le port correspond au comportement attendu du composant. Il faut s’assurer de la cohérence entre le comportement des ports et celui du calcul (Computation) par la notion de projection. Un port est une projection d’un composant si ce dernier agit de la même manière que le port quand nous ignorons tous les événements n’appartenant pas à l’alphabet de ce port.

Illustrons cette notion de projection en prenant le composant suivant :

**Component** Double
**Port** Input = read?x-> Input □ close -> §
**Port** Output = write!x-> Output ∏ close -> §
**Computation** = Input.read?x -> Output. write!(2*x) -> **Computation** □ Input.close -> Output.close -> §

Si nous ignorons tous les événements qui n’appartiennent pas à l’alphabet du port Input, nous obtenons :

**Computation** = Input.read?x -> **Computation** □ Input.close -> §

[5] Elles sont appelées tests par les auteurs de Wright.

Le port Input est bien une projection du Calcul. De la même manière, il est facile de montrer que le port Output est une projection du Calcul.

Le second aspect du port correspond à l'interaction avec l'environnement. Si la spécification des ports ne tient pas compte d'un événement, alors le composant (partie Calcul) n'a pas à s'en occuper. Mais d'un autre côté, il y a des situations dans lesquelles il peut être approprié d'avoir des spécifications d'un composant qui décrivent des comportements qui n'auront pas forcément lieu. Ceci est notamment le cas lors de la réutilisation d'une spécification d'un Calcul plus générale que nos besoins. Pour toutes ces raisons, la spécification des ports peut ne couvrir qu'un sous-ensemble des situations que le composant peut effectivement gérer. Illustrons nos propos par l'exemple du composant Double qui décrit un comportement d'échec indiqué par l'événement fail si nous souhaitons utiliser ce composant sans se préoccuper de cet événement, nous le spécifions par :

**Component** Double
**Port** Input = read?x-> Input □ close -> §
**Port** Output = $\overline{\text{write!}}$x-> Output ∏ $\overline{\text{close}}$ -> §
**Computation** = Input.read?x -> $\overline{\text{Output. write!(2*x)}}$ -> **Computation** □ Input.close -> Output.close -> □ Input.fail -> §

Si nous ignorons l'hypothèse du port Input qui est que l'événement fail ne va pas avoir lieu et nous projetons le port Output, nous obtenons :

**Computation** = $\overline{\text{Output. write!(2*x)}}$ -> **Computation** □ $\overline{\text{Output.close}}$ -> **§** □ **§**

Le port Output n'est plus une projection du Calcul. Ainsi la propriété permettant de vérifier la cohérence d'un composant Wright est formulée par :

**Propriété 1 : Cohérence Port / Calcul**

> *La spécification d'un port doit être une projection du Calcul, sous l'hypothèse que l'environnement obéisse à la spécification de tous les autres ports*

Intuitivement, la propriété 1 stipule que le composant ne se préoccupe pas des événements non traités par les ports (ici événement fail).

#### 2.5.3.1.1.2 Cohérence d'un connecteur

La description du connecteur doit vérifier que la coordination des rôles par la *Glu* est cohérente avec le comportement attendu des composants. Prenons comme exemple le connecteur suivant :

**Connector** Bogus
**Role** User1 = set -> User1 ∏ get -> User1 ∏ §
**Role** User2 = set -> User2 ∏ get -> User2 ∏ §
**Glue =** User1. set-> Continue □ User2. set-> Continue □ §
**where** {
Continue = User1. set-> Continue □ User2. set-> Continue □ User1. get-> Continue □User2. get-> Continue □ § }

La spécification du connecteur Bogus semble raisonnable, la *Glu* (Glue en anglais) du connecteur exige qu'un de ses deux participants initialise l'événement *set* mais n'indique pas lequel, si chacun commence par *set*, alors que l'événement se produira d'abord et la communication peut continuer sans aucun problème, si cependant chaque participant essaye légalement d'exécuter initialement l'événement *get* alors le connecteur aboutira à un interblocage. Sachant qu’un processus CSP est dit en situation d’interblocage quand il peut refuser de participer à tout événement, mais n’a pas pour autant terminé correctement (en participant à l’événement §). Inversement, un processus est sans interblocage s’il ne peut jamais être en situation d’interblocage. Ainsi, Wright propose la propriété 2.

**Propriété 2 : Connecteur sans interblocage**

> *La glu d’un connecteur interagissant avec les rôles doit être sans interblocage.*

Une autre catégorie d’incohérences est détectable comme une situation d’interblocage, lorsque la spécification d’un rôle est elle-même incohérente. Dans une spécification d’un rôle complexe, il peut y avoir des erreurs qui mènent à une situation dans laquelle aucun événement n’est possible pour ce participant, même si la *Glu* était prête à prendre tout événement.

**Propriété 3 : Rôle sans interblocage**

> *Chaque rôle d’un connecteur doit être sans interblocage.*

Pour empêcher le conflit de contrôle, un événement ne doit être initialisé que par un unique processus, tous les autres processus ne faisant que l’observer.

**Propriété 4 : Un initialiseur unique**

> *Dans une spécification de connecteur, tout événement ne doit être initialisé que par un rôle ou la glu. Tous les autres processus doivent soit l’observer, soit l’oublier (grâce à leur alphabet).*

La dernière propriété pour les connecteurs vérifie que les notations pour l’initialisation et l’observation des événements sont utilisées correctement. Ceci est illustré par l’exemple du port Output dont la spécification est la suivante :

**Port** Output = $\overline{\text{write!(w,i)}}$ -> Output □ $\overline{\text{close}}$ -> Output

Le port Output envoie des couples (mot, numéro de ligne) jusqu’à ce qu’il n’y en ait plus, ensuite il ferme son port et s’arrête.

Cette spécification n’est pas cohérente du fait que c’est le composant qui décide s’il lui reste des couples à envoyer et non pas l’environnement.

**Propriété 5 : Engagement de l’initialiseur**

> *Si un processus initialise un événement alors il doit s’engager dans cet événement sans être influencé par l’environnement.*

#### 2.5.3.1.1.3 Cohérence d’une configuration

Au niveau de la déclaration d’instances, la cohérence s’applique aux deux points suivants :

- Le nom de l’instance est-il unique?
- Des paramètres raisonnables ont-ils été donnés?

Dans l’exemple ci-dessous nous avons paramétré le nombre d’instances du port Output du composant Filtre_Texte.

**Component** Filtre_Texte (nout : 1 ..)
> **Port** Input = DataInput
> **Port** Output $_{1..\,nout}$ = DataOutput
> Computation = lire des données du port Input. Envoyer ces données successivement sur les ports Output $_1$ , Output $_2$ ,… , Output $_{nout}$

Le nombre d’instances du port Output sera déterminé au moment de l’instanciation du filtre.

**Propriété 6 : Substitution des paramètres**

> *Une déclaration d’instance paramétrant un type doit résulter d’une validation de ce type après avoir substitué tous les paramètres formels manquant.*

Dans le cas de paramètres numériques, il faut s’assurer que les paramètres entrent dans les bornes données dans la description du type.

**Propriété 7 : Test des valeurs sur leur intervalle donné**

> *Un paramètre numérique ne doit pas être plus petit que la limite inférieure (si elle est déclarée), et pas plus grande que la limite supérieure (si elle est déclarée).*

Au niveau des liens la question suivante se pose :

Quels ports peuvent être utilisés pour ce rôle?

La vérification sur le fait que les protocoles du port et du rôle soient identiques n’est pas suffisante. En effet, nous voulons avoir la possibilité d’attacher un port qui n’a pas un protocole identique au rôle.

Considérons le rôle suivant :

**Role** Source = $\overline{\text{write !x}}$ ->Source □ $\overline{\text{close}}$ -> §

Le rôle Source peut être attaché au port suivant :

**Port** Output3 = $\overline{\text{write!1}}$ -> $\overline{\text{write!2}}$ -> $\overline{\text{write!3}}$ -> $\overline{\text{close}}$ -> §

Le rôle Source et le port Output3 ne sont pas identiques. Le rôle Source qui émet des suites de x a une description plus générale que le port Output3 qui émet la suite 1 2 3.

D’autre part, il faut toujours vérifier qu’il n’existe pas une incompatibilité entre le rôle et le port qui lui est attaché. Par exemple, nous ne voulons pas accepter le fait qu’un port comme BadOutput (sans l’événement close) puisse être attaché au rôle Source.

**Port** BadOutput = $\overline{\text{write!x}}$ -> BadOutput ∏ §

Ainsi nous avons la propriété suivante :

**Propriété 8 : Compatibilité port / rôle**

> *Tout port attaché à un rôle doit toujours continuer son protocole dans une direction que le rôle peut avoir.*

Les deux propriétés suivantes concernent le concept de Style d’architectures. Une configuration d’un système est cohérente avec ses styles déclarés si elle obéit à chacune de leurs contraintes.

**Propriété 9 : Contraintes de Style**

> *Les prédicats d’un style doivent être vrais pour une configuration déclarée être dans ce style.*

Les contraintes d’un Style doivent être cohérentes entre elles.

- **Exemple**

  ∀ c : Components ; p : Ports (c) • Type (p) = DataOutput

  ∃ c : Components ; p : Ports (c) • Type (p) = DataOutput

Ces deux contraintes sont en contradiction, donc le Style contenant ces deux contraintes est incohérent.

**Propriété 10 : Cohérence de Style**

*Au moins une configuration doit satisfaire les contraintes de style.*

#### 2.5.3.1.2 Complétude

Une catégorie de complétude importante que vérifie Wright concerne la configuration :

- Au niveau des liens, si un lien est omis alors un composant va dépendre des événements qui ne vont jamais avoir lieu, ou une interaction va échouer car il manque un participant,
- D'autre part, il existe des ports de composants qui n'ont pas besoin d'être attachés et il y a des interactions qui peuvent continuer même si un participant manque.

Pour ces deux raisons, il n'est pas suffisant de contrôler que tous les ports et rôles soient bien attachés.

**Propriété 11 : la complétude des liens**

*Chaque port (respectivement rôle) non attaché dans la configuration doit être compatible avec le rôle (respectivement port)* avec le processus de terminaison avec succès noté §

La Figure 2.26 donne la liste complète des propriétés effectuées par Wright sur la cohérence et la complétude.

### 2.5.3.2 Techniques de vérification des propriétés Wright

Ici, nous nous interrogeons sur les techniques potentielles permettant de prouver les propriétés Wright présentées précédemment.

1. **Cohérence des ports avec le Calcul (composant)**
2. **Absence d'interblocage sur les connecteurs (connecteur)**
3. **Absence d'interblocage sur les rôles (rôle)**
4. **Initialiseur unique (connecteur)**
5. **Engagement de l'initialiseur (n'importe quel processus)**
6. **Substitution de paramètres (instance)**
7. **Bornes d'un intervalle (instance)**
8. **Compatibilité port / rôle (lien)**
9. **Contraintes pour les styles (configuration)**
10. **Cohérence de style (style)**
11. **Complétude des liens (configuration)**

**Figure 2.26 :** Propriétés définies par Wright

#### 2.5.3.2.1 Utilisation du raffinement CSP

Le raffinement CSP permet le développement incrémental des systèmes CSP. Un processus CSP peut être raffiné progressivement jusqu'à son implémentation (raffinement ultime). Par exemple, si une composition parallèle de deux processus P et Q raffine une spécification abstraite décrite par le processus S, alors nous écrivons :

$$S \sqsubseteq P \parallel Q.$$

Ensuite, nous pouvons développer la spécification S en raffinant d'une façon séparée P et Q : si $P \sqsubseteq P'$ et $Q \sqsubseteq Q'$, alors la composition de P' et Q' raffine aussi S : $S \sqsubseteq P' \parallel Q'$. Egalement le raffinement CSP peut être utilisé pour vérifier des propriétés de sûreté ou de vivacité. En effet, des propriétés Wright sont formalisées grâce au raffinement CSP. Ces propriétés sont les suivantes [Allen, 1997] :

- Propriété 1 : Cohérence des ports avec le Calcul,
- Propriété 2 : Absence d'interblocage sur les connecteurs,
- Propriété 3 : Absence d'interblocage sur les rôles,
- Propriété 8 : Compatibilité port/rôle.

#### 2.5.3.2.2 Formalisation

Afin de formaliser les propriétés Wright en utilisant le raffinement CSP, nous définissons les ensembles suivants :

- $\alpha P$ : l'alphabet du processus P,
- $\alpha_i P$ : le sous-ensemble de $\alpha P$ correspondant aux événements initialisés,
- $\alpha_0 P$ : le sous-ensemble de $\alpha P$ correspondant aux événements observés,

**Propriété 1 : Cohérence Port/Calcul**

Comme nous l'avons noté, la spécification d'un port a deux aspects :

- Des exigences sur le comportement du composant (le composant accomplit le comportement décrit par le port),
- Des suppositions sur l'environnement (qu'est-ce que l'environnement, c'est-à-dire les rôles des connecteurs auxquels le composant peut être attaché, va exiger pendant l'interaction).

Ainsi, pour modéliser le processus du Calcul dans l'environnement indiqué par les ports :

1. Nous devons prendre les ports et construire un processus qui est restreint aux événements observés (ce qui extrait les suppositions de l'environnement).

**Définition 1**

Pour tout processus p = (A, F, D) et un ensemble d'événements E, $P \upharpoonright E = (A \cap E, F', D')$ où $F' = \{(t', r') \mid \exists (t, r) \in F \mid t' = t \upharpoonright E \wedge \forall r' = r \cap E\}$ et $D' = \{t' \mid \exists t \in D \mid t' = t \upharpoonright E\}$.

La projection d'une trace ($t \upharpoonright E$) est une trace qui contient tous les éléments de t qui sont dans E, dans le même ordre, sans tous les éléments qui ne sont pas dans E.

- **Exemple**

  $< acadbcabc > \upharpoonright \{a, b\} = < aabab >$

2. Nous devons rendre ce nouveau processus déterministe. Ainsi, nous assurons que les décisions prises dans l'interaction sont faites par le Calcul et non par les ports.

**Définition 2**

Pour tout processus P = (A, F, D), det(P) = (A, F', $\emptyset$) où $F' = \{(t, r) \mid t \in \text{Traces}(P) \wedge \forall e : r \bullet t \wedge < e > \notin \text{Traces}(P)\}$.

La fonction det(P) a les mêmes traces que P, mais avec moins de refus. Ainsi, n'importe quel événement qui a lieu à tout point est entièrement contrôlable par l'environnement : det(P) est déterministe.

3. Il ne nous reste plus qu'à faire interagir ce nouveau processus déterministe (det(P)) avec celui du Calcul(C) en les mettant en parallèle : C || det(P). Nous avons donc au moins les traces de P mais où les décisions sont prises par C.

En utilisant le raffinement, il est possible de vérifier que le Calcul respecte bien les exigences de ports.

**Propriété 1 : Cohérence Port/Calcul**

Pour un composant avec un processus de Calcul C et des ports P, $P_1$, … $P_n$ ; C est cohérent avec P si $P \sqsubseteq (C \,||\, \forall i : 1 \ldots n \,||\, det(P_i \upharpoonright \alpha_0 P_i)) \upharpoonright \alpha P$.

**Propriété 2 et Propriété 3 : Absence d'interblocage sur les connecteurs et Absence d'interblocage sur les rôles**

Ces deux propriétés reviennent à vérifier si un processus est sans interblocage. D'une façon formelle, un processus P = (A, F, D) est sans interblocage si pour toute trace t telle que (t, A) $\in$ F, last(t) =$\surd$. Mais ceci peut être exprimé par une relation de raffinement entre le processus $DF_A$ et P $DF_A \sqsubseteq$ P avec $DF_A$ est défini par :

$DF_A = (\prod e : A \bullet e \rightarrow DF_A) \prod §$.

Le processus $DF_A$ permet toutes les traces possibles sur l'alphabet A mais sans jamais avoir la possibilité de refuser tous les événements : il s'agit d'un processus sans interblocage.

**Propriété 8 : compatibilité port / rôle**

La distinction entre un port et un rôle est que le port décrit un comportement spécifique alors qu'un rôle décrit un pattern de comportements permettant le lien de plusieurs ports.

Par contre le lien d'un port à un rôle doit toujours respecter les contraintes de spécification de ce rôle. Ainsi, le comportement d'un port attaché à un rôle est le comportement de ce processus port restreint aux traces de ce processus rôle.

Comme la restriction d'une trace est effectuée par la version déterministe d'un processus, nous testons donc le processus P | | det(R) pour exprimer cette restriction au processus rôle. Pour pouvoir utiliser le raffinement dans le test de compatibilité, il faut que les alphabets des deux processus port et rôle soient identiques. Pour cela, nous définissons comment augmenter l'alphabet d'un processus.

**Définition 3**

Pour tout processus P et un ensemble d'événements A, $P_{+A} = P \,||\, STOP_A$

**Propriété 8 : compatibilité**

Un port P est compatible avec un rôle R, noté P compat R, si

$R_{+(\alpha P - \alpha R)} \sqsubseteq P_{+(\alpha R - \alpha P)} \,||\, det(R)$

### 2.5.3.2.3 Automatisation

Les auteurs de Wright proposent un outil appelé Wr2fdr [Wr2fdr, 2005] (cf. chapitre 6) permettant d'automatiser les quatre propriétés (1, 2, 3 et 8 cf. Figure 2.26) décrites précédemment. Pour y parvenir, l'outil Wr2fdr traduit une spécification Wright en une spécification CSP dotée des relations de raffinement à vérifier. La spécification CSP engendrée pour l'outil Wr2fdr est soumise à l'outil de Model checking FDR (Failure-Divergence Refinement) [FDR2, 2003]. Dans la suite nous présentons successivement FDR et Wr2fdr.

- **FDR**

FDR permet de vérifier de nombreuses propriétés sur des systèmes d'états finis. FDR s'appuie sur la technique de « model checking » [Schnoebelen, 1995]. Celle-ci effectue la vérification d'un modèle d'un système par rapport aux propriétés qui sont attendues sur ce modèle. Cette vérification est entièrement automatisée et consiste à explorer tous les cas possibles.

- **Wr2fdr**

Wr2fdr est un outil développé par l'université de Carnegie Mellon [Wr2fdr, 2005]. Il permet de traduire une spécification Wright en une spécification CSP acceptée par l'outil FDR. Hormis les fonctionnalités lexico-syntaxiques et de génération de code CSP, l'outil Wr2fdr assure les fonctionnalités communes suivantes :

- Correspondances entre les événements locaux de Wright et les événements globaux de CSP,
- Détermination d'un processus CSP : det(P). Ceci permet de traiter l'opérateur non déterministe (∏) de CSP,
- Calcul de l'alphabet d'un processus CSP : αP, car FDR exige explicitement lors de la composition parallèle des processus (| |) leurs alphabets,
- Calcul des relations de raffinement liées aux propriétés 1, 2, 3 et 8.

La version actuelle de l'outil Wr2fdr ne fait pas la distinction entre les événements initialisés et observés. De plus, les événements ne portent pas des informations ni d'entrée ni de sortie.

#### 2.5.3.2.4 Autres techniques

Les auteurs de Wright ne proposent aucune automatisation des sept propriétés restantes. A notre avis la vérification automatique des propriétés 4, 6, 7 et 11 à savoir Initialiseur unique, Substitution de paramètres, Bornes d'intervalle et Complétude des liens nécessitent l'implémentation du langage Wright : analyseur lexico-syntaxique et analyseur sémantique. Ainsi la vérification des propriétés 4, 6 et 7 peut être assurée totalement par l'analyseur sémantique. Après avoir identifié les ports et les rôles non attachés au sein d'une configuration Wright, la propriété 11 peut être traitée de la même manière que la propriété 8 (compatibilité port/rôle), c'est-à-dire confiée à FDR. La vérification automatique de la propriété 9 (contraintes pour les styles) peut être obtenue par l'implantation du langage de contraintes de Wright.

### 2.5.3.3 Bilan sur la vérification d'architecture logicielle Wright

Wright en tant que langage « généraliste » ne peut pas proposer des propriétés architecturales spécifiques c'est-à-dire liées à un domaine d'application ou à des applications particulières. L'architecte qui désire vérifier des propriétés architecturales spécifiques doit travailler sur des spécifications CSP produites par l'outil Wr2fdr. De plus, il faut que les propriétés à vérifier soient naturellement exprimables sous forme de raffinement CSP. Mais les propriétés spécifiques potentiellement vérifiables dans le cadre de CSP ne couvrent pas toutes les classes de propriétés. En effet, CSP ne peut pas traiter naturellement les propriétés d'équité[6], les propriétés orientées état (invariant du système) et les propriétés qui incluent des événements et en excluent d'autres. La technique de model checking appliquée sur des programmes – notamment sur des programmes concurrents – ouvre des perspectives intéressantes pour vérifier des propriétés spécifiques plus ou moins diversifiées. Mais ceci suppose le passage d'une représentation architecturale sous forme d'un modèle (architecture logicielle en Wright) vers une représentation architecturale sous forme d'un programme concurrent. Dans le chapitre 7, nous proposons un outil IDM permettant de transformer de Wright vers un programme concurrent en Ada.

---

[6] Une propriété d'équité (fairness en anglais) énonce que, sous certaines conditions, quelque chose aura lieu (ou n'aura pas lieu) un nombre infini de fois. On parle de vivacité répétée [Schnoebelen, 1995].

## 2.6 Spécification des propriétés non fonctionnelles

### 2.6.1 Aperçu sur les langages et méthodes de spécification des propriétés non-fonctionnelles

Les propriétés non fonctionnelles sont par exemple la sécurité, la fiabilite, le temps de réponse et la sécurité. De nombreux travaux de recherche menés sur la spécification des propriétés non fonctionnelles ont abouti à la création des langages et des méthodes formelles supportant ce type de propriétés. On parle aussi de qualité de services (QoS).

#### 2.6.1.1 Les méthodes formelles

Les méthodes formelles permettent de spécifier le comportement du système d'une façon précise et correcte. Elles sont basées sur des fondations mathématiques et des outils de vérification formelle tels que les prouveurs et model-checkers. Ces méthodes formelles peuvent être utilisées pour spécifier et prouver les propriétés des systèmes. On peut donc naturellement utilisées ces approches pour spécifier les propriétés non fonctionnelles. Cependant la plupart des approches existantes se focalisent sur les aspects temporels et, précisément, dans le domaine multimédia. Parmi les méthodes formelles nous citons :

- QTL (Quality of service Temporal Logic) : La logique temporelle QTL [Blair, 1993], [Blair, 1997] permet de spécifier les exigences temporelles ainsi que les suppositions de performances du système. QTL est basée sur les événements à temps réel linéaire,
- SDL : le langage SDL [Ellsberger, 1997] est un langage principal pour la spécification des propriétés non fonctionnelles dans le domaine des télécommunications. SDL est basé sur des fondations mathématiques. Ce langage possède une représentation graphique basée sur les machines à états dont les processus sont représentés comme états de machines et les messages asynchrones sont représentés par des transitions entre eux.

#### 2.6.1.2 Les langages de spécification

Plusieurs langages de spécification de qualité de services existent [Samuel, 2008], tels que ODL [TINA, 1996], QDL [Pal, 2000], QIDL [Becker, 1999], QML [Frolund, 1998] et CQML [Aagedal, 2001].

- **TINA ODL** : TINA ODL [TINA, 1996] est un sur-ensemble de CORBA IDL [OMG, 1996]. Il permet la spécification des objets via leurs interfaces. TINA ODL supporte la spécification de QdS en utilisant une paire nom-valeur directement liée à une opération ou à un flot de données. L'inconvénient majeur de cette approche est qu'elle n'offre pas la possibilité d'associer plusieurs spécifications de QdS à la même interface. Par conséquent l'utilisation d'une même interface avec différents QdS doit nécessiter l'héritage de cette interface et l'ajout des spécifications de QdS différents à chaque nouvelle interface héritée.

- **QIDL :** QIDL [Becker, 1999] est une extension du langage OMG IDL qui supporte la spécification de QdS en fournissant la possibilité de spécifier des interfaces de QdS toute en leur assignant des interfaces fonctionnelles. QIDL apporte deux mots clés au langage IDL : "qos" pour la spécification d'une interface de QdS et "withQoS" pour attacher une interface fonctionnelle à une interface de QdS.

**- QDL :** QDL [Pal, 2000] est un langage de description de QdS basée sur IDL CORBA. Ce langage définit les relations de qualité en définissant des objets de QdS. Un objet de QdS contient une attente et une obligation de qualité. Chaque obligation contient un nombre de propriétés (simples ou complexes). Une propriété simple est une paire nom-valeur alors qu'une propriété complexe dépend d'autres propriétés provenant d'autres objets de QdS. Un besoin de qualité est spécifié comme une contrainte sur les propriétés des autres objets de QdS. QDL utilise OCL [Warmer, 2003] pour spécifier les relations de QdS.

Par la suite, nous allons présenter en détails QML et CQML. QML [Frolund, 1998] est le premier langage générique de spécification de QdS. Il sépare la spécification de QdS de la spécification des aspects fonctionnels (en IDL). CQML [Aagedal, 2001] est défendu comme étant le plus approprié pour la description des propriétés non fonctionnelles des composants logiciels. De plus, il est intégrable en UML.

#### 2.6.1.3 QML

QML [Frolund, 1998] est le premier langage générique de spécification de QdS. Il sépare la spécification de QdS de la spécification des aspects fonctionnels (en IDL). On distingue trois concepts fondamentaux dans QML : *contrat type*, *contrat* et *profil*.

**1- Contrat type** : un contrat type représente une catégorie de qualité de services comme par exemple la performance ou la fiabilité. Il décrit toutes les dimensions possibles de cette catégorie qui vont être utilisées pour caractériser un aspect particulier de QdS. Une dimension est définie par son nom et son domaine. Le domaine est constitué d'une direction (*increasing* ou *decreasing*), d'un ensemble de valeurs possibles (*numeric, set ou enum*) et peut avoir une unité.

a- Le mot-clé «*decreasing*» signifie que la diminution de la valeur de la dimension permet l'augmentation de la qualité du service.

b- Le mot-clé «*increasing*» signifie que l'augmentation de la valeur de la dimension permet l'augmentation de la qualité du service.

Ces mots-clés sont utilisés lors de la phase de vérification de conformité entre une qualité offerte et une qualité requise.

La Figure 2.27 présente une définition d'un contrat type appelé *Fiabilite*. Ce dernier est composé de deux dimensions (MTBF et MTTR).

```
Type Fiabilite = contrat
{     MTBF :  increasing numeric heure ;
      MTTR :  decreasing numeric  min ;
}
```

**Figure 2.27** : Description d'un contrat type en QML

**2- Contrat** : C'est une instance de contrat type qui représente une spécification particulière de QdS. Un contrat définit des contraintes sur les valeurs des dimensions de son contrat type. La Figure 2.28 présente une instance du contrat type *Fiabilite*.

```
BonneFiabilite =  Fiabilite contrat
{     MTBF >= 48 heure ;
      MTTR <  30 min ;
}
```

**Figure 2.28** : Description d'un contrat en QML

**3- Profil** : permet d'attacher un ensemble de contrats aux éléments d'une interface. QML sépare une qualité offerte d'une autre requise. Une qualité offerte est présentée par un contrat précédé du mot-clé *provide* tandis que la qualité requise est décrite par un contrat précédé du mot-clé *require*.

La Figure 2.29 montre que le service S1 de l'interface I1 exige de son environnement la qualité BonneFiabilité.

#### 2.6.1.4 CQML

CQML [Aagedal, 2001] est un langage lexical de spécification de QdS. Il reprend les concepts de QML et les étend pour les modèles à composants. La différence la plus notable à QML se présente dans l'utilisation d'OCL [Warmer, 2003] pour la spécification des invariants de caractéristiques de qualité et pour le calcul des valeurs de qualité.

```
FiabiliteI1 for I1 = profile
{
  From S1 require BonneFiabilite ;
}
```

**Figure 2.29 :** Description d'un *profile* en QML

Les caractéristiques les plus importantes du langage CQML peuvent être décrites par :

1- sa généralité,

2- sa compatibilité à UML,

3- son utilisation à différents niveaux d'abstraction,

4- sa précision dans la spécification des QdS suite à l'utilisation d'OCL,

5- son pouvoir de séparation entre l'aspect qualitatif et l'aspect fonctionnel.

CQML repose sur trois concepts clés : la caractéristique de qualité, la qualité et le profil. Les concepts de CQML sont très similaires à ceux de QML.

##### 2.6.1.4.1 Le concept caractéristique

La caractéristique de qualité (*quality_characteristic*) est la construction de base d'une spécification CQML. Cette caractéristique (dimension dans QML) représente un aspect non fonctionnel tel que performance, fiabilité, disponibilité, etc. Chaque caractéristique possède un nom et un domaine. Le domaine est constitué d'une direction (*increasing* ou *decreasing*), d'un ensemble de valeurs possibles (*numeric, set ou enum*) et peut avoir une unité.

Par ailleurs CQML permet de spécifier plus finement les caractéristiques que QML. En effet, les caractéristiques CQML peuvent être paramétrées. Les paramètres admis peuvent être des opérations, des classes ou des interfaces aux sens d'UML. En outre, une caractéristique CQML peut inclure dans sa définition un invariant exprimé à l'aide des prédicats OCL comme elle peut avoir aussi une clause «*Values*» exprimant la formule de calcul de la valeur de la qualité.

La Figure 2.30 issue de [Aagedal, 2001] montre une spécification en CQML de deux caractéristiques de qualité appelées respectivement TempsDeReponse et TauxDeTransfert. La caractéristique TempsDeReponse permet de mesurer le temps de réponse d'un composant. Cette caractéristique prend en paramètre un élément de type Flow [Aagedal, 2001]. La caractéristique TauxDeTransfert permet de mesurer le taux de transfert de données (ici images) d'un composant. Cette caractéristique prend en paramètre un élément de type Flow. La valeur de cette caractéristique est calculée à partir de la fonction OCL «eventsInRange» qui permet de calculer le nombre d'événements par seconde.

#### 2.6.1.4.2 Le concept qualité

Le concept qualité (*quality*) permet de spécifier une catégorie de qualité d'un service ou d'un ensemble de services. Toute qualité CQML est identifiée par un nom et un ensemble de sous-qualités et peut avoir des paramètres. Chaque sous-qualité est définie par une contrainte exprimée en OCL. Cette contrainte présente une restriction du domaine d'une caractéristique de qualité.

```
quality_characteristic TempsDeReponse (flow : Flow)
{
 domain : decreasing numeric milliseconds;
 values :  //Formule OCL
         if flow.SE->isEmpty then invalid
         else flow.SE->first.time() - flow.initiate.time()
         endif;
 invariant :   //Prédicat OCL
     flow.initiate =invalid implies flow.SE ->isEmpty ;
}
quality_characteristic TauxDeTransfert (flow : Flow)
{
 domain : increasing numeric Image/sec ;
 values :         //Formule OCL
           flow.SE -> eventsInRange (1000) ;
//avec  eventsInRange  est  une  opération  OCL  qu'on  doit  définir.  Elle
//permet de calculer le nombre d'événements //(ici image) par seconde
}
```

**Figure 2.30 :** Spécification des caractéristiques en CQML

La Figure 2.31 montre une spécification en CQML de deux qualités appelées respectivement Performant et TresPerformant. Ces deux qualités sont liées à la performance d'un service *S1* de type *Flow*. La qualité Performant spécifie que le taux de transfert du service *S1* est supérieur à 25 images par seconde et que son temps de réponse est inférieur ou égal à 20 msec. La qualité TresPerformant spécifie que le taux de transfert du service *S1* est supérieur à 30 images par seconde et que son temps de réponse est inférieur ou égal à 15 msec.

```
quality Performant (S1 : Flow)
{    TauxDeTransfert (S1) >= 25   ;
     TempsDeReponse  (S1) <= 20   ;
}
quality TresPerformant (S1 : Flow)
{
      TauxDeTransfert (S1) >= 30   ;
      TempsDeReponse  (S1) <= 15   ;
}
```

**Figure 2.31 :** Spécification des qualités en CQML

#### 2.6.1.4.3 Le concept *profil*

Après avoir défini les deux concepts qui permettent la spécification du QdS, nous allons présenter maintenant le concept du profil (*profile*). De la façon du QML, un *profil* permet d'attacher à chaque composant ses qualités qui peuvent être requises et/ou offertes. La spécification des profils CQML est moins fine que dans QML, CQML ne descend pas jusqu'au niveau de la méthode (ou service).

De la même façon, CQML sépare une qualité offerte d'une autre requise :

1- le mot-clé «*provides*» indique que toutes les qualités qui suivent ce mot sont de type qualité offerte.

2- le mot-clé «*uses*» indique que toutes les qualités qui suivent ce mot sont de type qualité requise.

La Figure 2.32 montre un composant UML2.0 qui propose deux interfaces (VideoStream et VideoPresented). L'interface VideoStream exige les trois services suivants : Lire, Avancer et Arreter. L'interface VideoPresented propose un seul service appelé Presenter.

La Figure 2.33 présente un profil CQML qui attache des qualités au composant VideoPlayer. Ce profil modélise que le composant VideoPlayer exige la qualité Performant sur le service Lire et la qualité TresPerformant sur le service Avancer. En outre, ce *profile* modélise que ce composant offre la qualité TresPerformant sur le service Presenter.

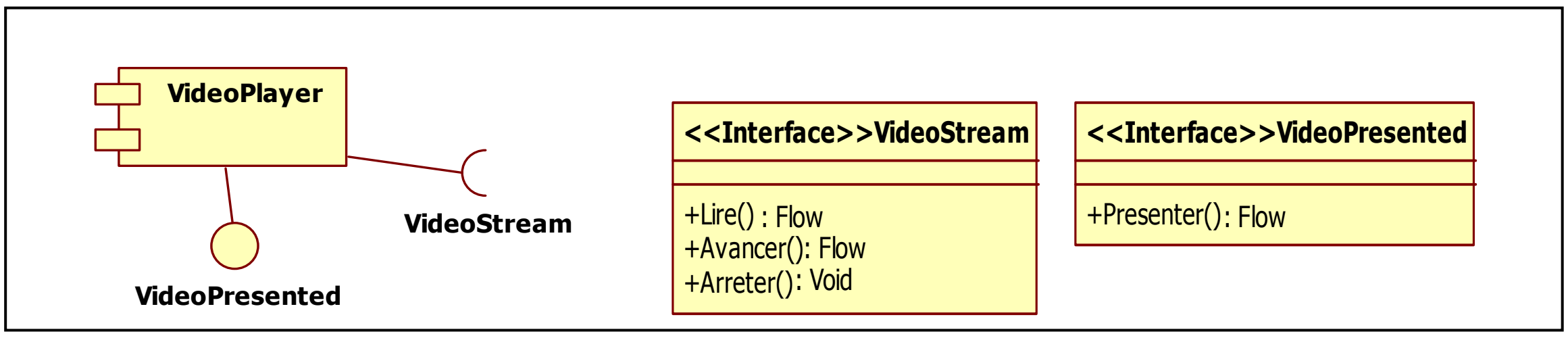

**Figure 2.32 :** Description en UML2 du composant VideoPlayer

## 2.6.2 Évaluation des langages et des méthodes de spécification des propriétés non-fonctionnelles

Bien que les méthodes formelles soient les plus précises dans la spécification de QdS et les plus performantes pour la comparaison des propriétés, leur utilisation reste limitée à certains domaines.

```
Profile P1 for VideoPlayer
 {
  Uses
    Performant (VideoStream.Lire) and
    TresPerformant (VideoStream.Avancer);
 Provides
    TresPerformant (VideoPresented.Presenter);
}
```

**Figure 2.33** : Spécification d'un *profil* de qualité en CQML

Dans [Aagedal, 2001], Jan Oyvind Aagedal propose un tableau d'évaluation des différents langages de spécification de QdS selon 25 critères dont on cite en particulier : la généralité, la séparation entre les spécifications fonctionnelles et les spécifications qualitatives, la composition des QdSs et la compatibilité à UML. Une autre constatation a été dégagée de cette étude : aucun de ces langages ne peut être considéré comme parfait, vis-à-vis de critères déjà mentionnés. Devant ces exigences, Jan Oyvind Aagedal a proposé le langage CQML dont il l'approuvait comme étant le langage le plus performant dans la description des QdS. Comme conséquence de cette étude, l'auteur a pu conclure que CQML est le langage le plus complet et le plus satisfaisant. Il est générique, sépare l'aspect qualitatif de l'aspect fonctionnel et intégrable à UML.

## 2.7 Conclusion

Dans ce chapitre nous avons présenté les différents formalismes qui seront utilisés tout au long de cette thèse, à savoir le modèle de composants UML2.0, le modèle de composants Ugatze, l'ADL Acme/Armani, l'ADL Wright et le langage CQML.

Dans le chapitre suivant, nous proposons deux démarches de vérification d'assemblage de composants: *VerifComponentUML2.0 et VerifComponentUgatze*. La démarche VerifComponentUML2.0 permet de vérifier la cohérence d'un assemblage de composants UML2.0 vis-à-vis des contrats syntaxiques, de QdS et synchronisation. Quant à la démarche VerifComponentUgatze, elle offre un cadre permettant la vérification de l'assemblage de composants vis-à-vis des contrats syntaxiques.

# **Chapitre 3** : Démarche de vérification d'assemblages de composants : cas d'UML2.0 et Ugatze

## 3.1 Introduction

Dans ce chapitre, nous proposons deux démarches : *VerifComponentUML2.0 et VerifComponentUgatze*. La démarche *VerifComponentUML2.0* permet de vérifier la cohérence d'un assemblage de composants UML2.0 vis-à-vis des contrats syntaxiques, de QdS et de synchronisation. Quant à la démarche *VerifComponentUgatze*, elle offre un cadre permettant la vérification de l'assemblage de composants Ugatze vis-à-vis des contrats syntaxiques. Ce chapitre comporte trois sections. La première section aborde la problématique du passage des notations semi-formelles vers des notations formelles. Nous justifions les notations formelles retenues pour les deux démarches *VerifComponentUML2.0* et *VerifComponentUgatze*. Les deux sections trois et quatre présentent respectivement les deux démarches *VerifComponentUML2.0* et *VerifComponentUgatze*.

## 3.2 D'une modélisation semi-formelle vers une modélisation formelle

D'une façon générale, les notations semi-formelles et formelles son complémentaires.
Les notations semi-formelles sont souvent à base de modèles graphiques et permettent une vue synthétique, structurante et intuitive du système modélisé [Dupuy, 2000], Ainsi, elles fournissent de bons vecteurs de communication tant entre concepteurs et utilisateurs qu'entre concepteurs et développeurs. En outre, elles entraînent un coût de formation peu élevé vis-à-vis des notations formelles. Mais ces notations manquent de sémantique précise. Ceci ne favorise pas la construction des outils permettant l'automatisation, même en partie, de la vérification des modèles semi-formels.
Les notations formelles sont basées sur des notations mathématiques. Elles apportent la précision et concision manquant aux modélisations semi-formelles. Ces notations formelles sont dotées des outils de vérification formelle tels que prouveurs et model-checkers.
Plusieurs travaux liés au couplage de notations semi-formelles et formelles sont proposés. Un tel couplage est souvent réalisé par une stratégie de traduction des modèles semi-formels en modèles formels. Ceci concerne aussi bien les modèles semi-formels orientés objets qu'à base de composants [Meyer, 1999], [Dupuy, 2000], [Graeme, 2000], [Laleau, 2002], [Ledang, 2001], [Ledang, 2002], [Marcano, 2002a], [Marcano, 2002b], [Rasch, 2003], [Idani, 2009] : les notations semi-formelles concernées sont à base d'UML (objet et composant). Tandis que les notations formelles ciblées sont Z, object-Z, B et CSP. Mais la traduction du semi-formel vers le formel se heurte aux problèmes suivants :

- difficulté d'obtenir une spécification formelle conservant le plus possible la structure de la spécification semi-formelle,
- difficulté d'obtenir une spécification formelle suffisamment claire pour être facilement lisible et exploitable.

Ces deux problèmes deviennent problématiques lorsque les écarts sémantiques entre les deux notations semi-formelle et formelle sont importants : par exemple entre le modèle de composants UML2.0 et la méthode orientée modèle comme B. Pour faire face à ces problèmes, nous préconisons une approche favorisant la continuité au niveau paradigme entre les deux notations semi-formelle et formelle. Ainsi, nous proposons deux stratégies de traduction du modèle de composants UML2.0 respectivement vers les modèles de composants formels Acme/Armani et Wright. De même, nous apportons une approche de traduction du modèle de composants semi-formel Ugatze vers Acme/Armani.

## 3.3 Vérification d'assemblages de composants UML2.0

Dans cette section, nous proposons une démarche appelée VerifComponentUML2.0 permettant de vérifier la cohérence d'un assemblage de composants UML2.0 vis-à-vis des contrats syntaxiques, de qualité de services et de synchronisation. Les aspects structuraux d'un assemblage de composants UML2.0 sont décrits à l'aide d'un diagramme de composants en se servant notamment des concepts composant primitif, interface offerte, interface requise et connecteur d'assemblage (cf. 2.2.1). Quant aux aspects non fonctionnels, ils sont décrits en utilisant le langage de modélisation CQML (cf. 2.6). Enfin, les aspects comportementaux sont décrits en utilisant une extension de PSM appelé PoSM (Port State Machine) (cf. 2.2.2.2). La démarche VerifComponentUML2.0 comporte trois étapes permettant respectivement de décrire et de vérifier les contrats syntaxiques, QdS et de synchronisation. La vérification des contrats syntaxiques et QdS est confiée à l'évaluateur de prédicats Armani supporté par la plate-forme AcmeStudio [ABLE, 2009]. Tandis que la vérification des contrats de synchronisation est confiée au model checker FDR [FDR2, 2003].

### 3.3.1 Etape1 : Vérification des contrats syntaxiques

L'enchaînement des opérations permettant à terme la vérification des contrats syntaxiques d'un assemblage de composants UML2.0 est fourni par la Figure 3.1

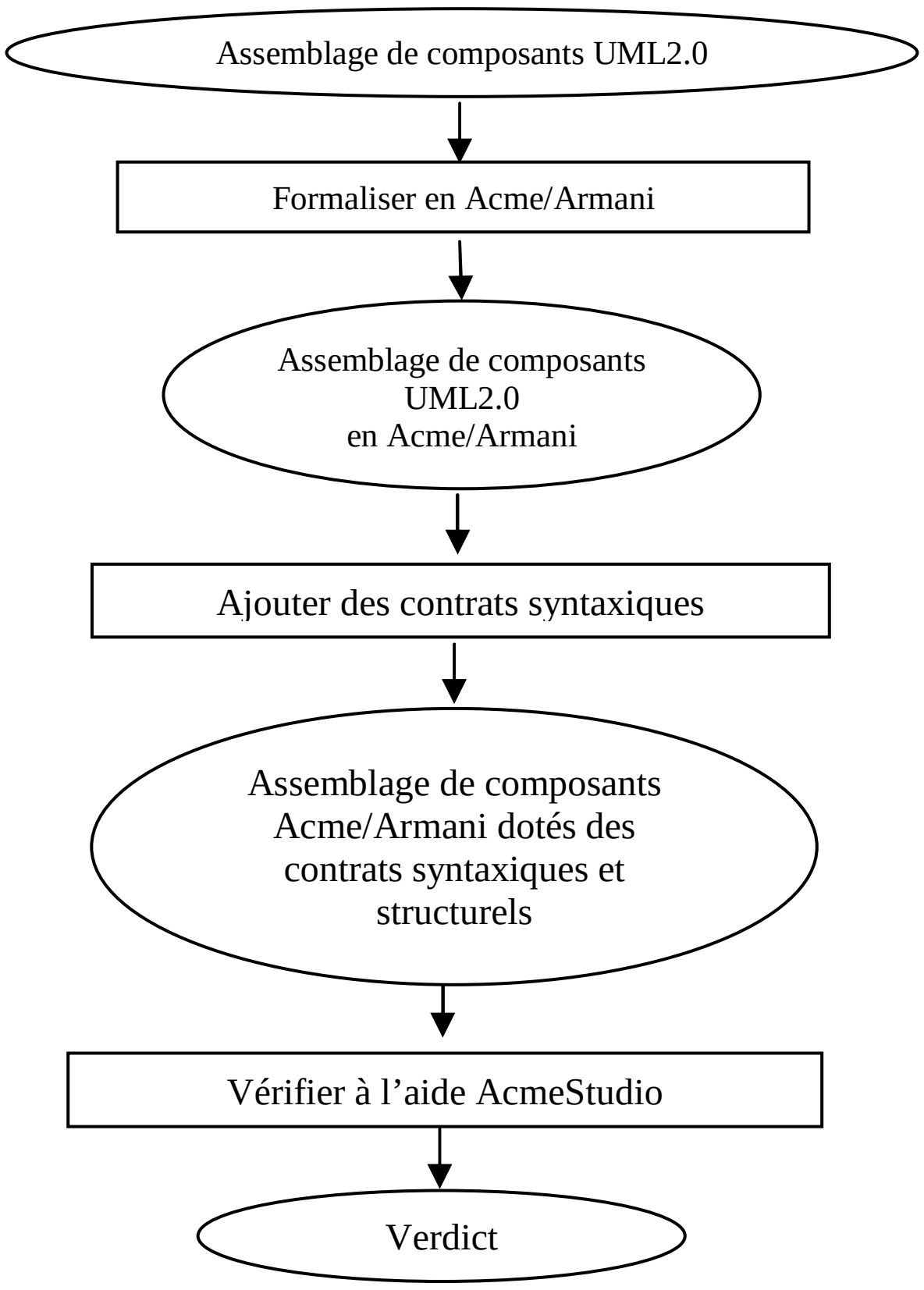

**Figure 3.1 :** Etape 1 de la démarche VerifComponentUML2.0 : Vérification des contrats syntaxiques

Dans un premier temps, nous proposons une formalisation en Acme/Armani des principaux concepts issus du modèle de composants UML2.0 tels que composant, interface offerte, interface requise, signature d'une méthode et connecteur d'assemblage. Dans un deuxième temps, nous ajoutons au produit obtenu des contrats d'assemblage décrits sous forme des propriétés invariantes permettant de vérifier la compatibilité entre une opération offerte et une opération requise, et les règles de composition des composants UML2.0. Enfin, nous utilisons la plateforme AcmeStudio afin de vérifier l'assemblage de composants en Acme/Armani dotés des contrats syntaxiques obtenus. La localisation des erreurs est facilitée par les possibilités de traitement des erreurs offertes par AcmeStudio. Le chapitre 4 présente d'une façon détaillée notre processus de vérification des contrats syntaxiques et structurels d'un assemblage de composants UML2.0.

### 3.3.2 Etape 2 : Vérification des contrats de QdS

Le processus permettant à terme de vérifier les contrats de QdS d'un assemblage de composants UML2.0 dotés des PNF spécifiées en CQML est fourni par la Figure 3.2. Dans un premier temps, nous formalisons en Acme/Armani les principaux concepts issus du langage de modélisation des PNF CQML. Ceci a donné naissance à un style Acme/Armani qui regroupe les constructions de base CQML telles que qualité caractéristique, qualité et profil. Dans un deuxième temps, nous combinons le style obtenu précédemment et l'assemblage de composants UML2.0/CQML en réutilisant la formalisation des aspects structuraux d'un assemblage de composants UML2.0 fournie lors de l'étape 1 de la démarche VerifComponentUML2.0. Ainsi, nous obtenons un assemblage de composants Acme/Armani traduisant un assemblage de composants UML2.0/CQML.

Dans un troisième temps, nous ajoutons les contrats syntaxiques –établis lors de l'étape 1- et les contrats de QdS élaborés lors de cette étape. Enfin, nous vérifions ces contrats en utilisant la plate-forme AcmeStudio. Le chapitre 4 décrit cette étape 2 de la démarche VerifComponentUML2.0.

### 3.3.3 Etape 3 : Vérification des contrats de synchronisation

Le processus de vérification des contrats de synchronisation d'un assemblage de composants UML2.0/PoSM est fourni par la Figure 3.3. Sachant que le langage PoSM [Samek, 2005] est une extension du langage PSM d'UML2.0. Dans un premier temps, nous proposons une traduction d'un assemblage de composants UML2.0/PoSM vers Wright. Ensuite, l'assemblage de composants Wright obtenu est traduit en CSP de Hoare grâce à l'outil Wr2fdr maintenu et amélioré par nous même (cf. chapitre 6). L'outil Wr2fdr génère des contrats sous forme d'assertions visant la compatibilité des connexions port et rôle, la cohérence d'un composant Wright et d'un connecteur Wright. De tels contrats sont à vérifier par le model-checker FDR. Egalement, nous pouvons traduire l'assemblage de composants Wright obtenu vers un programme concurrent Ada en utilisant notre outil IDM Wright2Ada (cf. chapitre 7). Ceci autorise l'utilisation des outils d'analyse statique et dynamique liés à Ada.

## 3.4 Vérification d'assemblages de composants Ugatze

Au cours de ces dernières années, notre équipe de recherche a développé et expérimenté (dans le cadre du projet européen ASIMIL [ASIMIL, 2002]) un méta-modèle de composants dénommé Ugatze, adapté à la réutilisation de composants logiciels autonomes, hétérogènes et distribués [Aniorté, 2004], [Seyler, 2004]. Les composants logiciels visés par le modèle Ugatze ne sont pas forcément conçus pour être réutilisés a priori : c'est la réutilisation a posteriori.

Afin d'enrichir le (méta)modèle de composants Ugatze, nous proposons une démarche appelée *VerifComponentUgatze* permettant, de vérifier la cohérence d'un assemblage de composants Ugatze vis-à-vis des contrats applicatifs : contrats syntaxiques, contrats sémantiques, contrats de synchronisation et contrats de QdS. Actuellement, *VerifComponentUgatze* comporte une seule étape liée à la vérification des contrats syntaxiques et structurels.

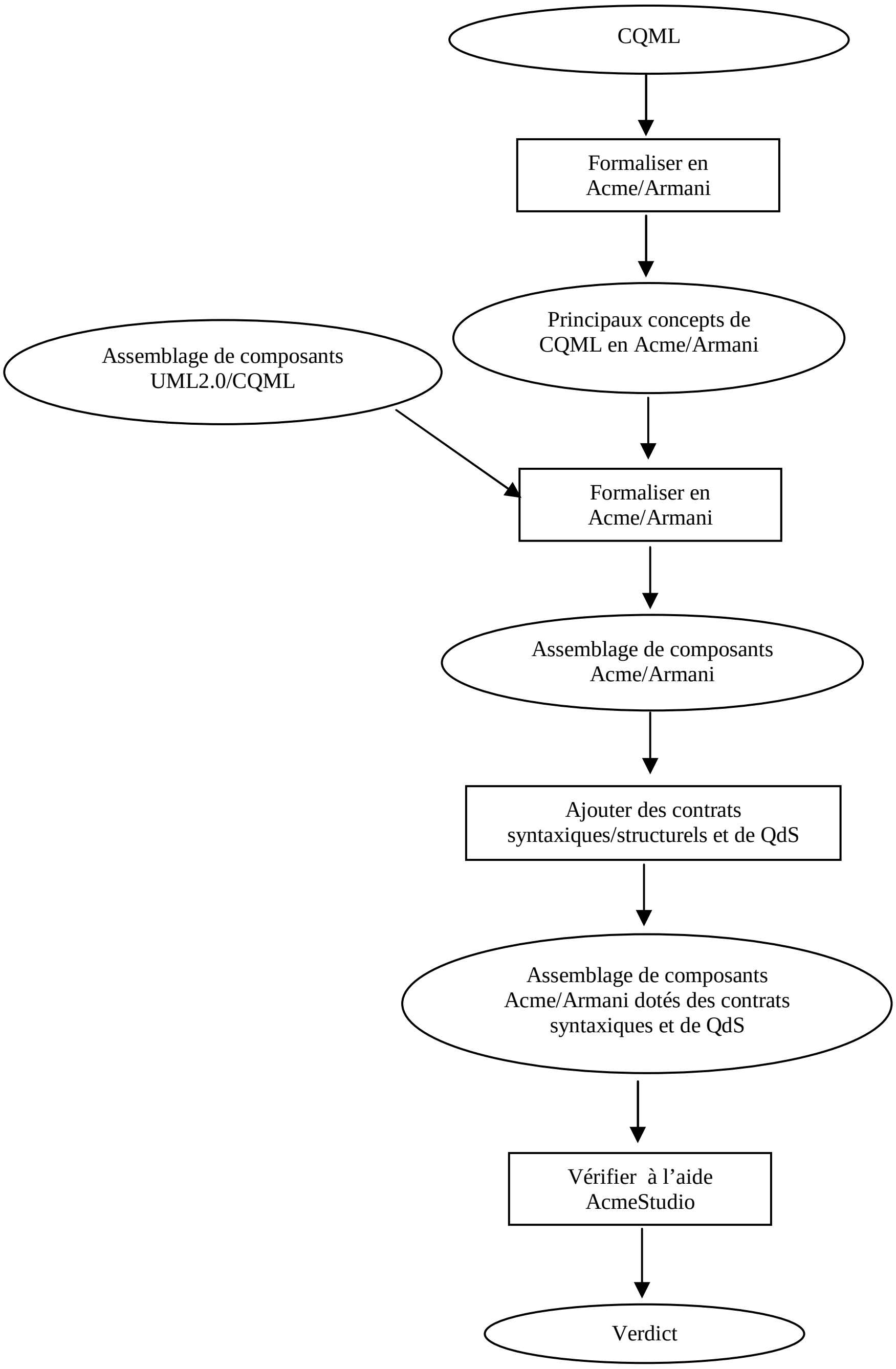

**Figure 3.2 :** Etape 2 de la démarche VerifComponentUML2.0 : vérification des contrats de QdS

Le processus permettant de vérifier les aspects syntaxiques et structurels d'un assemblage de composants Ugatze est fourni dans la Figure 3.4. Dans un premier temps, nous formalisons en Acme/Armani les aspects structuraux d'Ugatze tels que : composant, interaction, points d'interaction, etc. Dans un deuxième temps, nous ajoutons au produit obtenu des contrats syntaxiques et structurels décrits comme des propriétés invariantes en

Armani. Enfin, nous vérifions ces contrats à l'aide d'AcmeStudio. Le chapitre 8 est consacré à la démarche *VerifComponentUgatze*.

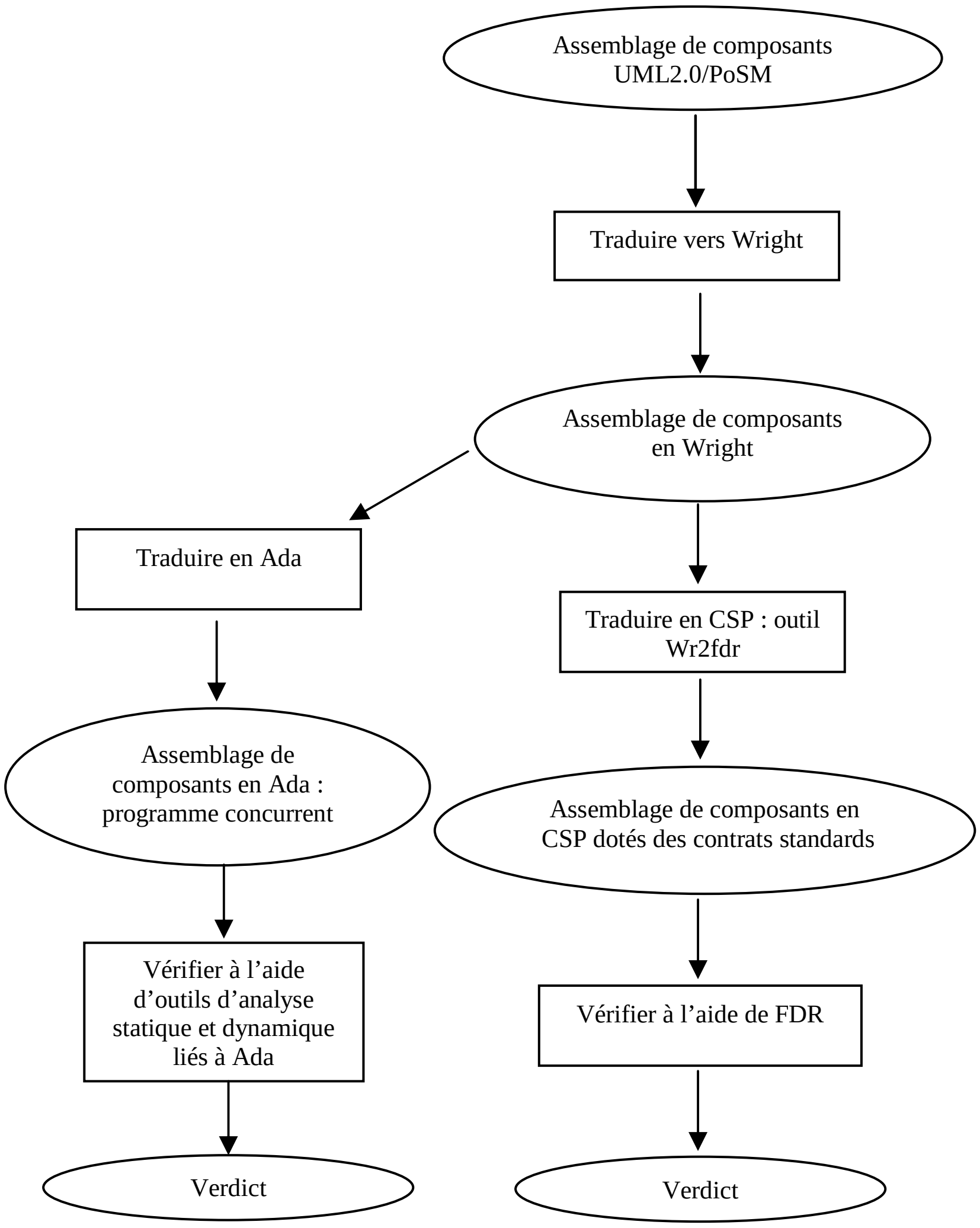

**Figure 3.3 :** Etape 3 de la démarche VerifComponentUML2.0 et ouverture sur les outils d'analyse statique et dynamique relatifs à Ada

## 3.5 Conclusion

Dans ce chapitre, nous avons proposé deux démarches *VerifComponentUML2.0 et VerifComponentUgatze* permettant de vérifier respectivement la cohérence d'assemblages de compossants UML2.0 et Ugatze. Les deux démarches s'appuient sur un socle des modèles de composants formels : Acme/Armani et Wright. Ceci favorise la continuité entre les modèles de composants semi-formels (UML2.0 et Ugatze) et ces deux modèles formels. En outre, la démarche *VerifComponentUML2.0* propose deux outils : Wr2fdr et Wright2Ada. L'outil Wr2fdr permet de traduire un assemblage de composants Wright vers une spécification CSP de Hoare acceptable par le model-checker FDR. L'outil Wright2Ada est un outil IDM permettant de transformer une spécification Wright vers un programme concurrent en Ada. Ces deux outils permettent l'analyse statique et dynamique d'un assemblage de composants décrits initialement par UML2.0

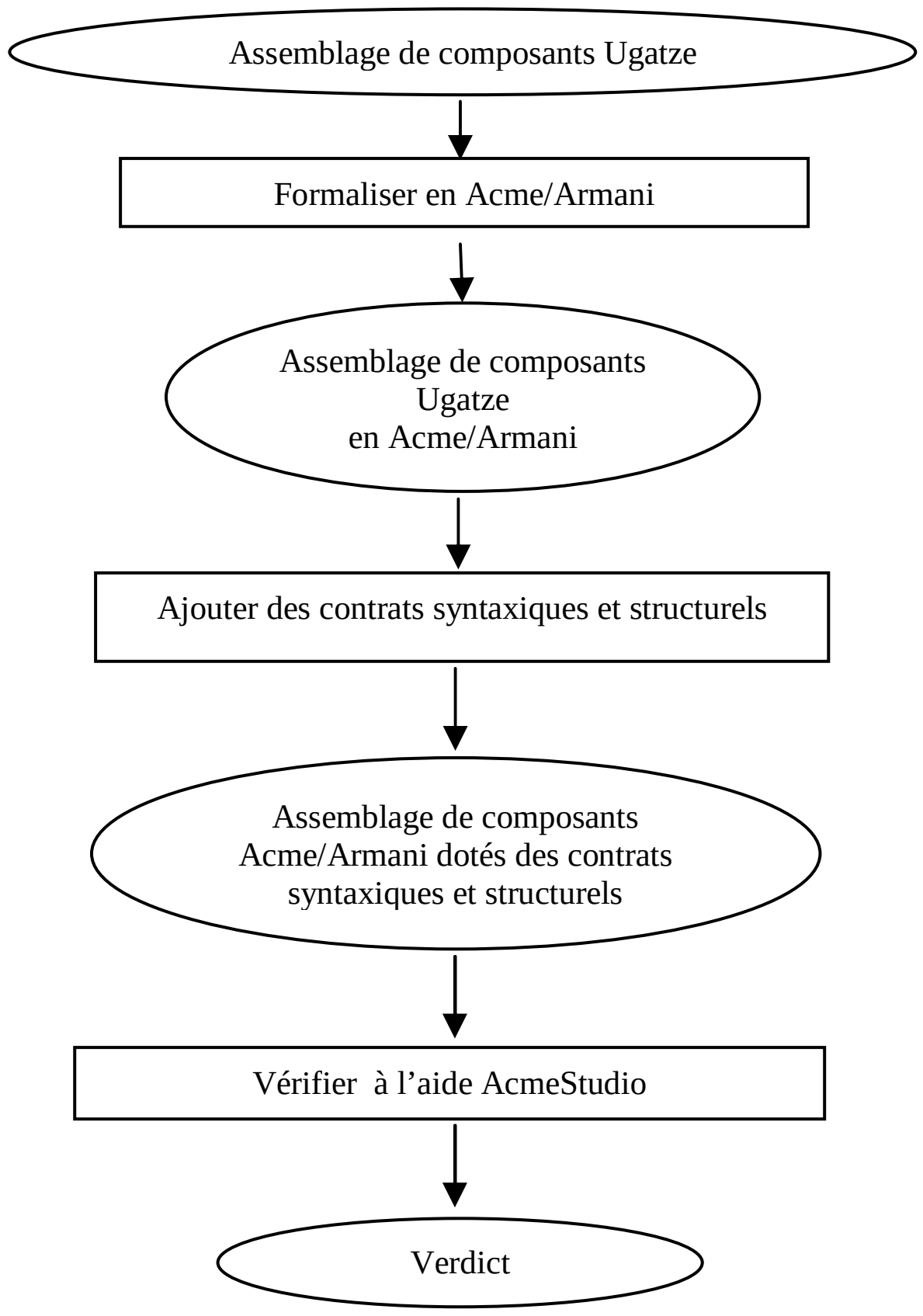

**Figure 3.4 :** Vérification des aspects syntaxiques et structurels d'un assemblage de composants Ugatze

Dans la suite de ce mémoire, nous allons présenter d'une façon approfondie respectivement les constituants des deux démarches *VerifComponentUML2.0 et Verif ComponentUgatze.*

# Chapitre 4 : Vérification des contrats syntaxiques d'assemblages de composants UML2.0

## 4.1 Introduction

Dans ce chapitre, nous proposons une traduction du modèle de composants UML2.0 en Acme afin de vérifier la cohérence d'assemblages de composants UML2.0 [Kmimech, 2009a], [Kmimech, 2009d], [Kmimech, 2009e]. La vérification des contrats syntaxiques et structurels est confiée à l'évaluateur des prédicats supporté par la plateforme AcmeStudio [ABLE, 2009]. Pour y parvenir, nous décrivons les principaux concepts issus du modèle de composants UML2.0 (niveau M2) en utilisant le concept style d'architectures d'Acme. Un assemblage de composants UML2.0 (niveau M1) est décrit à l'aide de la notion *system* d'Acme. Le niveau M1 est dit conforme au niveau M2 s'il vérifie les règles de cohérence décrites au niveau M2 en plus des règles spécifiques décrites au niveau M1.

Ce chapitre comporte deux sections. La première section est consacrée à la formalisation des concepts relatifs au modèle de composants UML2.0 en Acme/Armani ainsi qu'une expérimentation de cette formalisation. La deuxième section présente une formalisation en Acme d'un assemblage de composants UML2.0 d'une application « Réservation de chambres d'hôtels ». La particularité de cette application est qu'elle exige des types de données définis par le concepteur.

## 4.2 Le méta-modèle de composants UML2.0 en Acme

### 4.2.1 Motivation

UML2.0 [OMG, 2005a] propose un modèle de composants englobant des concepts tels que : composant, port, structure composite, connecteur, interface offerte, interface requise et protocol state machine. Ainsi UML2.0 favorise le développement basé sur le paradigme composant (cf. chapitre 2). Des règles de cohérence liées à la bonne façon de constituer un assemblage de composants UML2.0 sont décrites et formalisées en OCL 2.0 [OMG, 2005b] au niveau du méta-modèle UML2.0. Mais ces règles sont loin d'être complètes. En effet, le travail décrit dans [Malgouyres, 2005] propose des nouvelles règles de cohérence liées à la bonne utilisation du diagramme de structures composites UML2.0. Nous avons retenu uniquement les règles relatives à la vérification de la cohérence structurelle. Ces règles sont :

- Tout composant a au moins une interface fournie,
- L'ensemble des services fournis par le port ou l'interface doit être un sur-ensemble de l'ensemble des services requis,
- Les chemins graphiques qui peuvent apparaître dans un diagramme de composants sont :
    - des connecteurs d'assemblage,
    - des connecteurs de délégation.
- Les nœuds contenus dans un diagramme de composants ne peuvent être que :
    - des composants,
    - des interfaces,
    - des ports.

Comparé à l'ADL Acme/Armani [Garlan, 2000], [Monroe, 2001], [Garlan, 2006], UML2.0/OCL 2.0 manque d'expressivité pour représenter un certain nombre de contraintes liées à la compatibilité d'un assemblage de composants. Par exemple, Acme/Armani a été utilisé avec succès afin de faciliter la détection des incompatibilités des architectures orientées web services [Gacek, 2008]. Le choix d'Acme se justifie aussi par les concepts fournis par celui-ci. En effet, l'ADL Acme [Garlan, 2000] offre des concepts architecturaux structuraux (cf. chapitre 2) tels que *component*, *connector*, *role*, *port*, *representation*, *system* et *family*. En outre, il fournit un langage de prédicats assez puissant appelé Armani [Monroe, 2001] avec des fonctions appropriées au domaine d'architectures logicielles. Le langage Armani permet de décrire des propriétés architecturales sous forme d'un invariant ou heuristique attachées à n'importe quel élément architectural (*component, family, system, connector,…*). De telles propriétés sont exécutables au sein de l'environnement AcmeStudio [ABLE, 2009]. De même, l'ADL Acme supporte la notion de type. On peut définir des types d'éléments architecturaux (composant, connecteur, rôle, port et style). Le concept *property* d'Acme utilisable au niveau type et instance permet d'attacher des propriétés non fonctionnelles aux éléments architecturaux. Enfin, Acme fournit des types de base (*int, float, boolean et string*) et des constructeurs de types (*enum, record, set et sequence*).

Nous proposons par la suite de décrire le méta-modèle de composants UML2.0 (niveau M2) en utilisant le concept de style d'architectures d'Acme. Un assemblage de composants UML2.0 (niveau M1) est décrit à l'aide de la notion *system* d'Acme. Le niveau M1 est dit conforme au niveau M2 s'il vérifie les règles de cohérence décrites au niveau M2 en plus des règles spécifiques décrites au niveau M1.

### 4.2.2 Formalisation du méta-modèle de composants UML2.0

Nous avons modélisé le méta-modèle de composants UML2.0 par un style Acme appelé *CUML* [Kmimech, 2009a], [Kmimech, 2009e] en utilisant la construction *family* (niveau M2). Le style *CUML* (cf. Figure 4.1) modélise les principaux concepts relatifs au modèle de composants UML2.0 à savoir signature d'une opération, interface offerte, interface requise, composant et connecteur d'assemblage. Pour y parvenir, nous avons utilisé avec profit les possibilités de typage offertes par Acme : *property type*, *port type*, *component type*, *connector type*, *role type*, *set*, *record*, *sequence*.

Par la suite nous détaillons la formalisation en Acme/Armani des différents concepts relatifs au modèle de composants UML2.0.

### 4.2.3 Formalisation d'une opération UML2.0

Les aspects syntaxiques d'une opération UML2.0 figurant au sein d'une interface UML2.0 regroupent : son nom, ses paramètres formels typés et sa nature (procédure ou fonction). Ces aspects sont formalisés en Acme/Armani en utilisant judicieusement les possibilités de typage offertes par Acme/Armani (voir Figure 4.2). Les types Acme/Armani proposés sont :

- « *UML_type_base*» : modélise les types de base offerts par UML2.0 à savoir Boolean, Real, Integer, String et Void,
- « *nature_logique* » : modélise la nature logique des paramètres formels d'une opération UML2.0 (in, out et in/out),
- « *parametre* » : regroupe au sein d'un enregistrement (constructeur de type Record) les deux caractéristiques d'un paramètre formel à savoir son type et sa nature logique,
- « *pls_parametre* » : regroupe au sein d'une séquence (constructeur de type Sequence), tous les paramètres formels,
- « *signature* » : regroupe au sein d'un enregistrement les éléments formant la signature d'une opération UML2.0 : nom de l'opération, paramètres formels et nature de l'opération (procédure ou fonction),
- Enfin, le type « *sous-programme* » : regroupe au sein d'un ensemble (constructeur de type Set) les signatures des opérations d'une interface UML2.0.

```
Family CUML=
{  //types de données pour définir la signature d'une interface UML
   Property Type UML_type_base=enum{Boolean_UML, Real_UML, Integer_UML,
String_UML, Void_UML};
   Property Type nature_logique=enum{in_UML, out_UML, inout_UML};
   Property Type
parametre=Record[type_parametre:UML_type_base;mode:nature_logique;];
   Property Type pls_parametre=Sequence<parametre>;
   Property Type
signature=Record[nom_sp:String;p:pls_parametre;resultat:UML_type_base;];
   Property Type sous_programme=Set{signature};

   //interface offerte modélisée par un type de port Acme
   Port Type InterfaceOfferte=
   {     Property services_offerts:sous_programme;
   }
   //Idem pour interface requise
   Port Type InterfaceRequise=
   {      Property services_requis:sous_programme;
   }
//Un type de composant UML est modélisé par un type de composant Acme
   Component Type ComposantUML=
   { //Un composant UML est doté d'une interface au moins
      rule aumoinsInterface=invariant size(self.PORTS)>=1;
//Si un composant UML est doté d'une seule interface alors celle-ci doit être
//une interface offerte
      rule uneseuleInterfaceOfferte=invariant size(self.PORTS)==1 ->
```

**Figure 4.1 :** Méta-modèle de composants UML2.0 formalisé en Acme/Armani

```
Property Type UML_type_base=enum{Boolean_UML, Real_UML, Integer_UML, String_UML,
Void_UML};
   Property Type nature_logique=enum{in_UML, out_UML, inout_UML};
   Property Type parametre=Record[type_parametre:UML_type_base;mode:nature_logique;];
   Property Type pls_parametre=Sequence<parametre>;
   Property Type
    signature=Record[nom_sp:String;p:pls_parametre;resultat:UML_type_base;];
   Property Type sous_programme=Set{signature};
```

**Figure 4.2 :** Formalisation d'une opération en Acme/Armani

### 4.2.4 Formalisation d'un composant UML2.0

Un composant UML2.0 est formalisé par un type de composant Acme/Armani (cf. Figure 4.3). Plusieurs règles de cohérence exprimées par des invariants Acme/Armani sont proposées :

- « *aumoinsInterface* » : cette règle stipule qu'un composant UML2.0 possède au moins une interface,
- « *uneseuleInterfaceOfferte* » : cette règle stipule que si un composant UML2.0 est doté d'une seule interface alors celle-ci doit être une interface offerte,
- « *interfaceRequiseOfferte* » : cette règle stipule qu'un composant UML2.0 est doté soit des interfaces offertes soit des interfaces requises.

```
 Component Type ComposantUML=
 {
    rule aumoinsInterface=invariant size(self.PORTS)>=1;
    rule uneseuleInterfaceOfferte=invariant size(self.PORTS)==1 ->
         forall p:Port in self.PORTS|declaresType(p,InterfaceOfferte);

    rule interfaceRequiseOfferte=invariant forall p:Port in
           self.PORTS|declaresType(p,InterfaceOfferte) OR
                  declaresType(p,InterfaceRequise);
}
```

**Figure 4.3 :** Formalisation d'un composant UML2.0 en Acme/Armani

### 4.2.5 Formalisation d'un connecteur d'assemblage UML2.0

Un connecteur d'assemblage est formalisé par un connecteur Acme/Armani (cf. Figure 4.4). Plusieurs règles de cohérence structurelle sont proposées :

- « `un_port_offert` » : cette règle stipule qu'un connecteur d'assemblage est rattaché à l'une de ses extrémités par un seul port,
- « `interface_offerte`» : cette règle stipule qu'un connecteur d'assemblage est rattaché à l'une de ses extrémités par un port de type *InterfaceOfferte*,
- « `un_port_requis` » : cette règle stipule qu'un connecteur d'assemblage est rattaché à l'une de ses extrémités par un seul port,
- « `interface_requise`» : cette règle stipule qu'un connecteur d'assemblage est rattaché à l'une de ses extrémités par un port de type *InterfaceRequise*,
- « `binaire`» : cette règle stipule qu'un connecteur d'assemblage est binaire, c'est-à-dire rattaché à exactement deux interfaces. Les deux règles définies précédemment permettent de garantir que les deux interfaces sont de types différents (une offerte et une requise).

```
   Connector Type AssemblageUML=
   {      Role serveur=
      {
          rule un_port_offert=invariant size(self.AttachedPorts)==1;
          rule interface_offerte=invariant forall p:Port in
self.AttachedPorts|declaresType(p,InterfaceOfferte);
      }
      Role client=
      {
       rule un_port_requis=invariant size(self.AttachedPorts)==1;
       rule interface_requise=invariant forall p:Port in self.AttachedPorts|
           declaresType(p,InterfaceRequise);
       }
      rule binaire=invariant size(self.ROLES)==2;
```

**Figure 4.4 :** Formalisation d'un connecteur d'assemblage UML2.0 en Acme/Armani

### 4.2.6 Formalisation d'une interface

Une interface (offerte ou requise) est formalisée par un port Acme/Armani. La Figure 4.5 illustre la formalisation des différents types d'interfaces.

```
  Port Type InterfaceOfferte=
{     Property services_offert:sous_programme;
}

  Port Type InterfaceRequise=
{      Property services_requis:sous_programme;
}
```

**Figure 4.5 :** Formalisation des interfaces UML2.0 en Acme/Armani

La construction *InterfaceOfferte* (resp. *InterfaceRequises*) comporte une propriété appelée *services_offert* (resp. *services_requis*) de type *sous_programme* (cf. 4.2.3). Celle-ci formalise les opérations UML2.0 offertes (resp. requises) par cette interface.

### 4.2.7 Formalisation des règles de cohérence d'un assemblage

Plusieurs règles de cohérence relatives à un assemblage de composants UML2.0 sont modélisées par des propriétés invariantes. Ces règles sont définies au niveau M2 (niveau style ou familly). De telles règles de cohérence permettent de vérifier des propriétés structurelles génériques telles que :

- « *composants_admis* » : cette règle stipule que seuls les composants de type *ComponentUML* sont admis dans un assemblage de composants UML2.0,
- « *connecteurs_admis* » : cette règle stipule que seuls les connecteurs d'assemblage de type *AssemblageUML* sont admis dans un assemblage de composants UML2.0,
- « *appelant_appele* » : cette règle stipule que l'appelant et l'appelé doivent être différents dans un assemblage de composants. C'est-à-dire que chaque connecteur d'assemblage est binaire d'une part et d'autre part les attachements se font entre une interface requise et interface offerte,
- « *interface_requise_satisfaite* » : cette règle stipule que chaque interface requise doit être satisfaite.

La Figure 4.6 illustre la formalisation en Acme/Armani de ces différentes contraintes.

```
rule composants_admis=invariant forall c:Component in
self.COMPONENTS|declaresType(c,ComposantUML) ;

rule connecteurs_admis=invariant forall con:Connector in
self.CONNECTORS|declaresType(con,AssemblageUML);

rule appelant_appele=invariant forall c:Component in self.COMPONENTS|forall
p1:Port in c.PORTS|forall p2:Port in c.PORTS| declaresType(p1,InterfaceOfferte)
and declaresType(p2,InterfaceRequise) -> (forall con:Connector in
self.CONNECTORS|forall r1:Role in con.ROLES|forall r2:Role in
con.ROLES|!(attached(p1,r1)and attached(p2,r2)) );

rule interface_requise_satisfaite=invariant forall c:Component in
self.COMPONENTS|forall p:Port in c.PORTS|declaresType(p,InterfaceRequise) ->
(exists con:Connector in self.CONNECTORS|exists r:Role in
con.ROLES|attached(p,r));
```

**Figure 4.6 :** Formalisation des règles de cohérence relatives à un assemblage de composants UML2.0 en Acme/Armani

### 4.2.8 Vérification d'un assemblage de composants UML2.0 en Acme/Armani

Un assemblage de composants UML2.0 est modélisé par une configuration Acme qui dérive du style *CUML*. Ceci permet, à terme, de vérifier les règles de cohérence appartenant au style *CUML* sur l'assemblage de composants proposé par l'environnement AcmeStudio [ABLE, 2009]. Une règle de cohérence violée (un invariant évalué à faux) traduit forcément une incohérence dans l'assemblage de composants traité.

Par la suite, nous allons tester notre style *CUML* sur deux modélisations (une invalide et une autre valide) d'un système bancaire simple contenant un client et un serveur appelé GAB (Guichet Automatique Bancaire).

#### 4.2.8.1 Assemblage de composants valide

##### 4.2.8.1.1 Modélisation en UML2.0 du système *GAB1*

La Figure 4.7 montre une description architecturale d'un système simplifié d'un GAB modélisé par un assemblage de composants UML2.0. Ce système est composé de deux composants dont le premier appelé *Serveur* est doté d'une interface offerte regroupant trois services (*crediter*, *debiter* et *solde*) alors que le second appelé *Client* est doté d'une interface requise qui exige deux services (*crediter* et *solde*).

##### 4.2.8.1.2 Formalisation en Acme/Armani du système *GAB1*

La Figure 4.8 donne la traduction de la description UML2.0 (Figure 4.7) du système *GAB1* sous forme d'un système Acme/Armani en passant par le style *CUML*. Les deux propriétés *service_offert* et *service_requis* attachées respectivement au port *I1* de *Serveur* et au port *I2* de *Client* utilisent avec profit le type *sous_programme* venant du style *CUML*. Ces deux propriétés mémorisent respectivement la signature de l'interface *I1* et *I2*. La règle *service_offert_requis* traduit la formalisation Acme de la compatibilité d'une interface offerte vis-à-vis de l'interface requise.

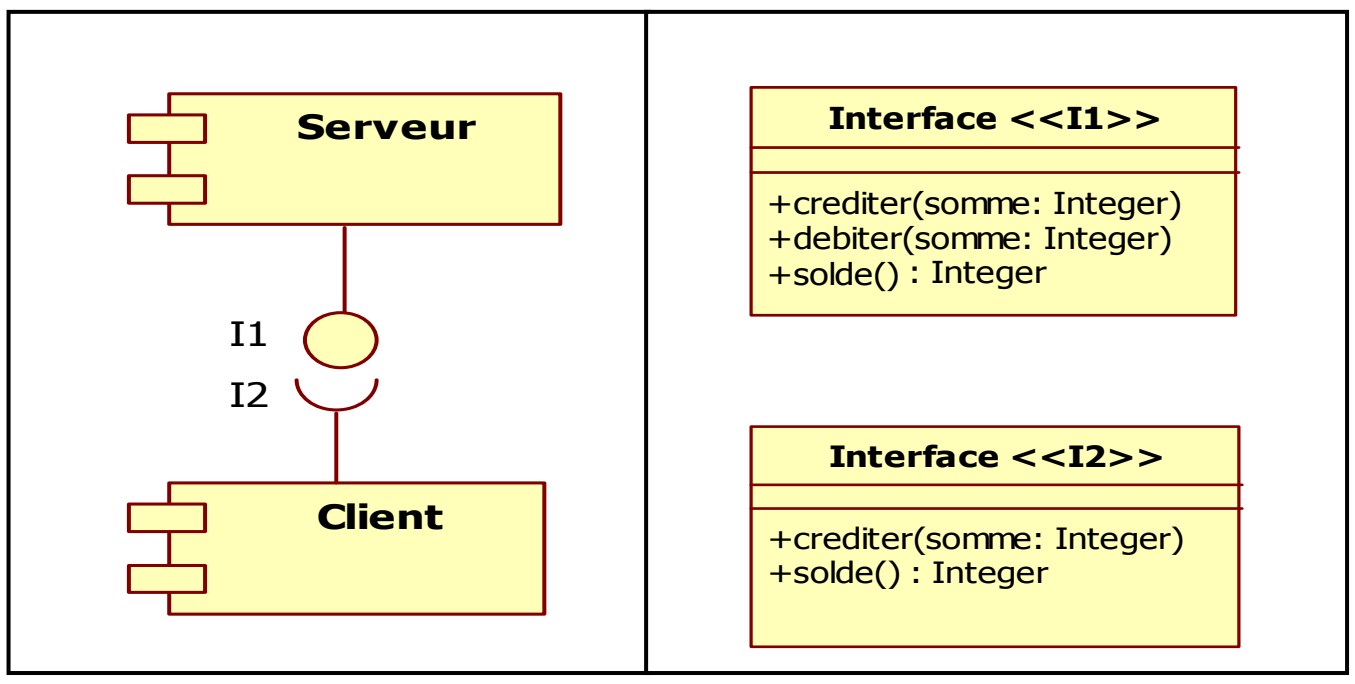

**Figure 4.7 :** Description Architecturale du système *GAB1* en UML2.0

```
import families/CUML.acme;

System GAB1 : CUML = new CUML extended with {

    Component Client : ComposantUML = new ComposantUML extended with {
        Port Ifictif : InterfaceOfferte = {
        }
        Port I2 : InterfaceRequise = new InterfaceRequise extended with {

            Property service_requis : sous_programme = {[nom_sp = "crediter";p =
<[type_parametre = Integer_UML;mode = in_UML;]>;resultat = Void_UML;],[nom_sp =
"solde";p = <>;resultat = Integer_UML;]};
        }
    }
    Component Serveur : ComposantUML = new ComposantUML extended with {
        Port I1 : InterfaceOfferte = new InterfaceOfferte extended with {

            Property service_offert : sous_programme = {[nom_sp = "crediter";p =
<[type_parametre = Integer_UML;mode = in_UML;]>;resultat = Void_UML;],[nom_sp =
"debiter";p = <[type_parametre = Integer_UML;mode = in_UML;]>;resultat =
Void_UML;],[nom_sp = "solde";p = <>;resultat = Integer_UML;]};
        }
    }
    Connector assemblage : AssemblageUML = new AssemblageUML extended with {

    }
    Attachment Client.I2 to assemblage.cli;
    Attachment Serveur.I1 to assemblage.serv;
    rule service_offert_requis = invariant isSubset(self.Client.I2.service_requis,
self.Serveur.I1.service_offert);
}
```

**Figure 4.8 :** Formalisation en Acme de l'exemple de la Figure 4.7

### 4.2.8.2 Assemblage de composants invalide

#### 4.2.8.2.1 Modélisation en UML2.0 du système *GAB2*

La Figure 4.9 montre une description architecturale d'un système simplifié d'un GAB modélisé par un assemblage de composants UML2.0. Ce système est composé de deux composants dont le premier appelé *Serveur* est doté d'une interface offerte regroupant trois services (*crediter*, *debiter* et *solde*) alors que le second *Client* est doté d'une interface requise qui exige trois services (*crediter*, *debiter* et *transferer*).

#### 4.2.8.2.2 Formalisation en Acme/Armani du système *GAB2*

La Figure 4.10 donne la traduction de la description UML2.0 (Figure 4.9) du système *GAB2* sous forme d'un système Acme/Armani en passant par le style *CUML*. Les deux propriétés *service_offert* et *service_requis* attachées respectivement au port *I2* de *Client* et au port *I1* de Serveur utilisent avec profit le type *sous_programme* venant du style *CUML*. Ces deux propriétés mémorisent respectivement la signature de l'interface *I1* et *I2*. La règle *service_offert_requis* traduit la formalisation Acme de la compatibilité d'une interface offerte vis-à-vis de l'interface requise.

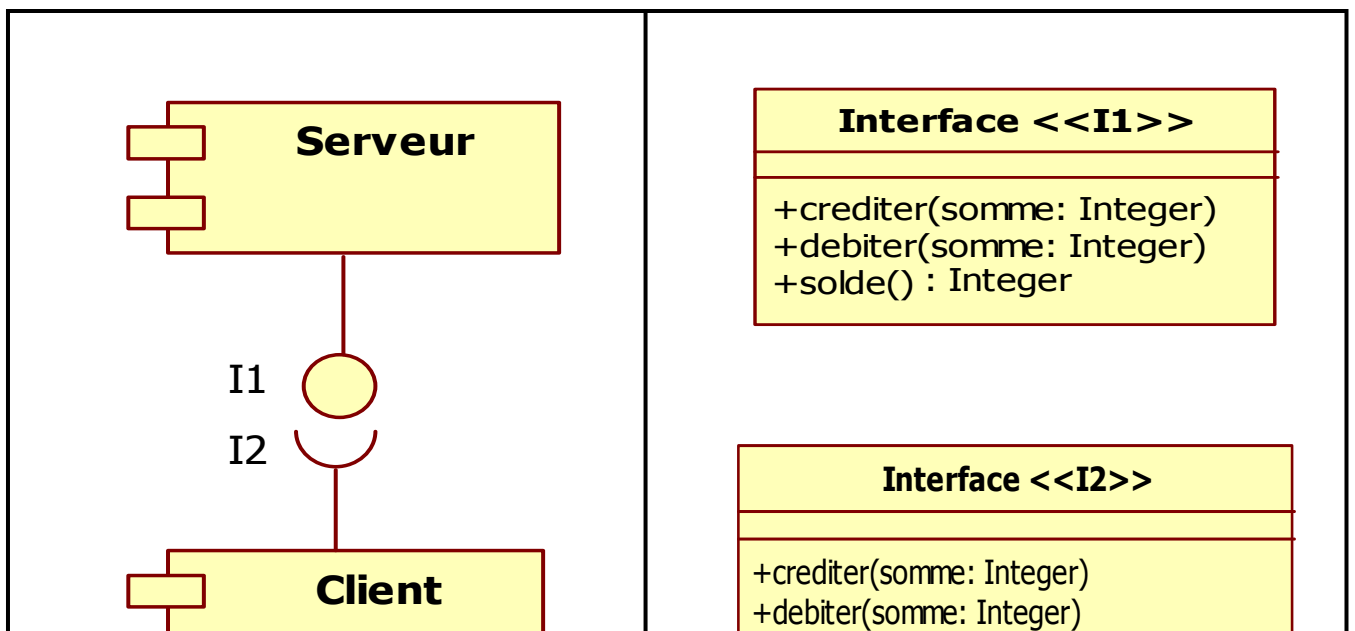

**Figure 4.9 :** Description Architecturale du système *GAB2* en UML2.0

```
import families/CUML.acme;

System GAB2 : CUML = new CUML extended with {

    Component Client : ComposantUML = new ComposantUML extended with {
        Port Ifictif : InterfaceOfferte = {

        }
        Port I2 : InterfaceRequise = new InterfaceRequise extended with {

            Property service_requis : sous_programme = {[nom_sp = "crediter";p =
<[type_parametre = Integer_UML;mode = in_UML;]>;resultat = Void_UML;],[nom_sp =
"debiter";p = <>;resultat = Integer_UML;],[nom_sp ="transferer";p = <[type_parametre =
Integer_UML;mode = in_UML;]>;resultat = Integer_UML;]};
        }
    }
    Component Serveur : ComposantUML = new ComposantUML extended with {
        Port I1 : InterfaceOfferte = new InterfaceOfferte extended with {

            Property service_offert : sous_programme = {[nom_sp = "crediter";p =
<[type_parametre = Integer_UML;mode = in_UML;]>;resultat = Void_UML;],[nom_sp =
"debiter";p = <>;resultat = Integer_UML;],[nom_sp = "solde";p = <[type_parametre =
Integer_UML;mode = in_UML;],[type_parametre = Real_UML;mode = in_UML;]>;resultat =
Integer_UML;]};
        }
    }
    Connector assemblage : AssemblageUML = new AssemblageUML extended with {

    }
    Attachment Client.I2 to assemblage.cli;
    Attachment Serveur.I1 to assemblage.serv;
    rule service_offert_requis = invariant isSubset(self.Client.I2.service_requis,
self.Serveur.I1.service_offert);
}
```

**Figure 4.10** Formalisation en Acme/Armani du système *GAB2*

L'environnement AcmeStudio montre (Figure 4.11) que la contrainte syntaxique «`service_offert_requis`» spécifiée au niveau du système *GAB2* est évaluée à faux. Ceci traduit forcément une incohérence dans l'assemblage des composants UML2.0 proposé par la Figure 4.9. En effet, le service *transferer* exigé par l'interface I2 n'est pas offert par l'interface I1.

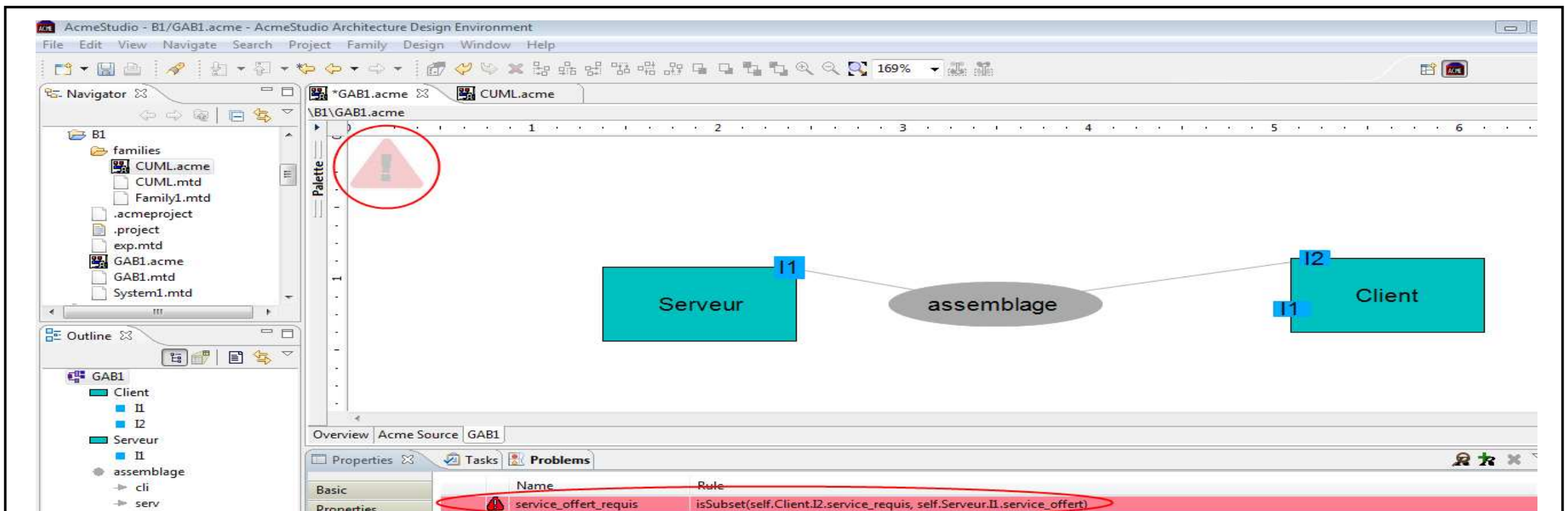

**Figure 4.11 :** Représentation graphique du système *GAB2* en Acme

## 4.3 Étude de cas : formalisation en Acme/Armani

Dans cette section, nous récupérons et adaptons en UML2 .0 une modélisation par composants d'un système de réservation de chambres d'hôtels [Cheesman, 2001]. Après avoir décrit d'une façon informelle le cahier des charges de notre application (cf. section 4.3.1), nous présentons une modélisation par composants assez détaillée de cette application (cf. section 4.3.2). Enfin, dans la section (cf. section 4.3.3) nous vérifions notre formalisation grâce à l'outil AcmeStudio [ABLE, 2009].

### 4.3.1 Cahier des charges

Le système de « Réservation de chambres d'hôtels » [Cheesman, 2001] souhaité doit autoriser des réservations dans n'importe quel hôtel appartenant à une chaîne d'hôtels. Une réservation peut être effectuée par téléphone en passant par un centre de réservation, par téléphone direct à un hôtel ou via Internet. Un avantage majeur de ce système de réservation est la possibilité d'offrir un hôtel alternatif lorsque l'hôtel désiré est complet. Chaque hôtel a un responsable permettant de contrôler les réservations dans cet hôtel. Afin de réduire la durée de réservation par téléphone, le système de réservation souhaité doit offrir un service permettant d'enregistrer et récupérer des informations liées aux clients antérieurs ou potentiels.

### 4.3.2 Modélisation en UML2.0

#### 4.3.2.1 Diagramme de composants

Le diagramme de composants associé à notre application est donné par la Figure 4.12. Un composant est doté des interfaces offertes et/ou requises. Une interface offerte propose un jeu de services à l'environnement. Par contre une interface requise exige des services venant de l'environnement.

Cette application exige des informations fournies par un utilisateur. Ainsi, nous avons eu recours à la définition de plusieurs types de données autres que les types de base (*real, integer, string, boolean*) définis dans le style *CUML*. Pour y parvenir, nous proposons une nouvelle formalisation d'un style appelé `DataType` (cf. section 4.3.1).

Nous détaillons par la suite les différents types de données et les signatures des services des interfaces de l'application « Réservation de chambres d'hôtels ».

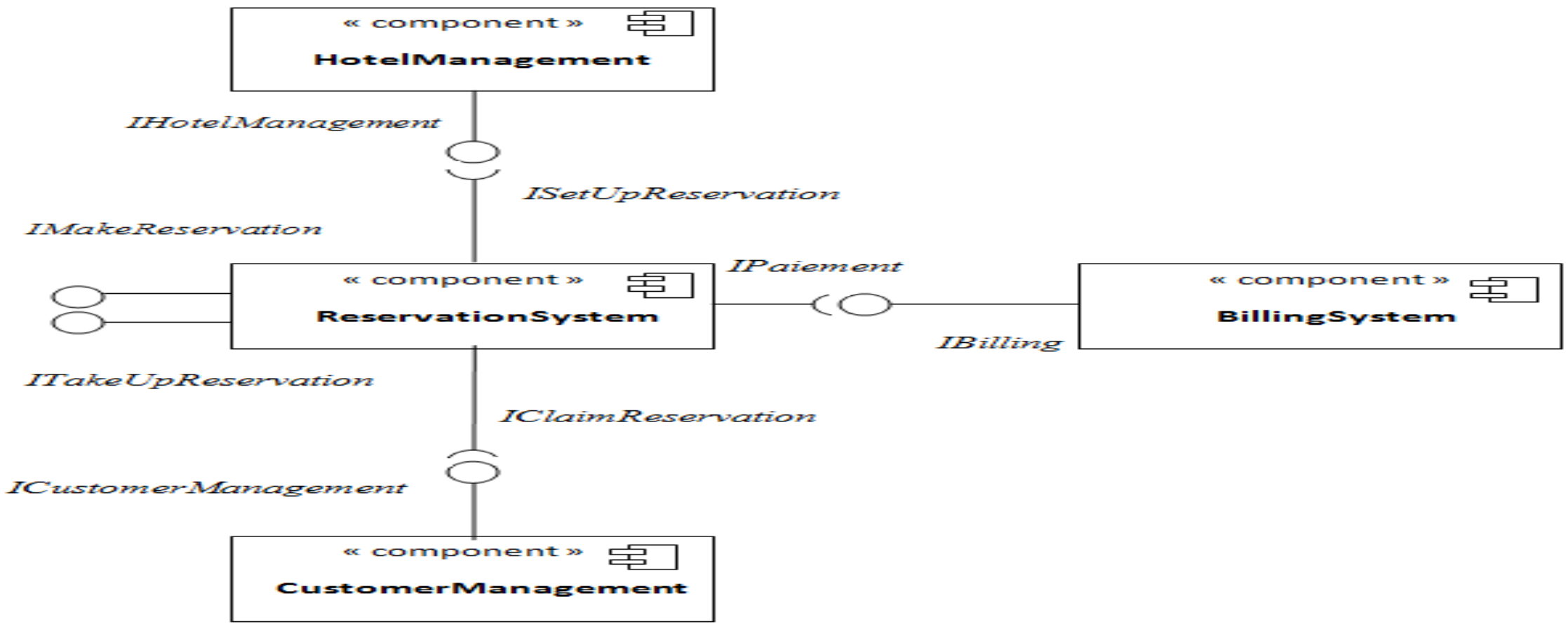

**Figure 4.12 :** Diagramme de composants avec identification des interfaces fournies /requises

### 4.3.2.2 Les types de données

Pour pouvoir décrire les services offerts/requis de l'application « Réservation de chambres d'hôtels », nous avons eu recours à plusieurs types de données autres que les types de base (*real*, *integer*, *string*, *boolean*). Un type de données peut avoir un ou plusieurs champs. Chaque champ est désigné par un identificateur et possède un type.

La Figure 4.13 présente les différents types de données de l'application « Réservation de chambres d'hôtels ».

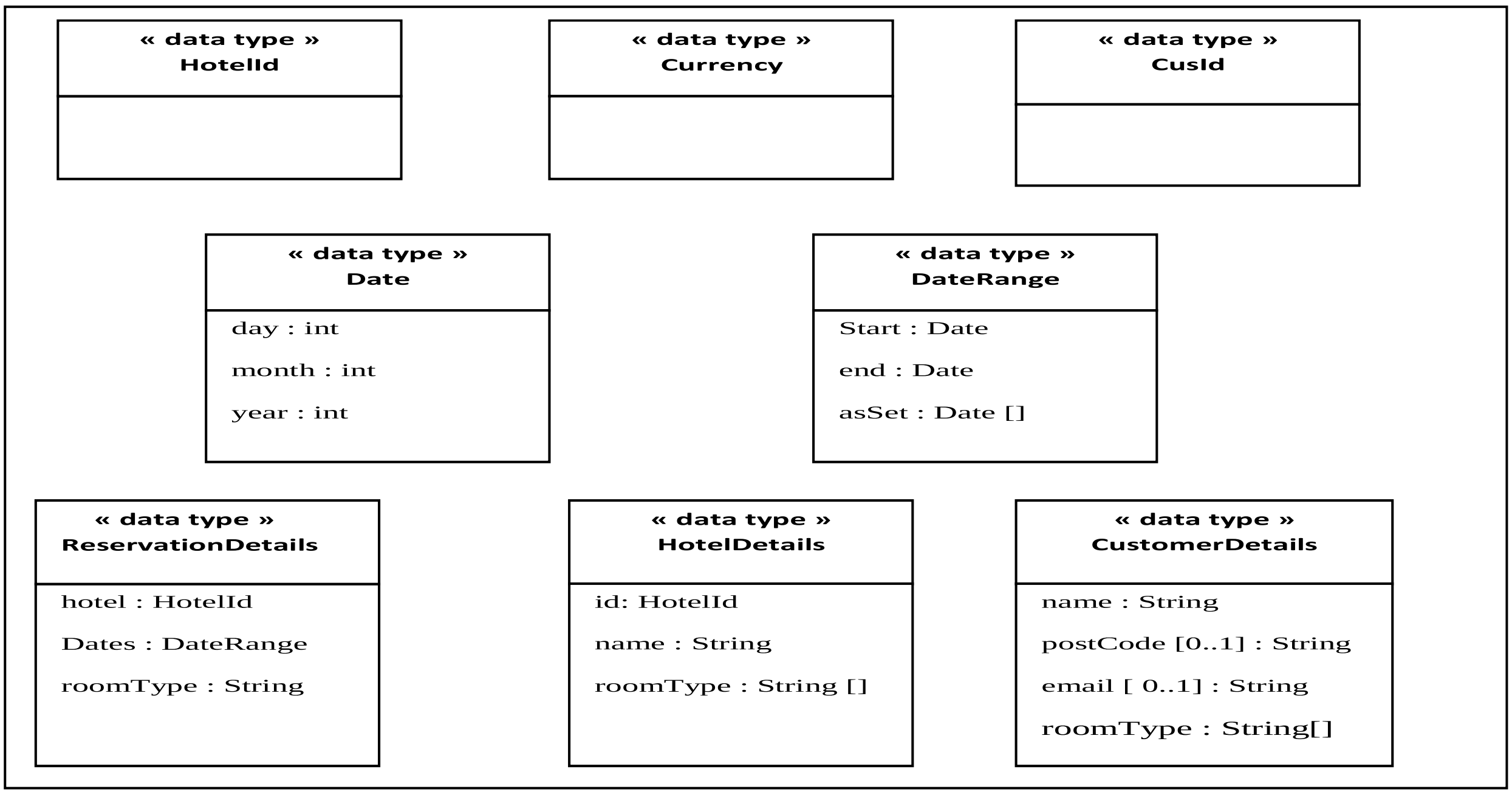

**Figure 4.13 :** Identification des types de données

**Explications :**

- Le type de données *CusId* fournit une structure d'accueil permettant d'identifier des clients,
- Le type de données *HotelId* fournit une structure d'accueil permettant d'identifier des hôtels,

- Le type de données *Date* permet de définir les caractéristiques d'une date à savoir : day, month, year,
- Le type de données *DateRange* regroupe trois champs permettant de mémoriser la date de début (start), la date de fin (end) d'une réservation ainsi que les jours de séjour (*asSet*),
- Le type de données *CustomerDetails* regroupe trois champs de type string permettant de définir respectivement *name*, *postCode* et *email* d'un client. Les deux derniers champs sont optionnels,
- Le type de données *HotelDetails* regroupe trois caractéristiques attachées à un hôtel : identifiant (*id*), nom (*name*) et les types de chambre (*roomType*). Ce dernier est un champ multivalué,
- Le type de données *ReservationDetails* regroupe des informations liées à une réservation : l'identifiant de l'hôtel (*hotel*), les dates proposées pour le séjour (*dates*) et la nature de chambre (*roomType*),
- Le type de données *Currency* fournit une structure d'accueil permettant de connaître l'unité monétaire utilisée pour les prix de chambres.

#### 4.3.2.3 Interfaces de l'application « Réservation de chambres d'hôtels »

Dans cette section, nous allons présenter les signatures des services des interfaces appartenant au diagramme de composants donné dans la Figure 4.12.

❖ **Interface IMakeReservation**

Cette interface (cf. Figure 4.14) a pour rôle de réserver des chambres d'hôtels. Elle offre trois services :

- **getHotelDetails ( )**

Cette opération fournit la liste des hôtels à partir de laquelle un client peut choisir son hôtel préféré. Le paramètre d'entrée (in par défaut) match est utilisé comme critère de sélection. Les noms d'hôtels sélectionnés doivent correspondre en partie ou totalement au nom (*match*) fourni comme paramètre in.

- **getRoomInfo ()**

Cette opération fournit deux grandeurs de sortie (deux paramètres out) liées à la disponibilité et au prix d'une réservation (*res*) éventuelle fournie comme paramètre in.

- **makeReservation ()**

Cette opération doit créer une réservation et avertir par email le client. La référence de la réservation créée est fournie dans le paramètre out (*resRef*). Elle rend explicitement une valeur nulle (0) en cas d'échec (informations manquantes ou redondantes) et une valeur non nulle en cas de succès.

| **« Interface »<br>IMakeReservation** |
|---|
| getHotelDetails (in match: String): HotelDetails [ ]<br>getRoomInfo (in res: ReservationDetails, out availability: Boolean, out price: Currency)<br>makeReservation (in res: ReservationDetails, in cus: CustomerDetails, out resRef: String): Integer |

**Figure 4.14 :** Interface IMakeReservation

## ❖ Interface ITakeUpReservation

Cette interface (cf. Figure 4.15) vise à confirmer la réservation en envoyant au client une fiche à remplir qui renferme tous les renseignements liés à cette réservation. L'interface fournit deux services qui sont les suivants :

- **getReservation ()**

Cette opération fournit deux grandeurs de sortie (*rd*) qui donne des informations liées à la réservation et (*cusId*) qui présente l'identificateur de client. La référence de la réservation doit correspondre à la référence indiquée comme paramètre in (*resRef*). L'opération retourne explicitement une valeur true si cette réservation est valide (*resRef* est *correcte*) et une valeur false si elle est invalide.

- **beginStay ()**

Cette opération annonce le début d'un séjour. Elle rend le numéro de la chambre (*roomNumber*). La référence de la réservation (*resRef*) est fournie comme paramètre in. Elle rend explicitement false en cas où *resRef* n'est pas valide.

| **« Interface »<br>ITakeUpReservation** |
|---|
| *getReservation (in resRef: String, out rd: ReservationDetails, out cus: CustomerDetails): Boolean*<br>*beginStay (in resRef: String, out roomNumber: String): Boolean* |

**Figure 4.15 :** Interface ITakeUpReservation

## ❖ Interface IHotelManagement

Cette interface (cf. Figure 4.16) a pour rôle de gérer les demandes de réservations de chambres d'hôtels. Elle comporte cinq services qui sont :

- **getHotelDetails ()**

Cette opération fournit la liste des hôtels à partir de laquelle un client peut choisir son hôtel préféré. Le paramètre d'entrée (in par défaut) match est utilisé comme critère de sélection. Les noms d'hôtels sélectionnés doivent correspondre en partie ou totalement au nom (*match*) fourni comme paramètre in.

- **getRoomInfo ()**

Cette opération fournit deux grandeurs de sortie (deux paramètres out) liées à la disponibilité et au prix d'une réservation (*res*) éventuelle fournie comme paramètre in.

- **makeReservation ()**

Cette opération doit créer une réservation et avertir par email le client. La référence de la réservation créée est fournie dans le paramètre out (*resRef*). Elle rend explicitement une valeur nulle (0) en cas d'échec (informations manquantes ou redondantes) et une valeur non nulle en cas de succès.

- **getReservation ()**

Cette opération fournit deux grandeurs de sortie (*rd*) qui donne des informations liées à la réservation et (*cusId*) qui présente l'identificateur de client. La référence de la réservation doit correspondre à la référence qui est indiquée comme paramètre in (*resRef*). L'opération retourne explicitement une valeur true si cette réservation est valide (*resRef* est correcte) et une valeur false si elle est invalide.

| **« Interface »<br>IHotelManagement** |
|---|
| *getHotelDetails (in match: String): HotelDetails [ ]*<br>*getRoomInfo (in res: ReservationDetails, out availability: Boolean, out price: Currency)*<br>*makeReservation (in res: ReservationDetails, in cus: CustomerDetails, out resRef: String): Integer*<br>*getReservation (in resRef: String, out rd: ReservationDetails, out cus: CustomerDetails): Boolean*<br>*beginStay (in resRef: String, out roomNumber: String): Boolean* |

**Figure 4.16 :** Interface IHotelManagement

- **beginStay ()**

Cette opération annonce le début d'un séjour. Elle rend le numéro de la chambre (*roomNumber*). La référence de la réservation est fournie comme paramètre in. Elle rend explicitement false en cas ou *resRef* n'est pas valide.

❖ **Interface ICustomerManagement**

Cette interface (cf. Figure 4.17) s'intéresse à gérer les clients. Elle offre quatre services qui sont les suivants :

- **getCustomerMatching ()**

Cette opération rend dans *CusId* l'identifiant d'un client correspondant aux informations fournies dans le paramètre in *CustD*. Elle rend explicitement le nombre des clients respectant *CustD*.

- **createCustomer ()**

Cette opération permet de créer un identifiant d'un client (*CusId*) en partant des informations stockées dans *CustD*. En cas de succès, elle rend true et false en cas d'échec.

- **getCustomerDetails ()**

Cette opération rend des informations liées à un client identifié par *cus* comme paramètre in.

- **notifyCustomer ()**

Cette opération permet d'envoyer le message (*msg*) au client (*cus*).

| **« Interface »<br>ICustomerManagement** |
|---|
| *getCustomerMatching (in custD: CustomerDetails, out cusId: CustId): Integer*<br>*createCustomer(in custD: CustomerDetails, out cusId: CustId): Boolean*<br>*getCustomerDetails (in cus: CustId): CustomerDetails*<br>*notifyCustomer (in cus: CustId, in msg: String)* |

**Figure 4.17 :** Interface ICustomerManagement

### ❖ Interface IBilling

L'interface IBilling (cf. Figure 4.18) permet de gérer la facturation. Elle comporte le service suivant :

- **openAccount ( )**

Cette opération permet de créer un compte pour le client qui vient de séjourner. Les informations liées à la réservation et au client sont fournies dans les deux paramètres in *res* et *cus*.

| **« Interface »**<br>**Interface IBilling** |
|---|
| *openAccount (in res: ReservationDetails, in cus: CustomerDetails)* |

**Figure 4.18 :** Interface IBilling

## 4.3.3 Formalisation en Acme/Armani

Dans cette section, nous proposons une formalisation Acme/Armani de la modélisation UML2 .0 [Kmimech, 2009d] de l'application « Réservation de chambres d'hôtels » décrite dans la section 4.3.2. Pour y parvenir, nous décrivons les types de données et les signatures des services offerts par l'application en utilisant judicieusement les possibilités de typage et la construction *family* offertes par Acme. Le diagramme de composants UML2.0 de notre application est formalisé en Acme en utilisant la construction *system*. Enfin, les règles de cohérence sont établies en utilisant la contrainte *invariant* supportée par Armani.

### 4.3.3.1 Les types de données et les signatures des opérations de l'application « Réservation de chambres d'hôtels »

Nous regroupons les types de données et les signatures des services offerts par notre application au sein d'un style Acme appelé *DataType*. Pour y parvenir, nous avons utilisé avec profit les possibilités de typage fournies par Acme : types simples prédéfinis (*int, float, boolean, string*), les constructeurs de types (*enum, renommage, property type, set, record* et *sequence*). Sachant que les deux types **set** et **sequence** sont génériques et modélisent respectivement un ensemble au sens mathématique (pas d'ordre et pas de doublons) et une collection avec ordre et doublons. La Figure 4.19 donne la formalisation Acme proposée. Par exemple, la signature1 correspond à la signature *getHotelDetails* offerte par l'interface *IMakeReservation*.

### 4.3.3.2 Assemblage de composants UML2.0 en Acme de l'application « Réservation de chambres d'hôtels »

L'assemblage de composants UML2 .0 de l'application est modélisé par un système appelé *Reservation* qui dérive de la famille *DataType*. Pour y parvenir, nous avons appliqué les règles suivantes :

***R1*** : Un composant UML2 .0 est traduit par un composant Acme.

***R1.1*** : Une interface attachée à un composant UML2.0 est traduite par un port Acme.

***R1.2*** : Un service déclaré au sein d'une interface UML2.0 est traduit par une propriété typée Acme attachée au port formalisant cette interface. Sachant que le type de la propriété modélise la signature du service.

***R2* :** Un connecteur d'assemblage UML2.0 reliant une interface offerte et une interface requise est modélisé par un connecteur binaire Acme ayant deux rôles.

***R3*:** les propriétés attachées à un rôle doivent être les mêmes que celles du port

```
Family DataType = {
    //Les types de données (data type) de l'application Réservation
Property Type CustId = int;
Property Type HoteId = int;
Property Type HotelDetails = Record [id: HotelId; name: string; room Types: Sequence <string>;];
Property Type Date = Record [day: int; month: int; year: int;];
Property Type Currency = Enum {euro, dollar, yen};
Property Type nature_logique = Enum {in_UML, out_UML, inout_UML};
Property Type ReservationDetails = Record [hotel: HotelId; dates: DateRange; roomType: string;];
Property Type DateRange = Record [start: date; end: date; asSet: Sequence <Date>;];
Property Type CustomerDetails = Record [name: string; postcode: Sequence <string>; email: Sequence <string>;];
Property Type signature1 = Record [name_service: string;
parametre: Record [name: string; nature: string; mode: nature_logique;]; result: Sequence <HotelDetails>;];

Property Type signature2 = Record [name_service: string;
parametre_1: Record [name: string; nature: ReservationDetails; mode: nature_logique;];
parametre_2: Record [name: string; nature: Boolean; mode: nature_logique;];
parametre_3: Record [name: string; nature: Currency; mode: nature_logique;];];
Property Type signature3 = Record [name_service: string;
parametre_1: Record [name: string; nature: ReservationDetails; mode: nature_logique;]; parametre_2: Record [name: string; nature: CustomerDetails; mode: nature_logique;]; parametre_3: Record [name: string; nature: string; mode: nature_logique;]; result: int;];
Property Type signature4 = Record [name_service: string;
parametre_1: Record [name: string; nature: string; mode: nature_logique;];
parametre_2: Record [name: string; nature: ReservationDetails; mode: nature_logique;];parametre_3: Record [name: string; nature: CustomerDetails; mode: nature_logique;]; result: boolean;];
Property Type signature5 = Record [name_service: string;
parametre_1: Record [name: string; nature: string; mode: nature_logique;];
parametre_2: Record [name: string; nature: string; mode: nature_logique;];
result: boolean;];
Property Type signature6 = Record [name_service: string;
parametre_1: Record [name: string; nature: string; mode: nature_logique;];
parametre_2: Record [name: string; nature: ReservationDetails; mode: nature_logique;];
parametre_3: Record [name: string; nature: CustomerDetails; mode: nature_logique;]; result: boolean;];
Property Type signature7 = Record [name_service: string;
parametre_1: Record [name: string; nature: CustomerDetails; mode: nature_logique;];
parametre_2: Record [name: string; nature: CustId; mode: nature_logique;];
result: int;];
Property Type signature8 = Record [name_service: string;
parametre_1: Record [name: string; nature: CustomerDetails; mode: nature_logique;];
parametre_2: Record [name: string; nature: CustId; mode: nature_logique;];
result: boolean;];
Property Type signature9 = Record [name_service: string;
parametre: Record [name : string; nature : CustId; mode : nature_logique;];];
Property Type signature10 = Record [name_service: string;

parametre_1: Record [name: string; nature: CustId; mode: nature_logique;];
parametre_2: Record [name: string; nature: string; mode: nature_logique;];];
Property Type signature11 = Record [name_service: string;
parametre_1 : Record [name : string; nature : ReservationDetails; mode : nature_logique;];
parametre_2 : Record [name : string; nature : CustomerDetails; mode : nature_logique;];];
```

**Figure 4.19 :** Types de données et signatures des services formalisés en Acme/Armani

La Figure 4.20 donne la spécification Acme de l'application « Réservation de chambres d'hôtels » issue de l'exécution des règles données ci-dessus.

```
import families/DataType.acme;
 System Reservation: DataType = new DataType extended with {

    Component ReservationSystem = {
   //interface offerte
        Port IMakeReservation = {
            Property getHotelDetails: signature1;
            Property getRoomInfo: signature2;
            Property makeReservation: signature3;
        }
  //interface offerte
        Port ITakeupReservation = {
            Property getReservation: signature4;
            Property beginStay: signature5 ;}
  //interface requise
        Port ISetupReservation = {
            Property getHotelDetails: signature1;
            Property getRoomInfo: signature2;
            Property makeReservation: signature3;
            Property getReservation: signature6;
            Property beginStay: signature5 ;}
 //interface requise
        Port IClaimReservation = {
            Property getCustomerMatching: signature7;
            Property createCustomer: signature8;
            Property getCustomerDetails: signature9;
            Property notifyCustomer: signature10;
        }
 //interface requise
       Port IPaiement = {
            Property openAccount: signature11 ;}}

Component BillingSystem = {
     //interface offerte
        Port IBilling = {
            Property openAccount: signature11 ;}}

Component CustomerManagement = {
    //interface offerte
        Port ICustomerManagement = {
            Property getCustomerMatching: signature7;
            Property createCustomer: signature8;
            Property getCustomerDetails: signature9;
            Property notifyCustomer: signature10 ;}}

Component HotelManagement = {
     //interface offerte
        Port IHotelManagement = {
            Property getHotelDetails: signature1;
            Property getRoomInfo: signature2;
            Property makeReservation: signature3;
            Property getReservation: signature6;
            Property beginStay: signature5;
              }
     }
Connector ReservationSystem_HotelManagement = {
        Role serveur_IHotelManagement = {
            Property getHotelDetails: signature1;
            Property getRoomInfo: signature2;
            Property makeReservation: signature3;
            Property getReservation: signature6;
            Property beginStay: signature5;
        }

       Role client_ISetupReservation = {
```

```
Connector ReservationSystem_CustomerManagement = {

        Role serveur_ICustomerManagement = {
            Property getCustomerMatching: signature7;
            Property createCustomer: signature8;
            Property getCustomerDetails: signature9;
            Property notifyCustomer: signature10 ;}

        Role client_IClaimReservation = {
            Property getCustomerMatching: signature7;
            Property createCustomer: signature8;
            Property getCustomer: signature9;
            Property notifyCustomer: signature10 ;}
}
Connector ReservationSystem_BillingSystem = {
        Role serveur_IBilling = {
            Property openAccount: signature11 ;}

        Role client_IPaiement = {
            Property openAccount: signature11 ;}}

Attachment ReservationSystem.IClaimReservation to
ReservationSystem_CustomerManagement.client_IClaimReservation;
Attachment CustomerManagement.ICustomerManagement to
ReservationSystem_CustomerManagement.serveur_ICustomerManagement;
Attachment ReservationSystem.IPaiement to
ReservationSystem_BillingSystem.client_IPaiement;
Attachment BillingSystem.IBilling to
ReservationSystem_BillingSystem.serveur_IBilling;
Attachment HotelManagement.IHotelManagement to
ReservationSystem_HotelManagement.serveur_IHotelManagement;
Attachment ReservationSystem.ISetupReservation to
ReservationSystem_HotelManagement.client_ISetupReservation;
}
```

**Figure 4.20 :** Formalisation de l'application « Réservation de chambres d'hôtels » par un système en Acme

### 4.3.4 Vérification

Dans cette section, nous allons proposer des règles de cohérence relatives au modèle de composants UML2.0. Ces règles sont modélisées par des propriétés invariantes en utilisant le concept *invariant* d'Acme. Elles concernent la vérification des attachements de la configuration Réservation.

Nous proposons deux règles de cohérence :

- ***Rc1*** : Un rôle et un port attachés ont le même nombre de propriétés,
- ***Rc2*** : Un port et un rôle attachés doivent avoir des propriétés compatibles ;

La Figure 4.21 donne une formalisation de ces deux règles en Acme/Armani.

```
// Un rôle et un port attachés ont le même nombre de propriétés
rule verifysizeproperty1 = invariant forall p: Port in
self.ReservationSystem_HotelManagement.serveur_IHotelManagement
.ATTACHEDPORTS|size (self. ReservationSystem_HotelManagement
.serveur_IHotelManagement. PROPERTIES) == size (p.PROPERTIES);

rule verifysizeproperty2 = invariant forall p : Port in
self.ReservationSystem_HotelManagement.client_ISetupReservation
.ATTACHEDPORTS|size(self. ReservationSystem_HotelManagement
.client_ISetupReservation. PROPERTIES) == size (p.PROPERTIES);

rule verifysizeproperty3 = invariant forall p : Port in
self.ReservationSystem_CustomerManagement.serveur_ICustomerManagement
.ATTACHEDPORTS|size(self. ReservationSystem_CustomerManagement
.serveur_ICustomerManagement. PROPERTIES) == size (p.PROPERTIES);

rule verifysizeproperty4 = invariant forall p : Port in
self.ReservationSystem_CustomerManagement.client_IClaimReservation
.ATTACHEDPORTS|size(self. ReservationSystem_CustomerManagement
.client_IClaimReservation. PROPERTIES) == size (p.PROPERTIES);

rule verifysizeproperty5 = invariant forall p : Port in
self.ReservationSystem_BillingSystem.serveur_IBilling.ATTACHEDPORTS |
size (self.ReservationSystem_BillingSystem.serveur_IBilling.PROPERTIES) ==
size(p.PROPERTIES);

rule verifysizeproperty6 = invariant forall p : Port in
self.ReservationSystem_BillingSystem.client_IPaiement.ATTACHEDPORTS |
size (self.ReservationSystem_BillingSystem.client_IPaiement.PROPERTIES) ==
size(p.PROPERTIES);

// En passant d'une façon explicite par les attachements: un port et un rôle
//attachés doivent avoir des propriétés compatibles. A défaut, le contrôleur
//de types signale une erreur liée au typage

rule verifyAttachment1 = invariant
self.ReservationSystem.ISetupReservation.getHotelDetails ==
self.ReservationSystem_HotelManagement.client_ISetupReservation.getHotelDetai
ls AND self.ReservationSystem.ISetupReservation.getRoomInfo ==
Self.ReservationSystem_HotelManagement.client_ISetupReservation.getRoomInfo
AND self.ReservationSystem.ISetupReservation.makeReservation ==
self.ReservationSystem_HotelManagement.client_ISetupReservation
.makeReservation AND self.ReservationSystem.ISetupReservation.getReservation
==
self.ReservationSystem_HotelManagement.client_ISetupReservation
.getReservation AND self.ReservationSystem.ISetupReservation.beginStay
==self.ReservationSystem_HotelManagement.client_ISetupReservation .beginStay;

rule verifyAttachment2 = invariant
self.HotelManagement.IHotelManagement.getHotelDetails ==
self.ReservationSystem_HotelManagement.serveur_IHotelManagement
.getHotelDetails AND self.HotelManagement.IHotelManagement.getRoomInfo ==
self.ReservationSystem_HotelManagement.serveur_IHotelManagement .getRoomInfo
AND self.HotelManagement.IHotelManagement.makeReservation
==self.ReservationSystem_HotelManagement.serveur_IHotelManagement
.makeReservation AND self.HotelManagement.IHotelManagement.getReservation ==
self.ReservationSystem_HotelManagement.serveur_IHotelManagement
.getReservation AND self.HotelManagement.IHotelManagement.beginStay ==
self.ReservationSystem_HotelManagement.serveur_IHotelManagement
.beginStay;
```

```
rule verifyAttachment3 = invariant
self.CustomerManagement.ICustomerManagement.getCustomerMatching ==
self.ReservationSystem_CustomerManagement.serveur_ICustomerManagement.getCust
omerMatching AND
self.CustomerManagement.ICustomerManagement.createCustomer ==
self.ReservationSystem_CustomerManagement.serveur_ICustomerManagement.createC
ustomer AND self.CustomerManagement.ICustomerManagement.getCustomerDetails ==
self.ReservationSystem_CustomerManagement.serveur_ICustomerManagement.getCust
omerDetails AND
self.CustomerManagement.ICustomerManagement.notifyCustomer ==
self.ReservationSystem_CustomerManagement.serveur_ICustomerManagement.notifyC
ustomer;
rule verifyAttachment4 = invariant
self.ReservationSystem.IClaimReservation.getCustomerMatching==
self.ReservationSystem_CustomerManagement.client_IClaimReservation.getCustome
rMatching AND self.ReservationSystem.IClaimReservation.createCustomer ==
self.ReservationSystem_CustomerManagement.client_IClaimReservation
.createCustomer AND self.ReservationSystem.IClaimReservation.getCustomer ==
self.ReservationSystem_CustomerManagement.client_IClaimReservation
.getCustomer AND
self.ReservationSystem.IClaimReservation.notifyCustomer ==
self.ReservationSystem_CustomerManagement.client_IClaimReservation
.notifyCustomer;

rule verifyAttachment5 = invariant self.BillingSystem.IBilling.openAccount ==
self.ReservationSystem_BillingSystem.serveur_IBilling.openAccount;

rule verifyAttachment6 = invariant
self.ReservationSystem.IPaiement.openAccount ==
self.ReservationSystem_BillingSystem.client_IPaiement.openAccount;
```

**Figure 4.21 :** Règles de cohérence relatives aux attachements de ports et rôles de notre application

La formalisation des règles de cohérence (***Rc1*** et ***Rc2***) en Acme/Armani proposées ci-dessus est dédiée à notre application. Une généralisation pourrait être envisagée (cf. Figure 4.22). L'idée consiste à proposer un type énuméré au niveau de la famille *DataType* regroupant :

- Les types de base d'UML : *Boolean_UML, Real_UML, Integer_UML et String_UML,*
- Les types de données propres à l'application traitée. Par exemple, dans notre application, nous avons : *ReservationDetails_app*, *CustomerDetails_app*, *Currency_app*, *CusId_app*, *HotelId_app* et *HotelDetais_app*.

Ensuite, nous initialisons dans la configuration concernée (*Reservation*) les champs formant la signature avec les valeurs adéquates (champ nature : *type_base_app*). Ainsi, la comparaison de deux propriétés devient possible car elle se fait sur les valeurs portées par les propriétés comparées.

```
Family Family1_essai = {
//Les types de données (data type) de l'application Reservation
Property Type CustId = int;
Property Type HoteId = int;
Property Type HotelDetails = Record [id: HotelId; name: string; room Types:
Sequence <string>;];
Property Type Date = Record [day: int; month: int; year: int;];
Property Type Currency = Enum {euro, dollar, yen};
Property Type nature_logique = Enum {in_UML, out_UML, inout_UML};
Property Type ReservationDetails = Record [hotel: HotelId; dates: DateRange;
roomType: string;];
Property Type DateRange = Record [start: date; end: date; asSet: Sequence
<Date>;];
Property Type CustomerDetails = Record [name: string; postcode: Sequence
<string>; email: Sequence <string>;];
Property Type type_base_app = Enum
{UML_string,UML_boolean,UML_integer,UML_Real,CustomerDetails_app,ReservationDe
tails_app,Currency_app,CustId_app,DateRange_app,HotelId_app,
sequence_HotelsDetails_app};
```

```
Property Type signature1 = Record [name_service : string;
    parametre  :  Record  [name  :  string;  nature  :  type_base_app;  mode  :
nature_logique;];
    result : type_base_app;];
Property Type signature2 = Record [name_service : string;
    parametre_1  :  Record  [name  :  string;  nature  :  type_base_app;  mode  :
nature_logique;];
    parametre_2  :  Record  [name  :  string;  nature  :  type_base_app;  mode  :
nature_logique;];
    parametre_3  :  Record  [name  :  string;  nature  :  type_base_app;  mode  :
nature_logique;];];
Property Type signature3 = Record [name_service : string;
    parametre_1  :  Record  [name  :  string;  nature  :  type_base_app;  mode  :
nature_logique;];
    parametre_2  :  Record  [name  :  string;  nature  :  type_base_app;  mode  :
nature_logique;];
    parametre_3  :  Record  [name  :  string;  nature  :  type_base_app;  mode  :
nature_logique;];
    result : type_base_app;];
Property Type signature4 = Record [name_service : string;
    parametre_1 : Record [name : string; nature : type_base_app;
    mode : nature_logique; ]; parametre_2 : Record [name : string;
    nature : type_base_app; mode : nature_logique; ];
    parametre_3 : Record [name : string; nature : type_base_app; mode       :
nature_logique;];
    result : type_base_app;];

 Property Type signature5 = Record [name_service : string;
    parametre_1 : Record [name : string; nature : type_base_app;
    mode : nature_logique;]; parametre_2 : Record [name : string;
    nature    :    type_base_app;    mode    :    nature_logique;];    result    :
type_base_app;] ;

Property Type signature6 = Record [name_service : string;
    parametre_1  :  Record  [name  :  string;  nature  :  type_base_app;  mode  :
nature_logique;];
    parametre_2  :  Record  [name  :  string;  nature  :  type_base_app;  mode  :
nature_logique;];
    parametre_3  :  Record  [name  :  string;  nature  :  type_base_app;  mode  :
nature_logique;];
    result : type_base_app;];

    Property Type signature7 = Record [name_service : string;
    parametre_1  :  Record  [name  :  string;  nature  :  type_base_app;  mode  :
nature_logique;];
    parametre_2  :  Record  [name  :  string;  nature  :  type_base_app;  mode  :
nature_logique;];
    result : type_base_app;];
Property Type signature8 = Record [name_service : string;
    parametre_1  :  Record  [name  :  string;  nature  :  type_base_app;  mode  :
nature_logique;];
    parametre_2  :  Record  [name  :  string;  nature  :  type_base_app;  mode  :
nature_logique;];
    result : type_base_app;];

Property Type signature9 = Record [name_service : string;
    parametre  :  Record  [name  :  string;  nature  :  type_base_app;  mode  :
nature_logique;];];

 Property Type signature10 = Record [name_service : string;
    parametre_1  :  Record  [name  :  string;  nature  :  type_base_app;  mode  :
nature_logique;];
    parametre_2  :  Record  [name  :  string;  nature  :  type_base_app;  mode  :
nature_logique;];];

Property Type signature11 = Record [name_service : string;
    parametre_1  :  Record  [name  :  string;  nature  :  type_base_app;  mode  :
nature_logique;];
    parametre_2  :  Record  [name  :  string;  nature  :  type_base_app;  mode  :
nature_logique;];];
```

**Figure 4.22 :** Types de données et signatures de services formalisés en Acme/Armani (version généralisée)

## 4.4 Conclusion

Dans ce chapitre, nous avons proposé une traduction du modèle de composants UML2.0 en Acme/Armani. L'objectif de cette traduction est de garantir les contrats syntaxiques en vérifiant la cohérence d'assemblages de composants UML2.0 [Kmimech, 2009a], [Kmimech, 2009d], [Kmimech, 2009e]. La vérification des contrats syntaxiques est confiée à l'évaluateur des prédicats supporté par la plateforme AcmeStudio [ABLE, 2009]. Pour y parvenir, nous décrivons les types de données et les signatures des services offerts par l'application en utilisant judicieusement les possibilités de typage et la construction *family* offertes par Acme. En outre, nous avons proposé une modélisation par composants en UML2.0 de l'application « Réservation de chambres d'hôtels ».

Le chapitre suivant sera consacré à la formalisation des propriétés non fonctionnelles en CQML (Component Quality Modeling Language). L'objectif recherché d'utiliser CQML est d'attacher des propriétés non fonctionnelles (PNF) aux composants UML2.0.

# **Chapitre 5 :** Vérification des contrats de qualité de services d'assemblages de composants UML2.0

## 5.1 Introduction

Dans le chapitre précédent, nous avons proposé un style d'architectures «*CUML*» permettant la formalisation et la vérification syntaxique des assemblages de composants UML2.0 en Acme/Armani. Dans ce chapitre, nous allons étendre ce style par des nouveaux concepts pour qu'il soit capable de formaliser les PNF des composants UML2.0 en Acme/Armani. Les nouveaux concepts présentés sont inspirés des langages de modélisation des PNF étudiés dans le chapitre 2 et principalement du langage CQML (cf. section 2.6).

Ce chapitre comporte cinq sections. La section 5.2 présente la modélisation en UML2.0/CQML d'une application appelée VideoCamera inspirée d'un exemple présenté dans [Blair, 1998]. L'architecture de cette application est modélisée en UML2.0 en se servant judicieusement des constructions composant, interface offerte, interface requise et connecteur d'assemblage. Quant aux aspects non fonctionnels, ils sont modélisés en CQML en utilisant les constructions *quality_characteristic*, *quality* et *profile*. La section 5.3 présente la formalisation en Acme/Armani des trois concepts fondamentaux décrivant des PNF issus d'un langage de type CQML : caractéristique de qualité, qualité et profil. La section 5.4 propose la formalisation en Acme/Armani d'un contrat de qualité permettant de vérifier la cohérence de qualité d'un assemblage de composants UML2.0. Enfin, dans la section 5.5 nous proposons des règles simples permettant la traduction d'une description non fonctionnelle décrite en CQML vers Acme/Armani en réutilisant notre style *CUML*. Ces règles de traduction proposées sont simples car *CUML* intègre les principaux concepts venant de CQML tels que : *quality_characteristic*, *quality* et *profile*. Tout au long de ce chapitre, nous fournissons des exemples illustratifs qui accompagnent les règles de traduction proposées.

## 5.2 Description en UML2.0/CQML : étude de cas d'un système VideoCamera

Le système CaméraVidéo numérique (VideoCamera) [Blair, 1998] est composé essentiellement par trois éléments : une Caméra qui capture des séquences vidéos, une Mémoire qui lui permet d'enregistrer ces séquences et d'un VideoPlayer permettant la présentation des séquences déjà enregistrées.

### 5.2.1 Spécification informelle des composants du système

Pour mieux comprendre le fonctionnement de ce système, nous allons essayer de détailler les trois composants qui le constituent :

1- Le premier composant appelé *Camera*, propose une interface appelée *Memorization*. Cette interface comporte deux services (*Stored* et *Stoped*).

2- Le second composant appelé *Memory*, propose à son tour deux interfaces. La première exige les services proposés par l'interface *Memorization* du composant *Camera*. La seconde, appelée *VideoStream*, fournit les quatre services (*Play*, *Forward*, *Rewind* et *Stop*).

3- Le troisième composant appelé *VideoPlayer* exige une interface *VideoStream* et fournit une autre interface appelée *VideoPresented*. Cette interface définit le service *Presented*.

### 5.2.2 Spécification informelle des propriétés non fonctionnelles

Pour garantir le bon fonctionnement de ce système, on décide d'attacher quelques propriétés non fonctionnelles aux services de ces composants. On propose de traiter la fiabilité, la disponibilité et quelques propriétés de performance telles que le débit et le temps de réponse. La spécification suivante illustre la signification détaillée de ces PNF.

**a- Fiabilité** (Reliability) : la probabilité qu'un composant soit en état de fonctionnement (sans panne). Cette propriété assure la continuité du service ;

**b- Disponibilité** (Availability) : la probabilité qu'un composant soit en état de marche à un instant donné. Cette propriété assure que le service soit prêt à l'emploi ;

**c- Débit** : taux de transfert de données (ici image) par unité de temps ;

**d- Temps de réponse** : temps entre l'activation et la réponse d'un service ;

Le composant *Camera* offre un taux de disponibilité supérieur à 90 % et un niveau de fiabilité supérieur à 80%.

Le composant *Memory* exige un taux de disponibilité supérieur à 85 %, un niveau de fiabilité supérieur à 70 %. En outre, ce composant offre une bonne performance (c'est-à-dire le temps de réponse du service *Play* est inférieur ou égal à 15 msec et le taux de transfert des données lors de l'utilisation du service *Play* est supérieur ou égal à 30 image/sec).

Le composant *VideoPlayer* exige une performance acceptable (c'est-à-dire le temps de réponse du service *Play* est inférieur ou égal 20 à msec et le taux de transfert des données lors de l'utilisation du service *Play* est supérieur ou égal à 25 image/sec).

### 5.2.3 Modélisation de l'architecture d'un système de Caméra Vidéo en UML2.0

La Figure 5.1 présente une description architecturale du système « *CameraVideo* » en utilisant un assemblage de composants UML2.0. Cet assemblage est accompagné d'une représentation explicite (Figure 5.2) des différents types d'interfaces de composants constituant ce système.

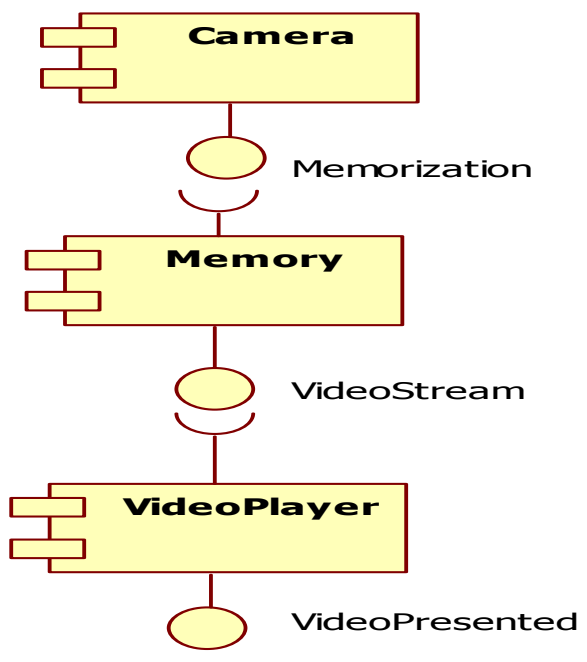

**Figure 5.1 :** Description architecturale en UML2.0 du système *VideoCamera*

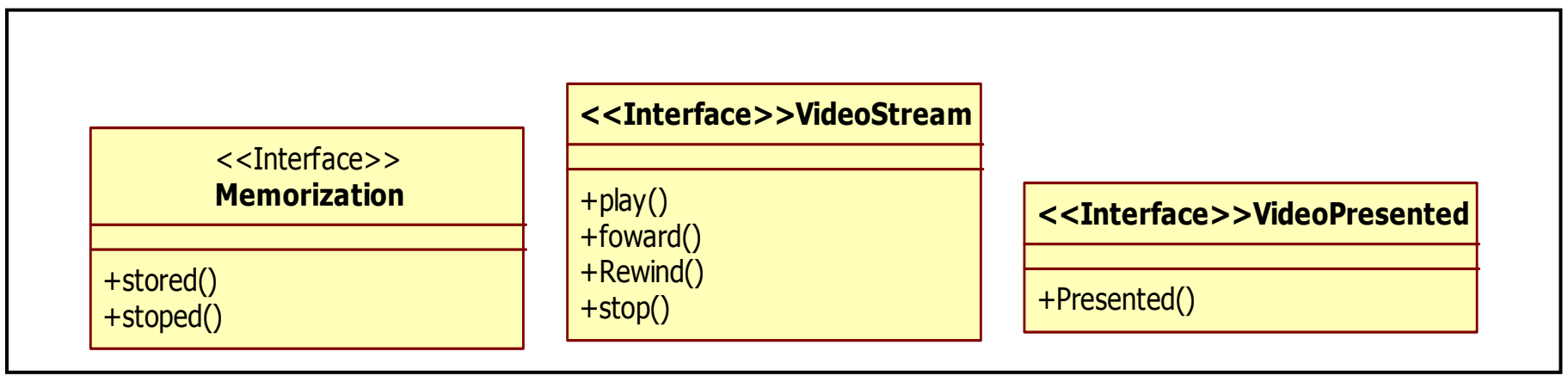

**Figure 5.2 :** Types d'interfaces des composants de *VideoCamera* en UML2.0

## 5.2.4 Formalisation des propriétés non fonctionnelles de l'application VideoCamera en CQML

Dans cette partie, nous allons spécifier les PNF du système VideoCamera en CQML.

Rappelons que CQML est basé principalement sur les trois concepts suivants :

- *quality_characteristic* **:** permet de définir un type d'une PNF (fiabilité, disponibilité, temps de réponse…),
- *quality* **:** permet de définir une qualité (FiabilitéBonne, FiabilitéMoyenne…) par la restriction d'un ensemble de caractéristiques. Une telle restriction se fait par une contrainte sur une caractéristique de qualité, par exemple la qualité FiabilitéBonne peut être définie par la contrainte suivante (Fiabilité > = 80 %),
- *profile* **:** permet d'attacher à chaque composant ses propres qualités.

### 5.2.4.1 Spécification des caractéristiques de qualité

#### a- Fiabilité

La spécification informelle définit le niveau de fiabilité comme un pourcentage. Ceci implique que les valeurs possibles de cette caractéristique doivent être de type réel dans l'intervalle [0,100]. De plus, la bonne fiabilité est attribuée aux valeurs les plus élevées. Ceci implique que la variance de cette caractéristique doit être « increasing ». La Figure 5.3 montre une formalisation de cette caractéristique en CQML.

```
quality_characteristic Fiabilite
  {         domain : increasing numeric real [0..100]  %;
  }
```

**Figure 5.3 :** Formalisation de la caractéristique Fiabilité en CQML

#### b- Disponibilité

Le taux de disponibilité est aussi défini comme un pourcentage. Il doit être représenté par une valeur réelle entre 0 et 100 et son unité doit être le « % ». L'élévation du pourcentage de cette caractéristique se trouve positivement corrélée à la qualité. La variation est donc « increasing ». La Figure 5.4 montre une formalisation de la Disponibilité en CQML.

```
quality_characteristic Disponibilite
{
      domain : increasing numeric real [0..100]  % ;
 }
```

**Figure 5.4 :** Formalisation de la caractéristique Disponibilité en CQML

**c- Temps de réponse**

L'unité de la caractéristique temps de réponse est définie dans la spécification informelle par «msec» conserver en CQML. Plus le temps de réponse est petit, plus la qualité de service est bonne : on constate que la variance de cette caractéristique doit être «decreasing». La Figure 5.5 présente une formalisation de cette caractéristique en CQML.

```
quality_characteristic TempsDeReponse
{
      domain : decreasing numeric msec;
}
```

**Figure 5.5 :** Formalisation de la caractéristique Temps de réponse en CQML

**d- Taux de transfert**

Le taux de transfert (débit) représente le nombre d'images transmises par seconde. Cette caractéristique doit être de variance « increasing » ayant une valeur entière, dont l'unité est image/sec. La Figure 5.6, représente une formalisation de cette caractéristique en CQML.

```
quality_characteristic TauxDeTransfert
{
      domain : increasing numeric integer image/sec ;
}
```

**Figure 5.6 :** Formalisation de la caractéristique Taux de transfert en CQML

## 5.2.4.2 Spécification des qualités des propriétés non fonctionnelles

Dans cette partie, nous allons spécifier en CQML les différentes contraintes exercées sur les caractéristiques déjà définies.

**a- Qualités liées à la caractéristique Fiabilité**

La spécification informelle des PNF propose deux contraintes sur la fiabilité :

- Fiabilité $\geq$ 70 % ➔ on associe la qualité FiabiliteAcceptable,
- Fiabilité $\geq$ 80 % ➔ on associe la qualité FiabiliteBonne.

La Figure 5.7 représente une formalisation de ces deux qualités en CQML en se servant du type Fiabilite déjà défini (cf. Figure 5.3).

```
quality FiabiliteAcceptable  {
      Fiabilite  >= 70  ;
};
quality FiabiliteBonne {
      Fiabilite >= 80  ;
};
```

**Figure 5.7 :** Formalisation en CQML des qualités liées à la Fiabilité

**b- Qualités liées à la caractéristique Disponibilité**

La spécification informelle des PNF propose deux contraintes sur la disponibilité :

- Disponibilité $\geq$ 85 % ➔ on associe la qualité DispBonne,
- Disponibilité $\geq$ 90 % ➔ on associe la qualité DispTresBonne.

La Figure 5.8 représente une formalisation de ces qualités en CQML en utilisant le type Disponibilité déjà défini (cf. Figure 5.4).

```
quality DispBonne {
      Disponibilite  >= 85 ;
};
quality DispTresBonne {
      Disponibilite  >= 90 ;
};
```

**Figure 5.8 :** Formalisation en CQML des qualités liées à la disponibilité

**c- Qualités liées à la caractéristique Performance**

La spécification informelle des PNF propose deux qualités liées à la performance.

La qualité *PerformanceAcceptable* est rapportée aux caractéristiques temps de réponse et taux de transfert. Elle modélise les deux contraintes suivantes :

- Temps de réponse $\leq$ 20 msec ,
- Taux de transfert >= 25 image/sec.

La qualité *PerformanceBonne* est aussi rapportée aux caractéristiques temps de réponse et taux de transfert. Elle modélise les deux contraintes suivantes :

- Temps de réponse $\leq$ 15 msec,
- Taux de transfert >= 30 image/sec.

La Figure 5.9 représente une formalisation de ces qualités en CQML en se servant des types *TempsDeReponse* et *TauxDeTransfert* déjà définis (cf. Figures 5.5 et 5.6).

```
quality PerformanceAcceptable  {
      TempsDeReponse  <= 20  ;
      TauxDeTransfert >= 25  ;
};
quality PerformanceBonne {
      TempsDeReponse  <= 15  ;
      TauxDeTransfert >= 30  ;
};
```

**Figure 5.9 :** Formalisation en CQML des qualités liées au Temps de réponse

### 5.2.4.3 Attachement des qualités aux composants

Dans cette partie, nous allons utiliser le concept *profil* offert par CQML pour associer à chaque composant du système *VideoCamera* ses qualités (requises et/ou offertes).

La Figure 5.10 présente trois *profils* CQML. Le premier appelé *QoSCamera* modélise les qualités proposées par le composant *Camera*, le second appelé *QoSMemory* attache au composant *Memory* ses qualités et le troisième appelé *QoSVideoPlayer* décrit les qualités exigées par le composant *VideoPlayer*.

## 5.3 Formalisation et vérification des propriétés non fonctionnelles des composants UML2.0 en Acme/Armani

### 5.3.1 Formalisation des propriétés non fonctionnelles des composants UML2.0

Les possibilités de typage d'Acme/Armani (cf. chapitre 2) sont utilisées avec profit afin de définir quatre nouveaux types de propriétés (*CaracteristiqueNumerique*, *CaracteristiqueOrdinaire*, *Qualite* et *Profile*) permettant la formalisation des PNF en Acme/Armani. Dans la suite, nous allons présenter ces nouveaux types de propriétés.

```
profile QoSCamera for Camera {
      provides DispTresBonne   and
             FiabiliteBonne  ;
}

profile QoSMemory for Memory {
      uses  DispBonne          and
          FiabiliteAcceptable ;

      provides PerformanceBonne  ;
}

profile QoSVideoPlayer for VideoPlayer {
      uses    PerformanceAcceptable  ;
}
```

**Figure 5.10 :** Formalisation des *profils* associés aux composants du système *VideoCamera*

#### 5.3.1.1 Formalisation des deux concepts «*CaracteristiqueNumerique*» et «*CaracteristiqueOrdinaire*»

La caractéristique de qualité est la construction de base de toute spécification non fonctionnelle. Cette caractéristique représente un aspect non fonctionnel tels que la performance, la fiabilité, la disponibilité, etc.

On peut formaliser une caractéristique de qualité par une propriété Acme/Armani. Cette propriété doit être de type enregistrement (*record*) composé de cinq champs :

1- *Nom* : qui représente le nom de la caractéristique (Performance, Disponibilité…). Ce champ peut être modélisé par une propriété de type chaîne de caractères.

2- *Parametres* : qui représente les paramètres de la caractéristique. Ce champ peut être modélisé par une séquence d'éléments de type composé. A ce niveau, chaque élément de cette séquence peut être modélisé par un enregistrement composé de deux champs :

   a- *Nom_Par* : qui représente le nom d'un paramètre de la caractéristique. Ce champ peut être modélisé par une chaîne de caractères,

   b- *Type_Par* : qui représente le type d'un paramètre de la caractéristique. Ce champ peut être modélisé par une chaîne de caractères.

3- *Valeur* : qui représente la formule de calcul de la valeur réelle de cette caractéristique. A ce niveau, ce champ est modélisé par une chaîne de caractères.

4- *Invariant* : qui représente une contrainte sur la caractéristique (exemple: la valeur de la disponibilité est toujours positive). A ce niveau, ce champ peut être modélisé par une chaîne de caractères.

5- *Domaine* : qui représente le domaine de la caractéristique. Ce champ peut être modélisé par un enregistrement composé de trois champs :

**5.1**- *Direction* : qui modélise la direction (increasing ou decreasing) de la caractéristique. Ce champ doit être de type énuméré (enum{ *increasing*, *decreasing*})

**5.2-** *Dom* : qui modélise l'ensemble de valeurs possibles de la caractéristique. Ce champ doit être de type :

**A-** *enum*{ *numeric_real*, *numeric_integer*} si le domaine est quantitatif c'est-à-dire si les valeurs de la caractéristique sont de type numérique

**B-** *set*{*string*} si le domaine est ordinaire c'est-à-dire si les valeurs de la caractéristique sont de type non numérique (exemple : {*Good, Medium*})

**5.3**- *Unite* : qui modélise l'unité de la caractéristique si elle existe. Ce champ peut être de type chaîne de caractères.

Nous avons constaté que les caractéristiques numériques et celles ordinaires ne peuvent pas être décrites par un même type de propriétés. Pour cela, nous avons proposé de formaliser deux types de propriétés. Le premier «*CaracteristiqueNumerique*» (cf. Figure 5.11) pour les caractéristiques numériques et le type «*CaracteristiqueOrdinaire*» (cf. Figure 5.12) pour celles de type ordinaire.

```
property type CaracteristiqueNumerique = record
       [Nom       : string;
      Parametres: sequence <record [Nom_Par : string;
                                   Type_Par: string;]>;
      Domaine   : record
               [ direction: enum {increasing, decreasing};
               dom:    enum{numeric_real, numeric_integer};
               unite:     string;  ];
      Valeur   : string ;
              Invar: string ;  ];
```

**Figure 5.11 :** Formalisation du concept «CaracteristiqueNumerique» en Acme/Armani

```
Property type CaracteristiqueOrdinaire = record
       [Nom     : string;
      Parametres: sequence <record [Nom_Par : string;
                                   Type_Par: string;]>;
      Domaine  : record
                  [ direction: enum {increasing, decreasing};
                  dom:       set{ string };
                  unite:     string ;  ];
      Valeur: string ;
      Invar: string ;  ];
```

**Figure 5.12 :** Formalisation du concept «CaracteristiqueOrdinaire» en Acme/Armani

❖ **Exemple :**

La Figure 5.13 représente la formalisation en Acme/Armani d'une caractéristique de qualité appelée *TempsDeReponse*. Cette caractéristique est de type quantitatif, elle permet de calculer le temps de réponse d'un composant. L'ensemble de valeurs possibles de cette caractéristique est l'ensemble des entiers dont la variance (ou direction) est croissante. Cette caractéristique est exprimée en milliseconde.

La modélisation de trois champs *Parametres*, *Valeur* et *Invariant* peut être renforcée par des contraintes Armani décrites au sein de la configuration (*system*) qui dérive de notre style *CUML*. En effet, une contrainte Armani peut être attachée au champ *Type_Par*. Une telle contrainte stipule que ce champ désigne un type dans l'assemblage de composants UML2.0 considéré. En outre, une formule décrite en Armani peut être attachée au champ *Valeur* montrant comment calculer la valeur d'une caractéristique de qualité. Enfin, une propriété invariante peut être attachée au champ *Invar* moyennant une approche de traduction OCL – cas de CQML– vers Armani.

```
Property QoSChar : CaracteristiqueNumerique =
      [Nom       = "TempsDeReponse";
      Parametres=  <>;
       [ Domaine =
              [ direction = decreasing;
               dom =  numeric_integer;
               unite ="milliseconde"; ];
      Valeur   = "" ;
      Invar    =""  ;];
```

**Figure 5.13 :** Formalisation de la caractéristique « *TempsDeReponse* » en Acme/Armani

### 5.3.1.2 Formalisation du concept «qualite»

Une qualité (ou QoS) spécifie un ensemble de PNF proposées par un composant. Chaque PNF représente une restriction du domaine d'une caractéristique de qualité. Cette restriction est généralement de la forme suivante : "*CaractéristiqueNF Operateur Valeur*" avec *CaractéristiqueNF* correspond à une caractéristique de qualité, *Operateur* est un simple opérateur de comparaison (<, <=, > ou >=) et *Valeur* correspond à la valeur permettant de restreindre le domaine de la caractéristique. Une qualité peut avoir un nom et des paramètres.

On peut formaliser une qualité par une propriété Acme. Cette propriété doit être de type enregistrement composé de quatre champs :

**1-** *Nom* : qui représente le nom de la qualité (GoodAvailability, MediumAvailability…). Ce champ peut être modélisé par une chaîne de caractères.

**2-** *Parametres* : qui représente les paramètres de la qualité. Ce champ peut être modélisé par une séquence d'éléments de type *Parametre*. A ce niveau, ce dernier peut être modélisé par un enregistrement composé de deux champs :

a- *Nom_Par* : qui représente le nom du paramètre. Ce champ peut être modélisé par une chaîne de caractères,

b- *Type_Par* : qui représente le type du paramètre. Ce champ peut être modélisé par une chaîne de caractères.

**3-** *SetPNFNum* : qui correspond à l'ensemble des PNF de type numérique. Ce champ peut être modélisé par un ensemble d'éléments (PNF) de type enregistrement composé de trois champs :

**3**.1- *CaractéristiqueNF* : qui modélise la caractéristique d'une PNF de cette qualité. Ce champ doit être de type «CaracteristiqueNumerique»,

**3.2-** *Operateur* : qui modélise l'opérateur de la contrainte appliquée sur cette caractéristique. Ce champ doit être de type enum {<, <=, >, >= },

**3.3-** *Valeur* : qui modélise la valeur permettant de restreindre la caractéristique de cette PNF. Ce champ doit être de type réel puisque le type réel regroupe tous les types numériques.

**4-***SetPNFOrdi* : qui correspond à l'ensemble de PNF de type ordinaire. Ce champ peut être modélisé par un ensemble d'éléments de type enregistrement composé de trois champs :

**4.1-** *CaractéristiqueNF* : qui modélise la caractéristique d'une PNF de cette qualité. Ce champ doit être de type «*CaracteristiqueOrdinaire*»,

**4.2-** *Operateur* : qui modélise l'opérateur de la contrainte appliquée sur cette caractéristique. Ce champ doit être de type enum { <, <=, >, >= },

**4.3-** *Valeur* : qui modélise la valeur permettant de restreindre la caractéristique de cette PNF. Ce champ doit être de type chaîne de caractères.

La Figure 5.14 montre la formalisation en Acme/Armani du type propriété «*Qualite*» qui représente un type de base des qualités formalisées en Acme/Armani.

```
property type Qualite = record [
     Nom : string ;
     Parametres : sequence <record [Nom_Par : string; Type_Par: string;]>;
     SetPNFNum : set { PNFNum };
     SetPNFOrdi : set { PNFOrdi };
     ];
property type PNFNum =record
               [CaracteristiqueNF : CaracteristiqueNumerique;
                Operateur : OperComparaison;
                Valeur : float; ];
property type PNFOrdi =record
              [CaracteristiqueNF : CaracteristiqueOrdinaire ;
               Operateur :OperComparaison ;
               Valeur : string; ];
property type OperComparaison= enum
                {Inferieur, InferieurOuEgal,  Superieur, SuperieurOuEgal};
```

**Figure 5.14 :** Formalisation du concept «Qualite» en Acme/Armani

❖ **Exemple :**

La Figure 5.15 représente la formalisation en Acme/Armani d'une qualité appelée Fiable. Le corps de cette qualité est spécifié par les deux propriétés non fonctionnelles suivantes :

- *PNF1* : le temps moyen de réparation (MTTR) est inférieur à 20 minutes,

- *PNF2* : le temps moyen entre deux défaillances consécutives (MTBF) est supérieur ou égal à 15 heures.

```
property  QoS : Qualite = [
     Nom ="Fiable";
     Parametres = < >;
    SetPNFNum ={
// PNF1 : MTTR < 20 minutes
   [ CaracteristiqueNF =
               [Nom = "MTTR";
          Parametres=< >;
          Domaine = [ direction = decreasing;
                      dom =  numeric_integer;
                      unite ="minute"; ];
          Valeur   = "" ;
          Invar    =""  ;];
       Operateur = Inferieur ;
       Valeur = 20 ; ] ,
// PNF2 : MTBF >= 15 heures
        [ CaracteristiqueNF =
               [Nom = "MTBF";
          Parametres=< >;
          Domaine = [ direction = increasing;
                      dom =  numeric_integer;
                      unite ="heure"; ];
          Valeur   = "" ;
          Invar    =""  ;];
       Operateur = SuperieurOuEgal ;
       Valeur = 15 ; ]
 };
     SetPNFOrdi ={ };
     ];
```

**Figure 5.15 :** Formalisation de la qualité Fiable en Acme/Armani

### 5.3.1.3 Formalisation du concept « *profile* »

Un composant peut avoir plusieurs qualités qui peuvent être requises et/ou offertes. De la même façon que QML et CQML, nous avons proposé de regrouper les qualités d'un composant dans un *profil*. Ce *profil* peut être modélisé en Acme par une propriété de type enregistrement composé de deux champs :

**1-** *QualitesExigees* **:** qui représente l'ensemble des qualités exigées par un composant. Ce champ doit être de type ensemble de propriétés de type «*Qualite*»,

**2-** *QualitesFournies* **:** qui représente l'ensemble des qualités fournies par un composant. Ce champ doit être de type ensemble de propriétés de type «*Qualite*».

La Figure 5.16 montre la formalisation en Acme/Armani du concept *profil* par un type de propriété.

```
property type Profile = record [
     QualitesExigees     : set { Qualite };
     QualitesFournies    : set { Qualite };
          ];
```

**Figure 5.16 :** Formalisation du concept *profile* en Acme

❖ **Exemple :**

Soit le composant UML2.0 *VideoPlayer* présenté par la Figure 5.17. Ce composant est enrichi par les propriétés non fonctionnelles suivantes :

- Le composant *VideoPlayer* exige la qualité Fiable (cf. Figure 5.15),
- Le composant *VideoPlayer* offre une bonne disponibilité (Disponibilité >= 85 %).

La formalisation en Acme/Armani de cette spécification est présentée par la Figure 5.18. Cette formalisation réutilise les types définis précédemment.

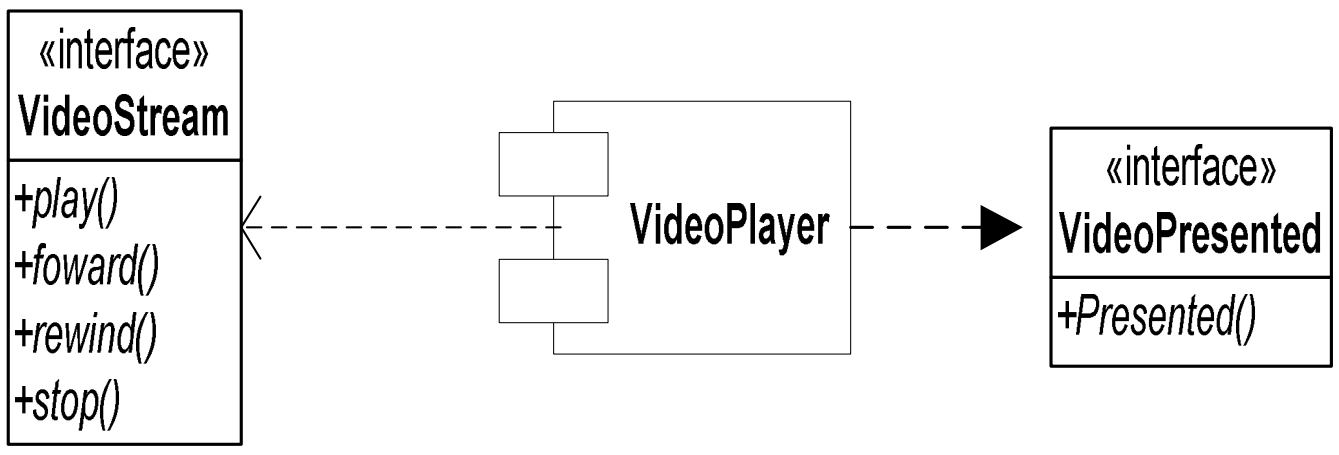

**Figure 5.17 :** Modélisation du composant *VideoPlayer* en UML2.0

```
  Component VideoPlayer: ComposantUML = new ComposantUML extended with {
//partie fonctionnelle
      Port VideoStream : InterfaceRequise = new InterfaceRequise extended with {
       property services_requis = {
               [nom_sp="Play"; p=<>; resultat= Void_UML;],
               [nom_sp="Forward"; p=<>; resultat= Void_UML;],
               [nom_sp="Rewind"; p=<>; resultat= Void_UML;],
               [nom_sp="Stop"; p=<>; resultat= Void_UML;]
       };
     } ;
   Port VideoPresented : InterfaceOfferte = new InterfaceOfferte extended with {
       property services_offerts = {
              [nom_sp="Presented"; p=<>; resultat= Void_UML;]
         };
  } ;
//partie non fonctionnelle
property Prof1 : Profile = [
//spécifiquation de l'ensemble des qualités requises
  QualitesExigees= {
     [Nom ="Fiable";
     Parametres = < >;
     SetPNFNum ={
// PNF1 : MTTR < 20 minutes
   [ CaracteristiqueNF =
          [Nom = "MTTR";
                    Parametres=< >;
                     Domaine = [ direction = decreasing;
                                          dom =  numeric_integer;
                                         unite ="minute"; ];
                      Valeur   = "" ;
                      Invar    =""  ;];
       Operateur = Inferieur ;
       Valeur = 20 ; ] ,
// PNF2 : MTBF >= 15 heures
      [ CaracteristiqueNF =
           [Nom = "MTBF";
                    Parametres=< >;
                    Domaine = [ direction = increasing;
                                        dom =  numeric_integer;
                                         unite ="heure"; ];
                     Valeur   = "" ;
                     Invar    =""  ;];
       Operateur = SuperieurOuEgal ;
       Valeur = 15 ; ]
 };
     SetPNFOrdi ={ };
     ]
};//fin de l'ensemble des qualités exigées par le composant VideoPlayer
//spécification de l'ensemble des qualités offertes
   QualitesFournies= {
            [ Nom ="BonneDisponibilité";
```

```
            Parametres = < >;
            SetPNFNum ={
// PNF3 : Disponibilité  >= 25 %
              [ CaracteristiqueNF =
              [Nom = "Disponibilité";
                     Parametres= < >;
                     Domaine = [ direction = increasing;
                                         dom =  numeric_real;
                                         unite ="%"; ];
                    Valeur   = " " ;
                    Invar    =""  ;];
          Operateur = SuperieurOuEgal;
          Valeur = 85 ; ]
    };
     SetPNFOrdi ={ };
     ] //fin de la qualité BonneDisponibilité
};//fin de l’ensemble des qualités offertes par le composant
 ];//fin du profile
}// Fin du composant VideoPlayer
```

**Figure 5.18 :** Formalisation du composant *VideoPlayer* en Acme/Armani

## 5.4 Vérification des contrats de qualité d’assemblages des composants UML2.0 en Acme/Armani

Afin de vérifier la cohérence de qualité d’un assemblage de composants UML2.0, nous proposons le contrat de qualité *CQualite* défini d’une façon informelle.

- *CQualite* **:** toutes les qualités exigées par un composant doivent être assurées par les composants connectés à ce dernier. Une Qualité requise *QRequise* est assurée par un composant *C* si et seulement si ce dernier propose une qualité offerte *QOfferte* répondant à la qualité requise *QRequise*. Une qualité *QRequise* est satisfaite par une qualité offerte *QOfferte* si et seulement si toutes les PNF formant la qualité *QRequise* sont assurées par celles formant la qualité *QOfferte*.

La Figure 5.19 montre une formalisation de cette contrainte par un invariant Armani.

```
rule CQualite = invariant
//lignes A1, B1 et C1 traitent les qualités exigées par chaque composant
//considéré : Comp1
 forall Comp1: ComposantUML in self.Components|                      //A1
     forall prof1:Profile in  Comp1.properties|                      //B1
        forall QoSExigee:Qualite in prof1.QualitesExigees|           //C1
//lignes A2, B2 et C2 ont pour objectif de vérifier s’il existe un composant Comp2
//attaché à Comp1 offrant la qualité requise par Comp1
   exists Comp2: ComposantUML in {select C: ComposantUML in self.Components|
   connected (Comp1, C) } |                                        //A2
   exists prof2:Profile in Comp2.properties|                       //B2
         exists QoSFournie:Qualite in prof2.QualitesFournies|      //C2
//toutes les PNF numériques de  la qualité  exigée sont assurées par celles de la
//qualité fournie
  ( forall PNF1 :PNFNum in QoSExigee.SetPNFNum|
      exists PNF2 :PNFNum in QoSFournie.SetPNFNum|
       PNF1.CaracteristiqueNF == PNF2.CaracteristiqueNF and
       PNF1.Operateur == PNF2.Operateur and
       ( (PNF1.Operateur == SuperieurOuEgal -> PNF2.Valeur >= PNF1.Valeur) and
         (PNF1.Operateur == Superieur -> PNF2.Valeur > PNF1.Valeur) and
        (PNF1.Operateur == InferieurOuEgal -> PNF2.Valeur <= PNF1.Valeur) and
        (PNF1.Operateur == Inferieur -> PNF2.Valeur < PNF1.Valeur)      )   )
  and
 //toutes les PNF ordinaires de  la qualité  exigée sont assurées par celles de la
```

```
 //qualité fournie
  ( forall PNF1 :PNFOrdi in QoSExigee.SetPNFOrdi|
  exists PNF2 :PNFOrdi in QoSFournie.SetPNFOrdi|
     PNF1.CaracteristiqueNF  == PNF2.CaracteristiqueNF and
     PNF1.Operateur==PNF2.Operateur
// la vérification des valeurs  des caractéristiques ordinaires peuvent être
//spécifiées au niveau du système
)  ;
```

**Figure 5.19 :** Formalisation du contrat de qualité en Acme/Armani

## 5.5 De CQML vers Acme/Armani

Dans cette section, nous allons proposer des règles permettant la traduction d'une description non fonctionnelle décrite en CQML vers Acme/Armani en réutilisant notre style CUML. Les règles de traduction proposées sont simples car CUML intègre les principaux concepts venant de CQML tels que : *quality_characteristic*, *quality* et *profile*. Tout au long de cette section, nous fournissons des exemples illustratifs qui accompagnent les règles de traduction proposées.

### 5.5.1 Traduction du concept *quality_characteristic*

Une caractéristique CQML est identifiée par un **nom**, des **paramètres**, une clause **Values**, une clause **Invariant** et un **domaine**.

On peut formaliser une caractéristique CQML par une propriété Acme de type:

A- «*CaracteristiqueNumerique*» si le domaine de la caractéristique est numérique (**numeric** en CQML),

B- «*CaracteristiqueOrdinaire*» si le domaine de la caractéristique est non numérique (**enum** ou **set** en CQML).

La correspondance entre chaque clause d'une caractéristique CQML et les champs de ces types de propriétés Acme est présentée par:

i. le champ *Nom* de la propriété prend le nom de la caractéristique CQML,

ii. le champ *Parametre*s de la propriété prend l'ensemble des paramètres de la caractéristique CQML. Les champs *Nom_Par* et *Type_Par* de chaque paramètre prennent respectivement le nom et le type de chaque paramètre CQML,

iii. le champ *Domaine* de la propriété prend la clause *Domain* de la caractéristique CQML tout en conservant l'ensemble des valeurs possibles, la direction et l'unité,

iv. le champ *Valeur* prend la clause *Values* de la caractéristique CQML,

v. le champ *Invar* prend la clause *Invariant* de la caractéristique CQML.

❖ **Exemple :**

La Figure 5.20 (source [Aagedal, 2001]) présente une modélisation CQML d'une caractéristique ordinaire appelée «*Resolution*». La traduction en Acme de cette caractéristique est présentée par la Figure 5.21. Cette formalisation est obtenue par une propriété de type «*CaractéristiqueOrdinaire*».

```
quality_characteristic Resolution (flow: Video) = {
doamin: increasing enum {160x120, 176x144, 320x240} pixel;
values: flow.DateReceiver.putDataUnit.SE-> Last.unit.resolution;
}
```

**Figure 5.20 :** Modélisation de la caractéristique «Resolution» en CQML

## 5.5.2 Traduction du concept *quality*

Une qualité CQML est identifiée par un **nom**, des **paramètres**, un **ensemble de PNF numériques** et un **ensemble de PNF ordinaires**. Chaque PNF est modélisée *par une contrainte de la forme suivante : «quality_characteristic* Opérateur Valeur» avec quality_characteristic qui correspond à une caractéristique non fonctionnelle, Opérateur est un simple opérateur de comparaison (<, >, >= ou >=) et Valeur correspond à la valeur permettant de restreindre le domaine de la caractéristique.

```
Property QoSChar1 : CaracteristiqueOrdinaire =
  [Nom       = "Resolution";
   Parametres = <[Nom_Par = "flow"; Type_Par= "Video";]>;
   Domaine  = [    direction = increasing ;
                       dom= {"160x120", "176x144", "320x240"};
                   unite = "pixel"; ];
   Valeur    = "flow.DateReceiver.putDataUnit.SE->Last.unit.resolution";
   Invar     = " "; ] ;
```

**Figure 5.21 :** Traduction de la caractéristique «Resolution» en Acme/Armani

On peut formaliser une qualité CQML par une propriété Acme de type «*Quality*». La correspondance entre chaque clause d'une qualité CQML et les champs de ce type de propriétés Acme est présentée sous forme de quatre règles :

i. le champ *Nom* de la propriété prend le nom de la qualité CQML,

ii. le champ *Parametres* de la propriété prend l'ensemble des paramètres de la qualité CQML. Les champs *Nom_Par* et *Type_Par* de chaque paramètre prennent respectivement le nom et le type de chaque paramètre CQML,

iii. le champ *SetPNFNum* prend l'ensemble des PNF numériques de la qualité CQML. Chaque élément de ce champ est composé de:

   A- *CaracteristiqueNF* : matérialisant la caractéristique de la PNF numérique, B- *Operateur* : matérialisant l'opérateur de la PNF numérique, C- *Valeur* : matérialisant la valeur de la PNF numérique.

iv. le champ SetPNFOrdi prend l'ensemble des PNF ordinaires de la qualité CQML. Chaque élément de ce champ est composé de :

   A- *CaracteristiqueNF* : matérialisant la caractéristique de la PNF ordinaire, B- *Operateur* : matérialisant l'opérateur de la PNF ordinaire, C- *Valeur* : matérialisant la valeur de la PNF ordinaire.

❖ **Exemple :**

La Figure 5.22 (source [Aagedal, 2001]) présente une modélisation CQML d'une qualité permettant de restreindre une caractéristique numérique. La traduction en Acme de cette qualité est présentée par la Figure 5.23. Cette formalisation est obtenue par une propriété de type «*Qualite*».

```
quality Fast ( flow : Flow ; initiator :Operation)
{      startUpTime ( flow, initiator.SR->last) <= 10 ;
 }
quality_ characteristic startUpTime  (flow : Flow ; initiatingEvent : Event)
{ domain : deceasing numeric integer milliseconds ;
   values : if flow.SE-> isEmpty then invalid
              else flow.SE->first.time ()-flow.initiate.time ()
                   endif ;
   invariant: flow.initiate= initiatingEvent ;
}
```

**Figure 5.22 :** Modélisation de la qualité Fast en CQML

## 5.5.3 Traduction du concept *profile*

Un profil CQML est identifié par ses deux clauses *Uses* et *Provides***.**

On peut formaliser un profil CQML par une propriété Acme de type «*Profile*». La correspondance entre chaque clause d'une qualité CQML et les champs de ce type de propriétés Acme est présentée par:

i. le champ QualitesExigees prend la clause Uses du profil CQML,

ii. le champ QualitesOffertes prend la clause Provides du profil CQML.

**- Exemple :**

La Figure 5.24 présente une modélisation CQML d'un profil qui modélise les qualités d'un composant appelé *Client*. La traduction en Acme de ce profil est présentée par la Figure 5.25. Cette formalisation est obtenue par une propriété de type «*Profile*».

```
property  QoS : Qualite = [
     Nom ="Fast";
     Parametres = <[Nom_Par = "flow"; Type_Par= "Flow";],
                              [Nom_Par = "initiator"; Type_Par= "Operation";] >;
    SetPNFNum ={
// PNF : startUpTime ( flow, initiator.SR->last) <= 10 ;
   [ CaracteristiqueNF =
               [Nom = " startUpTime";
          Parametres=<[Nom_Par="flow"; Type_Par= "Flow";]>;
          Domaine = [ direction = decreasing;
                      dom =  numeric_integer;
                      unite ="milliseconds"; ];
          Valeur   = " if flow.SE-> isEmpty then invalid
                       else flow.SE->first.time () - flow.initiate.time ()
                        endif " ;
          Invar    ="flow.initiate= initiatingEvent ";];
       Operateur = InferieurOuEgale ;
       Valeur = 10 ; ]
};
     SetPNFOrdi = { };
 ];
```

**Figure 5.23 :** Traduction de la qualité Fast en Acme/Armani

```
profile P1 for Client {
Uses  Fiable         ;
Provides  Performant  ;
} ;
```

```
quality Fiable
{  Fiabilite >= 90 ;
 };
quality_characterisique Fiabilite
{  Domain : increasing numeric integer  % ;
 };
quality Performant
{  MTBF >= 50 ;
   MTTR <= 30 ;
 };
quality_characterisique MTBF
{  Domain : increasing numeric integer  Jour ;
 };
quality_characterisique MTTR
{  Domain : decreasing numeric integer  Heure ;
 };
```

**Figure 5.24 :** Modélisation d'un *profil* CQML attaché à un composant UML2.0

```
Component Client : ComposantUML = new ComposantUML extended with {
//spécification de la partie fonctionnelle
//spécification de la partie non fonctionnelle
property P1 : Profile = [
//spécifiquation de l'ensemble des qualités requises
  QualitesExigees= {
            [ Nom ="Fiable";
             Parametres = < >;
             SetPNFNum ={
// Fiable : Fiabilité >=90 %
               [ CaracteristiqueNF =
               [Nom = "Fiable";
          Parametres=< >;
          Domaine = [ direction = increasing;
                      dom =  numeric_integer;
                      unite ="%"; ];
          Valeur   = "" ;
          Invar    =""  ;];
           Operateur = Inferieur ;

           Valeur = 90 ; ]  };
  SetPNFOrdi ={ };
     ]  //fin de la qualité Fiable
};//fin de l'ensemble des qualités exigées par le composant Client

//spécification de l'ensemble des qualités offertes
   QualitesFournies= {
            [ Nom ="Performant";
             Parametres = < >;
             SetPNFNum ={
// PNF1 : MTBF  >= 50 jours
               [ CaracteristiqueNF =
               [Nom = "MTBF";
          Parametres= < >;
          Domaine = [ direction = increasing;
                      dom =  numeric_integer;
                      unite ="Jour"; ];
          Valeur   = " " ;
          Invar    =""  ;];
           Operateur = SuperieurOuEgal;
           Valeur = 50 ; ] ,
//PNF2 : MTTR <= 30 heures
               [ CaracteristiqueNF =
               [Nom = "MTTR";
          Parametres= < >;
          Domaine = [ direction = decreasing;
                      dom =  numeric_integer;
```

```
                       unite ="Heur"; ];
           Valeur   = " " ;
           Invar    =""  ;];
            Operateur = InferieurOuEgal;
            Valeur = 30 ; ]
     };
      SetPNFOrdi ={ };
      ] //fin de la qualité Performant
};//fin de l'ensemble des qualités offertes par le composant
 ];//fin du profile
};// fin du composant
```

**Figure 5.25 :** Traduction du profil attaché au composant Client en Acme/Armani

## 5.6 Conclusion

Dans ce chapitre, nous avons modélisé le système *VideoCamera* en UML2.0/CQML. Les composants retenus de l'application sont modélisés en UML2.0. Chaque composant proposé offre et/ou exige des interfaces. Après avoir identifié les PNF souhaitées pour cette application, nous avons modélisé ces PNF en CQML en utilisant les constructions *quality_characteristic*, *quality* et *profile*. Nous avons proposé une approche permettant de formaliser une description architecturale UML2.0 dotée de PNF (décrites en CQML) en Acme/Armani afin de vérifier sa cohérence : chaque PNF exigée doit avoir sa réciproque (PNF offerte) dans l'assemblage des composants. Par la suite nous avons utilisé judicieusement les possibilités de typage et d'expression des contraintes fournies par Acme/Armani afin de formaliser les principaux concepts décrivant des PNF : caractéristique de qualité, qualité et profil. En outre, nous avons établi un contrat de qualité formalisé sous forme d'un invariant Acme/Armani permettant de vérifier la cohérence de qualité (non fonctionnelle) d'un assemblage de composants UML2.0. Des exemples illustratifs exhibant les intérêts de notre formalisation des PNF en Acme/Armani ont été fournis.

Nous avons proposé des règles simples permettant de traduire des expressions CQML vers Acme/Armani en passant par notre style *CUML*. Ces règles concernent la traduction en Acme/Armani des concepts *quality_characteristic*, *quality* et *profile* issus du langage CQML. Des exemples illustratifs montrant l'application des règles proposées ont été fournis dans ce chapitre.

Dans le chapitre suivant, nous allons proposer notre contribution à l'outil Wr2fdr qui accompagne l'ADL Wright. Un tel outil permet de traduire une spécification Wright en CSP Hoare acceptable par le model checker FDR afin de vérifier des propriétés standards telles que la cohérence des composants, des connecteurs et la compatibilité port/rôle. Notre contribution consiste à maintenir l'outil Wr2fdr aussi bien sur le plan correctif qu'évolutif.

# Chapitre 6 : Maintenance corrective et évolutive de l'outil Wr2fdr

## 6.1 Introduction

Les auteurs de Wright proposent un outil appelé Wr2fdr [Wr2fdr, 2005] permettant d'automatiser les quatre propriétés décrites dans le chapitre 2 (cf. section 2.5.4): la propriété 1 (cohérence des ports avec le Calcul), la propriété 2 (absence d'interblocage sur les connecteurs), la propriété 3 (absence d'interblocage sur les rôles) et la propriété 8 (compatibilité port/rôle). Pour y parvenir, l'outil Wr2fdr traduit une spécification Wright en une spécification CSP dotée des relations de raffinement à vérifier. La spécification CSP engendrée pour l'outil Wr2fdr est soumise à l'outil de Model checking FDR (Failure-Divergence Refinement) [FDR2, 2003]. Cependant la version actuelle Wr2fdr comporte des erreurs et elle est limitée en possibilités. En effet, suite à l'utilisation de l'outil Wr2fdr, nous avons remarqué que l'outil génère des erreurs liées aux propriétés 2 et 3. En plus, les propriétés 1 et 8 ne sont pas traitées par cette version de l'outil. Vu l'importance de cet outil, nous avons contacté les auteurs de Wright, expliqué les problèmes rencontrés et récupéré le source de cet outil pour une tâche de maintenance corrective et évolutive.

Ce chapitre comporte cinq sections. La première section présente les fonctionnalités souhaitées de l'outil Wr2fdr. La deuxième section propose une évaluation de l'outil Wr2fdr vis-à-vis des fonctionnalités présentées dans la première section en suivant une approche de vérfication orientée tests syntaxiques. La troisième section aborde les caractéristiques techniques de l'outil Wr2fdr. La quatrième section propose une correction des anomalies détectées lors de l'utilisation de Wr2fdr. Enfin, la cinquième section enrichit l'outil Wr2fdr par un analyseur de la sémantique statique de Wright.

## 6.2 Fonctionnalités souhaitées de l'outil Wr2fdr

Wr2fdr est un outil développé par l'université de Carnegie Mellon et il accompagne l'ADL Wright. Il permet de traduire une spécification Wright en une spécification CSP acceptée par l'outil FDR. L'outil Wr2fdr est censé assurer les fonctionnalités suivantes :

- Analyse lexico-syntaxique d'une spécification Wright,
- Génération de code CSP,
- Correspondances entre les événements locaux de Wright et les événements globaux de CSP,
- Déterminisation d'un processus CSP : det(P). Ceci permet de traiter l'opération non déterministe (Π) de CSP,
- Calcul de l'alphabet d'un processus CSP : αP. En effet, FDR exige explicitement lors de la composition parallèle des processus (||) leurs alphabets,
- Calcul des relations de raffinement liées aux propriétés 1, 2, 3 et 8 permettant de vérifier respectivement la cohérence Port/Calcul, l'absence d'interblocage sur les connecteurs, l'absence d'interblocage sur les rôles et la compatibilité port/rôle (cf. 2.5.3.1).

La version actuelle de l'outil Wr2fdr ne fait pas la distinction entre les événements initialisés et observés. De plus, ces événements ne doivent pas porter des informations ni d'entrée ni de sortie.

## 6.3 Vérification de l'outil Wr2fdr

L'outil Wr2fdr (cf. Figure 6.1) accepte en entrée un fichier contenant une spécification Wright et produit en sortie un fichier contenant une spécification CSP acceptable par l'outil de model-checking FDR afin de vérifier les propriétés 1, 2, 3 et 8 (cf. 2.5.3.1). En effet, l'outil Wr2fdr est censé automatiser ces propriétés en utilisant le concept de raffinement de CSP.

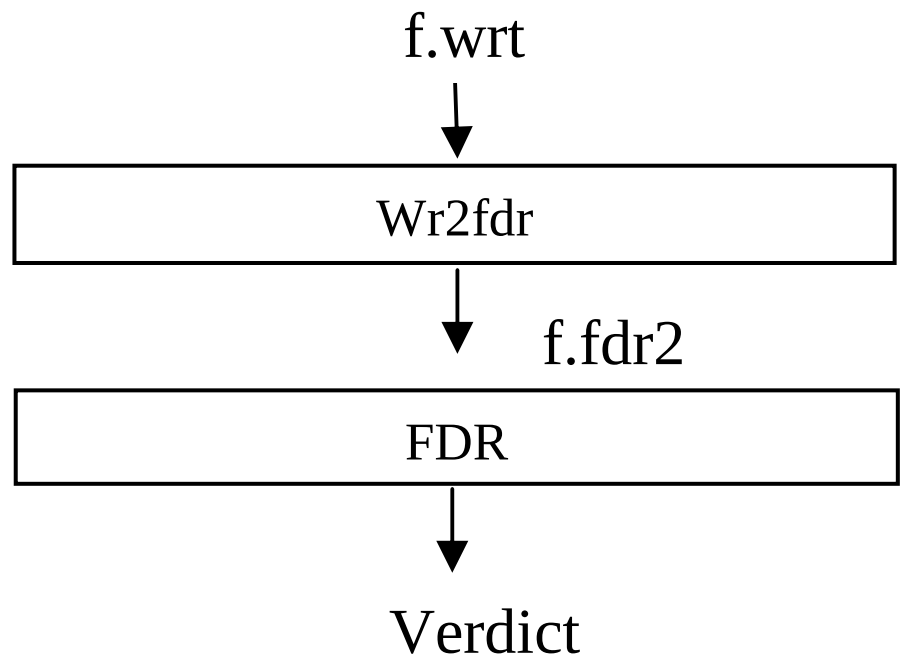

**Figure 6.1 :** Fonctionnement de l'outil Wr2fdr

En menant une activité de test fonctionnel orientée tests syntaxiques, nous avons constaté des écarts entre la spécification (cf. 6.1) et l'implémentation de l'outil Wr2fdr. Ainsi, nous pouvons dire que l'implémentation de l'outil Wr2fdr n'est pas conforme à sa spécification. En fait, l'outil Wr2fdr peut produire des spécifications CSP non acceptables par FDR. En outre, il peut s'arrêter brutalement en signalant une erreur à l'exécution.

Dans la suite, nous allons détailler les défaillances détectées lors du test de l'outil Wr2fdr. Ces défaillances concernent principalement le calcul des relations de raffinement liées aux propriétés 1, 2, 3 et 8.

### 6.3.1 Défaillances liées à la cohérence du connecteur

En Wright, la cohérence d'un connecteur est obtenue par la vérification des deux propriétés 2 et 3. Pour tester le comportement de l'outil Wr2fdr vis-à-vis de ces deux propriétés nous avons soumis l'entrée appelée *PipeConn.wrt* fournie par la Figure 6.2.

```
Style PipeConn
        Connector Pipe
                Role Writer = write -> Writer |~| close -> TICK
                Role Reader = DoRead |~| ExitOnly
                where {
                DoRead = read -> Reader [] readEOF -> ExitOnly
                ExitOnly = close -> TICK
        }
        Glue = Writer.write -> Glue [] Reader.read -> Glue
                [] Writer.close -> ReadOnly [] Reader.close -> WriteOnly
                where {
                ReadOnly = Reader.read -> ReadOnly
                [] Reader.readEOF -> Reader.close -> TICK
                [] Reader.close -> TICK
                WriteOnly = Writer.write -> WriteOnly [] Writer.close -> TICK
        }
        Constraints
        // no constraints
End Style
```

**Figure 6.2 :** Cas de test pour les propriétés 2 et 3

Puisque la construction syntaxique est un style comportant un seul connecteur Pipe, il est évident que les propriétés à générer seront seulement les propriétés 2 et 3.

L'outil Wr2fdr génère la spécification CSP PipeConn.wrt (cf. Figure 6.3). Lors de la vérification des trois relations de raffinement signalées par assert, l'outil FDR rencontre des problèmes visiblement d'ordre syntaxique (cf. Figure 6.4). Un examen du fichier *PipeConn.fdr2* montre que les identificateurs coloriés (ou marqués) ne sont pas définis.

```
- FDR compression functions
    transparent diamond
    transparent normalise
-- Wright defined processes
    channel abstractEvent
    DFA = abstractEvent -> DFA |~| SKIP
    quant_semi({},_) = SKIP
    quant_semi(S,PARAM) = |~| i:S @ PARAM(i) ; quant_semi(diff(S,{i}),PARAM)
    power_set({}) = {{}}
    power_set(S) = { union(y,{x}) | x <- S, y <- power_set(diff(S,{x}))}
-- Style PipeConn
-- Type declarations
-- events for abstract specification
     channel readEOF, read, close, write
-- Connector Pipe
-- generated definitions (to split long sets)
    ALPHA_Pipe = {|Reader.readEOF, Reader.read, Reader.close, Writer.write , Writer.close|}
ReadOnly = ((Reader.read -> ReadOnly) [] ((Reader.readEOF -> (Reader.close -> SKIP)) [] (Reader.close ->
SKIP))) WriteOnly = ((Writer.write -> WriteOnly) [] (Writer.close -> SKIP))
Glue = ((Writer.write -> Glue) [] ((Reader.read -> Glue) [] ((Writer.close
-> ReadOnly) [] (Reader.close -> WriteOnly))))
-- Rôle Writer
     ALPHA_Writer = {close, write}
     ROLEWriter = ((write -> Writer) |~| (close -> SKIP))
     WriterA = ROLEWriter [[ x <- abstractEvent | x <- ALPHA_Writer ]]
      assert DFA [FD= WriterA
-- Rôle Reader
     ALPHA_Reader = {readEOF, read, close}
     DoRead = ((read -> Reader) [] (readEOF -> ExitOnly))
     ExitOnly = (close -> SKIP)
     ROLEReader = (DoRead |~| ExitOnly)
     ReaderA = ROLEReader [[ x <- abstractEvent | x <- ALPHA_Reader ]]
     assert DFA [FD= ReaderA
     channel Writer: {close, write}
     channel Reader: {readEOF, read, close}
     Pipe = ( (ROLEWriter[[ x <- Writer.x | x <- {close, write } ]]
     [| diff({|Writer|}, {}) |]
   (ROLEReader[[ x <- Reader.x | x <- {readEOF, read, close } ]]
       [| diff({|Reader|}, {}) |] Glue)) )
       PipeA = Pipe [[ x <- abstractEvent | x <- ALPHA_Glue ]]
         assert DFA [FD= PipeA
-- No constraints
-- End Style
```

**Figure 6.3 :** Fichier CSP PipeConn.fdr2

### 6.3.2 Défaillances liées à la propriété 1 : Cohérence Port/Calcul

Nous avons exécuté le programme Wr2fdr avec le cas de test fourni par la Figure 6.5. Un tel cas comporte un seul composant appelé Double et par conséquent la propriété visée est évidemment la propriété 1. L'exécution de Wr2fdr sur ce cas de test entraîne une erreur à l'exécution (cf. Figure 6.6) : erreur de segmentation traduisant souvent l'utilisation d'un pointeur qui pointe nulle part. Ceci est plausible car Wr2fdr est écrit en C++.

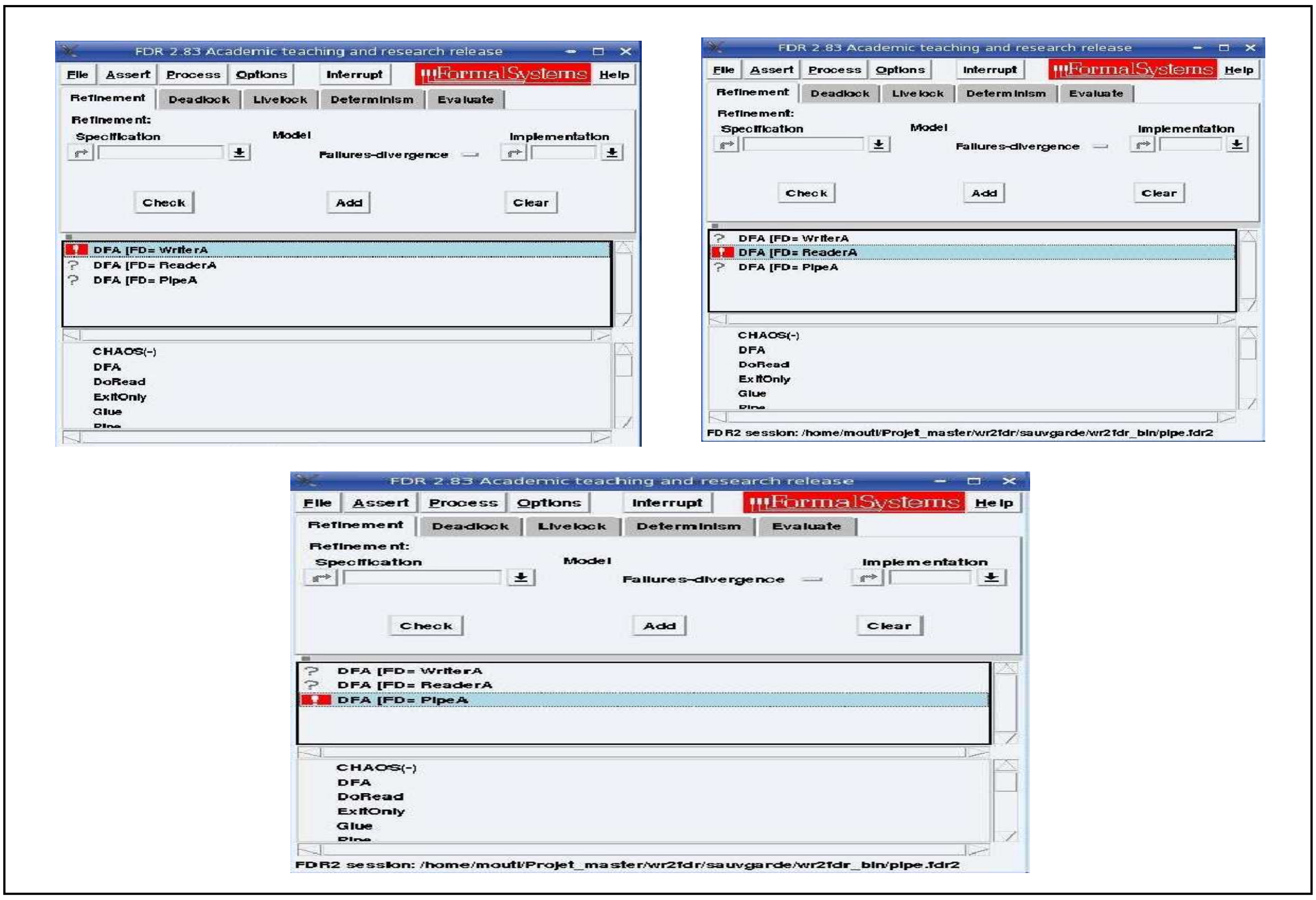

**Figure 6.4 :** Problèmes rencontrés par FDR

```
Style Double_Style
      Component Double
              Port In = read -> In [] close -> TICK
             Port Out = _write -> Out |~| _close -> TICK
             Computation = In.read -> _Out.write -> Computation [] In.close -> _Out.close -> TICK
       constraints
      //no constraints
End Style
```

**Figure 6.5 :** Cas de test pour la propriété 1

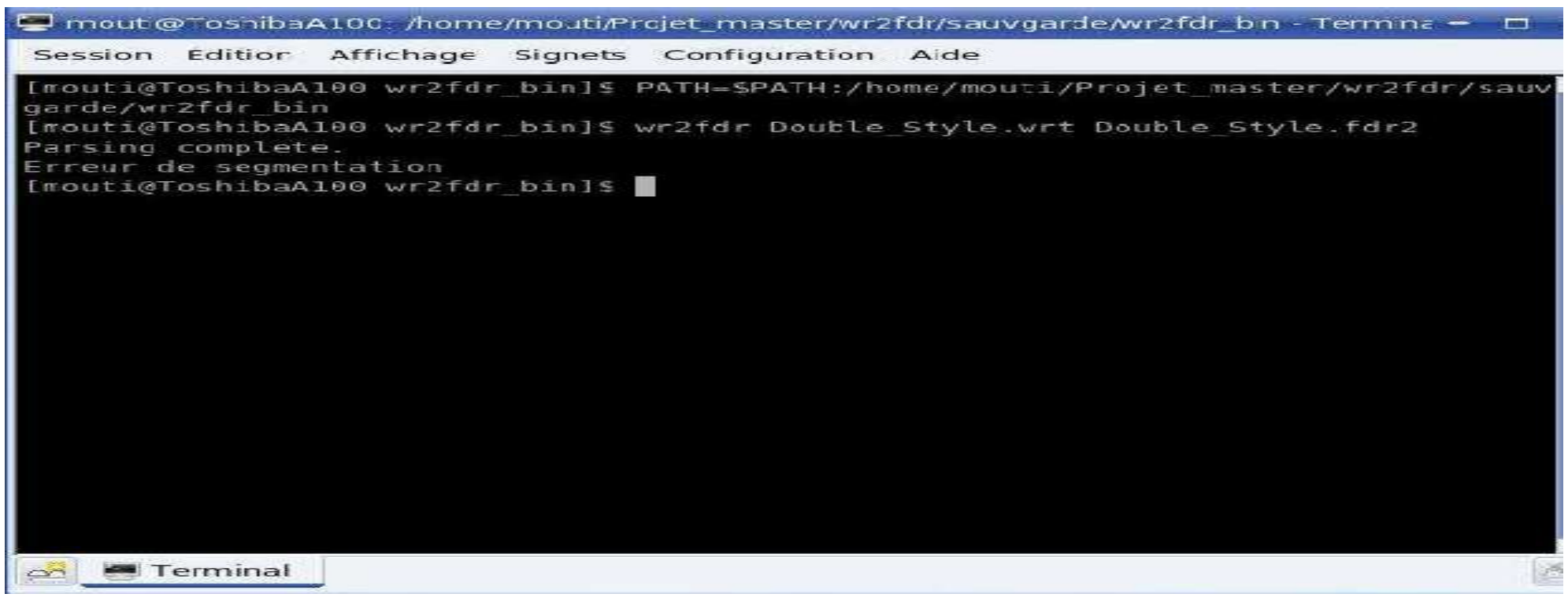

**Figure 6.6 :** Arrêt brutal de l'outil Wr2fdr

### 6.3.3 Défaillances liées à la propriété 8 : Compatibilité Port/Rôle

Pour pouvoir tester le comportement de l'outil Wr2fdr vis-à-vis de la propriété 8, il faut faire appel à la construction *Configuration* avec notamment les clauses *Instances* et *Attachments*. L'exécution du programme Wr2fdr avec le cas de test fourni par la Figure 6.7 entraîne le même arrêt brutal rencontré précédemment.

```
Configuration ABC
   Component Atype
       Port Output = _a -> Output |~| TICK
        Computation = _Output.a -> Computation |~| TICK
   Component Btype
       Port Input = c -> Input [] TICK
       Computation = Input.c -> _b -> Computation [] TICK
   connector Ctype
       Role Origin = _a -> Origin |~| TICK
       Role Target = c -> Target [] TICK
       Glue = Origin.a -> _Target.c -> Glue [] TICK
    Instances
      A : Atype
      B : Btype
      C : Ctype
   Attachments
     A.Output As C.Origin
     B.Input As C.Target
End Configuration
```

**Figure 6.7 :** Cas de test pour la propriété 8

## 6.4 Caractéristiques techniques de l'outil Wr2fdr

L'outil Wr2fdr est écrit en C++. Son code source est réparti physiquement sur plusieurs fichiers : trois fichiers «.hpp» et huit fichiers «.cpp». La complexité textuelle de l'outil Wr2fdr est de l'ordre de 16000 lignes C++. L'outil Wr2fdr englobe un analyseur lexico-syntaxique de Wright développé en utilisant les deux générateurs d'analyseurs lexicaux et syntaxiques célèbres Lex et Yacc. Le fonctionnement général de l'outil Wr2fdr est décrit par une séquence d'opérations. Dans un premier temps, l'opération *parse_result* est exécutée afin d'analyser syntaxiquement le fichier d'entrée contenant une spécification Wright. En cas de succès, cette opération produit un arbre syntaxique abstrait (Structure de données *astNode*). En cas d'échec, des erreurs lexico-syntaxiques sont signalées. Dans un deuxième temps, l'opération *fdrprint* applicable sur un objet de type *astNode* est exécutée afin de produire la traduction CSP correspondante.

L'outil Wr2fdr offre plusieurs structures de données non génériques considérées comme des classes C++. Parmi ces classes, nous citons : *astNode*, *SList*, *Set*, *SymEntry*, *Relation* et *LookupTable*. La classe *astNode* joue un rôle fondamental. Elle permet de matérialiser une spécification Wright sous forme d'un arbre syntaxique abstrait. La classe *astNode* offre plusieurs attributs et méthodes virtuelles. Elle admet plusieurs classes descendantes permettant de modéliser les constructions syntaxiques offertes par Wright telles que : Style, Configuration, Component, Connector, event, binaryop, unitaryop et declaration.

Nous avons fourni des efforts importants pour étudier le code source de l'outil Wr2fdr (16000 lignes C++). Nous avons extrait un diagramme de classes UML modélisant l'architecture orientée objet de l'outil Wr2fdr. En fait, nous avons appliqué des règles simples de traduction de C++ vers UML. Par exemple, un fichier «.hpp» est traduit par un package UML. Également, une classe C++ est traduite par une classe UML. En outre, les relations client et héritage de C++ sont traduites respectivement par une association et généralisation UML. Nous avons enrichi le diagramme de classes obtenu par des contraintes OCL : invariant, précondition et postcondition. Ces contraintes sont issues d'un examen approfondi des implémentations des classes formant l'outil Wr2fdr. Le diagramme de classes obtenu de code source de l'outil Wr2fdr nous a permis d'avoir une vue d'ensemble, d'identifier les abstractions principales et de connaître les choix techniques de Wr2fdr. Ceci nous a autorisé à lancer l'activité de maintenance de Wr2fdr.

## 6.5 Correction des erreurs et nouvelles fonctionnalités apportées à l'outil Wr2fdr

### 6.5.1 Localisation et correction des erreurs liées aux propriétés 2 et 3

La cohérence d'un connecteur Wright est régie par la vérification des deux propriétés 2 et 3 présentées dans (cf. 2.5.3.1). Mais l'outil Wr2fdr ne traite pas convenablement ces deux propriétés (cf. 6.3.1). Dans la suite, nous allons expliquer les anomalies constatées lors de l'exécution de Wr2fdr, localiser et corriger les erreurs détectées.

#### 6.5.1.1 Anomalies constatées

L'exécution de Wr2fdr avec l'entrée fournie par la Figure 6.8 produit, entre-autres, les deux processus CSP relatifs aux deux rôles Client et Serveur donnés ci-dessous :

```
ROLEClient = ((request -> (result -> Client)) |~| SKIP)
ROLEServer = ((invoke -> (return -> Server)) [] SKIP)
```

```
Style ClientServer
   Connector CSconnector
      Role Client = (request -> result -> Client) |~| TICK
      Role Server = (invoke -> return -> Server) [] TICK
      Glue = (Client.request -> Server.invoke -> Server.return -> Client.result -> Glue) [] TICK
  Constraints
   // no constraints
End Style
```

**Figure 6.8 :** Entrée ClientServer

On remarque que les auteurs de Wr2fdr ont décidé de nommer le processus CSP relatif à chaque rôle du connecteur en ajoutant le préfixe Role au nom initial du rôle mais ils ont oublié de traiter la récursivité. Ainsi les sorties attendues doivent être :

```
ROLEClient = ((request -> (result -> ROLEClient)) |~| SKIP)
ROLEServer = ((invoke -> (return -> ROLEServer)) [] SKIP)
```

En outre, l'exécution de Wr2fdr pour la même entrée produit les sorties observées :

```
ALPHA_CSconnector = {|Server.invoke, Server.return, Client.result, Client.request|}
CSconnectorA = CSconnector [[ x <- abstractEvent | x <- ALPHA_Glue ]
```

On remarque que l'identificateur ALPHA_Glue non défini a été utilisé à la place de ALPHA_CSconnector défini précédemment et regroupant l'alphabet du processus CSP associé au connecteur CSconnector. Ainsi, la sortie attendue doit être :

```
CSconnectorA = CSconnector [[ x <- abstractEvent | x <- ALPHA_CSconnector]]
```

### 6.5.1.2 Localisation et correction

Nous avons pu localiser les erreurs entraînant les anomalies signalées dans le paragraphe précédent en se servant du diagramme de classes UML que nous avons extrait du code source Wr2fdr.

La première erreur liée à la génération du processus CSP associé à chaque rôle d'un connecteur Wright est localisée dans la classe *name* qui dérive da la classe fondatrice *astNode*. C'est la méthode *fdrprint* de la classe *name* qui est responsable de l'erreur. Afin de corriger cette erreur, nous avons ajouté à la méthode *fdrprint* la séquence d'instructions suivante :

```
switch (higherScope_effectif->gtype){
    case ROLE_T:
            //dans le cas où c'est un rôle
            if( this->eq(((declaration*)higherScope_effectif)->n))
            //si c'est le même nom du rôle
            doPrint("ROLE");//ajouter le mot 'ROLE'
            doPrint(n);
            ….
    break;
```

La deuxième erreur liée à la génération du processus CSP associé au connecteur est localisée dans la méthode *fdrprint* appartenant à la classe *connector* qui dérive de *asNode*. La séquence d'instructions provoquant l'anomalie est :

```
conn_name->fdrprint();
doPrint("A = ");
conn_name->fdrprint();
doPrint(" [[ x <- abstractEvent | x <- ALPHA_");
glue->n->fdrprint();//affiche le mot 'Glue'
doPrint(" ]]");
```

Cette séquence d'instructions est corrigée par :

```
conn_name->fdrprint();
doPrint("A = ");
conn_name->fdrprint();
doPrint(" [[ x <- abstractEvent | x <- ALPHA_");
conn_name->fdrprint();//affiche le nom du connecteur
doPrint(" ]]");
```

Nous avons validé les corrections apportées par une série de tests représentatifs.

## 6.5.2 Localisation et correction des erreurs liées à la propriété 1

La cohérence d'un composant Wright est régie par la vérification de la propriété 1 (cf. 2.5.3.1). L'exanen approfondi de la méthode *fdrprint* de la classe *component* qui dérive d'*astNode* nous a permis de conclure que la propriété 1 n'est pas implantée par la version actuelle de l'outil Wr2fdr. L'erreur d'exécution de Wr2fdr présentée dans 6.3.2 est causée par un pointeur qui pointe nulle part (ayant la valeur NULL). Nous avons fourni un travail important afin d'implémenter la propriété 1. En effet, celle-ci nécessite :

- l'adaptation des processus CSP associés aux différents ports d'un composant Wright pour qu'ils soient acceptables par FDR,
- l'adaptation du processus CSP modélisant le comportement global d'un composant Wright (partie computation) pour qu'il soit acceptable par FDR,
- l'implantation de la déterminisation des processus det(P),

- l’implantation de la restriction des processus à un ensemble d’événements,
- la génération des relations de raffinement.

En ce qui concerne l’implémentation de la détermination des processus (det), nous avons établi un algorithme simple basé sur la substitution textuelle de l’opérateur du choix non déterministe Π par l’opérateur du choix déterministe □. Nous lançons un traitement préalable afin de conditionner le processus à déterminiser. Un tel traitement consiste à remplacer un processus de la forme P= e→Q Π e→S (où e est un événement, P, Q et S des processus) par P=e→ (Q Π S).

Nous avons adopté une approche orientée tests syntaxiques afin de valider les fonctionnalités ajoutées à l’outil Wr2fdr. En outre, nous avons complété le logiciel Wr2fdr par des assertions internes (en utilisant la macro-instruction assert de *C*) afin de localiser les erreurs. Les trois Figures 6.9, 6.10 et 6.11 donnent respectivement un cas de test pour la propriété 1, le fichier CSP engendré et la vérification des relations de raffinement par FDR.

```
Style CalculFormule
        Component calcul
            Port In = read -> In [] close -> TICK
            Port Out = _write -> Out |~| _close -> TICK
            Computation = In.read -> _Out.write -> Computation [] In.close -> _Out.close ->
            TICK
        Constraints
        //no constraints
End Style
```

**Figure 6.9 :** Cas de test pour la propriété 1 : CalculFomule.wrt

```
-- Style CalculFormule
-- Types declarations
-- events for abstract specification
        channel write, close, read
-- Component Calcul
        ALPHA_Calcul = {|Out.close, Out.write, In.read, In.close|}
        ComputationCalcul = ((In.read -> (Out.write -> ComputationCalcul)) [] (In.close
        -> (Out.close -> SKIP)))
        --Port Process
        ALPHA_In = {close, read}
        ALPHA_InI = {}
        PORTIn = ((read -> PORTIn) [] (close -> SKIP))
        InG = PORTIn[[ x <-In.x | x <- ALPHA_In ]]
        ALPHA_Out = {close, write}
-- no events observed!
        PORTOut = ((write -> PORTOut) |~| (close -> SKIP))
        OutG = PORTOut[[ x <-Out.x | x <- ALPHA_Out ]]
        channel In: {close, read}
        channel Out: {close, write}
--Deterministic Process restricted to the observed event
        PORTInDETR = ((read -> PORTInDETR) [] (close -> SKIP))
        CompIn = ((In.read -> CompIn) [] (In.close -> SKIP))
        PORTOutDETR = SKIP
        CompOut = ((Out.write -> CompOut) [] (Out.close -> SKIP))
        COMPIn = (( PORTOutDETR
        [| diff({}, {}) |]
        ComputationCalcul))\ diff(ALPHA_Calcul, {|In|})
```

```
            assert InG [FD= COMPIn
            COMPOut = (( PORTInDETR [[ x <- In.x | x <- {close, read } ]]
            [| diff({In.close, In.read}, {}) |]
            ComputationCalcul))\ diff(ALPHA_Calcul, {|Out|})
            assert OutG [FD= COMPOut
-- No constraints
            -- End Style
```

**Figure 6.10 :** Fichier CSP CalculFomule.fdr2

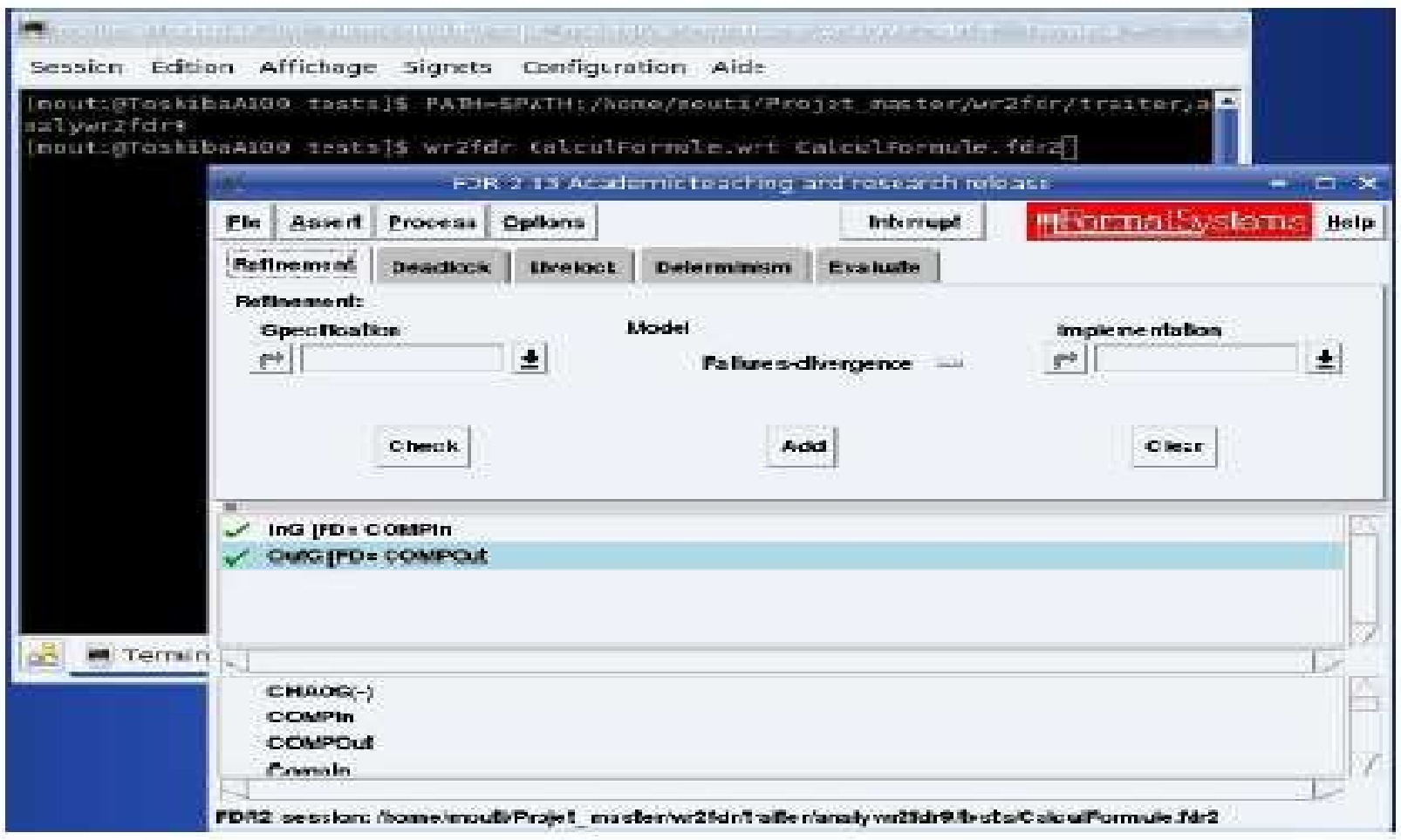

**Figure 6.11 :** Vérification à l'aide de FDR

### 6.5.3 Localisation et correction des erreurs liées à la propriété 8

La compatibilité Port/Rôle d'un assemblage de composants Wright est régie par la vérification de la propriété 8 (cf. 2.5.3.1). Après avoir étudié la méthode *fdrprint* de la classe *configuration* qui dérive *astNode*, nous avons conclu que la propriété 8 n'est pas implémentée par l'outil Wr2fdr. Nous avons réalisé cette propriété dans l'outil Wr2fdr par :

- l'augmentation de l'alphabet de processus : $P_{+A}$

- la généralisation des relations de raffinement adéquates.

Nous avons testé soigneusement l'implémentation proposée de la propriété 8. Les Figures 6.12, 6.13 et 6.14 donnent respectivement un cas de test pour la propriété 8, le fichier CSP produit et sa vérification à l'aide de FDR. Bien entendu, les relations de raffinement générées signalées par assert dans la Figure 6.14 concernent les propriétés 1, 2, 3 et 8. Car la structure syntaxique traitée est une configuration Wright.

## 6.6 Un analyseur de la sémantique statique de Wright

Nous avons enrichi l'outil Wr2fdr par un analyseur sémantique statique de Wright. Ceci permet d'avoir des constructions cohérentes aussi bien sur le plan syntaxique que sémantique. Nous avons établi et implémenté les six règles suivantes :

- **Règle 1** : Un identificateur doit désigner un seul élément architectural (component, connector, port, role, configuration et style),

- **Règle 2 :** Le type d'une instance (component, connector) doit être précédemment déclaré,

- **Règle 3 :** Toute instance doit être déclarée (clause *instances*) avant d'être utilisée dans les attachements (clause *attachments*),

- **Règle 4 :** Une interface d'un composant ou d'un connecteur doit être de la forme instance.port ou instance.role. Chaque port (respectivement role) doit figurer au sein du type composant (respectivement connecteur) utilisé pour définir l'instance,

- **Règle 5 :** Un attachement (clause *attachments*) doit être de la forme instance.port as instance.role,

- **Règle 6 :** Chaque port (respectivement rôle) d'un composant (respectivement connecteur) doit être attaché à un et un seul rôle (respectivement port) d'un connecteur (respectivement composant).

Nous avons implémenté ces règles relatives à la sémantique statique de Wright en augmentant l'analyseur lexico-syntaxique de Wr2fdr par des actions sémantiques appropriées. Également, nous avons testé avec succès notre analyseur sémantique.

```
Configuration ABC
    Component Atype
            Port Output = _a -> Output |~| TICK
            Computation = _Output.a -> Computation |~| TICK
    Component Btype
            Port Input = c -> Input [] TICK
            Computation = Input.c -> _b -> Computation [] TICK
     connector Ctype
            Role Origin = _a -> Origin |~| TICK
            Role Target = c -> Target [] TICK
            Glue = Origin.a -> _Target.c -> Glue [] TICK
     Instances

      A : Atype
      B : Btype
      C : Ctype
Attachments
     A.Output As C.Origin
     B.Input As C.Target
End Configuration
```

**Figure 6.12 :** Cas de test pour la propriété 8 : ABC.wrt

```
-- Configuration ABC
-- Types declarations
-- events for abstract specification
channel b, c, a
-- Component Atype
ALPHA_Atype = {|Output.a|}
ComputationAtype = ((Output.a -> ComputationAtype) |~| SKIP)
--Port Process
ALPHA_Output = {a}
-- no events observed!
PORTOutput = ((a -> PORTOutput) |~| SKIP)
OutputG = PORTOutput[[ x <-Output.x | x <- ALPHA_Output ]]
channel Output: {a}
--Deterministic Process restricted to the observed event
PORTOutputDETR = SKIP
COMPOutput = (ComputationAtype)\ diff(ALPHA_Atype, {|Output|})
assert OutputG [FD= COMPOutput
-- Component Btype
ALPHA_Btype = {|Input.c, b|}
ComputationBtype = ((Input.c -> (b -> ComputationBtype)) [] SKIP)
--Port Process
```

```
ALPHA_Input = {c}
ALPHA_InputI = {}
PORTInput = ((c -> PORTInput) [] SKIP)
channel Input: {c}
--Deterministic Process restricted to the observed event
PORTInputDETR = ((c -> PORTInputDETR) [] SKIP)
COMPInput = (ComputationBtype)\ diff(ALPHA_Btype, {|Input|})
assert InputG [FD= COMPInput
-- Connector Ctype
-- generated definitions (to split long sets)
ALPHA_Ctype = {|Target.c, Origin.a|}
GlueCtype = ((Origin.a -> (Target.c -> GlueCtype)) [] SKIP)
ALPHA_Origin = {a}
ROLEOrigin = ((a -> ROLEOrigin) |~| SKIP)

OriginA = ROLEOrigin [[ x <- abstractEvent | x <- ALPHA_Origin ]]
assert DFA [FD= OriginA
ALPHA_Target = {c}
ROLETarget = ((c -> ROLETarget) [] SKIP)
TargetA = ROLETarget [[ x <- abstractEvent | x <- ALPHA_Target ]]
assert DFA [FD= TargetA
ROLEOriginDET = ((a -> ROLEOriginDET) [] SKIP)
ROLETargetDET = ((c -> ROLETargetDET) [] SKIP)
channel Origin: {a}
channel Target: {c}
Ctype = (( ROLEOrigin[[ x <- Origin.x | x <- {a } ]]
[| diff({|Origin|}, {}) |]
( ROLETarget[[ x <- Target.x | x <- {c } ]]
[| diff({|Target|}, {}) |]
GlueCtype)))
CtypeA = Ctype [[ x <- abstractEvent | x <- ALPHA_Ctype ]]
assert DFA [FD= CtypeA
--Attachment Test
A_OutputPLUS = PORTOutput
[| diff( ALPHA_Origin , ALPHA_Output ) |] STOP
C_OriginPLUS = ROLEOrigin
[| diff( ALPHA_Output , ALPHA_Origin )|] STOP
A_OutputPLUSDET = A_OutputPLUS
[| union(ALPHA_Output , ALPHA_Origin ) |]
ROLEOriginDET
assert C_OriginPLUS [FD= A_OutputPLUSDET
B_InputPLUS = PORTInput
[| diff( ALPHA_Target , ALPHA_Input ) |] STOP
C_TargetPLUS = ROLETarget
[| diff( ALPHA_Input , ALPHA_Target )|] STOP
B_InputPLUSDET = B_InputPLUS
[| union(ALPHA_Input , ALPHA_Target ) |]
ROLETargetDET
assert C_TargetPLUS [FD= B_InputPLUSDET
-- End Configuration
```

**Figure 6.13 :** Fichier CSP ABC.fdr2

## 6.7 Conclusion

Dans ce chapitre, nous avons réalisé une activité de maintenance de l'outil Wr2fdr qui accompagne Wright. Nous avons corrigé les erreurs liées aux deux propriétés 2 et 3. En outre, nous avons proposé une implémentation des deux propriétés 1 et 8. Enfin, nous avons enrichi l'outil Wr2fdr avec un analyseur sémantique de Wright.

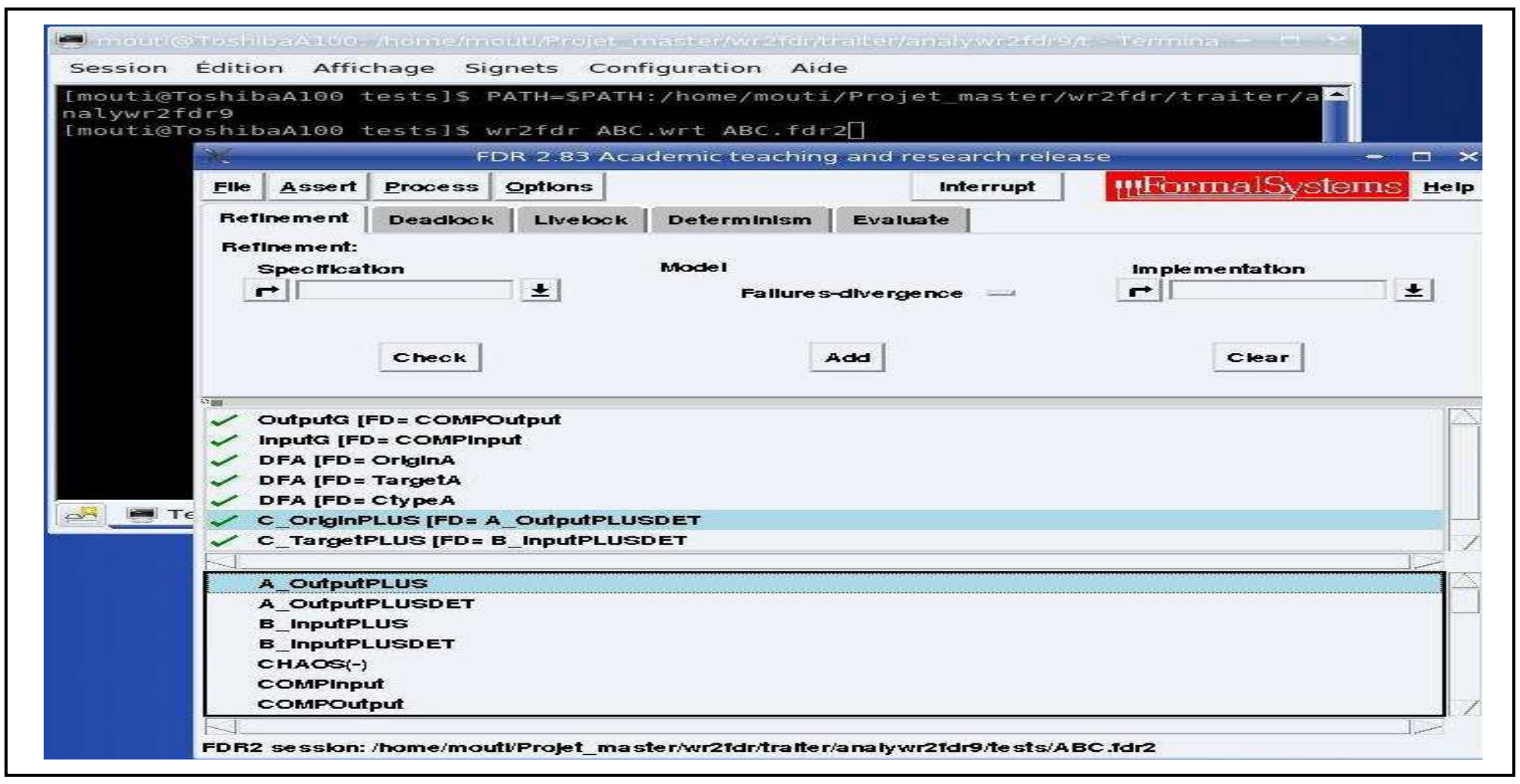

**Figure 6.14 :** Vérification à l'aide de FDR

Dans le chapitre suivant, nous allons proposer un outil IDM permettant de transformer un assemblage de composants Wright vers un programme concurrent Ada. Ceci favorise l'utilisation des outils d'analyse statique et dynamique associés à Ada.

# **Chapitre 7 :** De Wright vers Ada

## 7.1 Introduction

L'ADL formel Wright permet de décrire les aspects structuraux et comportementaux d'une architecture logicielle abstraite. Les aspects comportementaux sont décrits en CSP et vérifiés avec le model-checker FDR moyennant une traduction de Wright vers CSP acceptable par FDR en utilisant notre outil Wr2fdr (cf. chapitre 6). Mais l'ADL Wright n'offre aucun moyen de concrétiser de telles architectures abstraites. L'objectif de ce chapitre est d'ouvrir l'ADL Wright sur Ada en suivant une approche d'automatisation de type IDM (Ingénierie Dirigée par les Modèles). Pour y parvenir, nous avons élaboré deux méta-modèles en Ecore : le méta-modèle Wright et le méta-modèle partiel d'Ada. De plus, nous avons conçu, réalisé et testé notre outil Wright2Ada permettant de transformer une architecture logicielle décrite en Wright vers un programme concurrent en Ada en utilisant les langages IDM : ATL [Jouault, 2006], Xtext [Haase, 2007], Xpand [Klatt, 2007] et Check [Haase, 2007].
Ce chapitre comporte sept sections. La section 7.2 présente les principes généraux de l'IDM. Dans la section 7.3, nous proposons un méta-modèle Wright représentant la plupart des concepts issus de ce langage. Ce méta-modèle joue, dans notre contexte, le rôle de méta-modèle source dans notre approche IDM de transformation d'une architecture logicielle décrite en Wright vers un programme concurrent Ada. La section 7.4 présente la traduction de Wright vers Ada venant de [Bhiri, 2008]. La section 7.5 présente un méta-modèle partiel Ada issu de description BNF de ce langage [BNF-Ada] en se limitant aux constructions d'Ada utilisées dans la transformation de Wright vers Ada. La section 7.6 décrit d'une façon assez détaillée le programme Wright2Ada en traitant respectivement les aspects structuraux et comportementaux de traduction de Wright vers Ada. La section 7.7 propose des transformations IDM permettant d'avoir des interfaces conviviales afin d'utiliser notre programme Wright2Ada dans un contexte réel. Enfin, la section 7.8 préconise une approche basée sur le test fonctionnel permettant d'augmenter la confiance dans notre programme Wright2Ada.

## 7.2 L'ingénierie dirigée par les modèles

Dans cette section, nous commençons par une présentation des principes généraux de l'IDM (Ingénierie Dirigée par les Modèles) [Bézivin, 2004], [Diaw, 2009] ou MDE (Model Driven Engineering). Ensuite, nous donnerons un aperçu sur les origines de l'IDM, qui est l'architecture dirigée par les modèles.

### 7.2.1 Principes généraux de l'IDM

L'ingénierie dirigée par les modèles se base sur le principe « tout est modèle ». Un modèle est une abstraction de la réalité (le système). Il aide à répondre aux questions que l'on peut se poser sur le système modélisé. Pour qu'un modèle soit productif, il doit pouvoir être manipulé par une machine. Le langage de modélisation a pris la forme d'un modèle, appelé méta-modèle. Un méta-modèle est un modèle qui définit le langage d'expression d'un modèle [OMG, 2006]. Autrement dit, un méta-modèle est un modèle d'un ensemble de modèles. La Figure 7.1 inspirée de [Jouault, 2006] [Bézivin, 2004] représente la relation entre le système et le modèle, ainsi que, la relation entre le modèle et le méta-modèle.

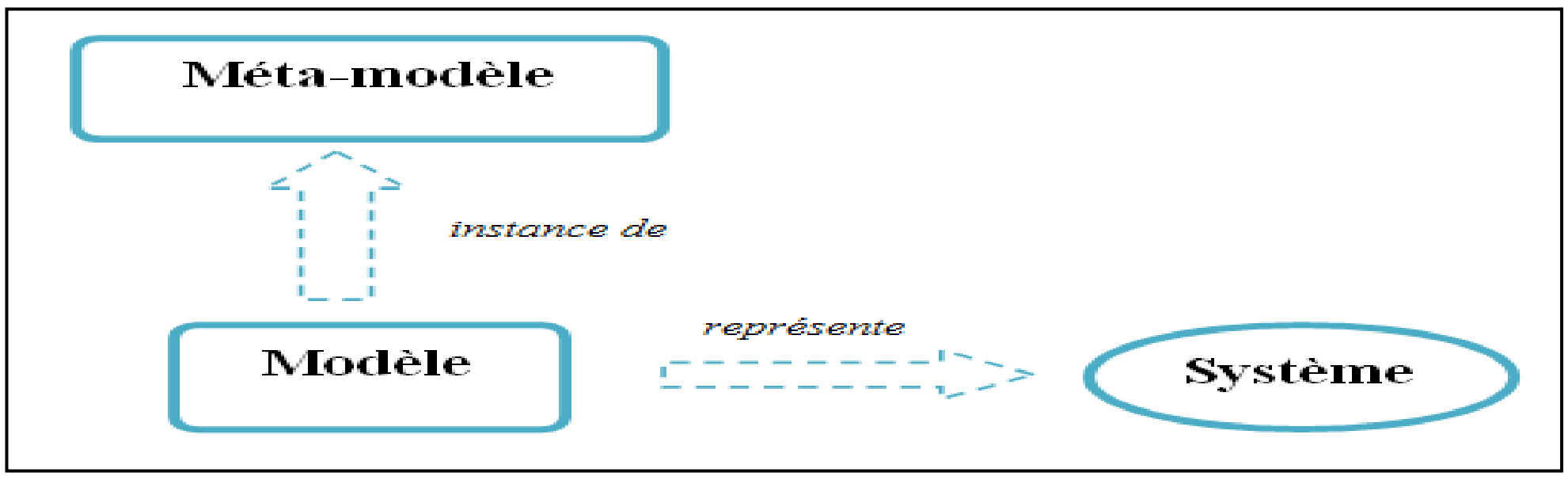

**Figure 7.1:** Relations de bases dans l'IDM

Dans la Figure 7.1, la relation « représente » dénote qu'un modèle est une représentation d'un système, tandis que la relation « instance » dénote qu'un modèle est conforme à un méta-modèle si ce modèle appartient à l'ensemble modélisé par ce méta-modèle.

### 7.2.2 Architecture dirigée par les modèles

Après l'acceptation du concept clé de méta-modèle comme langage de description de modèles, de nombreux méta-modèles ont émergé afin d'apporter chacun leurs spécificités dans un domaine particulier. Devant le danger de voir émerger indépendamment et de manière incompatible cette grande variété de méta-modèles, il y avait un besoin urgent de donner un cadre général pour leur description. La réponse logique fut donc d'offrir un langage de définition de méta-modèles qui prit lui-même la forme d'un modèle : ce fut le méta-méta-modèle MOF (Meta-Object Facility) [OMG, 2006]. En tant que modèle, il doit également être défini à partir d'un langage de modélisation. Pour limiter le nombre de niveaux d'abstraction, il doit alors avoir la propriété de méta-circularité, c'est-à-dire la capacité de se décrire lui-même [Combemale, 2008].

C'est sur ces principes que se base l'organisation de la modélisation de l'OMG généralement décrite sous une forme pyramidale représentée par la Figure 7.2 [Bézivin, 2003].

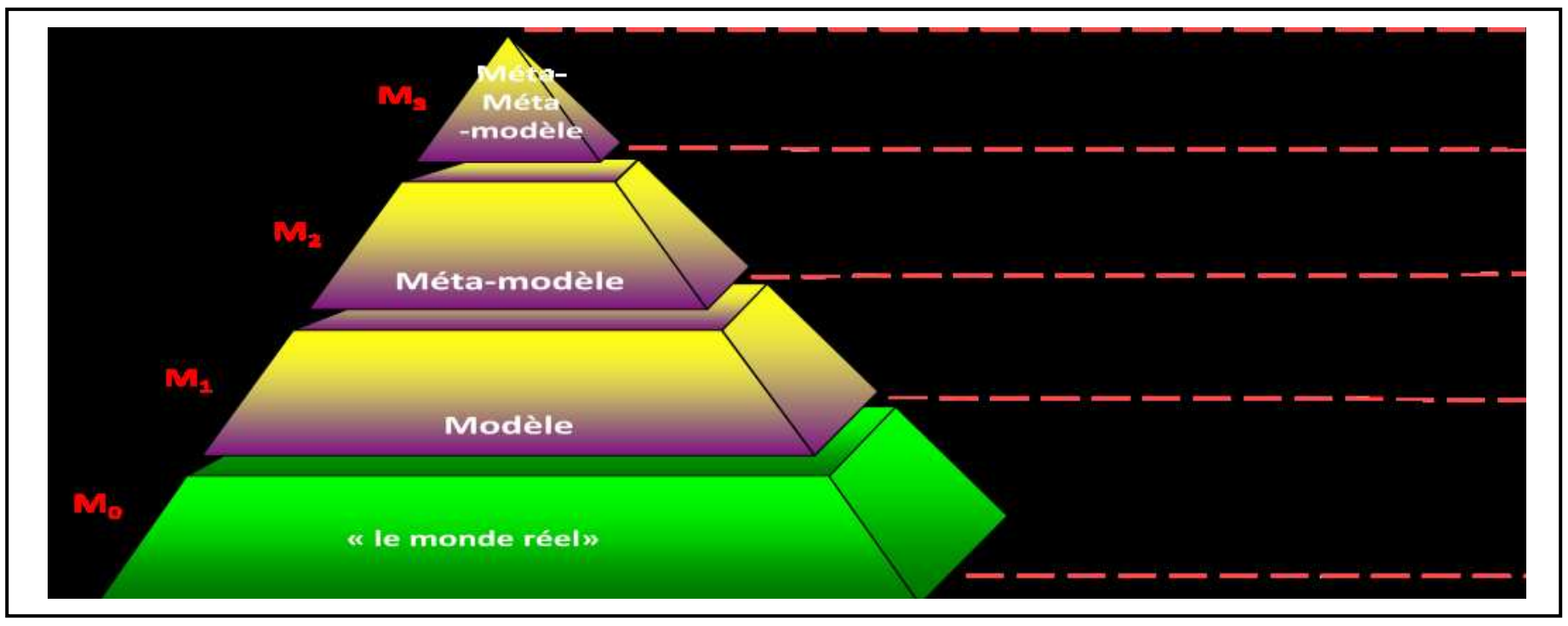

**Figure 7.2 :** Pyramide de modélisation de l'OMG

Le monde réel est représenté à la base de la pyramide (niveau M0). Les modèles représentant cette réalité constituent le niveau M1. Les méta-modèles permettant la définition de ces modèles constituent le niveau M2. Enfin, le méta-méta-modèle, unique et méta-circulaire, est représenté au sommet de la pyramide (niveau M3).
L'idée de base de MDA est de séparer les spécifications fonctionnelles d'un système des détails de son implémentation sur une plate-forme donnée. Pour cela, MDA définit une architecture de spécification structurée en plusieurs types de modèles.

- CIM (Computational Independent Model): aussi connu sous le nom modèle métier, il s'agit des modèles indépendants de l'informatisation. Un CIM modélise les exigences d'un système, son but étant d'aider à la compréhension du problème ainsi que de fixer un vocabulaire commun pour un domaine particulier (par exemple le diagramme des cas d'utilisation d'UML),
- PIM (Platform Independent Model): aussi connu sous le nom de modèle d'analyse et de conception. C'est un modèle abstrait indépendant de toute plate-forme d'exécution. Il a pour but de décrire une vue fonctionnelle du système,
- PDM (Platform Description Model) : pour les modèles de description de la plate-forme sur laquelle le système va s'exécuter. Il définit les différentes fonctionnalités de la plate-forme et précise comment les utiliser,
- PSM (Platform Specific Model) : pour les modèles spécifiques à une plate-forme donnée. En général il est issu de la combinaison du PIM et du PDM. Il représente une vue détaillée et opérationelle du système.

La Figure 7.3 donne une vue générale d'un processus MDA appelé couramment cycle de développement en Y en faisant apparaître les différents niveaux d'abstraction associés aux modèles.

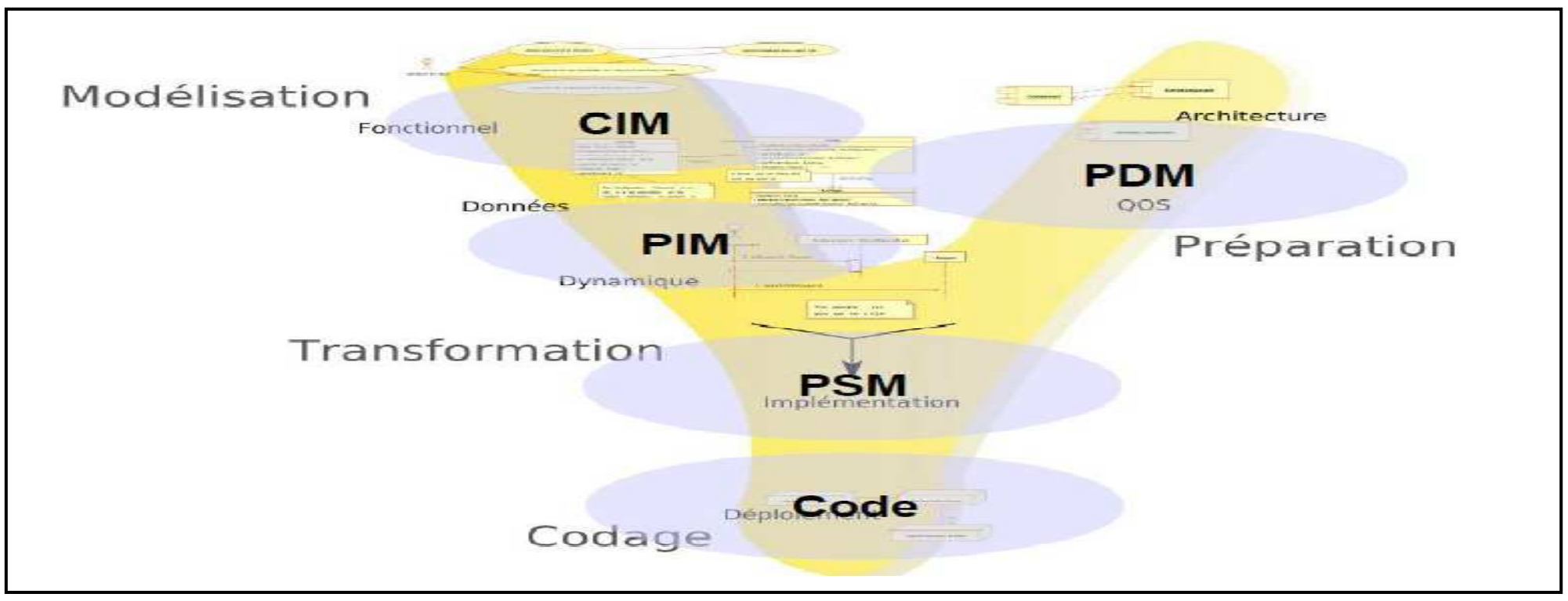

**Figure 7.3 :** Processus en Y de l'approche MDA

### 7.2.3 La transformation des modèles

Les transformations sont au coeur de l'approche MDA. Elles permettent d'obtenir différentes vues d'un modèle, de le raffiner ou de l'abstraire, de plus elles permettent de passer d'un langage vers un autre. Elles assurent le passage d'un ou plusieurs modèles d'un niveau d'abstraction donné vers un ou plusieurs autres modèles du même niveau (transformation horizontale) ou d'un niveau différent (transformation verticale). Les transformations horizontales sont de type PIM vers PIM ou bien PSM vers PSM. Les transformations verticales sont de type PIM vers PSM ou bien PSM vers code. Les transformations inverses verticales (rétro-ingénierie) sont type PSM vers PIM ou bien code vers PSM.

La Figure 7.4 [Piel, 2007] donne une vue d'ensemble sur la transformation des modèles.

Les règles de transformation sont établies entre les méta-modèles source et cible, c'est-à-dire entre l'ensemble des concepts des modèles source et cible. Le processus de transformation prend en entrée un modèle conforme au méta-modèle source et produit en sortie un ou plusieurs autre(s) modèle(s) conforme(s) au méta-modèle cible, en utilisant les règles préalablement établies.

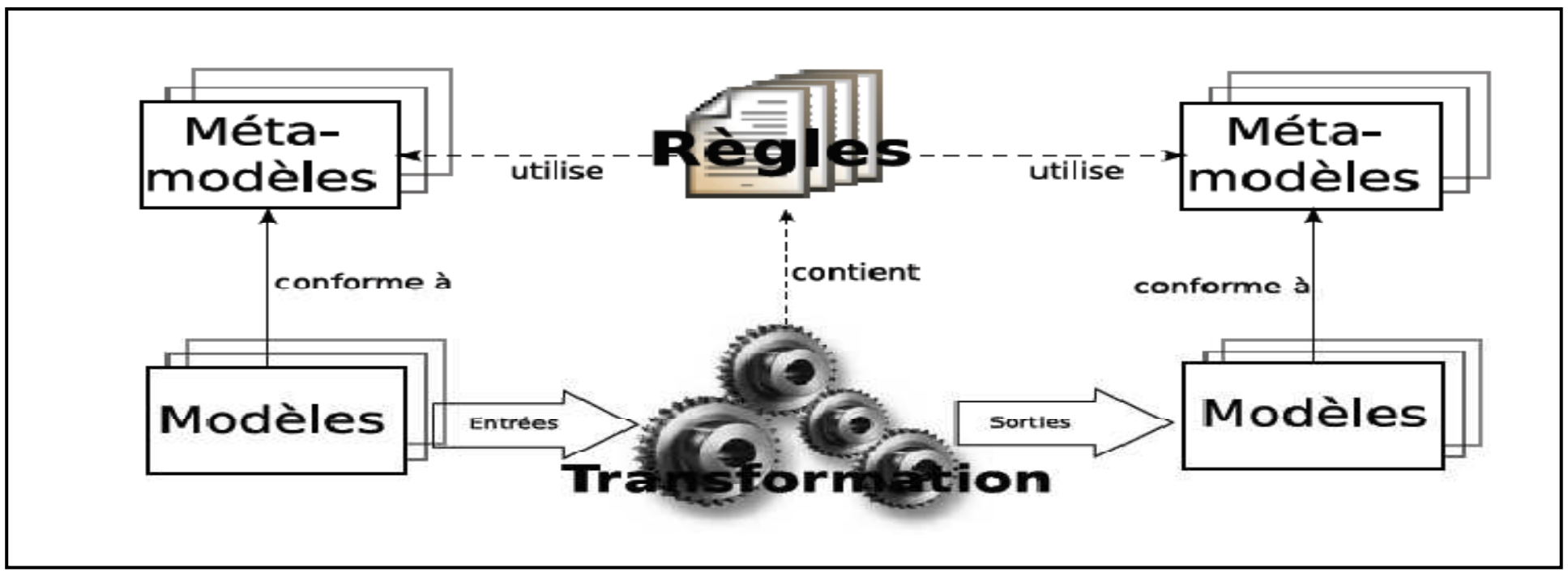

**Figure 7.4 :** Architecture de la transformation des modèles

## 7.3 Un méta-modèle du langage de description d'architectures Wright

Dans cette section, nous proposons un méta-modèle Wright représentant la plupart des concepts issus de ce langage à savoir : composant, connecteur, configuration et processus CSP. Ce méta-modèle joue, dans notre contexte, le rôle de méta-modèle source dans notre approche IDM de transformation d'une architecture logicielle décrite en Wright vers un programme concurrent Ada.

### 7.3.1 La partie structurelle

Cette section présente le fragment du méta-modèle Wright consacré aux aspects structuraux couvrant les concepts composant, connecteur et configuration

#### 7.3.1.1 Aspects syntaxiques

L'ADL Wright repose essentiellement sur les concepts composant, connecteur et configuration. La Figure 7.5 donne le fragment du méta-modèle Wright permettant de représenter ces concepts.

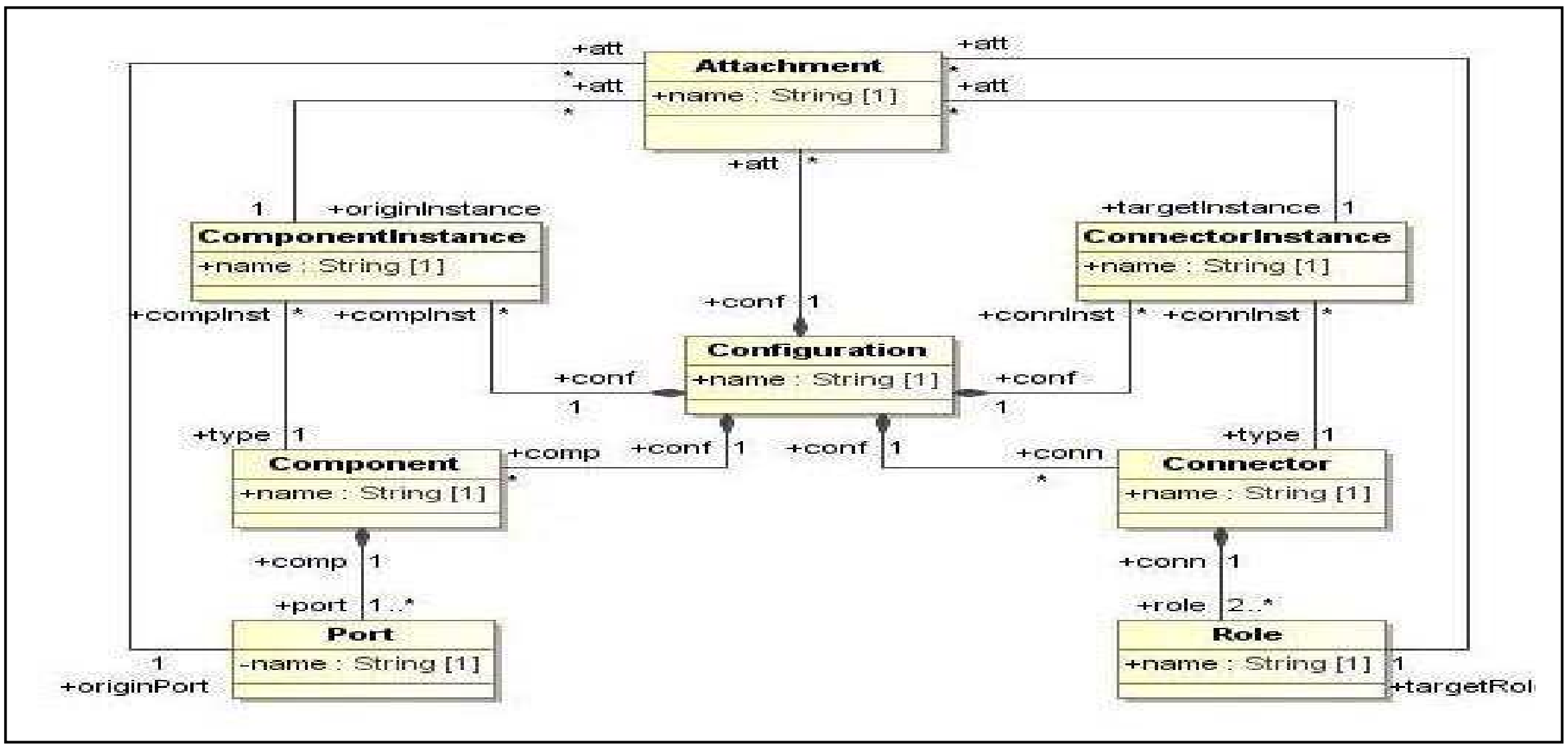

**Figure 7.5 :** Fragment du méta-modèle Wright : Partie structurelle

Un tel fragment comporte huit méta-classes et treize méta-associations. La méta-classe Configuration occupe une position centrale. Elle englobe des composants, des instances de composants, des connecteurs, des instances de connecteurs et des attachements. Ceci est traduit par une méta-composition entre Configuration et respectivement Component, ComponentInstance, Connector, ConnectorInstance et Attachment. À un composant Wright -respectivement connecteur- est attaché plusieurs ports –respectivement plusieurs rôles-. Ceci est traduit par une méta-composition entre Component et Port –respectivement entre Connector et Role-. Une instance de composant doit avoir un type de Composant. Ceci est traduit par la méta-association entre Component et ComponentInstance. De même, une instance de connecteur doit avoir un type de connecteur. Ceci est traduit par la méta-association entre Connector et ConnectorInstance. Un attachement concerne un port appartenant à une instance de composant et un rôle appartenant à une instance de connecteur. Ceci est traduit par les méta-associations entre *Attachment* et respectivement *ComponentInstance*, *Port*, *ConnectorInstance* et *Role*.

Dans la suite, nous décrivons les contraintes OCL attachées au fragment du méta-modèle relatif aux aspects structuraux de Wright.

#### 7.3.1.2 Les contraintes OCL

Nous avons établi plusieurs propriétés décrivant des contraintes d'utilisation des constructions structurelles de Wright. De telles propriétés sont décrites d'une façon informelle et formelle en se servant d'OCL.

- Propriété 1 :

Les noms désignant des composants, des instances de composants, des connecteurs, des instances de connecteurs, des ports, des rôles et des configurations doivent être des identificateurs valides au sens de Wright.

```
context Component
 inv identifier_card: name.size()>0
 inv letter: --le premier caractère de name doit être une lettre
 majuscule --ou miniscule.
 inv tail: --les autres caractères doivent être lettres majuscules, ou
 --miniscules, ou des chiffres.
```

- Propriété 2 :

Tous les ports attachés à un composant doivent avoir des noms deux à deux différents.

```
context Component
  inv different_port_names : self.port-> forAll( p1, p2 : Port | p1<>p2
  implies p1.name<>p2.name)
```

- Propriété 3 :

Tous les rôles attachés à un connecteur doivent avoir des noms deux à deux différents.

```
context Connector
 inv different_role_names : self.role-> forAll( r1, r2 : Role | r1<>r2
 implies r1.name<>r2.name)
```

- Propriété 4 :

Dans une même configuration un composant, une instance de composant, et une instance de connecteur doivent avoir des noms deux à deux différents.

```
context Configuration
 inv different_names_component : self.comp->forAll(c1, c2 : Component |
 c1<>c2 implies c1.name<>c2.name)
```

```
inv different_names_connector : self.conn->forAll(c1, c2 : Connector |
c1<>c2 implies c1.name<>c2.name)
inv different_names_componentInstance : self.compInst->forAll(c1, c2 :
ComponentInstance | c1<>c2 implies c1.name<>c2.name)
inv different_names_component : self.connInst->forAll(c1, c2 :
ConnectorInstance | c1<>c2 implies c1.name<>c2.name)
inv different_names_in_configuration : self.comp->
collect(self.comp.name)->excludesAll(self.compInst->
collect(self.compInst.name))
and self.comp->collect(self.comp.name)->excludesAll(self.conn->
collect(self.conn.name))
and self.comp->collect(self.comp.name)->excludesAll(self.connInst->
collect(self.connInst.name))
and self.compInst->collect(self.compInst.name)->excludesAll(self.conn->
collect(self.conn.name))
and self.compInst->collect(self.compInst.name)->
excludesAll(self.connInst->collect(self.connInst.name))
and self.conn->collect(self.conn.name)->excludesAll(self.connInst->
collect(self.connInst.name))
```

- Propriété 5 :

Une configuration privée de composants n'admet ni instance de composant ni attachement. De même, une configuration privée de connecteurs n'admet ni instance de connecteur ni attachement.

```
context Configuration
 inv component_without : self.comp -> size()= 0 implies ( self.compInst -
 > size()= 0 and self.att ->size()= 0)
 inv connector_without : self.conn -> size()= 0 implies ( self.connInst -
 > size()= 0 and self.att ->size()= 0)
```

- Propriété 6 :

Chaque instance déclarée au sein d'une configuration doit utiliser un type déclaré au sein de la même configuration.

```
context Configuration
 inv declared_component : self.compInst -> forAll( i : ComponentInstance|
 self.comp ->includes( i.type))
 inv declared_connector : self.connInst -> forAll( i : ConnectorInstance|
 self.conn ->includes( i.type))
```

- Propriété 7 :

Tous les attachements utilisent des instances déclarées au sein de la même configuration.

```
context Configuration
 inv declared_instance : self.att -> forAll( a : Attachment
 |self.compInst -> includes(a.originInstance) and self.connInst->i
 ncludes(a.targetInstance))
```

- Propriété 8 :

Un attachement est valide si et seulement si le port et le rôle concernés sont bel et bien attachés respectivement à l'instance concernée de type composant et l'instance concernée de type connecteur.

```
context Attachment
 inv attachment_port_concerns_component : self.originInstance.type.port -
 > includes( self.originPort)
```

```
inv attachment_role_concerns_connector : self.targetInstance.type.role -
> includes( self.targetRole)
```

- Propriété 9 :

Les instances de composants reliées à un composant donné doivent être de même type.

```
context Component
 inv instance_type_component: self.compInst ->
 forAll(i:ComponentInstance|i.type=self)
```

- Propriété 10 :

Les instances de connecteurs reliées à un connecteur donné doivent être de même type.

```
context Connector
inv instance_type_connector: self.connInst ->forAll(i:ConnectorInstance|i.type=self)
```

### 7.3.2 La partie comportementale

Cette section présente le fragment du méta-modèle Wright relatif aux aspects comportements couvrant le concept de Processus CSP.

#### 7.3.2.1 Les aspects syntaxiques

Le langage CSP de Hoare repose sur deux concepts essentiels: événement et processus. Il offre plusieurs opérateurs permettant d'enchaîner des événements et par conséquent de construire des processus CSP tels que: préfixage (ou séquencement), récursion, choix déterministe et choix non déterministe. En outre, Wright augmente le langage CSP en distinguant entre événement initialisé et observé.

La Figure 7.6 résume le fragment du méta-modèle Wright lié à ses aspects comportementaux.

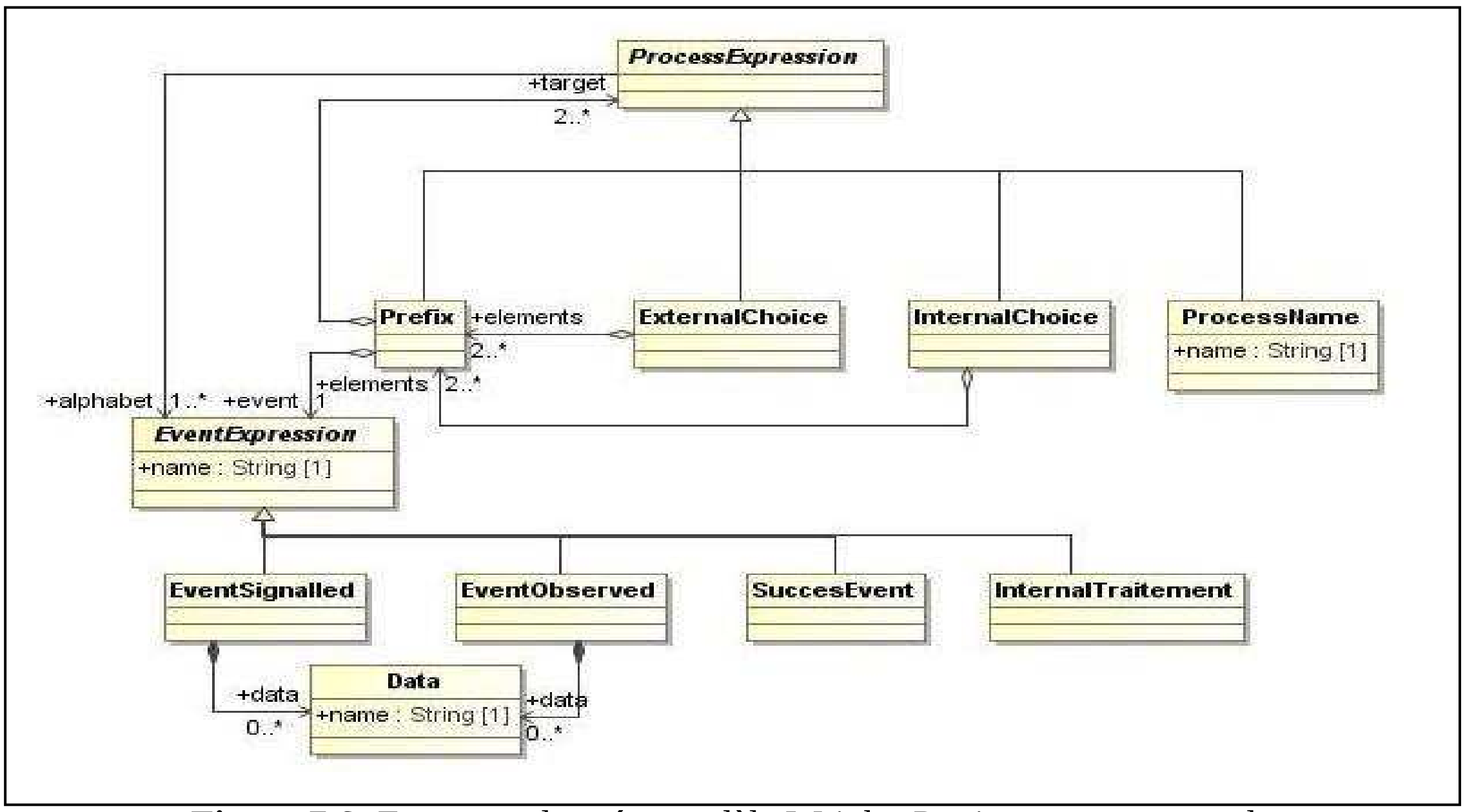

**Figure 7.6:** Fragment du méta-modèle Wright: Partie comportementale

Un tel fragment comporte deux hiérarchies. La hiérarchie ayant comme méta-classe fondatrice *ProcessExpression* modélise le concept de processus en CSP. Les méta-classes descendantes *Prefix*, *ExternalChoice*, *InternalChoice* et *ProcessName* représentent respectivement les opérateurs préfixage, choix externe (ou déterministe), choix interne (ou non déterministe) et le nommage d'un processus (favorisant la récursion) fournis par CSP. L'autre hiérarchie ayant comme méta-classe fondatrice *EventExpression* représente le concept d'événement en CSP Wright. Les méta-classes descendantes *EventSignalled*, *EventObserved*, *InternalTraitment* et *SuccesEvent* représentent respectivement événement initialisé, événement observé, traitement interne et événement succès fournis par CSP de Wright. Les liens entre ces deux hiérarchies sont traduits par les deux méta-agrégations entre *Prefix* et *EventExpression* et *ProcessExpression* et *EventExpression* qui exprime l'alphabet d'un processus. Les deux méta-agrégations entre *Prefix* et respectivement *EventExpression* et *ProcessExpression* traduisent fidèlement la structure d'un opérateur de préfixage (e → P): il s'engage dans l'événement e puis se comporte comme P. La structure de l'opérateur de choix déterministe est traduite par la méta-agrégation entre *ExternalChoice* et *Prefix*. De même, la struture de l'opérateur de choix non déterministe est traduite par la méta-agrégation entre *InternalChoice* et *Prefix*.

#### 7.3.2.2 Les contraintes OCL

Les propriétés attachées au fragment du méta-modèle décrivant les aspects comportementaux de Wright sont :

- Propriété 11 :

Le méta-attribut name de la méta-classe *ProcessName* doit stocker un identificateur valide au sens Wright.

```
context ProcessName
 inv identifier_card: name.size()>0
 inv letter: --le premier caractère de name doit être une lettre
 majuscule --ou miniscule.
 inv tail: --les autres caractères doivent être lettres majuscules, ou
--miniscules, ou des chiffres.
```

- Propriété 12 :

Le méta-attribut name de la méta-classe *EventExpression* doit stocker un identificateur valide au sens Wright –possibilité d'utiliser la notation qualifiée . -.

```
context ProcessName
 inv identifier_card: name.size()>0
 inv letter: --le premier caractère de name doit être une lettre
 majuscule --ou miniscule.
 inv tail: --les autres caractères doivent être lettres majuscules, ou
 --miniscules, ou des chiffres ou le caractère '.'.
```

- Propriété 13 :

Un choix externe doit être basé uniquement sur des événements observés et succès. Ceci peut être formalisé en OCL par :

```
context ExternalChoice
 inv event_observed_in_EC: self.elements -> forAll( e : Prefix |
 e.event.oclIsTypeOf(EventObserved) or e.event.oclIsTypeOf(SuccesEvent) )
```

### 7.3.3 Connexions entre les deux parties structurelle et comportementale

La Figure 7.7 donne les liens entre les deux fragments du méta-modèle Wright présentés ci-dessus.

Le comportement d’un port est décrit par un processus CSP. Ceci est traduit par la méta-agrégation entre Port et ProcessExpression. De même, le comportement d’un composant Wright est décrit par un processus CSP. Ceci est traduit par une méta-agrégation entre Component et ProcessExpression. D’une façon symétrique, les aspects comportementaux d’un rôle et d’un connecteur sont décrits respectivement par deux méta-agrégations entre Role et ProcessExpression et Connector et ProcessExpression.
Afin d’apporter plus de précisions à notre méta-modèle Wright, nous avons défini des nouvelles propriétés :

- Propriété 14 :

L’alphabet d’un processus associé à un port ne doit pas inclure des événements décrivant des traitements internes. Ceci peut être formalisé en OCL par :

```
context Port
 inv not_IT_behavior_port : self.behavior.alphabet -> forAll(
 a:EventExpression | not a.oclIsTypeOf(InternalTraitement))
```

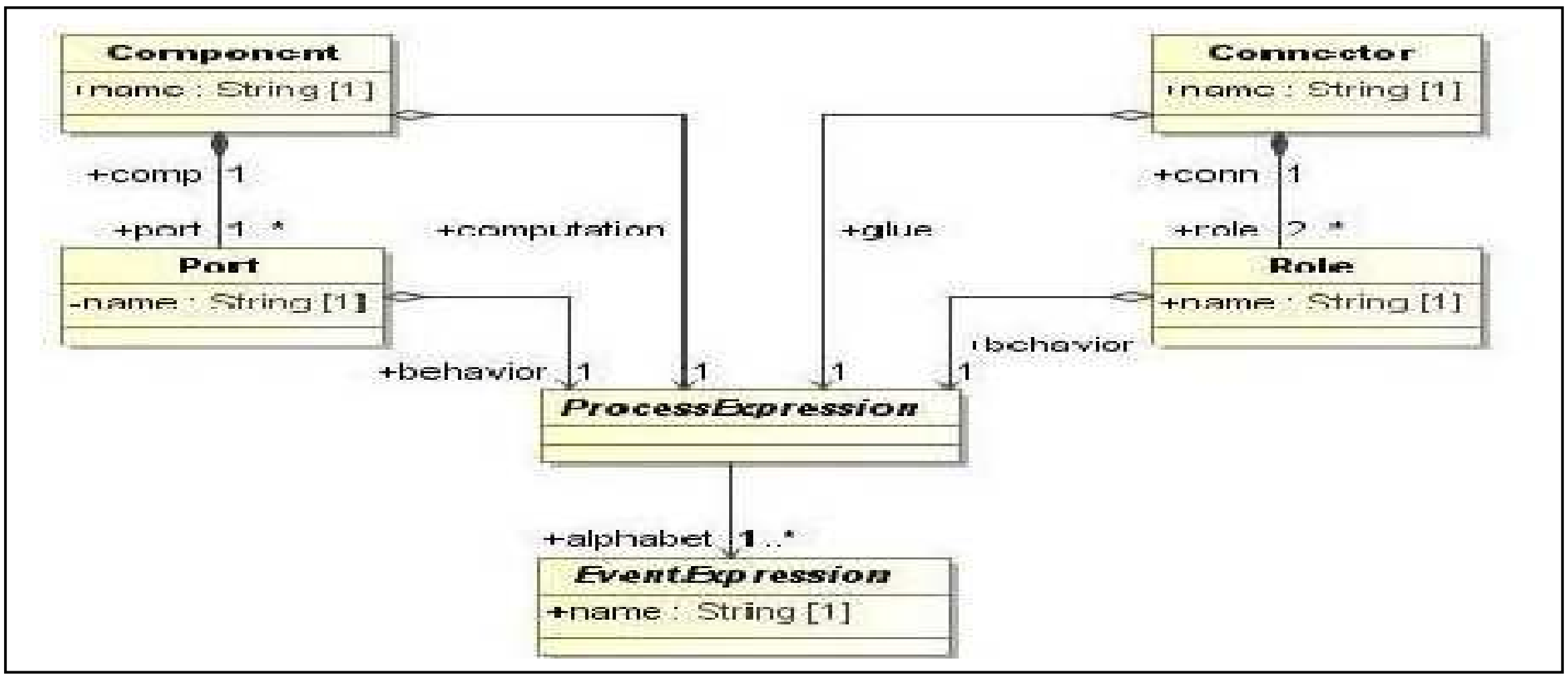

**Figure 7.7 :** Connexion entre les deux fragments du méta-modèle Wright

- Propriété 15 :

L’alphabet d’un processus associé à un rôle ne doit pas inclure des événements décrivant des traitements internes. Ceci peut être formalisé en OCL par :

```
context Role
 inv not_IT_behavior_role: self.behavior.alphabet -> forAll(
 a:EventExpression | not a.oclIsTypeOf(InternalTraitement))
```

- Propriété 16 :

Tous les alphabets des processus associés aux ports d’un composant doivent être inclus dans l’alphabet du processus associé au composant. Ceci peut être formalisé en OCL par :

```
context Component
 inv: self.computation.alphabet ->
 select(s:EventExpression|s.oclIsTypeOf(EventObserved) or
 s.oclIsTypeOf(EventSignalled)) -> collect(o:EventExpression| o.name)-
 >includesAll(self.port -> collect(p:Port|p.behavior.alphabet ->
 collect(a:EventExpression|p.name.concat('.').concat(a.name))))
```

- Propriété 17 :

Tous les alphabets des processus associés aux rôles d’un connecteur doivent être inclus dans l’alphabet du processus associé au connecteur. Ceci peut être formalisé en OCL par :

```
context Connector
 inv: self.glue.alphabet ->
 select(s:EventExpression|s.oclIsTypeOf(EventObserved) or
 s.oclIsTypeOf(EventSignalled)) -> collect(o:EventExpression| o.name) ->
 includesAll(self.role -> collect(r:Role|r.behavior.alphabet ->
 collect(a:EventExpression|r.name.concat('.').concat(a.name))))
```

### 7.3.4 Vue d’ensemble sur le méta-modèle Wright

Le méta-modèle de Wright utilisé comme méta-modèle source pour notre approche de transformation de Wright vers Ada est donné par la Figure 7.8.

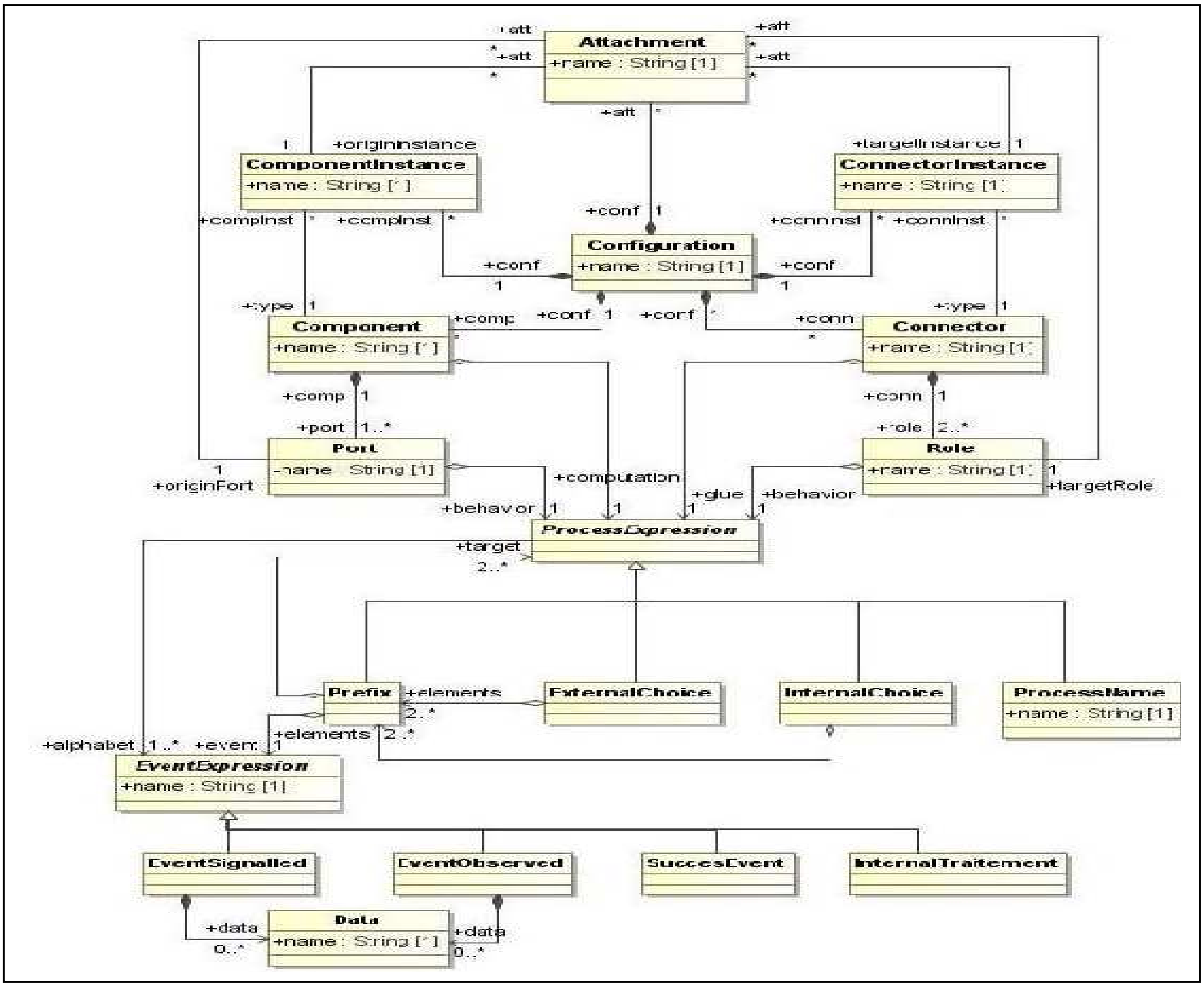

**Figure 7.8 :** Méta-modèle de Wright

## 7.4 Du langage de description d’architectures Wright vers le programme concurrent Ada

Dans cette section, nous présentons la contribution de [Bhiri, 2008] permettant de traduire d’une façon systématique une architecture logicielle formalisée en Wright vers Ada. Une telle contribution comporte un ensemble de règles permettant de traduire les constructions

Wright (configuration, composant, connecteur et processus CSP) en Ada. Le code Ada à générer correspond à l'architecture de l'application. Nous allons suivre une démarche descendante pour présenter le processus de traduction de Wright vers Ada.

### 7.4.1 Traduction d'une configuration

Une configuration Wright est traduite en Ada par un programme concurrent dans lequel :

– chaque instance de type composant est traduite par une tâche Ada,

– chaque instance de type connecteur est traduite également par une tâche Ada,

– les tâches de même type ne communiquent pas entre elles.

La Figure 7.9 illustre le principe de la traduction d'une configuration Wright en Ada. Pour des raisons de traçabilité, nous gardons les mêmes identificateurs utilisés dans la spécification Wright. En plus, pour favoriser des retours en arrière, – d'Ada vers Wright – nous transportons la nature de chaque instance soit Component, soit Connector.

| ***Spécification en Wright*** | ***Code Ada*** |
|---|---|
| **Configuration** ClientServeur<br>  **Component** Client<br>  **Component** Serveur<br>  **Connector CS**<br>   **Instances**<br>   c : Client<br>   s : Serveur<br>  cls : CS<br> **Attachments**<br> …<br>**End Configuration** | **procedure** ClientServeur **is**<br> **task** Component_c **is**<br>**end Component_c ;**<br> **task** Component_s **is**<br>**end Component_s;**<br>**task** Connector_cls **is**<br>**end Connector_cls;**<br>**task body** Component_c **is**<br>**end Component_c;**<br>**task body** Component_s **is**<br>**end Component_s;**<br>**task body** Connector_cls **is**<br>**end Connector_cls;**<br>  **begin**<br>   **null;**<br>**end ClientServeur;** |

**Figure 7.9 :** Traduction d'une configuration Wright

La traduction proposée possède un avantage majeur : elle permet de conserver la sémantique d'une configuration Wright. En effet, celle-ci est définie formellement en CSP comme la composition parallèle des processus modélisant les composants et les connecteurs formant cette configuration [Graiet, 2007]. De plus, un programme concurrent en Ada peut être modélisé en CSP comme la composition parallèle des tâches formant ce programme.

### 7.4.2 Traduction des événements

Nous ignorons les données portées par les événements CSP. De telles données seront introduites progressivement en appliquant un processus de raffinement sur le code Ada généré. Ainsi, nous distinguons :

– un événement observé de la forme e,

– une émission ou encore événement initialisé de la forme _e.

#### 7.4.2.1 Traduction d'un événement observé

Un événement observé de la forme e est traduit par une entrée (**entry**) et par une acceptation de rendez-vous (instruction **accept**).

La Figure 7.10 illustre le principe de la traduction d'une réception CSP en Ada.

| *Spécification en Wright* | *Code Ada* |
|---|---|
| **Component** Client<br>**Port** appelant =<br>_request → result → appelant Π §<br>**Instances**<br>c : Client<br>… | **task** Component_c **is**<br>**entry** result;<br>**end Component_c ;**<br>**task body** Component_c **is**<br>…<br>**accept** result;<br>…<br>**end Component_c;** |

**Figure 7.10 :** Traduction d'une réception

#### 7.4.2.2 Traduction d'un événement initialisé

Un événement initialisé de la forme _e est traduit par une demande de rendez-vous sur l'entrée e exportée par une tâche de type différent (seules les tâches de types différents communiquent) à identifier. Pour y parvenir, il faut analyser la partie **Attachments** de la configuration. La Figure 7.11 illustre le principe de la traduction d'une émission en Ada.

| *Spécification en Wright* | *Code Ada* |
|---|---|
| **Component** Client<br>**Port** appelant =<br>_request → result → appelant Π §<br>**Connector cs**<br>**Role** client = _request → result → client Π §<br>**Role** serveur = request →_result → serveur □ §<br>**Instances**<br>c : Client<br>cls: cs<br>**Attachments**<br>Client**.** appelant **As** cls**.**client | **task** Component_c **is**<br>**entry** result;<br>**end Component_c ;**<br>**task** Connector_cls **is**<br>**entry** request;<br>**entry** result;<br>**end Connector_cls;**<br>**task body** Component_c **is**<br>**begin**<br>Connector_cls**.**request;<br>**end Component_c;** |

**Figure 7.11:** Traduction d'une émission

### 7.4.3 Traduction de l'interface d'un composant

L'interface d'un composant Wright est traduite par une interface d'une tâche en Ada. Cette interface est obtenue de la manière suivante :

*Pour chaque port appartenant au composant Wright*
*Faire*
    *Pour chaque événement appartenant au port*
    *Faire*
        *Si événement est un événement observé de la forme e*
        *Alors créer une entrée ayant le nom suivant : port_e*
        *Finsi*
    *Finfaire*
*Finfaire*

La Figure 7.12 illustre le principe de la traduction de l'interface d'un composant Wright.

| *Spécification en Wright* | *Code Ada* |
|---|---|
| **Component** Client | **task** Component_c **is** |
| **Port** appelant =<br>_request → result → appelant Π §<br>**Instances**<br>c : Client<br>… | **entry** appelant_result;<br>**end Component_c ;** |

**Figure 7.12 :** Traduction de l'interface d'un composant

### 7.4.4 Traduction de l'interface d'un connecteur

L'interface d'un connecteur Wright est traduite par une interface d'une tâche Ada. Cette interface est obtenue de la manière suivante :

*Pour chaque rôle appartenant au connecteur Wright*
*Faire*
*Pour chaque événement appartenant au rôle*
*Faire*
*Si événement est un événement initialisé de la forme_e*
*Alors*
*Créer une entrée ayant le nom suivant : rôle_e*
*Finsi*
*Finfaire*
*Finfaire*

La Figure 7.13 illustre le principe de la traduction de l'interface d'un connecteur Wright.

### 7.4.5 De CSP Wright vers Ada

Dans cette section, nous décrivons les règles permettant de traduire en Ada les opérateurs CSP couramment utilisés en Wright.

| ***Spécification en Wright*** | ***Code Ada*** |
|---|---|
| **Connector cs**<br>**Role** client = _request → result → client Π §<br>**Role** serveur = request →_result → serveur □ §<br>**Instances**<br>cls: cs | **task** Connector_cls **is**<br>**entry** client_request;<br>**entry** serveur_result;<br>**end Connector_cls;** |

**Figure 7.13:** Traduction de l'interface d'un connecteur

#### 7.4.5.1 Traduction de l'opérateur de préfixage

La Figure 7.14 illustre la traduction en Ada de l'opérateur de préfixage. Nous distinguons les deux cas :

| *CSP* | *Traduction Ada* |
|---|---|
| Cas 1 : a → P | accept a ;<br>traiter P |
| Cas 2 : _a → P | nom_ tache.a;<br>traiter P |

**Figure 7.14 :** Traduction de l'opérateur de préfixage

#### 7.4.5.2 Traduction de l'opérateur de récursion

La récursion en CSP permet la description des entités qui continueront d'agir et d'interagir avec leur environnement aussi longtemps qu'il le faudra. La Figure 7.15 illustre la traduction de l'opérateur de récursion. Nous distinguons les cas suivants :

| *CSP* | *Traduction Ada* |
|---|---|
| Cas 1 : P= a → Q → P | loop<br>accept a;<br>traiter Q<br>end loop; |
| Cas 2 : P= _a → Q → P | loop<br>nom_ tache.a;<br>traiter Q<br>end loop; |
| Cas 3 : P= a → Q → P Π § | loop<br>exit when condition_interne ;<br>accept a;<br>traiter Q<br>end loop; |

**Figure 7.15 :** Traduction de l'opérateur de récursion

### 7.4.5.3 Traduction de l'opérateur de choix non déterministe

La notation P Π Q avec P ≠ Q, dénote un processus qui se comporte soit comme P soit comme Q, la sélection étant réalisée de façon arbitraire, hors du contrôle ou de la connaissance de l'environnement extérieur. Nous distingons les cas fournis par la Figure 7.16.

| *CSP* | *Traduction Ada* |
|---|---|
| Cas 1 : a → P Π b → Q<br>avec a et b quelconques. | **if** condition_interne **then**<br>accept a;<br>traiter P<br>**else**<br>accept b;<br>traiter Q<br>**end if;** |
| Cas 2 : _a → P Π § | **if** condition_interne **then**<br>nom_tache.a;<br>traiter P<br>**else**<br>exit;<br>**end if;** |
| Cas 3 : _a → P Π _b → Q | **if** condition_interne **then**<br>nom_tache.a;<br>traiter P<br>**else**<br>nom_tache.b;<br>traiter Q<br>**end if;** |
| Cas 4 : _a → P Π b → Q | **if** condition_interne **then**<br>nom_tache.a;<br>traiter P<br>**else**<br>accept b;<br>traiter Q<br>**end if;** |
| Cas 5 : _a → P Π _a → Q | nom_tache.a;<br>**if** condition_interne **then** |
| | traiter P<br>**else**<br>traiter Q<br>**end if;** |

**Figure 7.16:** Traduction de l'opérateur de choix non déterministe

**Remarque:** Le traitement modélisé par condition_interne utilisé dans la traduction précédente traduit un non déterminisme lié à l'opérateur Π. Il cache souvent des fonctionnalités à fournir par le futur logiciel. Ces fonctionnalités peuvent être précisées en réduisant progressivement le non déterminisme.

#### 7.4.5.4 Traduction de l'opérateur de choix déterministe

Le processus P □ Q avec P ≠ Q, introduit une opération par laquelle l'environnement peut contrôler celui de P ou de Q qui sera sélectionné, étant entendu que ce contrôle s'exerce sur la toute première action ou événement. Nous proposons (cf. Figure 7.17) la traduction suivante :

| *CSP* | *Traduction Ada* |
|---|---|
| a → P □ b → Q<br>avec a et b quelconques. | **select**<br>accept a;<br>traiter P<br>**or**<br>accept b;<br>traiter Q<br>**end select;** |

**Figure 7.17 :** Traduction de l'opérateur de choix déterministe

## 7.5 Méta-modèle partiel d'Ada

Dans cette section, nous proposons un méta-modèle partiel Ada issu de description BNF de ce langage [BNF-Ada] en se limitant aux constructions d'Ada utilisées dans la transformation de Wright vers Ada.

### 7.5.1 Concepts structurels retenus

#### 7.5.1.1 Sous-programmes Ada

En Ada, un sous-programme est une unité de programmation comportant deux parties : interface et implémentation. La partie implémentation possède deux parties : partie déclarative et partie exécutive. La partie interface correspond à la signature du sous-programme. En outre Ada distingue nettement les fonctions des procédures aussi bien sur le plan syntaxique que sémantique. En effet, l'appel d'une procédure est considéré comme instruction. Par contre, l'appel d'une fonction doit être inséré au sein d'une expression Ada.

La description BNF d'un sous-programme est donnée ci-dessous :

```
proper_body ::= subprogram_body | …
subprogram_body ::=
subprogram_specification "is"
      declarative_part
      "begin"
      handled_sequence_of_statements
      "end" [ designator ] ";"
subprogram_specification ::=
      ( "procedure" defining_program_unit_name [ formal_part ] )
|     ( "function" defining_designator [ formal_part ] "return" subtype_mark
)
declarative_part ::= { ( basic_declarative_item | body ) }
body ::= proper_body | …
basic_declarative_item ::= basic_declaration |…
basic_declaration ::= object_declaration | subprogram_declaration |…
handled_sequence_of_statements ::= sequence_of_statements [ …]
sequence_of_statements ::= statement { statement }
subprogram_declaration ::= subprogram_specification ";"
```

De cette description nous pouvons dériver le méta-modèle de la Figure 7.18.

La méta-classe *SubprogramBody* représente le concept de sous-programme ayant trois parties : en-tête, partie déclarative et partie exécutive. Ces trois parties sont traduites respectivement par trois méta-agrégations entre : *SubprogramBody* et *Declaration*, et, *SubprogramBody* et *Statement*.

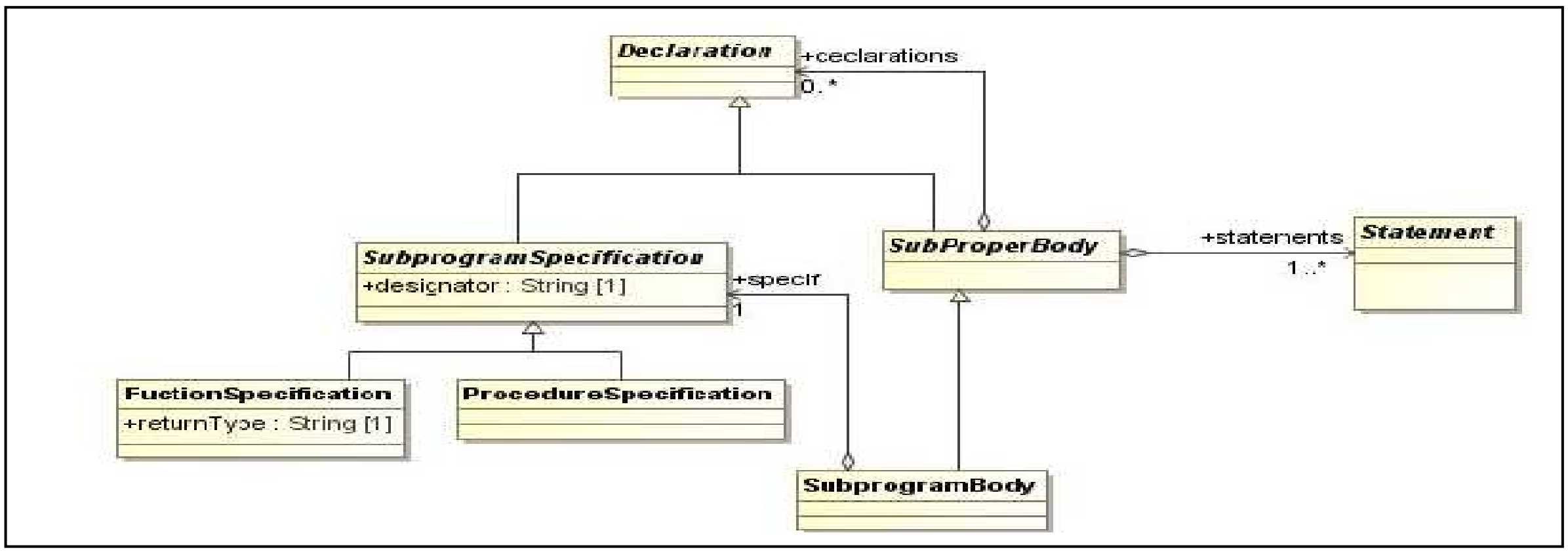

**Figure 7.18 :** Méta-modèle d’un sous-programme Ada

### 7.5.1.2 Tâches Ada

Une tâche en Ada est une unité de programmation comportant deux parties : interface et implémentation. La partie interface offre des services appelés entrées (entry).

Ces services indiquent des possibilités de rendez-vous fournis par la tâche. La partie implémentation comporte deux parties : partie déclarative et partie exécutive. La partie exécutive réalise la politique d’acceptation de rendez-vous par la tâche.

La description BNF d’une tâche Ada est donnée ci-dessous.

```
object_declaration ::= single_task_declaration |…
proper_body ::= subprogram_body | task_body |… single_task_declaration ::=
"task" defining_identifier [ "is" task_definition ] ";"
task_definition ::= { task_item } [ … ] "end" [ task_identifier ]
task_item ::= entry_declaration | …
entry_declaration ::= "entry" defining_identifier [ … ] ";"
task_body ::= "task" "body" defining_identifier "is" declarative_part
"begin"
handled_sequence_of_statements
"end" [ task_identifier ] ";"
```

Nous pouvons enrichir le méta-modèle de la figure 7.18 pour celui de la Figure 7.19. La méta-classe *TaskBody* représente le concept de tâche ayant trois parties : spécification (ou interface), partie déclarative et partie exécutive. Ces trois parties sont traduites respectivement par les deux méta-agrégations entre : *TaskBody* et *Declaration*, *TaskBody* et *Statement*.

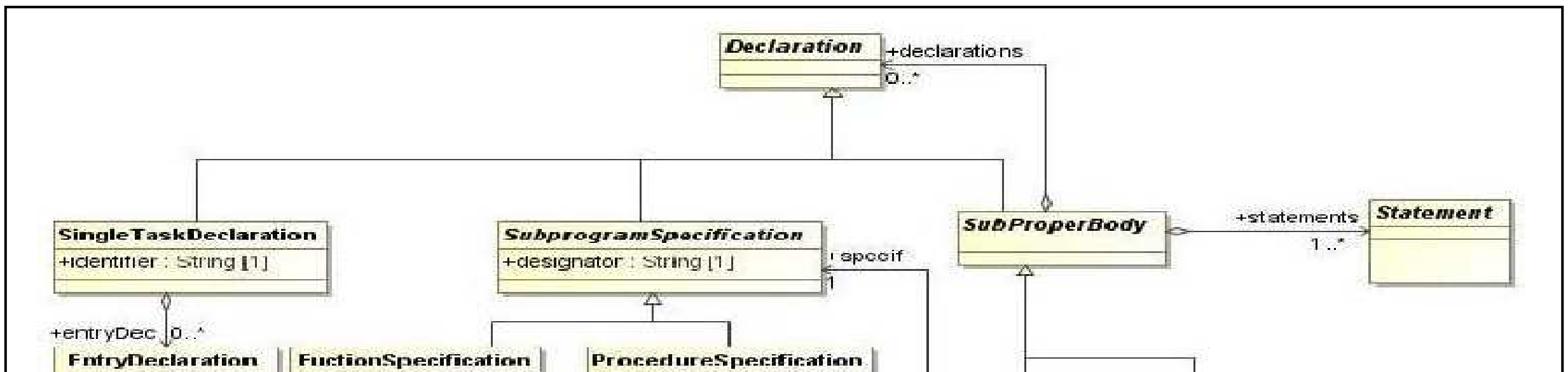

**Figure 7.19:** Méta-modèle représentant un sous-programme et une tâche Ada

### a) Instructions Ada

Les instructions concernées sont simples ou composées. Pour chaque instruction nous donnons son écriture BNF.

- **instructions simples**

  - L'instruction nulle :

L'instruction nulle est l'instruction qui ne fait rien.

```
null_statement ::= "null" ";"
```

- L'instruction exit :

L'instruction exit est utilisée pour achever l'exécution de l'instruction loop englobante; l'achèvement est conditionné si elle comprend une garde (une condition).

```
exit_statement ::= "exit" [ loop_name ] [ "when" condition ] ";"
condition ::= expression
```

- L'instruction return :

```
return_statement ::= "return" [ expression ] ";"
```

- L'invocation d'une procédure :

```
procedure_call_statement ::= ( procedure_name | prefix )
[ actual_parameter_part ] ";"
```

Dans notre transformation nous ne nous n'intéressons pas aux paramètres.

- Les appels d'entrée :

Les appels d'entrée ou encore demandes de rendez-vous peuvent apparaître dans divers contextes.

```
entry_call_statement ::= entry_name [ actual_parameter_part ] ";"
```

Dans notre transformation nous ne nous n'intéressons pas aux paramètres.

Le méta-modèle qui représente les instructions simples est présenté par la Figure 7.20.

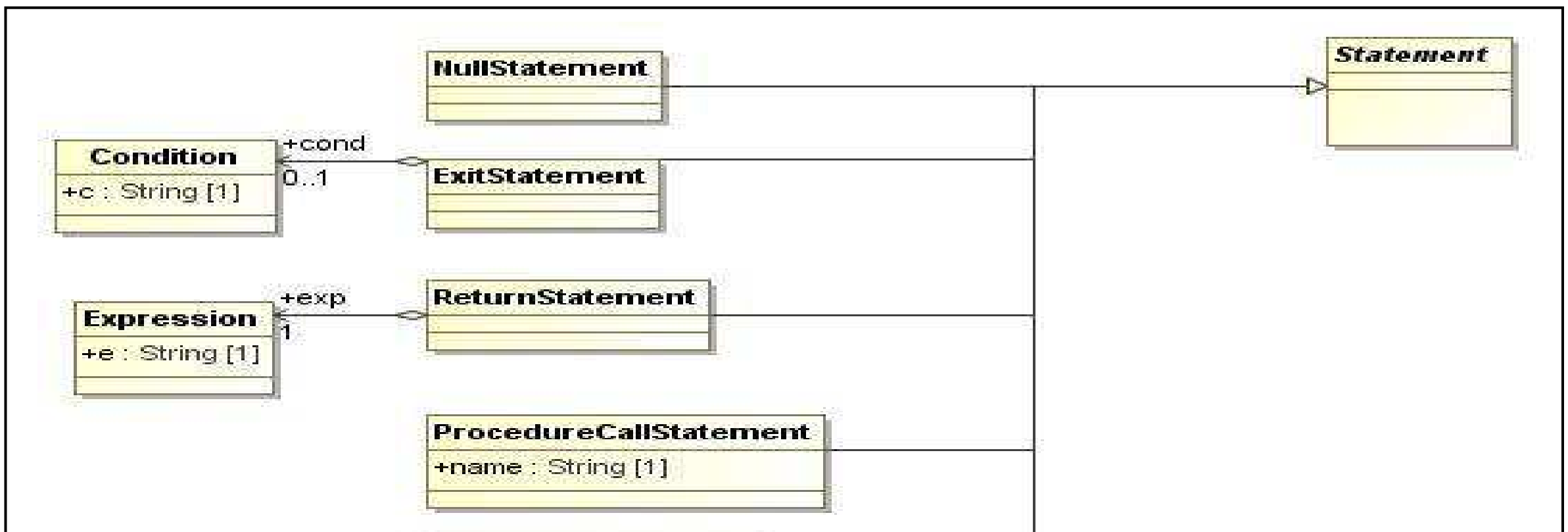

**Figure 7.20 :** Méta-modèle des instructions simples

Le méta-attribut *name* appartenant à la méta-classe *ProcedureCallStatement* mémorise l'identificateur de la procédure appelée. Egalement, le méta-attribut *entryName* stocke le nom de l'entrée appelé. Les deux méta-agrégations *ExitStatement* et *Condition*, et *ReturnStatement* et *Expression* modélisent respectivement la condition attachée à l'instruction *exit* et l'expression associée à *return*.

- **Les instructions composées**
    - L'instruction if :

Dans notre cas, nous nous intéressons à un simple if_then_else.

```
if_statement ::=
"if" condition "then"
      sequence_of_statements
{ "elsif" condition "then" sequence_of_statements }
[ "else" sequence_of_statements ]
"end" "if" ";"
condition ::= expression
```

- L'instruction case :

```
case_statement ::= "case" expression "is"
case_statement_alternative
{ case_statement_alternative }
"end" "case" ";"
case_statement_alternative   ::=   "when"   discrete_choice_list   "=>"
sequence_of_statements
discrete_choice_list ::= discrete_choice { "|" discrete_choice }
discrete_choice ::= expression | discrete_range | "others"
```

- L'instruction accept :

Il s'agit d'une instruction d'acceptation d'un rendez-vous. Elle est utilisée au sein de la partie exécutive d'une tâche. Dans notre cas nous nous intéressons à une simple instruction accept.

```
accept_statement ::= "accept" direct_name
[ "(" entry_index ")" ] parameter_profile
[ "do" handled_sequence_of_statements "end" [ entry_identifier ] ] ";"
```

- L'instruction select:

Il s'agit d'une instruction utilisée au sein de la partie exécutive d'une tâche. Elle favorise le non déterminisme lors de l'acceptation des rendez-vous éventuellement gardés.

Dans notre cas nous nous intéressons à un simple select_or sans garde et sans alternative d'attente.

```
selective_accept ::= "select"[ guard ] select_alternative
{ "or" [ guard ] select_alternative }
[ "else" sequence_of_statements ]
"end" "select" ";"
guard ::= "when" condition "=>"
select_alternative ::=
accept_alternative|delay_alternative|terminate_alternative
accept_alternative ::= accept_statement [ sequence_of_statements ]
terminate_alternative ::= "terminate" ";"
```

- L'instruction loop :

Il s'agit de l'instruction itérative de base offerte par Ada. Dans notre cas nous nous intéressons à une simple instruction loop.

```
loop_statement ::= [ statement_identifier ":" ]
[ ( "while" condition ) |
("for" defining_identifier "in" ["reverse"]
discrete_subtype_definition) ]
"loop" sequence_of_statements "end" "loop" [ statement_identifier] ";"
```

Le méta-modèle qui représente les instructions composées est présenté par la Figure 7.21.

La structure des instructions composées est définie d'une façon récursive. Par exemple, la méta-classe IfElse descend de Statement et regroupe plusieurs instructions dans les deux parties then et else. Ceci est traduit par les deux méta-agrégations orientées s1 et s2 entre IfElse et Statement.

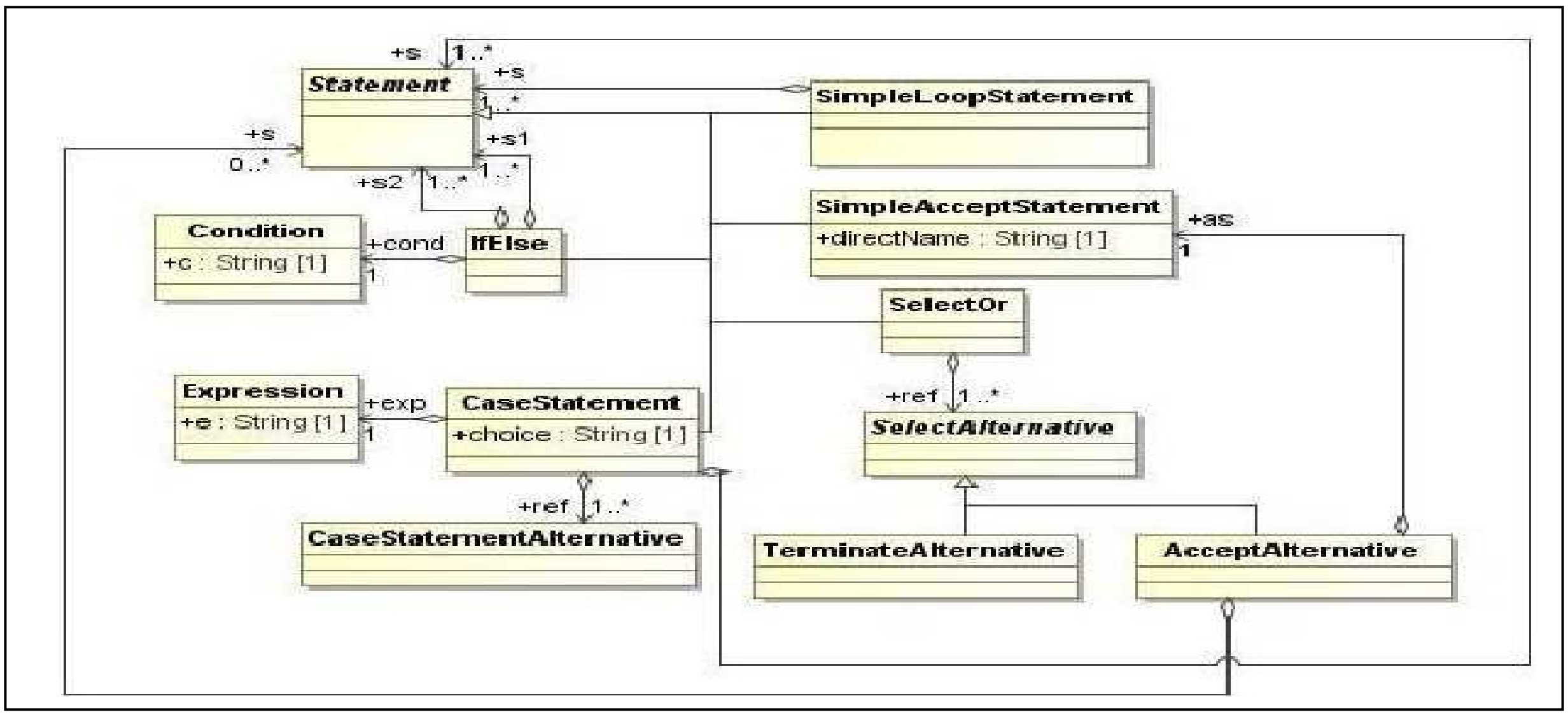

**Figure 7.21 :** Méta-modèle des instructions composées

Le méta-modèle partiel d'Ada utilisé comme méta-modèle cible pour notre approche de transformation de Wright vers Ada est donné par la Figure 7.22.

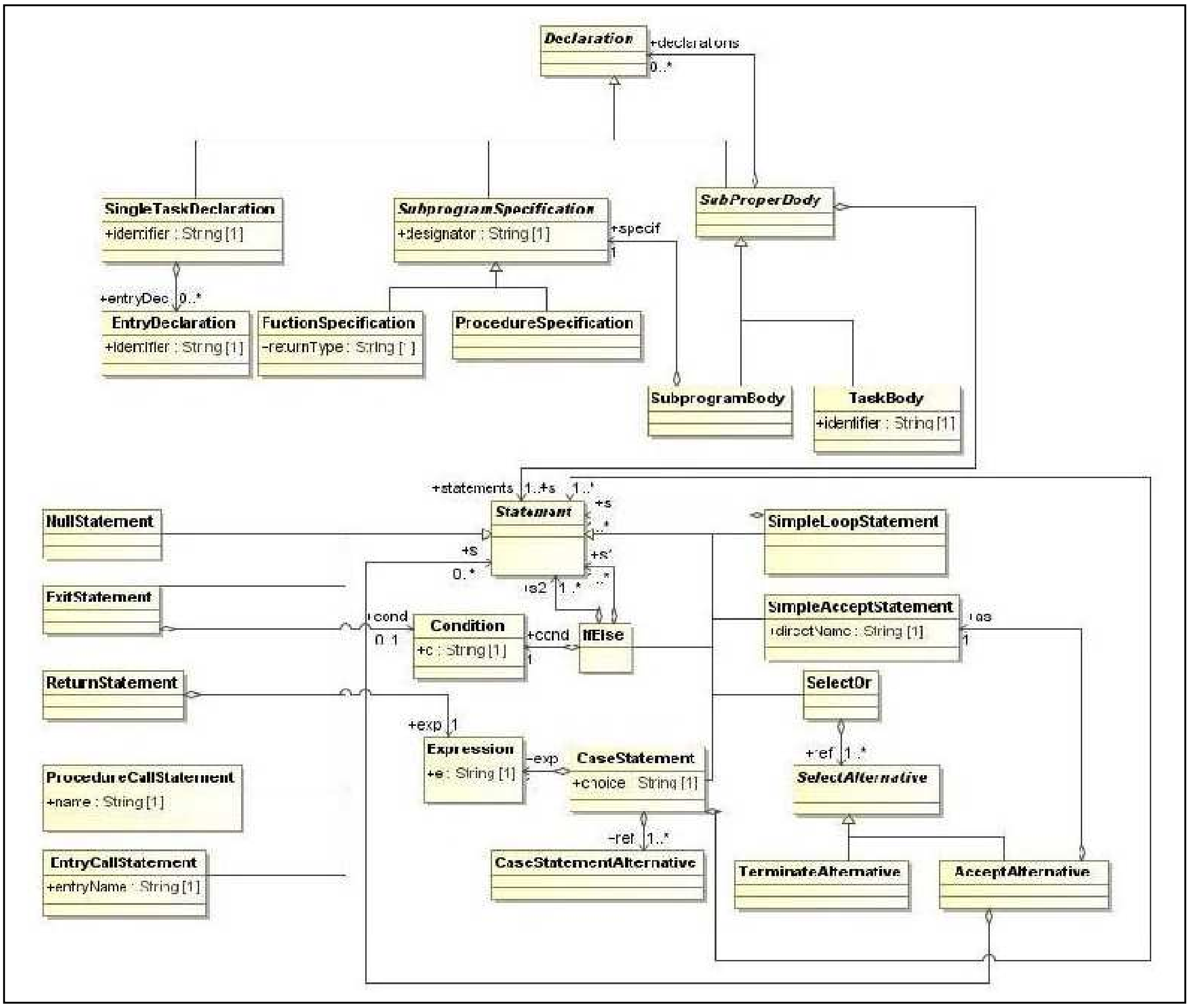

**Figure 7.22 :** Méta-modèle partiel d'Ada

## 7.5.2 Aspects sémantiques

Nous avons établi plusieurs propriétés décrivant des contraintes d'utilisation des constructions d'Ada. De telles propriétés sont décrites d'une façon informelle et formelle en se servant d'OCL.

### 7.5.2.1 Sémantique statique de la partie structurelle d'Ada

Nous proposons deux illustrations de propriétés liées à la sémantique statique de la partie structurelle d'Ada. Les autres propriétés sont présentées au niveau de l'annexe A.

- Propriété 1 :

Au sein de la partie déclarative d'un sous-programme, les noms des tâches (partie spécification et implémentation) et des sous-programmes (partie spécification et implémentation) doivent être deux à deux différents.

```
context SubprogramBody
```

```
def: col1:Sequence(String) = self.declarations ->
select(e:Declaration|e.oclIsKindOf(SubprogramSpecification)) ->
collect(e:SubprogramSpecification|e.designator)
def: col2:Sequence(String) = self.declarations ->
select(e:Declaration|e.oclIsTypeOf(SingleTaskDeclaration)) ->
collect(e:SingleTaskDeclaration|e.identifier)
def: col3:Sequence(String) = self.declarations ->
select(e:Declaration|e.oclIsTypeOf(TaskBody)) ->
collect(e:TaskBody|e.identifier)
def: col4:Sequence(String) = self.declarations ->
select(e:Declaration|e.oclIsTypeOf(SubprogramBody)) ->
collect(e:SubprogramBody|e.specif.designator)
inv: col1 -> excludesAll(col2)
inv: col1 -> excludesAll(col3)
inv: col2 -> excludesAll(col4)
inv: col3 -> excludesAll(col4)
inv: col2->includesAll(col3) and col2->size()=col3->size()
```

- Propriété 2 :

Au sein de la partie déclarative d'un sous-programme, les identificateurs des sous-programme doivent être différents.

```
context SubprogramBody
  inv: self.declarations ->
  select(e:Declaration|e.oclIsKindOf(SubprogramSpecification)) ->
  forAll(e1:SubprogramSpecification, e2:SubprogramSpecification| e1<>e2
  implies e1.designator<>e2.designator)
  inv: self.declarations->
  select(e:Declaration|e.oclIsTypeOf(SubprogramBody)) ->
  forAll(e1:SubprogramBody, e2:SubprogramBody| e1<>e2 implies
  e1.specif.designator<>e2.specif.designator)
```

### 7.5.2.2 Sémantique statique de la partie comportementale d'Ada

Nous proposons deux illustrations de propriétés (propriété 7 et 8) liées à la sématique statique de la partie comportementale d'Ada. Les autres propriétés sont présentées au niveau de l'annexe B.

- Propriété 7 :

Une fonction contient au moins une instruction return.

```
context SubprogramBody
  inv: specif.oclIsTypeOf(FunctionSpecification) implies statements ->
  collect(s:Statement|s.oclIsTypeOf(ReturnStatement)) -> size()>=1
```

- Propriété 8 :

Un sous-programme ne contient pas d'instruction accept.

```
context SubprogramBody
      inv: statements -> forAll(s:Statement | not
      s.oclIsTypeOf(SimpleAcceptStatement))
```

## 7.6 Transformation de Wright vers Ada : le programme Wright2Ada en ATL

### 7.6.1 Vue d’ensemble sur le programme Wright2Ada

La Figure 7.23 donne le contexte de notre programme Wright2Ada permettant de transformer une architecture logicielle décrite en Wright vers un programme concurrent Ada.

Les modèles source et cible (architecture logicielle en Wright et programme concurrent en Ada) ainsi que le programme Wright2Ada sont conforme à leurs méta-modèles Wright, Ada et ATL. Ces méta-modèles sont conformes au méta-modèle Ecore.

Le méta-modèle source de Wright, respectivement cible d’Ada, est représenté par un diagramme Ecore donné par la Figure 7.24, respectivement par la Figure 7.25.

L’en-tête du programme Wright2Ada stocké dans le fichier Wright2Ada.atl se présente par :

```
module WrightToAda;
create exampleAda : Ada from exampleWright : Wright;
```

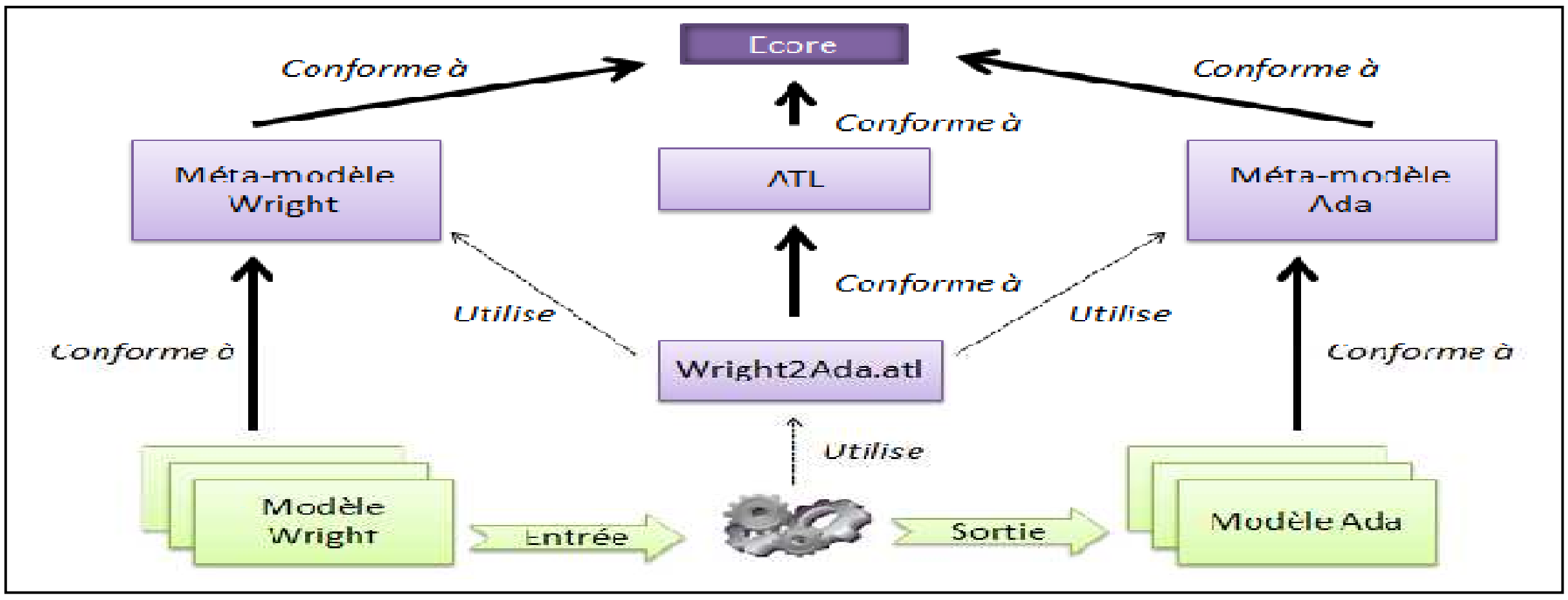

**Figure 7.23 :** Contexte général du programme Wright2Ada

Dans notre programme le modèle cible est représenté par la variable exampleAda à partir du modèle source représenté par exampleWright. Les modèles source et cible sont respectivement conformes aux méta-modèles Wright et Ada. Notre programme Wright2Ada opère sur le modèle source exampleWright en lecture seule et produit le modèle cible exampleAda en écriture seule.

Dans la suite, nous allons présenter progressivement les helpers et les règles standards et paresseuses formant notre programme Wright2Ada écrit en ATL. Notre transformation de Wright vers Ada est basée sur les règles issues de [Bhiri, 2008].

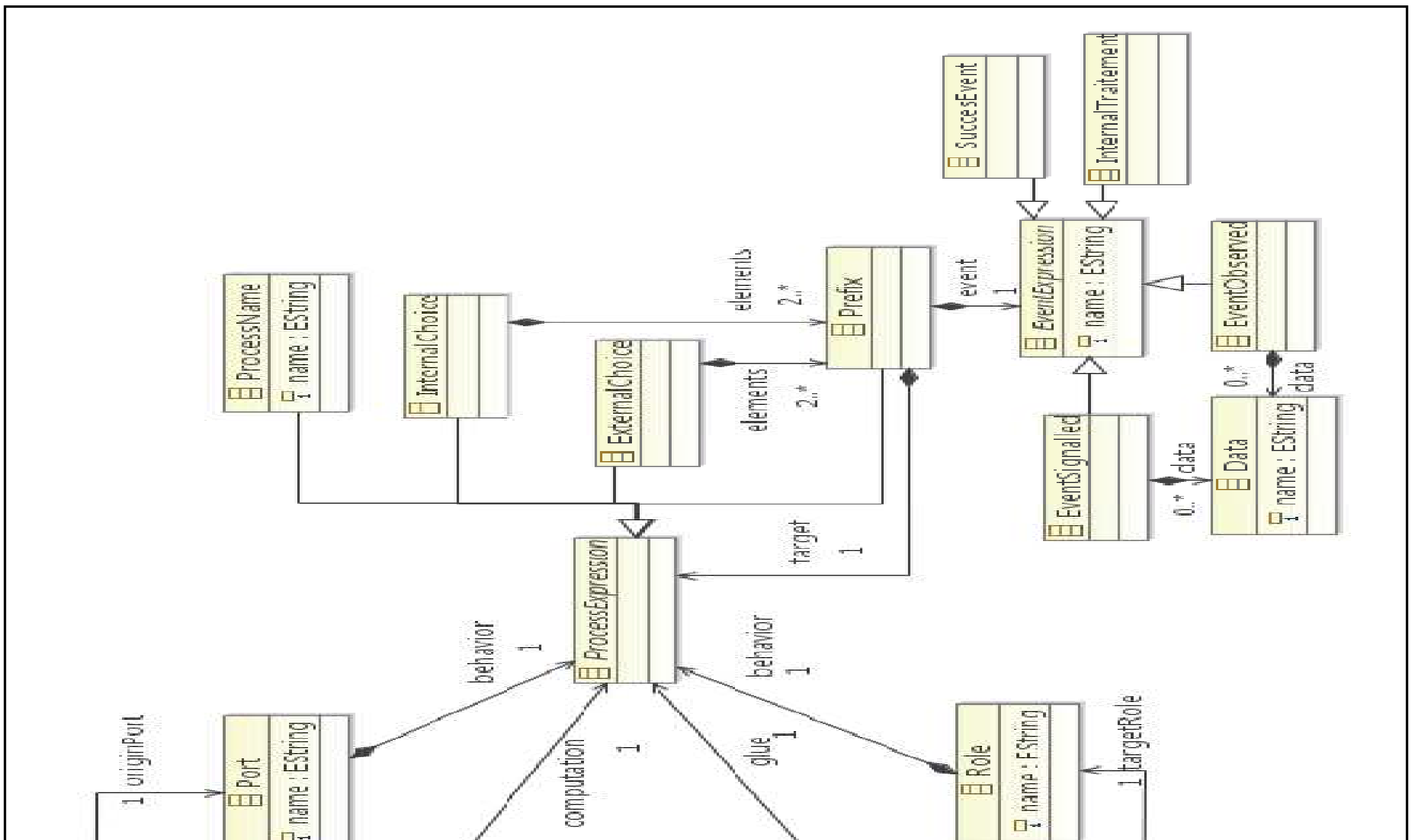

**Figure 7.24:** Méta-modèle Wright en diagramme Ecore

### 7.6.2 Traduction de la partie structurelle de l'ADL Wright

Dans cette section, nous présentons la traduction des aspects structuraux de Wright. Chaque règle de transformation est présentée informellement et illustrée sur un exemple avant de passer à sa formalisation en ATL. Les règles de transformation de la partie structurelle de Wright vers Ada sont illustrées sur l'architecture Client-serveur donnée dans la Figure 7.26.

Dans ce type d'architecture le composant *Client* envoie une requête au composant *Serveur* et attend sa réponse. Le composant *Serveur* quant à lui attend la requête pour répondre. Le connecteur *Lien_CS* joue le rôle d'intermédiaire entre le composant *Client* et le composant *Serveur*.

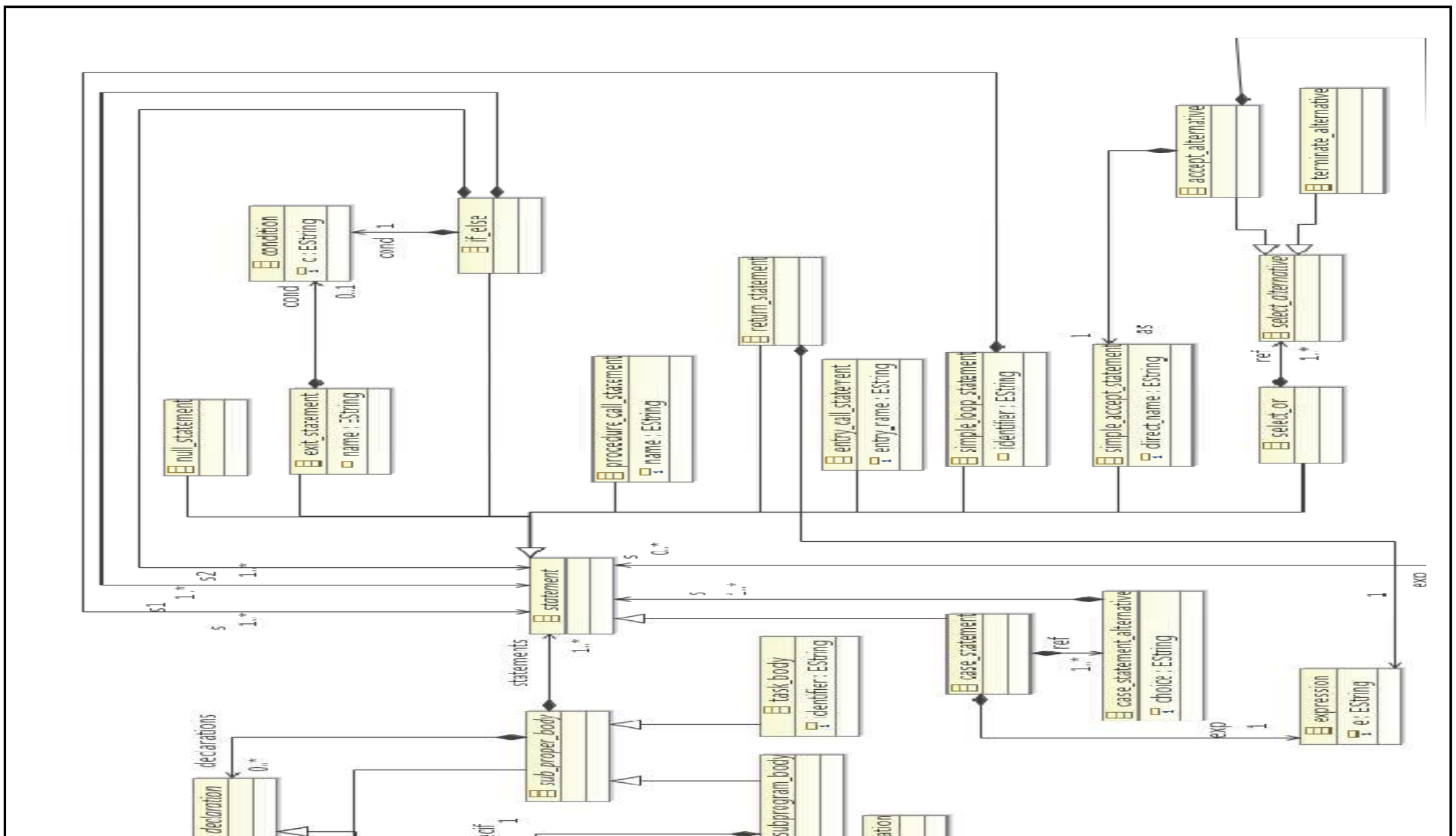

**Figure 7.25:** Méta-modèle partiel d'Ada en diagramme Ecore

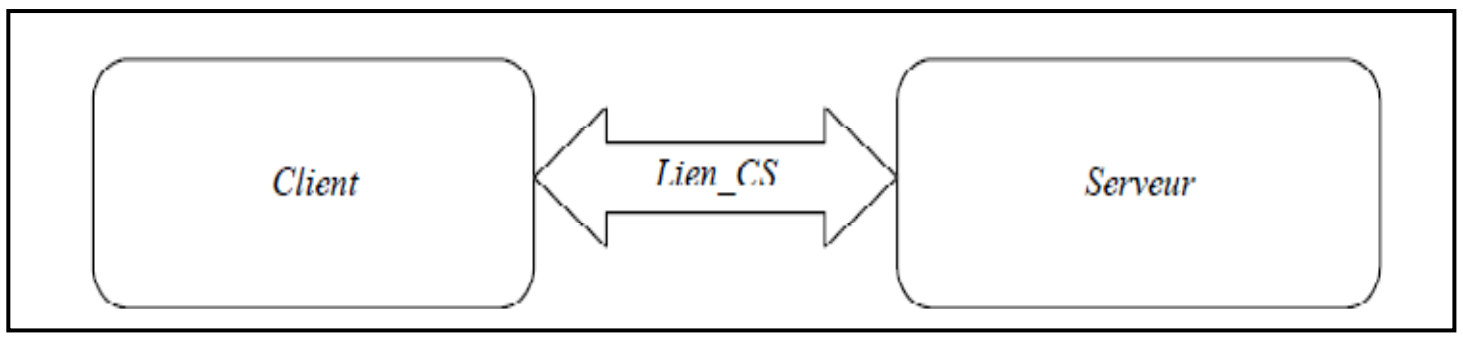

**Figure 7.26 :** Exemple Client-Serveur

**- Traduction d'une configuration Wright :**

Une configuration Wright est traduite en Ada par une procédure. Cette tâche ne fait rien (corps vide); elle constitue une structure d'accueil.

- Illustration sur l'exemple Client-Serveur :

| **Modélisation en Wright** | **Modélisation en Ada** |
|---|---|
| Configuration Client_Serveur<br>...<br>End Configuration | procedure Client_Serveur is<br>...<br>begin<br>null;<br>end Client_Serveur; |

Traduction en ATL:

```
rule Configuration2subprogram{
      from c: Wright!Configuration
      to sb: Ada!subprogram_body ( specif <- sp ,statements <- st ,
declarations <- ... ) , sp: Ada!procedure_specification( designator <-
c.name), st: Ada!null_statement
}
```

Dans cette règle nous créons la procédure qui constitue la structure d'accueil de notre configuration. Dans sa spécification, elle porte le nom de la configuration en question, soit *c.name*, et elle contiendra l'instruction nulle. Sa partie déclarative sera fournie ultérieurement.

### 7.6.2.2 Traduction de la partie structurelle d'une instance de composant et de connecteur

Chaque instance de type composant est traduite par une tâche Ada portant le nom *Component_nomInstanceComposant*.
Chaque instance de type connecteur est traduite par une tâche Ada portant le nom *Connector_nomInstanceConnecteur*.

Les noms sont conservés pour des raisons de traçabilité.

- Illustration sur l'exemple Client-Serveur :

| Modélisation en Wright | Modélisation en Ada |
|---|---|
| Configuration Client_Serveur<br>Component Client<br>...<br>Component Serveur<br>...<br>Connector Lien_CS<br>...<br>Instances<br>client1: Client<br>serveur1: Serveur<br>appel_cs: Lien_CS<br>Attachments<br>...<br>End Configuration | procedure Client_Serveur is<br>task Component_client1 is<br>...<br>end Component_client1;<br>task Component_seveur1 is<br>...<br>end Component_serveur1;<br>task Connector_ appel_cs is<br>...<br>end Connector_ appel_cs;<br>task body Component_client1 is<br>begin<br>...<br>end Component_client1;<br>task body Component_seveur1 is<br>begin<br>...<br>end Component_serveur1;<br>task body Connector_ appel_cs is<br>begin<br>...<br>end Connector_ appel_cs;<br>begin<br>null;<br>end Client_Serveur ; |

**Traduction en ATL :**

```
rule Configuration2subprogram{
   from c: Wright!Configuration
   to sb: Ada!subprogram_body (
specif <- sp , statements <- st , declarations <- c.compInst ->
collect(e|thisModule.ComponentInstance2single_task_declaration(e))
->union(c.connInst ->
collect(e|thisModule.ConnectorInstance2single_task_declaration(e)))
->union(c.compInst ->
collect(e|thisModule.ComponentInstance2task_body(e)))
e|thisModule.ConnectorInstance2task_body(e))) ... ) ,
sp: Ada!procedure_specification( designator <- c.name),
st: Ada!null_statement }
```

La partie déclarative et le corps des tâches font parties de la partie déclarative de la procédure qui joue le rôle de structure d'accueil. Cette règle déclenche les règles paresseuses correspondantes à la partie déclarative et le corps des tâches des instances de composants et de connecteurs.

```
lazy rule ComponentInstance2single_task_declaration{
      from ci:Wright!ComponentInstance
      to    std:Ada!single_task_declaration(identifier <-
      'Component_'+ci.name, entryDec <-... )
}
lazy rule ComponentInstance2task_body{
      from ci:Wright!ComponentInstance
```

```
      to    tb:Ada!task_body( identifier <-'Component_'+ ci.name,
            statements <- ... )
}
```

Dans la partie déclarative et dans le corps des tâches qui représentent les instances de composants nous préservons le nom de l'instance *ci.name* précédé par le préfixe *Component_*. Les instructions des tâches seront fournies ultérieurement.

```
lazy rule ConnectorInstance2single_task_declaration{
      from ci:Wright!ConnectorInstance
      to    std:Ada!single_task_declaration( identifier <-
            'Connector_'+ci.name, entryDec <-... )
}
lazy rule ConnectorInstance2task_body{
      from ci:Wright!ConnectorInstance
      to    tb:Ada!task_body( identifier <-'Connector_'+ ci.name,
            statements <- ... )
}
```

Dans la partie déclarative et dans le corps des tâches qui représentent les instances de connecteurs nous préservons le nom de l'instance *ci.name* précédé par le préfixe *Connector_*. Les instructions des tâches représentant les connecteurs seront fournies ultérieurement.

### 7.6.3 Traduction de la partie comportementale de l'ADL Wright

Cette section présente la traduction de certains aspects comportementaux de Wright décrits en CSP. L'annexe C donne la traduction de tous les concepts comportementaux de Wright vers Ada.

#### 7.6.3.1 Elaboration de la partie déclarative des tâches représentant les instances de composants et de connecteurs

Les événements observés de la partie calcul (Computation) d'un composant, ainsi que de la glu (Glue) d'un connecteur représentent les entrées des tâches qui les matérialisent. Afin d'identifier ces entrées (entry), nous nous inspirons des deux algorithmes décrits dans [Bhiri, 2008] mais pour plus de facilité dans l'automatisation, nous raisonnons sur la partie calcul au lieu des ports, respectivement glu au lieu des rôles.

✓ Algorithme d'élaboration de la partie déclarative des tâches représentant les instances de composants :

```
Pour chaque instance de type composant
 Faire
      Pour chaque événement appartenant à Computation
       Faire
            Si événement est un événement observé de la forme «nomPort.événement»
            Alors
                  Créer une entrée portant le nom nomPort_événement
                  Soit entry nomPort_événement ;
            Fin Si
      Fin Faire
Fin Faire
```

✓ Algorithme d'élaboration de la partie déclarative des tâches représentant les instances de connecteurs :

```
Pour chaque instance de type connecteur
Faire
      Pour chaque événement appartenant à Glue
      Faire
            Si événement est un événement observé de la forme «nomRôle.événement»
            Alors
            Créer une entrée portant le nom nomRôle_événement
            Soit entry nomRôle_événement ;
```

```
            Fin Si
        Fin Faire
Fin Faire
```

– Illustration sur l'exemple Client-Serveur :

| Modélisation en Wright | Modélisation en Ada |
|---|---|
| Configuration Client_Serveur<br>Component Client<br>...<br>Computation= traitement_interne -><br>_port_Client.requete -> port_Client.reponse -><br>Computation \|~\| §<br>Component Serveur<br>...<br>Computation= traitement_interne -><br>port_Serveur.requete -> _port_Serveur.reponse -><br>Computation \|~\| §<br>Connector Lien_CS<br>...<br>Glue= Appelant.requete -> _Appele.requete -> Glue<br>[] Appele.reponse -> _Appelant.reponse -> Glue<br>[] §<br>Instances<br>client1: Client<br>serveur1: Serveur<br>appel_cs: Lien_CS<br>Attachments<br>...<br>End Configuration | procedure Client_Serveur is<br>task Component_client1 is<br>entry port_Client_reponse;<br>end Component_client1;<br>task Component_seveur1 is<br>entry port_Serveur_requete;<br>end Component_serveur1;<br>task Connector_ appel_cs is<br>entry Appelant_requete;<br>entry Appele_reponse;<br>end Connector_ appel_cs;<br>task body Component_client1 is<br>begin<br>...<br>end Component_client1;<br>task body Component_seveur1 is<br>begin<br>...<br>end Component_serveur1;<br>task body Connector_ appel_cs is<br>begin<br>...<br>end Connector_ appel_cs;<br>begin<br>null;<br>end Client_Serveur ; |

**Traduction en ATL :**

Afin d'élaborer la partie déclarative des tâches représentant les instances de composants et de connecteurs, un parcours du processus CSP Wright représentant la partie calcul d'un composant et la glu d'un connecteur est indispensable. Le helper *getEventObserved* fourni ci-dessous permet de faire le parcours nécessaire du processus CSP Wright, à la recherche des événements observés. Il retourne à la règle appelante un ensemble *Set* contenant les événements observés rencontrés lors de son parcours.

```
helper context Wright!ProcessExpression def: getEventObserved():
Set(Wright!EventObserved) =
      if self.oclIsTypeOf(Wright!Prefix)then
         if self.event.oclIsTypeOf(Wright!EventObserved) then
            Set{self.event}->union(self.target.getEventObserved())
         else
            self.target.getEventObserved()
         endif
      else
        if self.oclIsTypeOf(Wright!InternalChoice) or
          self.oclIsTypeOf(Wright!ExternalChoice) then
          self.elements->iterate( child1 ; elements1 :
          Set(Wright!EventObserved) = Set{} | elements1->
          union(child1.getEventObserved()))
      else
            Set{}
      endif
```

```
endif;
```

La règle paresseuse *ComponentInstance2single_task_declaration* fournie ci-dessous, correspond à la traduction de la partie déclarative des tâches représentant les instances de composants. Elle comporte un appel au helper *getEventObserved*, qui retourne l'ensemble des événements observés dans la partie calcul du type composant de l'instance de composant, et déclenche la règle paresseuse qui transforme un événement observé en une entrée *EventObserved2entry_declaration*.

```
lazy rule ComponentInstance2single_task_declaration{
 from ci:Wright!ComponentInstance
 to   std:Ada!single_task_declaration(identifier <- 'Component_'+ci.name,
 entryDec <-ci.type.computation.getEventObserved()->
 collect(e|thisModule.EventObserved2entry_declaration(e)) ) }
```

La règle paresseuse *ConnectorInstance2single_task_declaration* fournie ci-dessous, correspond à la traduction de la partie déclarative des tâches représentant les instances de connecteurs. Cette règle est analogue à la précédente.

```
lazy rule ConnectorInstance2single_task_declaration{
   from ci:Wright!ConnectorInstance
   to std:Ada!single_task_declaration(identifier <- 'Connector_'+ci.name,
   entryDec <-ci.type.glue.getEventObserved()->
  collect(e|thisModule.EventObserved2entry_declaration(e)) )
}
```

La règle paresseuse qui transforme un événement observé en une entrée à la tâche se présente par :

```
lazy rule EventObserved2entry_declaration{
 from eo:Wright!EventObserved
 toed:Ada!entry_declaration( Identifier<- eo.name.replaceAll('.','_'))}
```

### 7.6.3.2 Traduction des événements internes

Les événements internes contenus dans une configuration, c'est-à-dire dans la description des comportements de ses composants ou de ses connecteurs, sont traduits par des procédures dont le corps est à raffiner. Dans cette traduction, le corps de ces procédures contiendra l'instruction nulle.

- Illustration sur l'exemple Client-Serveur :

| Modélisation en Wright | Modélisation en Ada |
|---|---|
| Configuration Client_Serveur<br>  Component Client<br>  ...<br>  Computation= traitement_interne -><br>  _port_Client.requete -> port_Client.reponse<br>  -> Computation \|~\| §<br> Component Serveur<br>  ...<br>  Computation= traitement_interne -><br>  port_Serveur.requete -><br>  _port_Serveur.reponse -> Computation \|~\|<br>  §<br> Connector Lien_CS<br>  ...<br>  Glue= Appelant.requete -><br>  _Appele.requete -> Glue<br>  [] Appele.reponse -> _Appelant.reponse -><br>  Glue<br>  [] §<br>Instances | procedure Client_Serveur is<br>procedure traitement_interne is<br>begin<br>  null;<br>end traitement_interne;<br>task Component_client1 is<br>  entry port_Client_reponse;<br>end Component_client1;<br>task Component_seveur1 is<br>  entry port_Serveur_requete;<br>end Component_serveur1;<br>task Connector_ appel_cs is<br>  entry Appelant_requete;<br>  entry Appele_reponse;<br>end Connector_ appel_cs;<br>task body Component_client1 is<br>begin<br>  ...<br>end Component_client1;<br>task body Component_seveur1 is |
| client1: Client<br>serveur1: Serveur<br>appel_cs: Lien_CS<br>Attachments<br>...<br>End Configuration | begin<br>...<br>end Component_serveur1;<br>task body Connector_ appel_cs is<br>begin<br>...<br>end Connector_ appel_cs;<br>begin<br>null;<br>end Client_Serveur ; |

**-Traduction en ATL :**

Pour ajouter les procédures représentant l'ensemble des événements internes contenus dans une configuration, un parcours des parties calcul (Computation) des composants et des parties glu (Glue) des connecteurs contenus dans cette configuration est indispensable.

Le helper *getInternalTrait* fait le parcours d'un processus CSP Wright à la recherche des événements internes.

```
helper context Wright!ProcessExpression def: getInternalTrait():
Set(Wright!InternalTraitement) =
      if self.oclIsTypeOf(Wright!Prefix)then
            if self.event.oclIsTypeOf(Wright!InternalTraitement) then
                  Set{self.event}->union(self.target.getInternalTrait())
            else
                  self.target.getInternalTrait()
            endif
      else
            if self.oclIsTypeOf(Wright!InternalChoice) or
              self.oclIsTypeOf(Wright!ExternalChoice) then
              self.elements->iterate( child1 ; elements1 :
              Set(Wright!InternalTraitement) = Set{} | elements1->
              union(child1.getInternalTrait()))
            else
                  Set{}
            endif
      endif;
```

Le helper *getInternalTraitement* permet de collecter les traitements internes contenus dans la partie *Computation* des composants et dans la partie *Glue* des connecteurs de la configuration. Pour y parvenir, ce helper fait appel au helper *getInternalTrait* décrit précédement.

```
helper context Wright!Configuration def: getInternalTraitement:
Set(Wright!InternalTraitement)=
      self.conn->iterate( child1 ; elements1 :
Set(Wright!InternalTraitement) = Set{} | elements1->
      union(child1.glue.getInternalTrait()))
      ->union(self.comp->iterate( child2 ; elements2 :
Set(Wright!InternalTraitement) = Set{} | elements2->
      union(child2.computation.getInternalTrait())));
```

Une mise à jour est apportée à la règle de transformation de la configuration en une procédure. Cette règle contiendra, de plus, un appel au helper getInternalTraitement qui collecte l'ensemble des événements internes dans la configuration pour déclencher ensuite la règle paresseuse qui transforme un traitement interne en une procédure. La mise à jour, ainsi que la règle paresseuse déclenchée sont présentées ci-dessous :

```
rule Configuration2subprogram{
```

```
        from c: Wright!Configuration
        to sb: Ada!subprogram_body (specif <- sp , statements <- st ,
              declarations <-c.getInternalTraitement ->
              collect(e|thisModule.InternalTraitement2subprogram(e))
              ->union(c.compInst ->
              collect(e|thisModule.ComponentInstance2single_task_declaration
              (e))) ->union(c.connInst ->
              collect(e|thisModule.ConnectorInstance2single_task_declaration
              (e))) ->union(c.compInst ->
              collect(e|thisModule.ComponentInstance2task_body(e)))
              ->union(c.connInst ->
              collect(e|thisModule.ConnectorInstance2task_body(e)))),
              sp: Ada!procedure_specification( designator <- c.name),
              st: Ada!null_statement
}
```

La règle paresseuse *InternalTraitement2subprogram* fournie ci-dessous traduit un événement interne en une procédure dont le corps est vide, a priori, et dont le nom est celui de l'événement interne en question.

```
lazy rule InternalTraitement2subprogram{
      from i:Wright!InternalTraitment to sb: Ada!subprogram_body
      ( specif <- ps, statements <-ns), ns:Ada!null_statement,ps:
      Ada!procedure_specification( designator <- i.name)
}
```

### 7.6.3.3 Traduction de l'opérateur de récursivité

Tous les processus relatifs à la description des composants et des connecteurs ont un aspect récursif. Dans notre cas, nous nous intéressons plus particulièrement au processus de description de la partie calcul d'un composant et de la glu d'un connecteur. L'opérateur de récursivité est traduit par l'instruction *loop* d'Ada.

- Traduction en ATL :

En tenant compte du fait que les processus représentant la *computation* d'un composant et la *glue* d'un connecteur sont délimités par l'opérateur de récursivité de CSP Wright, il en sera de même pour la traduction en Ada qui commencent par l'instruction *loop*. Ceci peut être traduit par les deux règles paresseuses suivantes :

```
lazy rule ComponentInstance2task_body{
      from ci:Wright!ComponentInstance
      to tb:Ada!task_body( identifier <-'Component_'+ ci.name,
       statements <- ls ), ls : Ada!simple_loop_statement(
      s<- ci.type.computation.transformation(ci.name)
      )
}
lazy rule ConnectorInstance2task_body{
 from ci:Wright!ConnectorInstance to tb:Ada!task_body(
 identifier <-'Connector_'+ ci.name, statements <- ls ),
 ls : Ada!simple_loop_statement(s<- ci.type.glue.transformation(ci.name))
}
```

L'élaboration du corps de la boucle *loop* se fait par l'intermédiaire du helper *transformation(instance : String)*. Celui-ci est redéfini plusieurs fois selon le contexte dans lequel il est appelé, il prend comme paramètre le nom de l'instance de composant ou de connecteur qui l'appelle. Le nom de l'instance passée en paramètre effectif est passé de ce niveau vers le niveau inférieur.

### 7.6.3.4 Traduction de l'opérateur de choix externe

L'opérateur de choix externe ou de choix déterministe est traduit en Ada par l'instruction *select*.

| Modélisation en CSP Wright | Modélisation en Ada |
|---|---|
| a ->P1 [] b->P2 [] V->STOP [] c->Q<br>Les a, b et c sont des événements observés.<br>Le « V » est l'événement succès.<br>Le Pi peut être : préfixe ou un opérateur de choix externe ou un opérateur de choix interne.<br>Le Q et le STOP sont des processus. | *Select*<br>traduction de a puis de P1<br>*or*<br>traduction de b puis de P2<br>*or*<br>traduction de c<br>*or*<br>terminate ;<br>*end select ;*<br>Nous commençons par la traduction des préfixes qui commencent par les événements observés suivis de la traduction du préfixe<br>qui commence par l'événement succès « V » s'il existe. |

**-Traduction en ATL :**

Le helper *getPrefixInOrder* permet de réordonner les préfixes contenus dans l'operateur de choix externe de façon à avoir les préfixes qui commencent par un événement observé suivi du préfixe qui commence par l'événement succès s'il existe. Ce helper retourne un ensemble ordonné contenant l'ensemble des préfixes directement accessibles par l'opérateur de choix externe.

```
helper context Wright!ExternalChoice def:
getPrefixInOrder():OrderedSet(Wright!Prefix) =
self.elements->select(c | c.event.oclIsTypeOf(Wright!EventObserved))
->union(self.elements->select(c |
c.event.oclIsTypeOf(Wright!SuccesEvent)));
```

Le helper redéfini, qui permet de déclencher la règle paresseuse responsable de la transformation d'un opérateur de choix externe en une instruction select est décrite ci-dessus.

```
helper context Wright!ExternalChoice def: transformation(instance :
String):Ada!select_or=
thisModule.ExternalChoice2select_or(self,instance);
```

Le déclenchement est délégué au helper transformation pour pouvoir profiter des facilités offertes par la propriété de redéfinition.

La règle paresseuse responsable de la traduction d'un opérateur de choix externe en une instruction *select* s'appelle `ExternalChoice` ci-dessus.

```
lazy rule ExternalChoice2select_or{
      from p:Wright!ExternalChoice,
            instance : String
            to s:Ada!select_or( ref <- p.getPrefixInOrder()->collect(e|
      if e.event.oclIsTypeOf(Wright!EventObserved)then
            if e.target.oclIsTypeOf(Wright!ProcessName) then
            thisModule.Prefix2accept_alternative1(e,instance)
            else
                  thisModule.Prefix2accept_alternative2(e,instance)
            endif
      else
            thisModule.SuccesEvent2terminate_alternative(e)
      endif ))
}
```

Cette règle paresseuse fait appel au helper *getPrefixInOrder* pour ordonner les préfixes directement accessibles puis déclenche la règle paresseuse adéquate. Si le préfixe commence par un événement observé suivi du nom d'un processus un déclenchement de la règle paresseuse *Prefix2accept_alternative1* aura lieu, si le préfixe commence par un événement observé suivi d'un opérateur de choix externe ou interne ou un autre préfixe, un déclenchement de la règle paresseuse *Prefix2accept_alternative2* aura lieu. Si le préfixe commence par l'événement succès un déclenchement de la règle paresseuse *SuccesEvent2terminate_alternative* aura lieu.

```
lazy rule Prefix2accept_alternative1{
      from p:Wright!Prefix,
            instance : String
          to a:Ada!accept_alternative(
          as <- thisModule.EventObserved2simple_accept_statement(p.event)
)
}
```

Cette règle paresseuse correspond à une alternative de l'instruction *select*. Elle déclenche la règle paresseuse permettant de traduire un événement observé.

```
lazy rule Prefix2accept_alternative2{
      from p:Wright!Prefix,
         instance : String
         to a:Ada!accept_alternative(
         as <- thisModule.EventObserved2simple_accept_statement(p.event),
         s<- p.target.transformation(instance)
)
}
```

Cette règle paresseuse admet le même comportement que la règle précédente, avec en plus, l'appel au helper *transformation* afin de traduire la cible du préfixe.

```
lazy rule SuccesEvent2terminate_alternative{
      from p:Wright!SuccesEvent
      to a:Ada!terminate_alternative
}
```

Cette règle paresseuse correspond à l'alternative *terminate* de l'instruction *select*.

Dans cette section, nous avons présenté d'une façon assez détaillée le programme Wright2Ada conçu et réalisé dans le cadre de cette thèse afin de transformer une architecture logicielle décrite en Wright vers un programme concurrent Ada. Notre programme Wright2Ada est purement déclaratif et comporte :

- 1 règle standard (ou matched rule),
- 19 règles paresseuses (ou lazy rules),
- 3 helpers attributs,
- 12 helpers opérations,
- 3 helpers polymorphiques.

Le programme Wright2Ada en ATL est fourni dans l'annexe D.

## 7.7 Interfaces conviviales d'utilisation de Wright2Ada

Dans cette section, nous apportons des interfaces conviviales afin d'utiliser notre programme Wright2Ada dans un contexte réel. Ces interfaces permettent d'utiliser le programme Wright2Ada en introduisant du code Wright et en produisant du code Ada. La transformation accomplie par le programme Wright2Ada, présentée dans la section précédente, suppose une compréhension des méta-modèles source et cible par l'utilisateur, un certain savoir-faire pour produire le modèle source et comprendre le modèle cible généré. De plus, il est souvent difficile de produire l'entrée de la transformation lorsque la spécification Wright est complexe. En effet, l'utilisateur est censé se servir du navigateur de modèles et de l'éditeur de propriétés afin d'introduire le texte Wright sous format XMI. Ceci est non convivial et sujet à des erreurs potentielles. En outre, il est censé transformer manuellement le modèle Ada au format XMI en code Ada afin de se servir des outils associés à Ada tels que: compilateur et model-checker.

Dans cette section, nous nous proposons de traiter les deux étapes d'injection et d'extraction. L'injection prend un modèle exprimé dans la syntaxe concrète textuelle de Wright et génère un modèle conforme au méta-modèle Wright dans l'espace technique de l'ingénierie des modèles. L'extraction travaille sur la représentation interne des modèles exprimés en Ada et crée la représentation textuelle (code Ada).

Pour ce faire nous nous proposons d'utiliser en premier lieu les possibilités fournies par l'outil Xtext de oAW (open Architecture Ware), afin de procéder à la transformation du texte Wright vers son modèle. Ensuite, nous proposons une transformation du modèle d'Ada vers son code avec l'outil Xpand de oAW. Enfin, nous présentons un exemple d'utilisation.

### 7.7.1 Texte Wright vers modèle Wright

Cette partie présente la validation et la transformation d'un texte Wright vers un modèle Wright conforme au méta-modèle Wright proposé dans la section 7.2. Le schèma général de cette transformation est donné dans la Figure 7.27.
La transformation proposée comporte trois étapes. La première étape a pour objectif de créer le méta-modèle Ecore appelé Grammaire Wright à partir d'une description en Xtext de la grammaire de Wright et de produire l'analyseur lexico-syntaxique de Wright via Xtext. La deuxième étape a pour objectif de valider (via Check) et transformer le modèle exprimé dans la syntaxe concrète textuelle de Wright en un modèle XMI conforme au méta-modèle Grammaire Wright. La troisième étape a pour objectif de transformer le modèle XMI conforme au méta-modèle Grammaire Wright vers un modèle XMI conforme au méta-modèle Wright via ATL.

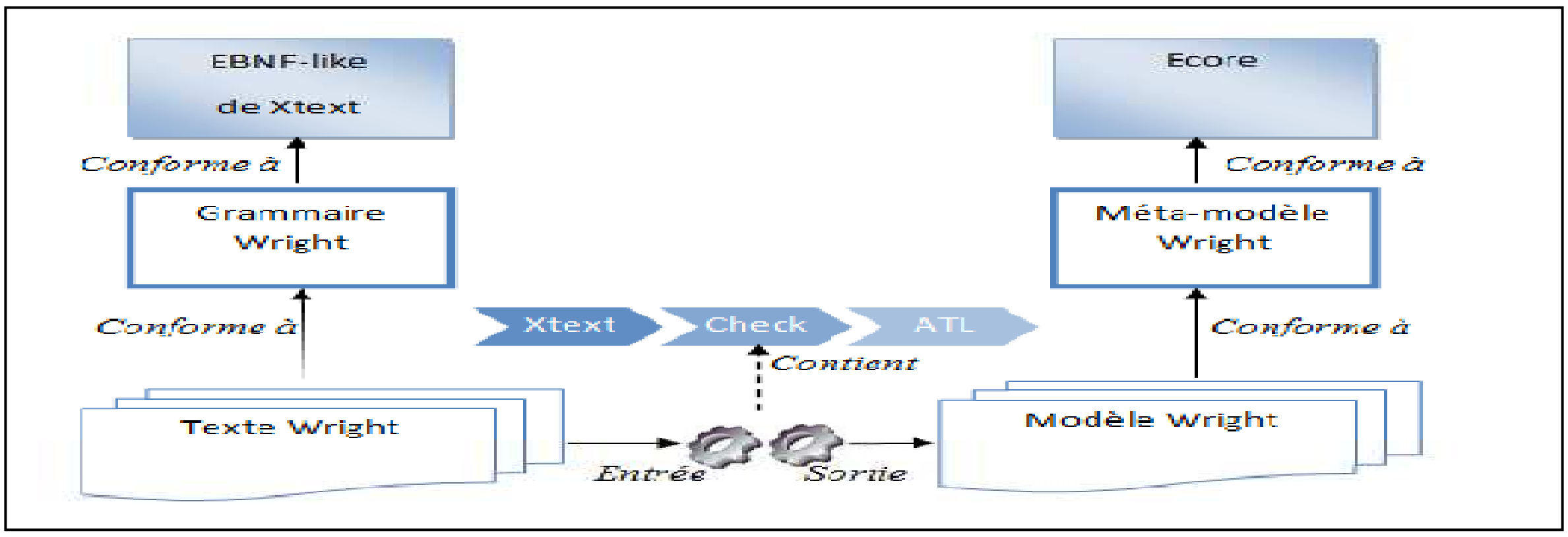

**Figure 7.27 :** Vue d'ensemble sur la transformation texte vers modèle Wright

### 7.7.1.1 Injection via Xtext

Dans cette sous partie, nous commençons par présenter la création du projet Xtext. Puis, nous présentons la grammaire du langage Wright avec Xtext. Enfin, nous donnons un aperçu sur le méta-modèle de Wright généré avec Xtext.

#### 7.7.1.1.1 Création du projet xtext

Pour commencer nous allons créer un nouveau projet Xtext avec l'option de création de générateur de projet, car nous allons utiliser ce générateur plus tard. Ceci est présenté par la Figure 7.28.

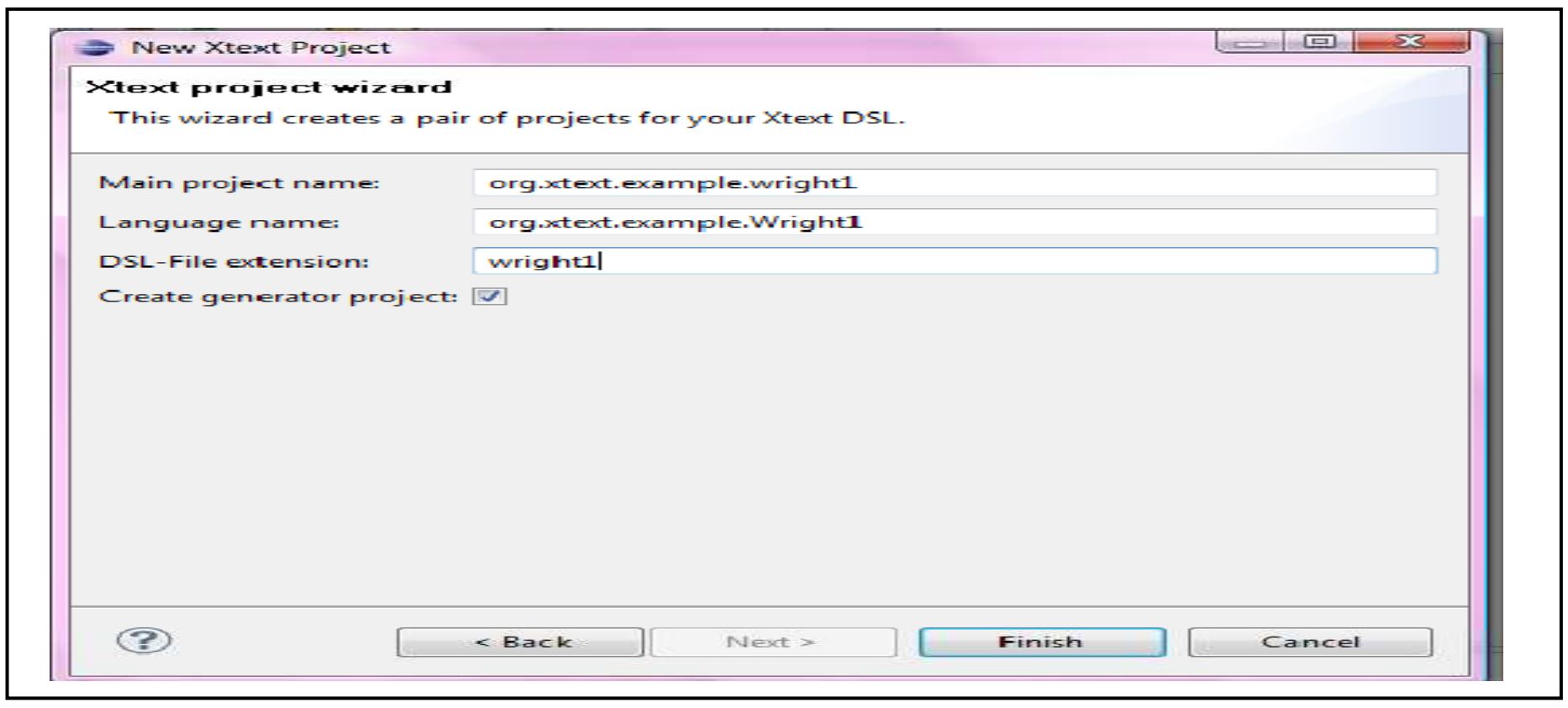

**Figure 7.28 :** Création du projet xtext

Nous devons avoir après cette étape de création, trois projets dans notre espace de travail. Le premier projet est le projet principal de Xtext où nous allons définir la grammaire de l'ADL Wright. Le second projet est l'éditeur de projet, il contiendra l'éditeur Xtext généré automatiquement à base de notre DSL Wright. Enfin, le troisième projet fournit l'interface d'utilisation (User Interface).

#### 7.7.1.1.2 Grammaire de l'ADL Wright

Dans le premier projet, et dans le fichier d'extension .xtext, nous allons créer la grammaire de notre DSL Wright. Ce fichier contient au premier niveau les deux lignes suivantes :

```
grammar org.xtext.example.Wright1 with org.eclipse.xtext.common.Terminals
generate wright1 "http://www.xtext.org/example/Wright1"
```

La première ligne déclare l'identificateur du modèle et la base des déclarations. La deuxième ligne est la directive de création du méta-modèle Ecore généré avec son emplacement.

- Création de l'entité Configuration :

La première entité de notre grammaire est la configuration. Une configuration Wright a le format suivant :

| |
|---|
| Configuration *nom_de _la configuration*<br>*L'ensemble des définitions de composants et de connecteurs*<br>Instances<br>*L'ensemble de déclaration des instances*<br>Attachment<br>*L'ensemble des attachements*<br>End Configuration |

La méta-classe Configuration dans le méta-modèle Wright est présentée par la Figure 7.29.

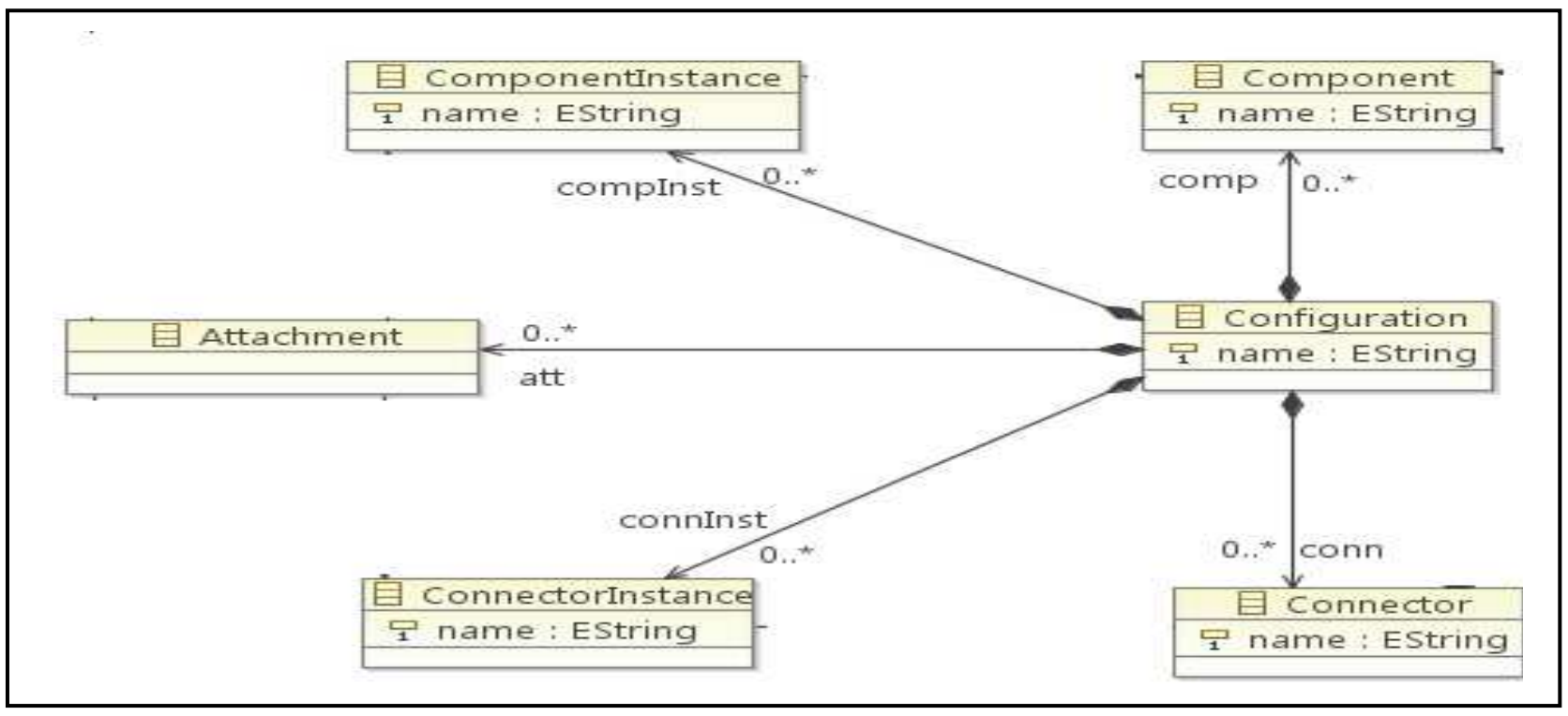

**Figure 7.29 :** Méta-classe Configuration

L'entité Configuration peut être traduite par la règle de production ci-dessous.

```
Configuration : "Configuration" name=ID
( TypeList+=Type )*
"Instances"
( InstanceList+=Instance)*
"Attachments"
( att+=Attachment )*
"End Configuration";
Instance: ComponentInstance | ConnectorInstance ;
Type: Component| Connector;
```

Nous venons de définir une configuration avec un nom, l'ensemble de types qui peuvent être des composants ou des connecteurs, l'ensemble d'instances qui peuvent être des instances de composants ou de connecteurs, et enfin, l'ensemble des attachements.

Le symbole « += » signifie que la variable contient un ensemble du type correspondant. Le symbole « * » signifie la cardinalité zéro ou plusieurs.

Il y a une différence entre le méta-modèle présenté et la grammaire de l'entité Configuration. Cette différence est dûe au fait que nous ne pouvons pas imposer un ordre de déclaration pour les composants et les connecteurs. Ceci est également vrai pour les instances de composants et de connecteurs.

- Création des entités Component et Port:

Un composant est défini selon le format suivant :

| |
|---|
| Component *nom_du _composant*<br>*L'ensemble des définitions de ces ports*<br>Computation =<br>*L'expression du processus compuation* |

Les méta-classes Component et Port sont présentées par la Figure 7.30.

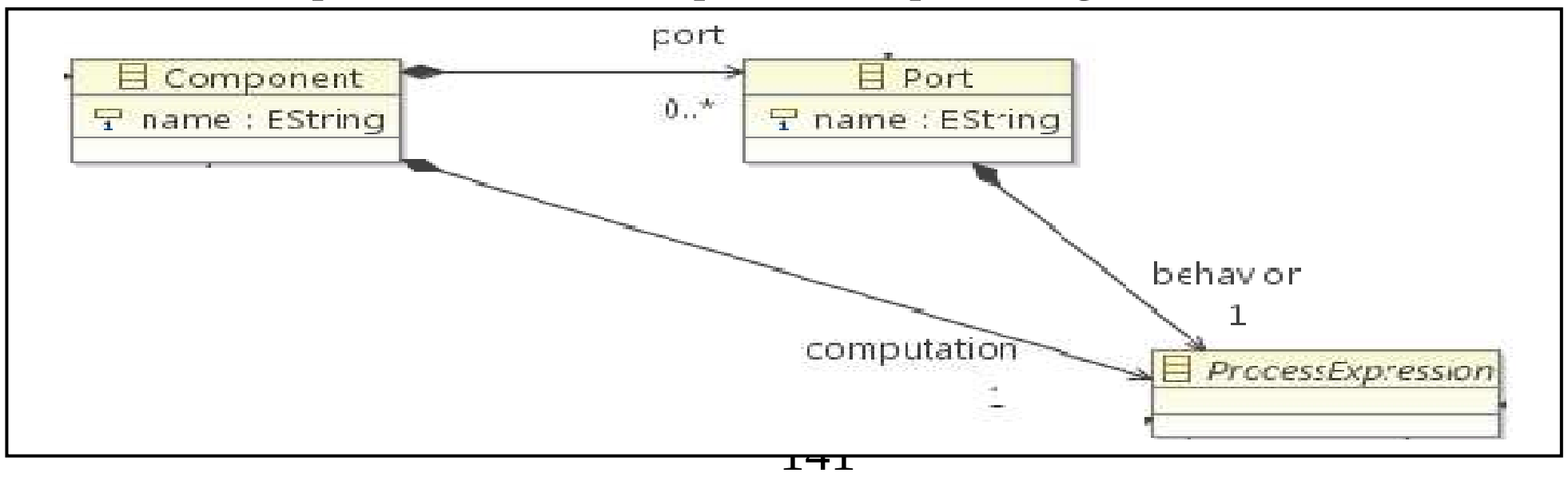

**Figure 7.30 :** Méta-classes Component et Port

L'entité Component peut être traduite par la règle de production ci-dessous.

```
Component : "Component" name=ID
( port+=Port )+
"Computation" '=' computation=ProcessExpression ;
```

Nous venons de définir un composant avec un nom, l'ensemble de ses ports et son processus compuation.

Un port est défini par :

| |
|---|
| Port *nom_du _port =*<br>*Le comportement du port ;l'expression du processus nom_port* |

L'entité port peut être traduite par la règle de production ci-dessous.

```
Port : "Port" name=ID '=' behavior=ProcessExpression;
```

Un port a un nom et un comportement décrit par l'expression d'un processus

- Création des entités Connector et Role:

Un connecteur est défini par :

| |
|---|
| Connector *nom_du _connecteur*<br>*L'ensemble des définitions de ces rôles*<br>Glue = *L'expression du processus glue* |

Les méta-classes Connector et Role sont présentées par la Figure 7.31.

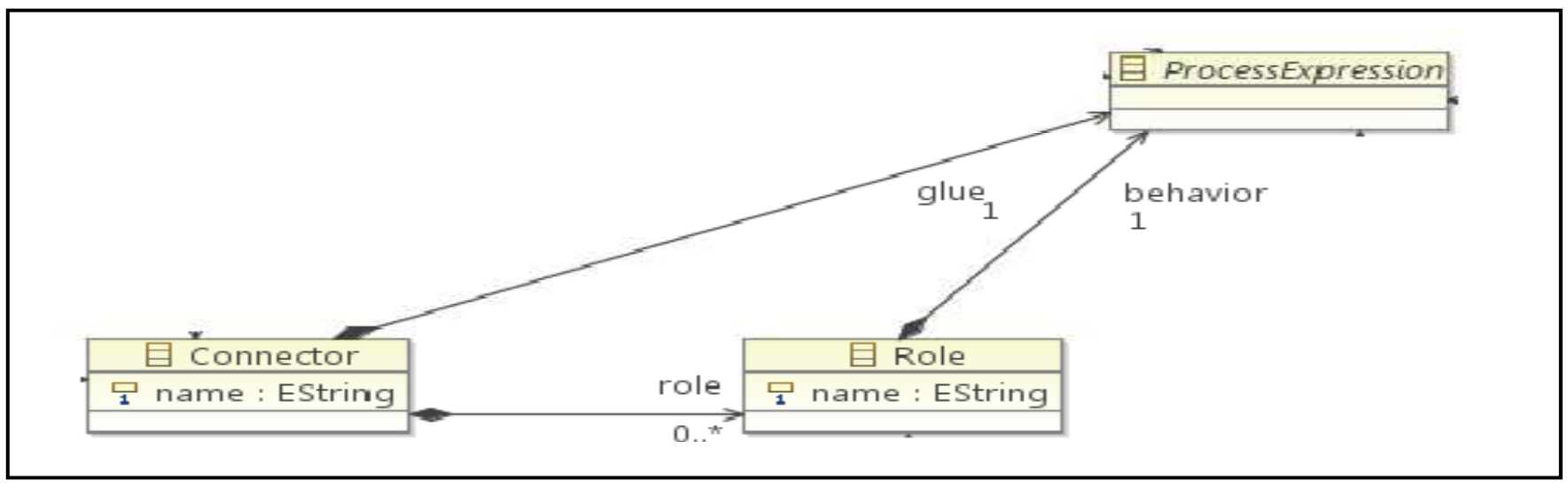

**Figure 7.31:** Méta-classes Connector et Role

L'entité Connector peut être traduite par la règle de production ci-dessous.

```
Connector : "Connector" name=ID
( role+=Role )+
"Glue" '=' glue=ProcessExpression ;
```

Nous venons de définir un connecteur avec un nom, l'ensemble de ses rôles et son processus glue.

Un rôle est défini par :

| |
|---|
| Role *nom_du _rôle =  Le comportement du rôle ;l'expression du processus nom_rôle* |

L'entité rôle peut être traduite par la règle de production ci-dessous.

```
Role : "Role" name=ID '=' behavior=ProcessExpression;
```

Un rôle a un nom et un comportement décrit par l'expression d'un processus

✓ Création des entités ComponentInstance et ConnectorInstance :

Une instance de composant est définie par :

| |
|---|
| *Nom_ de_l'instance_de_composant* : *nom_du_composant_type* |

La méta-classe ComponentInstance est présentée par la Figure 7.32.

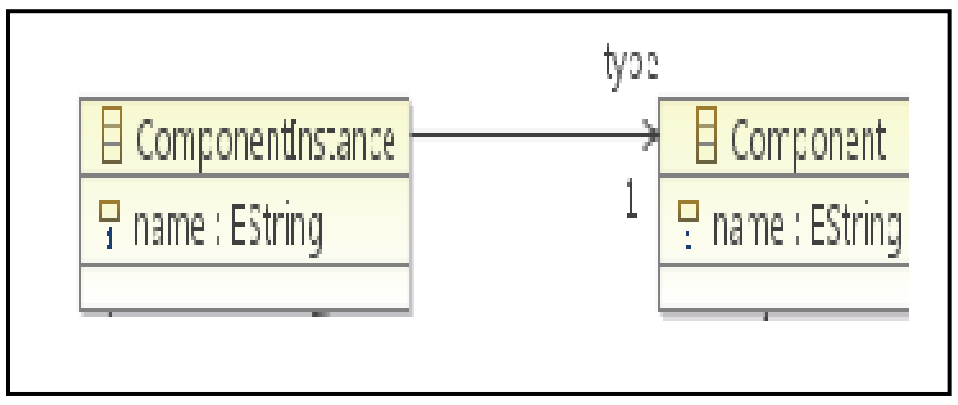

**Figure 7.32 :** Méta-classe ComponentInstance

Cette instance peut être traduite par la règle de production ci-dessous.

```
ComponentInstance : name=ID ':' type=[Component];
```

Une instance de composant a un nom et une référence vers le composant type.

Une instance de connecteur est définie par :

| |
|---|
| *Nom_ de_l'instance_de_connecteur* : *nom_du_connecteur_type* |

Un raisonnement similaire donne la règle de production de l'instance de connecteur :

```
ConnectorInstance : name=ID ':' type=[Connector];
```

Une instance de connecteur a un nom et une référence vers le connecteur type.

Les deux règles présentées ci-dessus posent un problème avec l'analyseur lexico-syntaxique de Xtext, car elles sont similaires. L'analyseur va avoir une confusion sur l'alternative qui va choisir ; celle de l'instance de composant ou de l'instance de connecteur.

Une solution pour remédier à ce problème, est de changer la règle de production de ces deux entités par :

```
ComponentInstance : name=ID ':' "Component" type=[Component];

ConnectorInstance : name=ID ':' "Connector" type=[Connector];
```

La définition de ces deux instances devient alors :

| |
|---|
| *Nom_ de_l'instance_de_composant* : Component *nom_du_composant_type* |

Création de l'entité Attachment :

Un attachement est défini par :

| |
|---|
| *Nom_ de_l'instance_de_connecteur* : Connector *nom_du_connecteur_type* |
| *le_nom_d'une_instance_de_composant*<br>'.'<br>*le_nom_du_port_d'origine*<br>"As"<br>*le_nom_d'une_instance_de_connecteur*<br>'.'<br>*le_nom_du_role_cible* |

La méta-classe Attachment est présentée par la Figure 7.33.

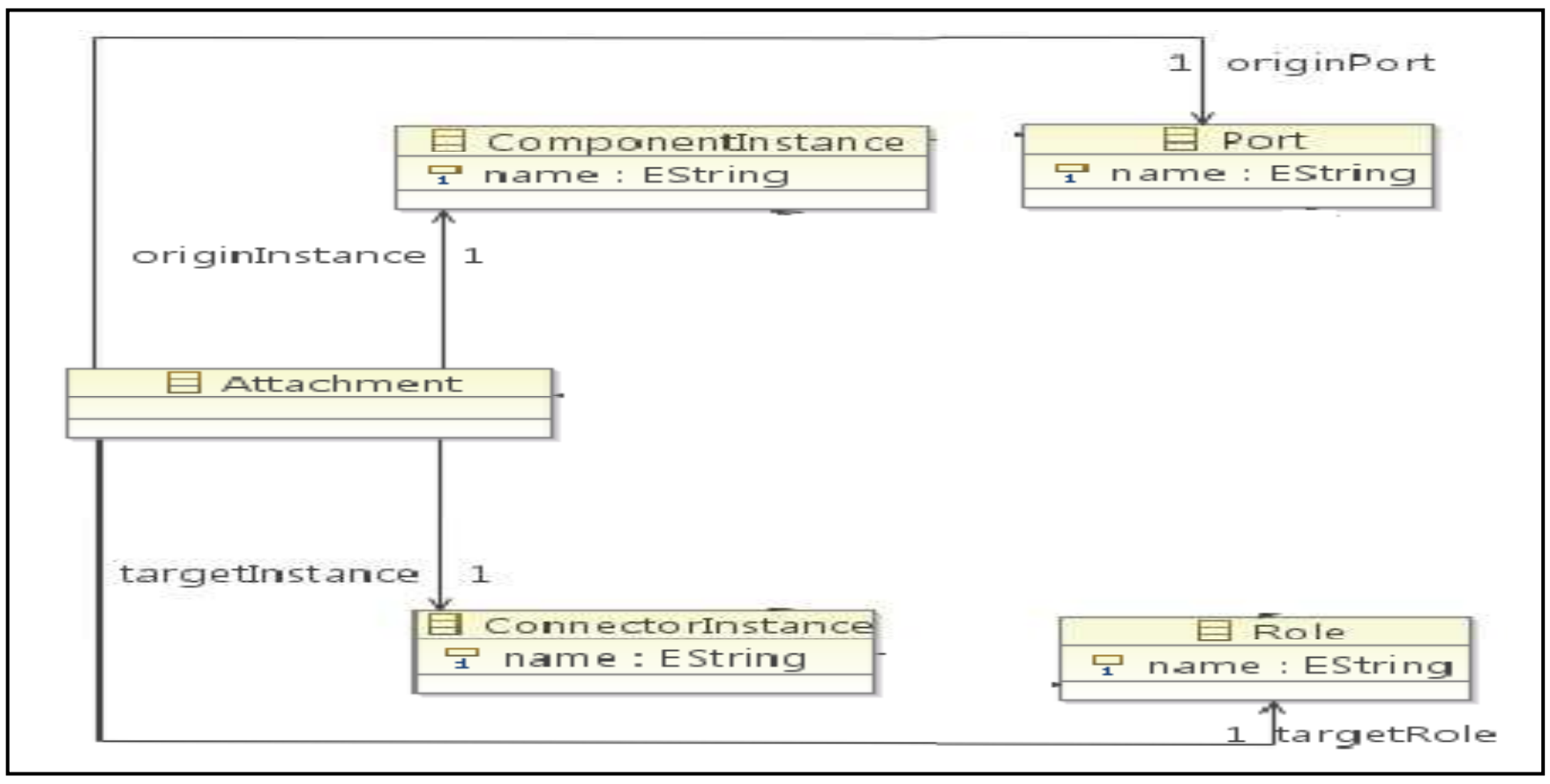

**Figure 7.33 :** Méta-classe Attachment

L'entité Attachment peut être traduite par la règle de production ci-dessous.

```
Attachment : originInstance=[ComponentInstance] '.' originPort=[Port]
"As" targetInstance=[ConnectorInstance] '.' targetRole=[Role] ;
```

Un attachement est composé de quatres références qui ont pour cible les méta-classes Component, Port, Connector et Role.

La règle présentée ci-dessus pose problème avec le « . » car il se trouve dans la règle terminale ID qui à son tour se présente par ::

```
terminal ID: ('a'..'z'|'A'..'Z') ('a'..'z'|'A'..'Z'|'_'|'.'|'0'..'9')*;
```

Une solution pour remédier à ce problème est de le remplacer par « - » .

La règle de production de l'entité Attachment devient :

```
Attachment : originInstance=[ComponentInstance] '-' originPort=[Port]
"As" targetInstance=[ConnectorInstance] '-' targetRole=[Role] ;
```

La définition d'un attachement devient :

*le_nom_d'une_instance_de_composant*
'-'
*le_nom_du_port_d'origine*
"As"
*le_nom_d'une_instance_de_connecteur*
'-'
*le_nom_du_role_cible*

Création des entités des événements Wright:
Les événements sont présentés par le méta-modèle de la Figure 7.34.

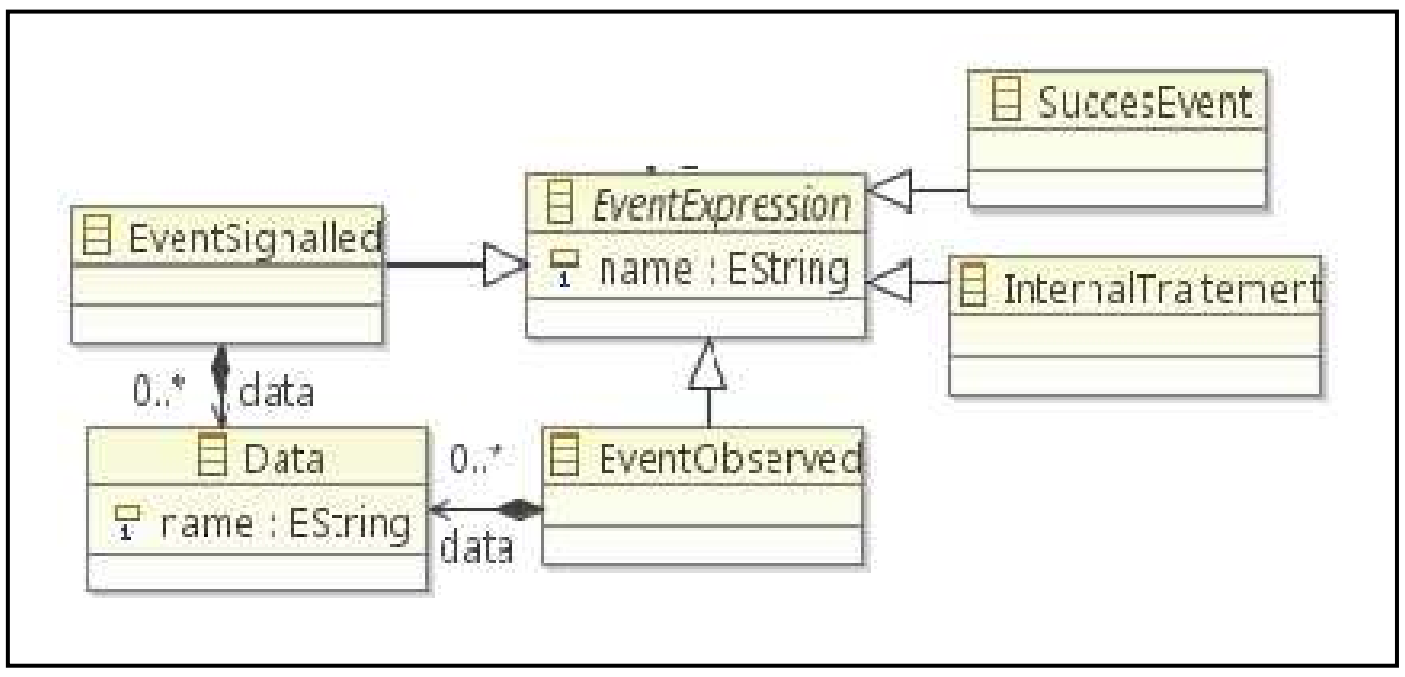

**Figure 7.34:** Méta-modèle des événements

Pour faire la distinction entre les événements observés, les événements initialisés et les traitements internes les événements initialisés doivent être obligatoirement préfixés par « _ » et le traitement interne par « - ». L'événement succès est toujours nommé √ soit « V ».
Ci-dessous nous présentons les entités des événements :

```
EventExpression : EventSignalled | EventObserved | InternalTraitement |
SuccesEvent;
EventSignalled: '_' name=ID (data+=Data)*;
EventObserved: name=ID (data+=Data)*;
InternalTraitement: '-' name=ID;
SuccesEvent: name='V';
```

Les événements observés et initialisés peuvent transporter des données préfixées par « ! » ou par « ? » ce qui représentent respectivement des données en sortie et en entrée.

```
Data : ('?' | '!') name=ID;
```

✓ Création des opérateurs du processus CSP Wright:
Le processus CSP Wright est décrit par le méta-modèle de la Figure 7.35.

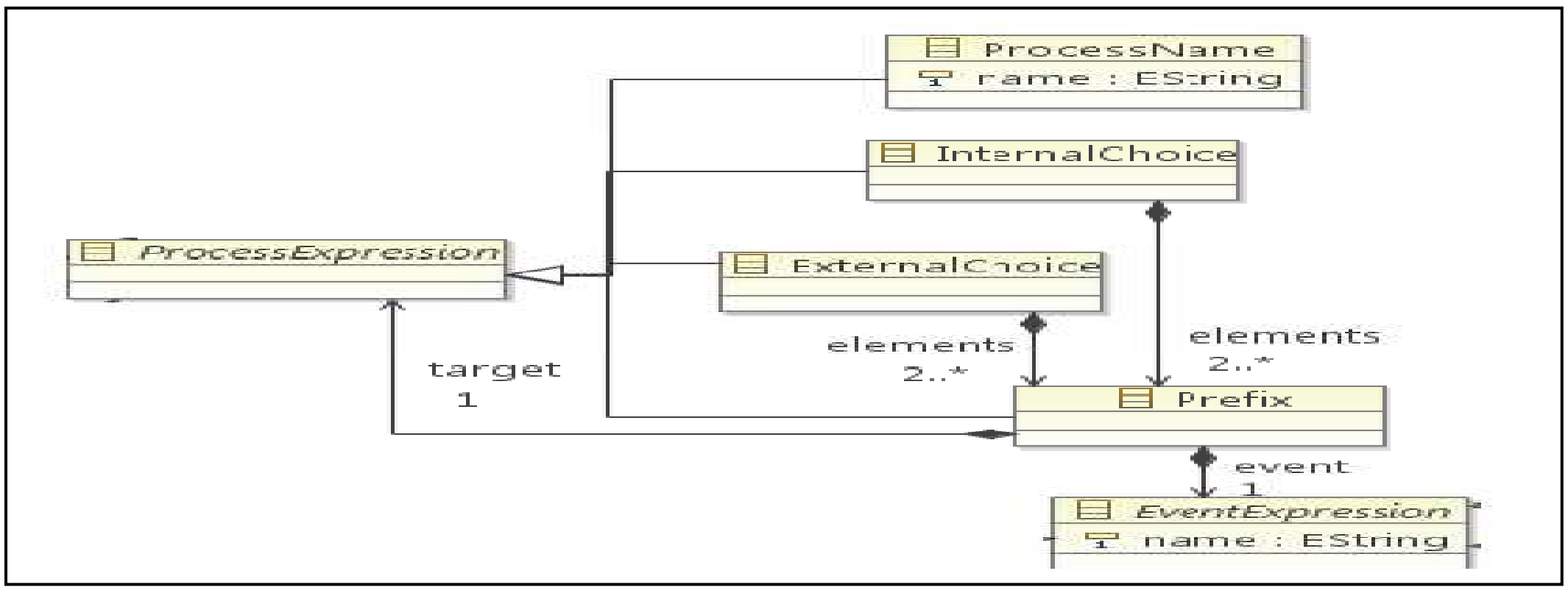

**Figure 7.35 :** Méta-modèle du processus CSP Wright

La méta-classe *ProcessName* est traduite par la règle de production ci-dessous.

```
ProcessName: name=ID ;
```

Un opérateur de préfixe peut être décrit par:

| | |
|---|---|
| *EventExpression* -> *ProcessExpression* | *EventExpression* -> (*ProcessExpression*) |

Un opérateur de choix externe est décrit par :

| |
|---|
| *Préfixe1* [] *Préfixe2* [] … |

Un opérateur de choix interne est décrit par :

| |
|---|
| *Préfixe1* \|~\| *Préfixe2* \|~\| … |

Ces derniers peuvent êtres traduits par les règles de production ci-dessous.

```
ProcessExpression: InternalChoice | ExternalChoice | ProcessName |Prefix
| Parentheses;
Parentheses: '(' p=ProcessExpression ')';
Prefix: event=EventExpression '->' target=ProcessExpression;
InternalChoice: p=Prefix ('|~|' e+=Prefix)+;
```

```
ExternalChoice: p=Prefix ('[]' e+=Prefix)+;
```

Mais cette solution pose malheureusement un problème, car elle n'est pas LL(*). Xtext fonctionne avec l'analyseur syntaxique ANTLR qui est basé sur les algorithmes LL(*). Le problème peut être résolu par une factorisation gauche. Nos règles de productions deviennent :

```
Prefix: event=EventExpression '->' target=TargetPrefix;
TargetPrefix: Parentheses | Prefix | ProcessName;
Parentheses: '(' p=ProcessExpression ')';
ProcessExpression : right=Prefix (('[]' ECLeft+=Prefix)+|('|~|'
ICLeft+=Prefix)+)?;
```

De plus, le symble § ou encore SKIP désigne V -> STOP, donc la règle de production de préfix devient :

```
Prefix: event=EventExpression '->' target=TargetPrefix | name='§'|
name='SKIP';
```

Dans les règles présentées ci-dessus l'opérateur de préfixe, l'opérateur de choix interne et externe sont traduits dans une même règle de grammaire nommée ici ProcessExpression.
La grammaire de l'ADL Wright décrite en Xtext est fournie dans l'annexe E.

##### 7.7.1.1.3 Méta-modèle de Wright généré avec Xtext

L'exécution du moteur workflow qui existe par défaut dans le premier projet permet, entre autre, de générer le diagramme Ecore présenté dans la Figure 7.36. Le diagramme Ecore généré correspond à la grammaire de l'ADL Wright en Xtext.

#### 7.7.1.2 Vérification et génération du modèle Wright en XMI

##### 7.7.1.2.1 Sémantique statique de Wright

La sémantique statique de Wright est décrite à l'aide des contraintes OCL attachées au méta-modèle Wright (cf. 7.3). Ces contraintes sont réécrites en Check et attachées au méta-modèle Grammaire Wright -appelé Wright1- généré par l'outil Xtext.

Les contraintes Check données ci-dessous seront évaluées sur les textes Wright. Ensuite, ces textes Wright seront transformés en XMI conformes au méta-modèle Grammaire Wright -appelé Wright1- moyennant l'utilisation des plugins : - org.xtext.example.wright1 que nous avons développé dans la section 7.8.1.1. et le plugin org.eclipse.xtext. Ses plugins permettent d'interpréter le texte Wright comme étant un modèle conforme au méta-modèle Grammaire Wright –appelé Wright1-.

##### 7.7.1.2.2 Le moteur de vérification et de génération

Le moteur workflow du deuxième projet doit être modifié par :

```
<workflow>
<property name="modelFile" value="classpath:/model/MyModel1.wright1"/>
<property name="targetDir" value="src-gen/example1"/>
<bean class="org.eclipse.emf.mwe.utils.StandaloneSetup"
platformUri=".."/>
<component class="org.eclipse.emf.mwe.utils.DirectoryCleaner"
directory="${targetDir}"/>
<component class="org.eclipse.xtext.MweReader" uri="${modelFile}">
<!—Cette classe est générée par le générateur de xtext -->
<register class="org.xtext.example.Wright1StandaloneSetup"/>
</component>
<!—valider le modèle -->
<component class="org.eclipse.xtend.check.CheckComponent">
<metaModel
class="org.eclipse.xtend.typesystem.emf.EmfRegistryMetaModel"/>
<checkFile value="model::CheckFile" />
<emfAllChildrenSlot value="model" />
</component>
<!—générer le modèle -->
<component class="org.eclipse.emf.mwe.utils.Writer">
<modelSlot value="model"/>
```

```
<uri value="${targetDir}/exampleWright1.xmi"/>
</component>
</workflow>
```

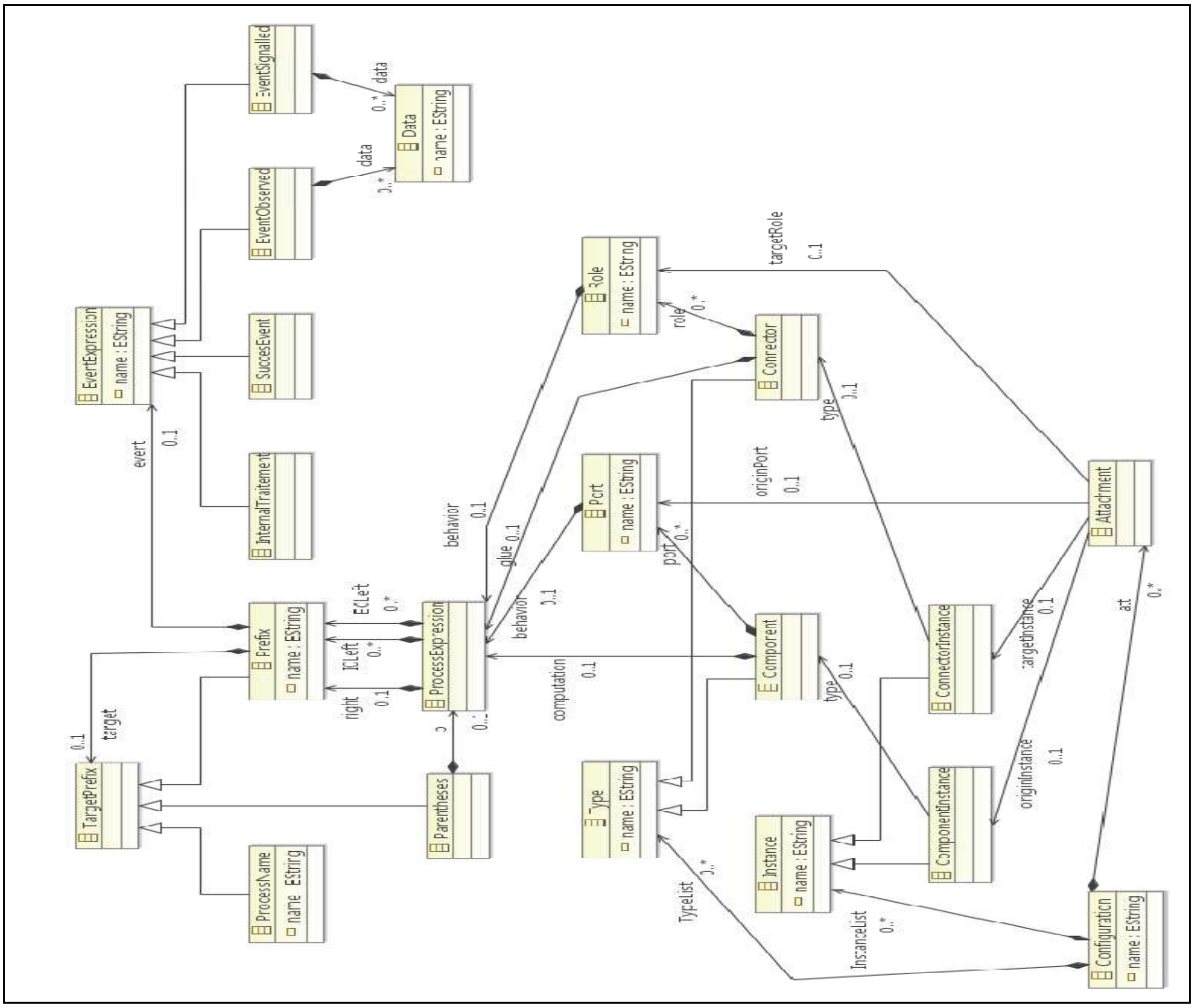

**Figure 7.36 :** Diagramme Ecore du méta-modèle Grammaire Wright généré -Wright1-

L'exécution de ce workflow permet la génération du modèle XMI conforme au méta-modèle Grammaire Wright –appelé Wright1- relatif au texte Wright écrit dans le fichier d'extension wright1. Le fichier d'extension wright1 se trouve dans le dossier src du deuxième projet. Le modèle XMI généré se trouve dans le dossier src-gen du deuxième projet. Cette étape est présentée par la Figure 7.37.

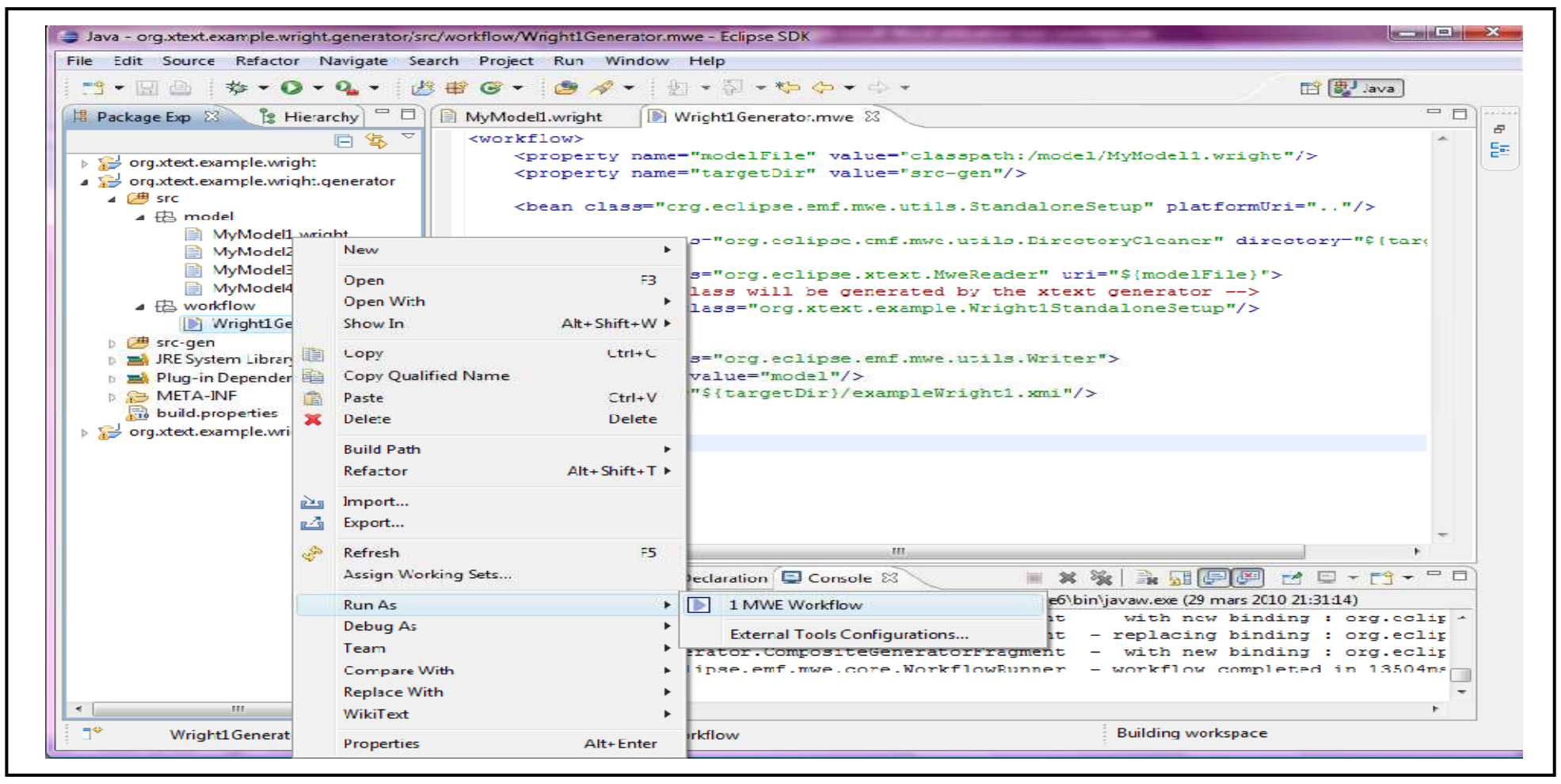

**Figure 7.37 :** Capture d'écran de l'exécution workflow du deuxième projet

### 7.7.1.2.3 Exemple Client-Serveur

Dans ce qui suit, nous allons donner une illustration sur l'exemple Client-Serveur fourni ci-dessous.

```
Configuration Client_Serveur
Connector Lien_CS
Role Appelant= _requete -> reponse -> Appelant |~| V -> STOP
Role Appele= requete -> _reponse -> Appele [] V -> STOP
Glue = Appelant.requete -> _Appele.requete -> glue
[] Appele.reponse -> _Appelant.reponse -> glue
[] V -> STOP
Component Client
Port port_Client= _requete -> reponse -> port_Client |~| V -> STOP
Computation= -traitement_interne1 -> _port_Client.requete ->
port_Client.reponse -> computation |~| V -> STOP
Component Serveur
Port port_Serveur= requete -> _reponse -> port_Serveur |~| V -> STOP
Computation= -traitement_interne2 -> port_Serveur.requete ->
_port_Serveur.reponse -> computation |~| V -> STOP
Instances
client1: Component Client
serveur1: Component Serveur
appel_cs: Connector Lien_CS

Attachments
client1-port_Client As appel_cs-Appelant
serveur1-port_Serveur As appel_cs-Appele
End Configuration
```

Après avoir vérifié les propriétés syntaxiques et sémantiques en passant par l'analyseur lexico-syntaxique généré par Xtext et l'évaluation des contraintes Check, le modèle correspondant à la configuration Client_Serveur est généré. Un tel modèle XMI est conforme au méta-modèle Grammaire Wright -appelé Wright1-.

## 7.7.1.3 Grammaire Wright vers Wright

Dans cette partie, nous allons présenter une partie du programme GrammaireWright2Ada écrit en ATL permettant la transformation des modèles sources conformes au méta-modèle Grammaire Wright –appelé Wright1- vers des modèles cibles conformes au méta-modèle Wright.

En-tête de ce fichier ATL :

```
module Wright1ToWright;
create exampleWright : Wright from exampleWright1 : Wright1;
```

✓ Transformation de la méta-classe *Configuration* de Wright1 vers la méta-classe *Configuration* de Wright:

La Figure 7.38 présente la partie du méta-modèle Wright1 qui modélise une configuration.

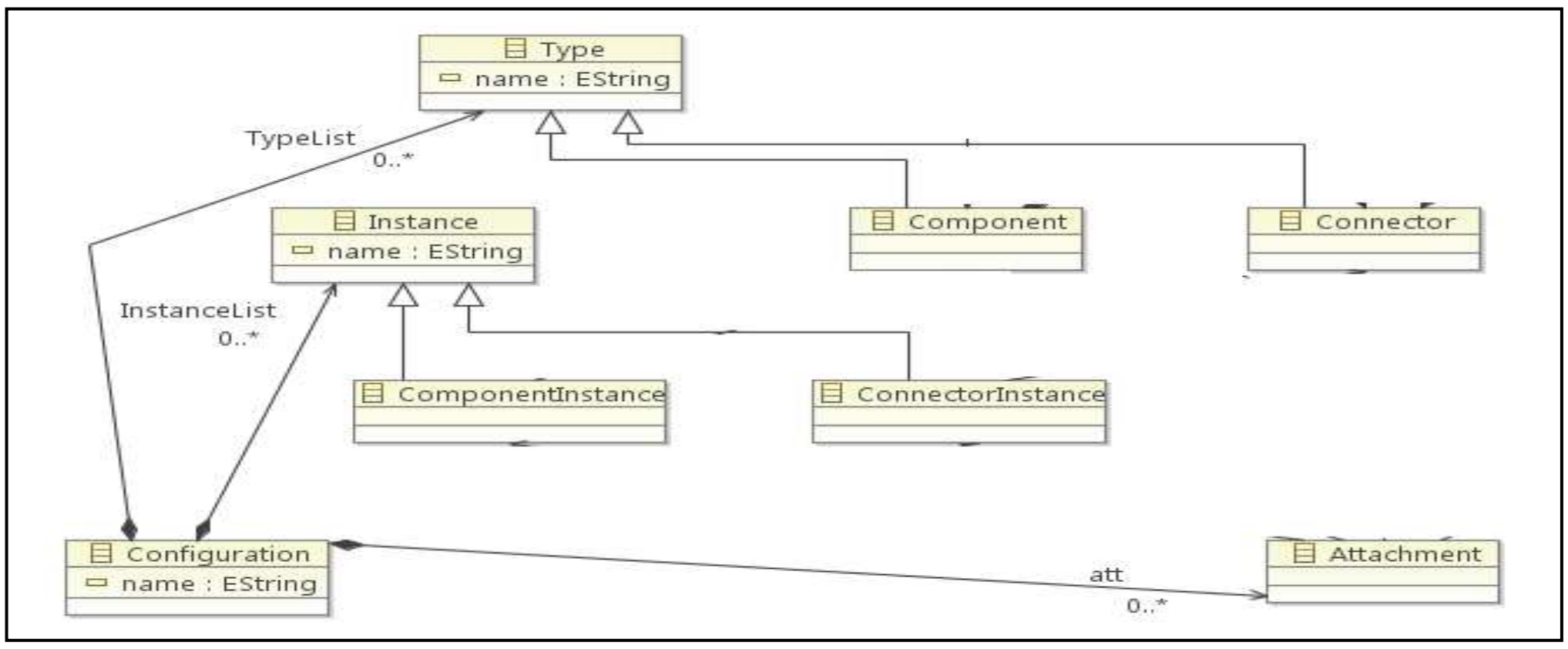

**Figure 7.38 :** Méta-classe Configuration du méta-modèle Wright1

La Figure 39 illustre la représentation d'une configuration dans le méta-modèle de Wright.

Règle de transformation de la configuration :

```
rule Configuration2Configuration{
 from c1:Wright1!Configuration
 to c:Wright!Configuration( name<-c1.name,comp<-c1.TypeList->
 select(e|e.oclIsTypeOf(Wright1!Component)),
 conn<-c1.TypeList->select(e|e.oclIsTypeOf(Wright1!Connector)),
 compInst<-c1.InstanceList->
 select(e|e.oclIsTypeOf(Wright1!ComponentInstance)),
 connInst<-c1.InstanceList->
 select(e|e.oclIsTypeOf(Wright1!ConnectorInstance)), att<-c1.att)
}
```

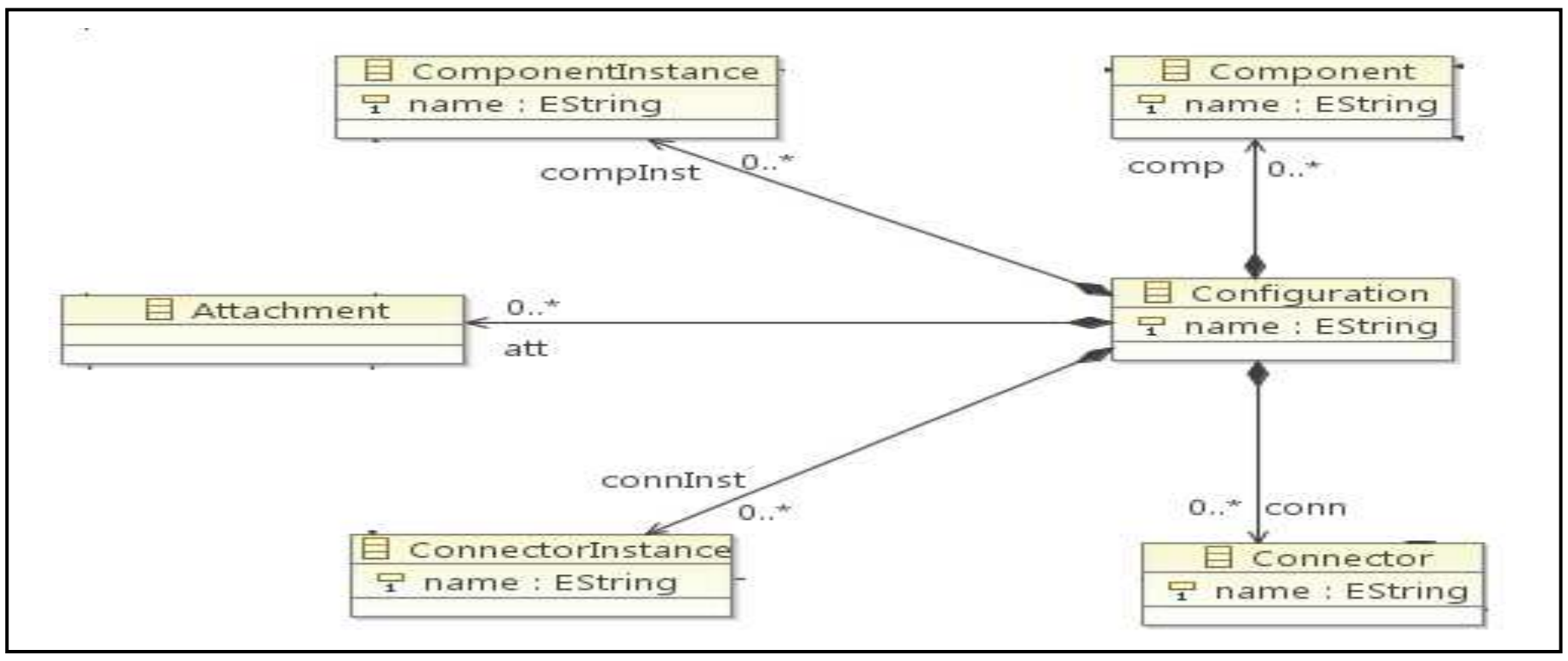

**Figure 7.39 :** Méta-classe Configuration du méta-modèle Wright

La référence *comp* prend l'ensemble des éléments de la méta-classe Component référencé par *TypeList*. Réciproquement, la référence *conn* prend l'ensemble des éléments de la méta-classe *Connector* référencé par *TypeList*. Et la référence *compInst* prend l'ensemble des éléments de la méta-classe *ComponentInstance* référencé par *InstanceList*. Réciproquement, la référence *connInst* prend l'ensemble des éléments de la méta-classe *ConnectorInstance* référencé par *InstanceList*. Le nom *name* et les attachements *att* restent inchangés.

✓ Transformation des méta-classes *ComponentInstance* et *ConnectorInstance* :

Pour les méta-classes *ComponentInstance* et *ConnectorInstance* : on ne signale aucun changement. Règles de transformation :

```
rule ComponentInstance2ComponentInstance{
from i1:Wright1!ComponentInstance
to i:Wright!ComponentInstance( name<-i1.name, type<-i1.type )
}
rule ConnectorInstance2ConnectorInstance{
from i1:Wright1!ConnectorInstance
to i:Wright!ConnectorInstance( name<-i1.name, type<-i1.type )
}
```

✓ Transformation de la méta-classe *Attachment* :

Pour la méta-classe *Attachment*, il ne va y avoir aucun changement. Règle de transformation :

```
rule Attachment2Attachment{
```

```
 from a1:Wright1!Attachment to a:Wright!Attachment( originInstance<-
  a1.originInstance,  targetInstance<-a1.targetInstance, originPort<-
  a1.originPort,  targetRole<-a1.targetRole )
}
```

✓ Transformation des méta-classes *Component* et *Connector* :

Pour les méta-classes *Component* et *Connector*, le seul changement est dans l'expression du processus CSP Wright référencé par *computation* respectivement *glue*. Règles de transformation :

```
rule Component2Component{
from c1:Wright1!Component to c:Wright!Component( name<-c1.name, port<-
c1.port, computation<-c1.computation.transformation() )
}
rule Connector2Connector{
from c1:Wright1!Connector to c:Wright!Connector( name<-c1.name,
 role<-c1.role,glue<-c1.glue.transformation()) }
```

Le helper *transformation* est un helper redéfini. Ce helper se charge de la transformation du processus CSP Wright selon le contexte dans lequel il est appelé.

✓ Transformation des méta-classes *Port* et *Role* :

Pour les méta-classes *Port* et *Role*, le seul changement est dans l'expression du processus CSP Wright référencé par *behavior*. Règles de transformation :

```
rule Port2Port{
from p1:Wright1!Port
to p:Wright!Port( name<-p1.name, behavior<-p1.behavior.transformation())
}
rule Role2Role{
from r1:Wright1!Role
to r:Wright!Role( name<-r1.name, behavior<-r1.behavior.transformation() )
}
```

Le helper *transformation* est un helper redéfini. Ce helper se charge de la transformation du processus CSP Wright selon le contexte dans lequel il est appelé.

✓ Transformation du processus CSP Wright:

la Figure 7.40 représente le processus CSP dans le méta-modèle Wright1.

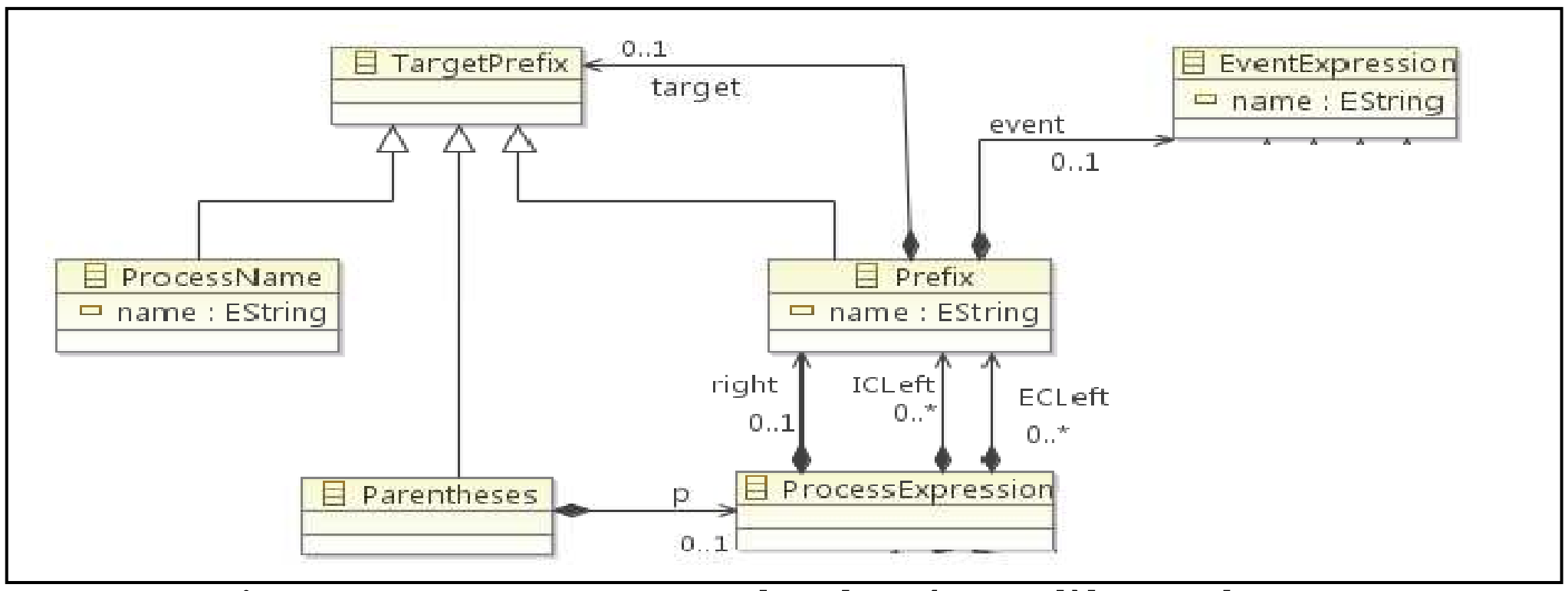

**Figure 7.40 :** Processus CSP dans le méta-modèle Wright1

La Figure 7.41 représente le processus CSP dans le méta-modèle Wright.

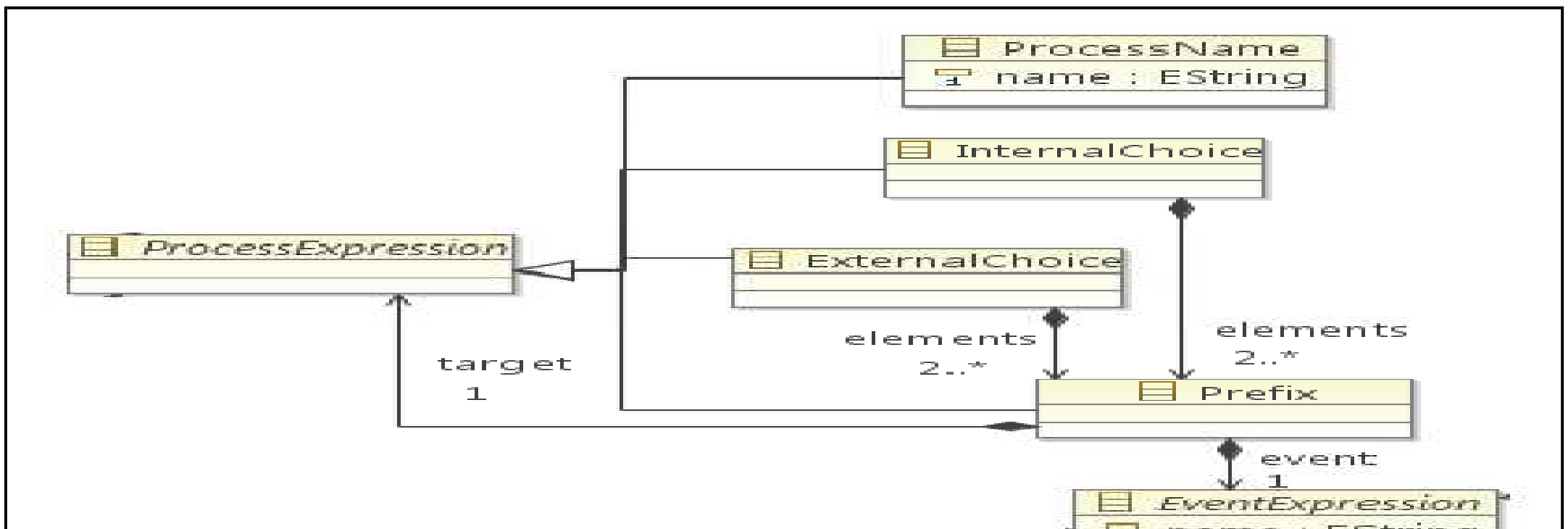

**Figure 7.41 :** Méta-modèle du processus CSP Wright

Nous avons dû écrire en ATL des règles standards, règles paresseuses et helpers afin de transformer un processus CSP conforme au méta-modèle Wright1 vers un processus CSP conforme au méta-modèle Wright.

### 7.7.2 Modèle Ada vers texte Ada : extraction via Xpand

Cette section présente la validation du modèle Ada conforme au méta-modèle partiel d'Ada et la transformation de ce modèle vers un texte Ada. Le principe d'extraction est fourni par la Figure 7.42. Pour y parvenir, nous avons utilisé avec profit les outils Check pour la validation et Xpand pour la transformation.

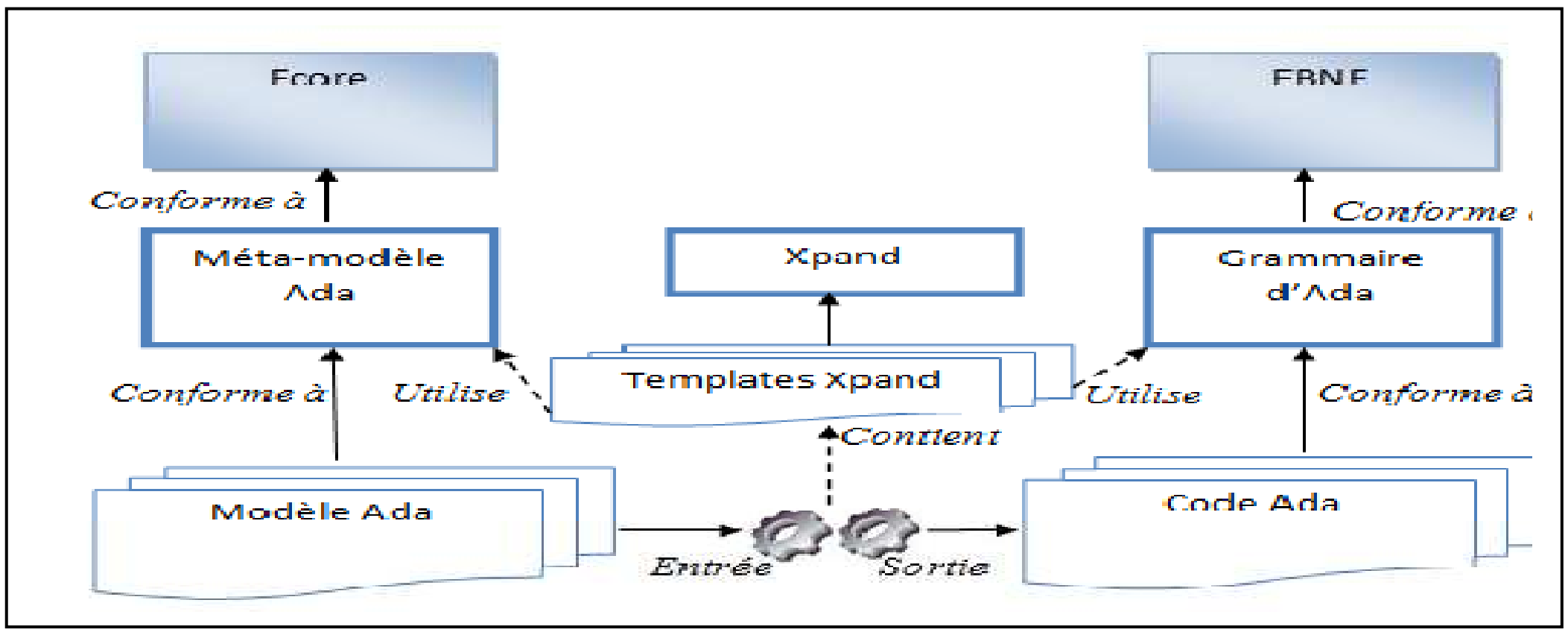

**Figure 7.42:** Schéma de transformation de modèle Ada vers texte Ada

#### 7.7.2.1 Sémantique statique d'Ada

La sémantique statique d'Ada est décrite à l'aide des contraintes OCL attachées au méta-modèle partiel d'Ada (cf. section 7.5). Ces contraintes sont réécrites en Check et attachées au méta-modèle partiel d'Ada. Les contraintes Check données ci-dessous sont évaluées sur les modèles Ada conformes au méta-modèle Ada. Ensuite, ces modèles sont transformés en code Ada moyennant le moteur workflow (cf. section 7.7.2.3) qui utilise les templates Xpand (cf. section 7.8.2.2).

#### 7.7.2.2 Génération de code d'un sous-programme Ada

Le sous-programme joue le rôle d'une fonction principale. Il est composé d'une spécification, d'un corps composé d'une partie déclarative et d'une partie exécutive. Ceci peut être traduit par le code Xpand suivant :

```
«DEFINE main FOR subprogram_body»
```

```
«FILE "adaCode.adb"»
«EXPAND specification FOR this.specif-»
«EXPAND declaration FOREACH this.declarations-»
begin
«EXPAND statement FOREACH this.statements-»
end «this.specif.designator»;
«ENDFILE»
«ENDDEFINE»
```

Les templates specification, declaration et statement seront redéfinis selon le contexte de leurs appels. Ceci permet de simplifier le code.

```
«DEFINE specification FOR subprogram_specification»
«ENDDEFINE»
«DEFINE declaration FOR declaration»
«ENDDEFINE»
«DEFINE statement FOR statement»
«ENDDEFINE»
```

#### 7.7.2.2.1 Spécification d'un sous-programme Ada

Il existe deux formes de spécification pour les sous-programmes: une procédure et une fonction. Ceci peut être traduit par le code suivant :

```
«DEFINE specification FOR procedure_specification»
procedure «this.designator» is
«ENDDEFINE»
«DEFINE specification FOR function_specification»
function «this.designator» return «this.returnType» is
«ENDDEFINE»
```

#### 7.7.2.2.2 Partie déclarative d'Ada

La partie déclarative d'un sous-programme Ada peut contenir la déclaration d'autres sous-programmes et des tâches.

a) Déclaration de sous-programme

La déclaration de sous-programme se fait par leurs prototypes. Ceci est traduit par le code suivant :

```
«DEFINE declaration FOR procedure_specification»
procedure «this.designator» ;
«ENDDEFINE»
«DEFINE declaration FOR function_specification»
function «this.designator» return «this.returnType» ;
«ENDDEFINE»
```

b) Sous-programmes

La déclaration d'autres sous-programmes se traduit par le code suivant :

```
«DEFINE declaration FOR subprogram_body»
«EXPAND specification FOR this.specif-»
«EXPAND declaration FOREACH this.declarations-»
begin
«EXPAND statement FOREACH this.statements-»
end «this.specif.designator»;
«ENDDEFINE»
```

c) Tâches Ada

Une tâche Ada est constituée d'une partie déclarative et d'un corps.

✓La partie déclarative d'une tâche :

La partie déclarative d'une tâche peut contenir les entrées de cette dernière.

```
«DEFINE declaration FOR single_task_declaration»
task «this.identifier» «IF this.entryDec.isEmpty» ; «ELSE» is
```

```
«EXPAND Entry FOREACH this.entryDec»
end «this.identifier»;
«ENDIF»
«ENDDEFINE»
«DEFINE Entry FOR entry_declaration»
entry «this.identifier» ;
«ENDDEFINE»
```

✓ Le corps d'une tâche :

Cette partie est constituée en deux parties : une partie décrivant les éventuelles déclarations de la tâche et une autre décrivant sa partie exécutive. Ceci peut être traduit par le code suivant :

```
«DEFINE declaration FOR task_body»
task body «this.identifier» is
«EXPAND declaration FOREACH this.declarations»
begin
«EXPAND statement FOREACH this.statements»
end «this.identifier»;
«ENDDEFINE»
```

Les deux parties déclarative et exécutive sont les mêmes que celles d'un sous-programme Ada.

#### 7.7.2.2.3 Partie exécutive d'Ada

Cette partie concerne les instructions Ada. Nous présentons deux illustrations, l'instruction « if » et « case ». L'annexe F présente l'ensemble des spécifications Xpand corespondant aux instructions de la partie exécutive.

✓ L'instruction if :

```
«DEFINE statement FOR if_else»
if «this.cond.c» then
«EXPAND statement FOREACH this.s1»
else
«EXPAND statement FOREACH this.s2»
end if;
«ENDDEFINE»
```

✓ L'instruction case :

```
«DEFINE statement FOR case_statement»
case «this.exp.e» is
«IF this.ref.notExists(e|e.choice=="others")»
«EXPAND Case FOREACH this.ref.reject(e|e.choice=="others")»
others => null;
«ELSE»
«EXPAND Case FOREACH this.ref.reject(e|e.choice=="others")»
«EXPAND Case FOREACH this.ref.select(e|e.choice=="others")»
«ENDIF»
end case;
«ENDDEFINE»
«DEFINE Case FOR case_statement_alternative»
when «this.choice» => «EXPAND statement FOREACH this.s»
«ENDDEFINE»
```

Le template de génération de code Ada, en entier, est fourni dans l'annexe G.

### 7.7.2.3 Moteur de vérification et de génération de code Ada

Le workflow donné ci-dessous permet de générer le code Ada relatif au modèle XMI conforme au méta-modèle partiel d'Ada en utilisant les templates Xpand fournis précédemment.

```
<workflow>
<property name="model"
value="my.generator.ada/src/example1/exampleAda.xmi" />
<property name="src-gen" value="src-gen/example1" />
<!-- set up EMF for standalone execution -->
<bean class="org.eclipse.emf.mwe.utils.StandaloneSetup" >
<platformUri value=".."/>
<registerEcoreFile
value="platform:/resource/my.generator.ada/src/metamodel/Ada.ecore" />
</bean>
<!-- load model and store it in slot 'model' -->
<component class="org.eclipse.emf.mwe.utils.Reader">
<uri value="platform:/resource/${model}" />
<modelSlot value="model" />
</component>
<!-- check model -->
<component class="org.eclipse.xtend.check.CheckComponent">
<metaModel
class="org.eclipse.xtend.typesystem.emf.EmfRegistryMetaModel"/>
<checkFile value="metamodel::CheckFile" />
<emfAllChildrenSlot value="model" />
</component>
<!-- generate code -->
<component class="org.eclipse.xpand2.Generator">
<metaModel
class="org.eclipse.xtend.typesystem.emf.EmfRegistryMetaModel"/>
<expand
value="template::Template::main FOR model" />
<outlet path="${src-gen}" />
</component>
</workflow>
```

### 7.7.2.4 Exemple d'utilisation

En exécutant le workflow sur le modèle d'Ada en XMI conforme au méta-modèle Wright, nous obtenons le code Ada suivant :

| | |
|---|---|
| **procedure** Client_Serveur **is** | **end if;** |
| **function** condition_interne **return** Boolean **is** | **end loop;** |
| **begin** | **end Component_client1;** |
| **return** true; | **task body** Component_serveur1 **is** |
| **end condition_interne;** | **begin** |
| **procedure** traitement_interne1 **is** | **loop** |
| **begin** | **if** condition_interne **then** |
| **null;** | **exit;** |
| **end traitement_interne1;** | **else** |
| **procedure** traitement_interne2 **is** | traitement_interne2; |
| **begin** | **accept** port_Serveur_requete; |
| **null;** | Connector_appel_cs.Appele_reponse; |
| **end traitement_interne2;** | **end if;** |
| **task** Component_client1 **is** | **end loop;** |
| **entry** port_Client_reponse ; | **end Component_serveur1;** |
| **end Component_client1;** | **task body** Connector_appel_cs **is** |
| **task** Component_serveur1 **is** | **begin** |
| **entry** port_Serveur_requete ; | **loop** |
| **end Component_serveur1;** | **select** |
| **task** Connector_appel_cs **is** | **accept** Appelant_requete; |
| **entry** Appele_reponse ; | Component_serveur1.port_Serveur_requete; |
| **entry** Appelant_requete ; | **or** |
| **end Connector_appel_cs;** | **accept** Appele_reponse; |
| **task body** Component_client1 **is** | Component_client1.port_Client_reponse; |
| **begin** | **or** |
| **loop** | **terminate;** |
| **if** condition_interne **then** | **end select;** |
| **exit;** | **end loop;** |
| **else** | **end Connector_appel_cs;** |
| traitement_interne1;<br>Connector_appel_cs.Appelant_requete;<br>**accept** port_Client_reponse; | **begin**<br>**null;**<br>**end Client_Serveur;** |

Nous avons compilé et exécuté ce programme concurrent Ada en utilisant l'environnement [ObjectAda]. Un tel programme traduisant l'architecture abstraite en Ada peut être raffiné step-by-step en prenant des décisions conceptuelles et techniques. La correction du raffinement est obtenue par l'utilisation des outils de vérification formelle associés à Ada tel que FLAVERS [Cobleigh, 2002].

## 7.8 Vérification

Dans cette section, nous nous penchons sur la vérification de notre programme Wright2Ada en utilisant une approche basée sur les tests syntaxiques (Syntax-Based Testing) (Xanthakis, 1999). L'objectif de cette section est de vérifier notre programme Wright2Ada écrit en ATL permettant de transformer une architecture logicielle décrite en Wright vers un programme concurrent Ada comportant plusieurs tâches (task). Pour y parvenir, nous préconisons une activité de vérification de ce programme basée sur le test fonctionnel ou encore boîte noire.

### 7.8.1 Tests syntaxiques

Nous considérons le programme Wright2Ada comme boîte noire. Ainsi, nous nous plaçons dans le cadre d'un test fonctionnel. Notre programme Wright2Ada nécessite des données d'entrée (des spécifications ou des descriptions en Wright) respectant une syntaxe rigide et bien définie : la syntaxe de Wright décrite en Xtext. Afin de couvrir l'espace de données du programme Wright2Ada, nous retenons les deux critères de couverture suivants :

**Critère 1** : Couverture des symboles terminaux. Ils sont au nombre de 79 unités lexicales couvrant des mots-clefs et des symboles utilisés par l'ADL Wright tels que : « Configuration », « Component », « Port », « Connector », « Role », « Instances », « Attachments », « As », « : », « ! », « ? », « § », « -> », « [] », « |~| », etc.
**Critère 2** : Couverture des règles de productions permettant de définir les constructions syntaxiques offertes par Wright. Elles sont au nombre de 20 règles telles que : ComponentInstance, ConnectorInstance, EventSignalled, InternalTraitement, SuccesEvent, Data, ProcessName, Prefix, ProcessExpression, Port, Role, Component, Connector, Configuration. Nous avons suivi une approche de prédiction des sorties attendues afin de tester notre programme Wright2Ada. La fonction d'oracle permettant de comparer la sortie observée par rapport à la sortie attendue pour une donnée de test : DT fournie est actuellement manuelle. Une automatisation de celle-ci peut être envisagée en s'inspirant de la commande *diff* offerte par un système d'exploitation de type Unix.

### 7.8.2 Les données de test

Afin de couvrir les deux critères proposés dans la section précédente, nous avons établi les données de test (DT) décrites ci-dessus :

#### 7.8.2.1 Exemple du dîner des philosophes

Nous testons ci-après l'exemple bien connu du dîner des philosophes. Notre exemple est tiré de (Déplanche, 2005). Nous nous limitons à une configuration de deux philosophes.
Cet exemple couvre toutes les règles de production exceptée la règle InternalTraitement. De même, il couvre tous les terminaux exceptés l'opérateur déterministe [].

| | |
|---|---|
| **Configuration Diner** | **Glue** = Mangeur.prendre -> _Outil.prend -> glue \|~\| |
| **Component Philo**<br>**Port** Gauche = _prendre -> _deposer -> Gauche \|~\| **§**<br>**Port** Droite = _prendre -> _deposer -> Droite \|~\| **§**<br>**Computation** = -penser -> _Gauche.prendre -> _Droite.prendre -> -manger -> _Gauche.deposer -> _Droite.deposer -> computation \|~\| **§**<br>**Component Fourchette**<br>**Port** Manche = prend -> depose -> Manche \|~\| **§**<br>**Computation** = Manche.prend -> Manche.depose -> computation \|~\| **§**<br>**Connector Main**<br>**Role** Mangeur = _prendre -> _deposer -> Mangeur \|~\| **V-**>STOP<br>**Role** Outil = prend -> depose -> Outil \|~\| **V**->STOP | Mangeur.deposer -> _Outil.depose -> glue \|~\| **SKIP**<br>**Instances**<br>p1: Component Philo<br>p2: Component Philo<br>f1: Component Fourchette<br>f2: Component Fourchette<br>m11: Connector Main<br>m12: Connector Main<br>m21: Connector Main<br>m22: Connector Main<br>**Attachments**<br>p1-Gauche As m11-Mangeur<br>p1-Droite As m12-Mangeur<br>p2-Gauche As m21-Mangeur<br>p2-Droite As m22-Mangeur<br>f1-Manche As m11-Outil<br>f1-Manche As m22-Outil<br>f2-Manche As m12-Outil<br>f2-Manche As m21-Outil<br>**End Configuration** |

Traduction Ada de l’exemple du dîner des philosophes :

| | |
|---|---|
| **procedure** Diner **is**<br>**function** condition_interne **return** Boolean **is**<br>**begin return** true;<br>**end condition_interne;**<br>**function** condition_interne1 **return** Integer **is**<br>**begin return** 1;<br>**end condition_interne1;**<br>**procedure** penser **is**<br>**begin** null; **end** penser;<br>**procedure** manger **is**<br>**begin** null; **end** manger;<br>**task** Component_p1;<br>**task** Component_p2;<br>**task** Component_f1 **is**<br>**entry** Manche_prend ;<br>**entry** Manche_depose ;<br>**end Component_f1;**<br>**task** Component_f2 **is**<br>**entry** Manche_prend ;<br>**entry** Manche_depose ;<br>**end Component_f2;**<br>**task** Connector_m11 **is**<br>**entry** Mangeur_prendre ;<br>**entry** Mangeur_deposer ;<br>**end Connector_m11;**<br>**task** Connector_m12 **is**<br>**entry** Mangeur_prendre ;<br>**entry** Mangeur_deposer ;<br>**end Connector_m12;**<br>**task** Connector_m21 **is**<br>**entry** Mangeur_prendre ;<br>**entry** Mangeur_deposer ;<br>**end Connector_m21;**<br>**task** Connector_m22 **is** | **task body** Component_f1 **is**<br>**begin loop**<br>**if** condition_interne **then**<br>**accept Manche_prend;**<br>**accept Manche_depose;**<br>**else** exit; **end if; end loop;**<br>**end Component_f1;**<br>**task body** Component_f2 **is**<br>**begin loop**<br>**if** condition_interne **then**<br>**accept Manche_prend;**<br>**accept Manche_depose;**<br>**else** exit; **end if; end loop;**<br>**end Component_f2;**<br>**task body** Connector_m11 **is**<br>**begin loop**<br>**case condition_interne1 is**<br>**when** 1 => exit;<br>**when** 2 => **accept** Mangeur_prendre;<br>Component_f1.Manche_prend;<br>**when** 3 => accept Mangeur_deposer;<br>Component_f1.Manche_depose;<br>**when others => null;**<br>**end case; end loop;**<br>**end Connector_m11;**<br>**task body** Connector_m12 **is**<br>**begin loop**<br>**case condition_interne1 is**<br>**when** 1 => exit;<br>**when** 2 => **accept** Mangeur_prendre;<br>Component_f2.Manche_prend;<br>**when** 3 => **accept** Mangeur_deposer;<br>Component_f2.Manche_depose;<br>**when others => null ;** |

```
entry Mangeur_prendre ;
entry Mangeur_deposer ;
end Connector_m22;
task body Component_p1 is
begin loop
if condition_interne then
penser;
Connector_m11.Mangeur_prendre;
Connector_m12.Mangeur_prendre;
manger;
Connector_m11.Mangeur_deposer;
Connector_m12.Mangeur_deposer;
else exit; end if; end loop;
end Component_p1;
task body Component_p2 is
begin loop
if condition_interne then
penser;
Connector_m21.Mangeur_prendre;
Connector_m22.Mangeur_prendre;
manger;
Connector_m21.Mangeur_deposer;
Connector_m22.Mangeur_deposer;
else exit; end if; end loop;
end Component_p2;
end case; end loop;
end Connector_m12;
task body Connector_m21 is
begin loop
case condition_interne1 is
when 1 => exit;
when 2 => accept Mangeur_prendre;
Component_f2.Manche_prend;
when 3 => accept Mangeur_deposer;
Component_f2.Manche_depose;
when others =>null;
end case; end loop;
end Connector_m21;
task body Connector_m22 is
begin loop
case condition_interne1 is
when 1 => exit;
when 2 => accept Mangeur_prendre;
Component_f1.Manche_prend;
when 3 => accept Mangeur_deposer;
Component_f1.Manche_depose;
when others =>null;
end case; end loop;
end Connector_m22;
begin
null;
end Diner;
```

### 7.8.2.2 Exemple de la gestion de places d'un parking

Nous testons ci-après un exemple d'une configuration pour la gestion de places d'un parking tiré de [Bhiri, 2008]. Cet exemple couvre toutes les règles de production exceptée les règles InternalTraitement et SuccesEvent. De même, il couvre tous les teminaux exceptés SKIP et §.

```
Configuration GestionParking
Component Acces
Port r1 = voitureArrive -> (_reservation -> (reponsePositive -> r1[] reponseNegative
-> r1) |~| _liberation -> r1)
Computation = r1.voitureArrive -> (_r1.reservation -> (r1.reponsePositive ->
computation []r1.reponseNegative -> computation) |~| _r1.liberation -> computation)
Component Afficheur
Port ecran = maj -> ecran
Computation = ecran.maj -> computation
Connector Parking
Role acces1 = voitureArrive -> (_reservation -> (reponsePositive -> Acces1 []
reponseNegative -> acces1) |~| _liberation -> acces1)
Role acces2 = voitureArrive -> (_reservation -> (reponsePositive -> Acces2 []
reponseNegative -> acces2) |~| _liberation -> acces2)
Role afficheur = maj -> afficheur
Glue = _acces1.voitureArrive -> (acces1.reservation ->(_acces1.reponsePositive
-> _afficheur.maj -> glue |~| _acces1.reponseNegative -> glue) [] acces1.liberation -
> _afficheur.maj -> glue)
|~| acces2.voitureArrive -> (acces2.reservation ->
(_acces2.reponsePositive -> _afficheur.maj -> glue |~| _acces2.reponseNegative ->
glue) [] acces2.liberation -> _afficheur.maj -> glue)
Instances
acces1: Component Acces
acces2 : Component Acces
afficheur1 : Component Afficheur
parking1 : Connector Parking
Attachments
acces1-r1 As parking1-acces1
acces2-r1 As parking1-acces2
afficheur1-ecran As parking1-afficheur
End Configuration
```

Traduction Ada de l’exemple de la gestion de places d’un parking :

| | |
|---|---|
| **procedure** GestionParking **is**<br>**function** condition_interne **return** Boolean **is**<br>**begin return** true;<br>**end condition_interne;**<br>**task** Component_acces1 **is**<br>**entry** r1_reponsePositive ;<br>**entry** r1_reponseNegative ;<br>**entry** r1_voitureArrive ;<br>**end Component_acces1;**<br>**task** Component_acces2 **is**<br>**entry** r1_reponsePositive ;<br>**entry** r1_reponseNegative ;<br>**entry** r1_voitureArrive ;<br>**end Component_acces2;**<br>**task** Component_afficheur1 **is**<br>**entry** ecran_maj ;<br>**end Component_afficheur1;**<br>**task** Connector_parking1 **is**<br>**entry** acces2_reservation ;<br>**entry** acces1_reservation ;<br>**entry** acces1_liberation ;<br>**entry** acces2_liberation ;<br>**end** Connector_parking1;<br>**task body** Component_acces1 **is**<br>**begin loop**<br>**accept** r1_voitureArrive;<br>**if** condition_interne **then**<br>Connector_parking1.acces1_liberation;<br>**else**<br>Connector_parking1.acces1_reservation; **select**<br>**accept** r1_reponseNegative;<br>**or accept** r1_reponsePositive;<br>**end select; end if; end loop;**<br>**end Component_acces1;** | **task body** Component_acces2 **is**<br>**begin loop**<br>**accept** r1_voitureArrive;<br>**if** condition_interne **then**<br>Connector_parking1.acces2_liberation;<br>**else**<br>Connector_parking1.acces2_reservation; **select**<br>**accept** r1_reponseNegative;<br>**or accept** r1_reponsePositive;<br>**end select; end if; end loop;**<br>**end Component_acces2;**<br>**task body** Component_afficheur1 **is**<br>**begin loop**<br>**accept** ecran_maj;<br>**end loop; end** Component_afficheur1;<br>**task body** Connector_parking1 **is**<br>**begin loop**<br>**if** condition_interne **then**<br>Component_acces1.r1_voitureArrive;<br>**select** accept acces1_reservation;<br>**if** condition_interne **then**<br>Component_acces1.r1_reponsePositive;<br>Component_afficheur1.ecran_maj;<br>**else**<br>Component_acces1.r1_reponseNegative; end if;<br>**or accept** acces1_liberation;<br>Component_afficheur1.ecran_maj;<br>**end select;**<br>**else**<br>Component_acces2.r1_voitureArrive;<br>**select accept** acces2_liberation;<br>Component_afficheur1.ecran_maj;<br>**or accept** acces2_reservation;<br>**if** condition_interne **then**<br>Component_acces2.r1_reponseNegative; **else**<br>Component_acces2.r1_reponsePositive;<br>Component_afficheur1.ecran_maj;<br>**end if; end select; end if; end loop;**<br>**end Connector_parking1;**<br>**begin** null; **end** GestionParking; |

### 7.8.2.3 Exemple d’architecture client-serveur

L’architecture Client-Serveur est déjà présentée dans la section 7.7.1.2.3. Le code en Ada correspondant est présenté dans la section 7.7.2.3.
Cet exemple couvre toutes les règles de production. De plus, il couvre tous les teminaux exceptés § et SKIP.

En conclusion, les trois exemples fournis ci-dessus couvrent les deux critères de générations des données de tests retenus dans 7.8.1.

## 7.9 Conclusion

Nous avons proposé une approche IDM permettant de transformer une architecture logicielle décrite à l’aide de l’ADL formel Wright vers un programme concurrent Ada comportant plusieurs tâches exécutées en parallèle. Pour y parvenir, nous avons élaboré deux méta-modèles en Ecore : le méta-modèle de Wright et le méta-modèle partiel d'Ada. De plus, nous avons conçu et réalisé un programme Wright2Ada permettant de transformer un modèle source Wright conforme à son méta-modèle Wright vers un modèle cible Ada conforme au méta-modèle partiel Ada. Notre programme est purement déclaratif et utilise avec profit les constructions déclaratives fournies par le langage ATL telles que : règle standard, règle paresseuse, helper (attributs, opérations). En outre, nous avons proposé des interfaces conviviales permettant de transformer du texte Wright vers du code Ada en utilisant les outils de modélisation Xtext, Xpand et Check. Enfin, nous avons testé notre programme Wright2Ada en adoptant une approche orientée tests syntaxiques.

Dans le chapitre suivant, nous proposons une approche permettant de vérifier les contrats syntaxiques et structurels d'un assemblage de composants Ugatze.

# **Chapitre 8 :** Vérification des contrats syntaxiques d'assemblages de composants Ugatze

## 8.1 Introduction

Dans ce chapitre, nous proposons une approche de traduction du modèle de composants semi-formel Ugatze vers le modèle de composants Acme/Armani. Ceci permet la vérification des contrats syntaxiques et structurels d'un assemblage de composants Ugatze.

Ce chapitre comporte deux sections. La section 8.2 formalise les principaux concepts issus d'Ugatze en Acme/Armani. La section 8.3 présente une étude de cas : diagnostic médical distribué modélisée en Ugatze, traduite en Acme/Armani et vérifiée à l'aide de la plate-forme AcmeStudio.

## 8.2 Formalisation du méta-modèle Ugatze

Le méta-modèle de composants Ugatze est modélisé par un style architectural Acme/Armani appelé *UGATZE* en utilisant la construction « *family* » [Kmimech, 2009b], [Kmimech, 2009c]. Le style *UGATZE* réutilise et adapte plusieurs styles d'architecture standards tels que pipe-and-filter, shared-variable et client-server [Shaw, 1996]. Le style *UGATZE* formalise tous les concepts relatifs au modèle de composants Ugatze. Il comporte la formalisation des types de données, signatures d'opérations, points d'interaction, composants et interactions supportés par le modèle Ugatze. Pour y parvenir, nous avons utilisé avec profit les possibilités de typage offertes par Acme/Armani. La Figure 8.1 présente le style UGATZE.

Les règles de cohérence (Well-Formedness Rules) relatives au modèle de composants Ugatze sont modélisées par des propriétés invariantes en utilisant le concept d'*invariant* d'Acme/Armani. Ces règles sont judicieusement réparties sur les éléments architecturaux définis dans le style *UGATZE*.

```
Family UGATZE = {
Property type
base_type_ugatze=enum{boolean_ugatze,real_ugatze,integer_ugatze,char_u
gatze,string_ugatze,void_ugatze};
  Property type logical_nature=enum{in_ugatze,
out_ugatze,inout_ugatze};
  Property type parameter
=Record[parameter_type:base_type_ugatze;mode:logical_nature;];
  Property type pls_parameter=sequence<parameter>;
  Property type
signature=Record[p:pls_parameter;resultat:base_type_ugatze;];
  Property type porttype=enum {provider,required,environment};

   Port Type DataPoint ={
     Property protocol : string ;
    }
   Port Type OperationPoint ={
     property porttype:port_type;
     Property operation:signature;
   }
   Port Type EnvironementPoint ={
     Property protocol : string ;
    }
```

```
Port Type OIP extends DataPoint with  {
        property porttype:port_type=provider;
    }
Port Type IIP extends DataPoint with {
        property porttype:port_type=required;
    }

Port Type PIOP extends OperationPoint  with {
       property porttype:port_type=provider;
    }

    Port Type UIOP extends OperationPoint  with {
       property porttype:port_type=required;
    }

    Component Type ComponentUgatze = {
      //un composant possède au moins un point d'interaction
      rule haveAtLeastone = invariant size(self.PORTS) >= 1;
      }

    Component Type ComponentFiltre extends ComponentUgatze with {
// un composant ComponentClient est défini par des points d'interactions
//de type IIP ou OIP
      rule PortType1 = invariant forall p : Port in self.PORTS
|declaresType(p,IIP) or declaresType(p, OIP);
      }

 Component Type ComponentFiltreClientServer extends ComponentUgatze with
{
// un composant ComponentFiltreClientServer est défini par au moins un
// point d'interaction de type OIP ou IIP et un point d'interaction de
// type de type UIOP ou PIOP
    rule PortType1 = invariant exists p : Port in self.PORTS
|declaresType(p,IIP) or declaresType(p, OIP);
    rule PortType2 = invariant exists p : Port in self.PORTS
|declaresType(p,UIOP) or declaresType(p, PIOP);
    }
 Component Type ComponentClientServer extends ComponentUgatze with {
// un composant ComponentClientServer est défini par des points
// d'interactions de type UIOP ou PIOP
    rule PortType3 = invariant exists  p : Port in self.PORTS
|declaresType(p, UIOP) or declaresType(p, PIOP);
       }

Role Type sourcePipe = {
        Property protocol : string;

//chaque rôle de type sourceT est rattaché à un seul point d'interaction
         rule oneAttachment = invariant size(self.ATTACHEDPORTS) == 1;
// chaque rôle de type sourceT est rattaché à un point d'interaction
//de type OIP
  rule attachedPortsAreOIP = invariant forall p : Port in
self.ATTACHEDPORTS|declaresType(p, OIP);
    }

Connector Type Pipe = {
        Property bufferSize : int <<  bufferSize : int = 0; >> ;

        Role source : sourcePipe = new sourcePipe extended with {
        }
        Role sink : sinkPipe = new sinkPipe extended with {
        }

      // chaque interaction Pipe est défini par deux rôles
         rule exactlyTwoRoles = invariant size(self.ROLES) == 2;

      // la taille du buffer ne peut être négative
         rule bufferpositive = invariant self.bufferSize >= 0;
```

```
//chaque rôle de type sourcePipe est rattaché à un port de type IIP
//et chaque rôle de type sinkPipe est rattaché à un port de type OIP
          rule precondition1 = invariant forall r1 : Role in self.ROLES |
            declaresType(r1, sinkPipe) -> forall p1 : Port in
r1.ATTACHEDPORTS |declaresType(p1, IIP) -> forall r2 : Role in self.ROLES
| declaresType(r2, sourcePipe) -> forall p2 : Port in r2.ATTACHEDPORTS |
declaresType(p2, OIP);

 // un rôle ne peut être non rattaché à un point d'interaction
            rule noDanglingRoles = invariant forall r : Role in
self.ROLES |attachedOrBound(r);
         }

    Role Type sinkPipe = {
        Property protocol : string;
 //chaque rôle de type sinkPipe est rattaché à un port de IIP
        rule attachedPortsAreIIP = invariant forall p : Port in
self.ATTACHEDPORTS | declaresType(p, IIP);
    }

// Toutes les interactions sont indépendantes. Elles ne partagent pas de
// points d'interactions
       rule connectorIndependence = invariant forall c : connector in
self.Connectors |forall r :  role in c.ROLES | forall p : Port in
r.ATTACHEDPORTS|size(p.ATTACHEDROLES) ==1;

Role Type serverT = {
// chaque rôle de type serverT est rattaché à un seul point d'interaction
        rule oneAttachment = invariant size(self.ATTACHEDPORTS) == 1;

// chaque rôle de type serverT est rattaché à un point d'interaction de
//type PIOP
        rule attachedPortsArePIOP = invariant forall p : Port in
self.ATTACHEDPORTS | declaresType(p, PIOP);
    }
 Connector Type InteractionOperation = {
        Role source : serverT = new serverT extended with {

        }
        Role sink : clientT = new clientT extended with {

        }
        // chaque interaction d'opération est défini par deux rôles
        rule exactlyTwoRoles = invariant size(self.ROLES) == 2;

        // un rôle doit être rattaché à un point d'interaction
              rule noDanglingRoles = invariant forall r : Role in
      self.ROLES | attachedOrBound(r);

// Toutes les interactions sont indépendantes. Elles ne partagent pas de
// points d'interactions
    rule connectorIndependence = invariant forall r : Role in self.ROLES
| size(r.ATTACHEDPORTS) == 1;
   }

Role Type clientT = {
      //chaque rôle de type sinkT est rattaché à un seul port de type IIP
        rule oneAttachment = invariant size(self.ATTACHEDPORTS) == 1;

      //chaque rôle de type clientT est rattaché à un port de type IIP
        rule  attachedPortsAreUIOP  =  invariant  forall  p  :  Port  in
self.ATTACHEDPORTS | declaresType(p, UIOP);
   }
    Role Type sourceShared = {
  //  chaque  rôle  de  type  sourceShared  est  rattaché  à  au  moins
//un point d'interaction
   rule atLeastOneAttachment = invariant size(self.ATTACHEDPORTS) >= 1;
```

```
//  chaque  rôle  de  type  sourceShared  est  rattaché  à  un  point
//d'interaction de type OIP
         rule  attachedPortsAreOIP  =  invariant  forall  p  :  Port  in
self.ATTACHEDPORTS | declaresType(p, OIP);
         }

    Role Type sinkShared = {
 // chaque rôle de type sinkShared est rattaché à au moins un point
//d'interaction
   rule atLeastOneAttachment = invariant size(self.ATTACHEDPORTS) >= 1;

//  chaque  rôle  de  type  sinkShared  est  rattaché  à  un  point
//d'interactions de type IIP
         rule attachedRolesAreSourceT  =  invariant  forall  p  :  Port  in
self.ATTACHEDPORTS | declaresType(p, IIP);
    }
```

**Figure 8.1:** Style Ugatze

### 8.2.1 Formalisation des types de données

La Figure 8.2 illustre la formalisation des types de données et de signatures d'opérations Ugatze. Pour y parvenir, nous avons utilisé avec profit les possibilités de typage offertes par Acme/Armani. En effet, l'ADL Acme/Armani supporte la notion de type : des types de base (*int*, *float*, *boolean* et *string*) et des constructeurs de types (*enum*, *record*, *set et sequence*). Les types Acme/Armani proposés sont:

- « *base_type_ugatze* » : modélise les types de base fournis par Ugatze. Il s'agit d'un type énuméré,
- « *logical_nature* » : modélise la nature logique des paramètres formels d'une opération Ugatze. Il s'agit d'un type énuméré,
- « *pls_parameter* » : regroupe au sein d'un enregistrement les deux caractéristiques d'un paramètre formel d'une opération Ugatze à savoir son type et sa nature logique,
- « *parameter* » : regroupe au sein d'une séquence (*sequence)* les paramètres d'une opération Ugatze, en tenant compte de l'ordre de ces paramètres,
- « *signature* » : regroupe au sein d'un enregistrement la signature d'une opération Ugatze en utilisant les types définis précédemment *pls_paramet*er et *base_type_ugatze*,
- « *porttype* » modélise les différents types de ports supportés par Ugatze. Il s'agit d'un type énuméré.

```
Property type base_type_ugatze=enum
{boolean_ugatze,real_ugatze,integer_ugatze, char_ugatze,
string_ugatze,void_ugatze};
//Formalisation des signatures des opérations
Property type logical_nature=enum{in_ugatze, out_ugatze,inout_ugatze};
Property type parameter=Record[parameter_type:base_type_ugatze;mode:
logical_nature;];
Property type pls_parameter=sequence<parameter>;
Property type signature=Record[p:pls_parameter;resultat:base_type_ugatze;];
Property type porttype=enum {provider,required,environment};
```

**Figure 8.2 :** Formalisation des types de données en Acme/Armani

### 8.2.2 Formalisation des points d'interaction

Un point d'interaction Ugazte (*IIP*, *OIP*, *UIOP*, *PIOP*) est formalisé par un port Acme/Armani. La Figure 8.3 illustre la formalisation des différents types des points d'interaction Ugatze. Les types proposés sont :

- «*DataPoint*» : fondateur des points d'interaction et admet comme descendant les types «*OIP*» et «*IIP*». Le type «*DataPoint*» englobe une propriété «*protocol*» de type chaîne de caractères,
- «OperationPoint» : fondateur des points d'interaction d'opérations admet comme descendants les types «UOIP» et «PIOP». Le type «OperationPoint» englobe deux propriétés « porttype » de type «*port_type*» et «*operation*» de type «*signature*»,
- «*OIP*» : définit un point d'interaction de données. Le type «*OIP*» hérite de «DataPoint» et englobe une propriété « *porttype* » de valeur « *provider* »,
- «*IIP*» : définit un point d'interaction de données. Le type «*IIP*» hérite de «DataPoint» et englobe une propriété « *porttype* » de valeur « *required* »,
- «*PIOP*» : définit un point d'interaction d'opération fourni. Il hérite de «OperationPoint» et englobe une propriété «*porttype*» de valeur «*provider*»,
- «*UIOP*» : définit un point d'interaction d'opération requis. Il hérite de «*OperationPoint*» et englobe une propriété «*porttype*» de valeur «*required*».

### 8.2.3 Formalisation des composants Ugatze

Un composant Ugatze est formalisé par un composant Acme/Armani (cf. Figure 8.4). Une règle « *haveAtLeastone* » relative à un composant Ugatze est modélisée par une propriété invariante en utilisant le concept d'invariant d'Acme/Armani. Cette règle stipule qu'un composant Ugatze de type *componentUgatze* possède au moins un point d'interaction.

Par ailleurs, un composant Ugatze est formalisé par une hiérarchie de composants dont le fondateur est *componentUgatze* ayant comme descendants *componentFilter*, *componentFilterClientServer* et *componentClientServer*. La Figure 8.1 illustre les différents descendants.

```
Port Type DataPoint ={
      Property protocol : string ;
      }

Port Type OperationPoint ={
      property porttype:port_type;
       Property operation:signature;
      }

Port Type EnvironementPoint ={
      Property protocol : string ;
      }

Port Type OIP extends DataPoint with{property porttype:port_type=provider;}

Port Type IIP extends DataPoint with {property porttype:port_type=required;}

Port Type PIOP extends OperationPoint with {property porttype:port_type=provider;}

Port Type UIOP extends OperationPoint
with{property porttype:port_type=required;}
```

**Figure 8.3 :** Formalisation des points d'interaction en Acme/Armani

```
Component Type ComponentUgatze = {
      rule haveAtLeastone = invariant size(self.PORTS) >= 1;
}
```

**Figure 8.4:** Formalisation d'un composant Ugatze en Acme/Armani

Un *componentFilter* (cf. Figure 8.5) englobe une règle « *PortType1*». Cette règle stipule qu'un composant Ugatze de type *ComponentFilter* est défini par des points d'interaction de type « *IIP* » ou « *OIP* ».

```
Component Type ComponentFilter extends ComponentUgatze with {
      rule  PortType1  =  invariant  forall  p  :  Port  in  self.PORTS
|declaresType(p,IIP) or declaresType(p, OIP); }
```

**Figure 8.5:** Formalisation d'un ComponentFilter en Acme/Armani

Un *ComponentClientServer* (cf. Figure 8.6) englobe une règle «*PortType2*». Cette règle stipule qu'un composant Ugatze de type *ComponentClientServer* est défini par des points d'interaction de type de « *UIOP* » ou « *PIOP* ».

```
Component Type ComponentClientServer extends ComponentUgatze with {
     rule  PortType2  =  invariant  exists   p  :  Port  in  self.PORTS
|declaresType(p, UIOP) or declaresType(p, PIOP);
}
```

**Figure 8.6 :** Formalisation d'un ComponentClientServer en Acme/Armani

Un *ComponentFilterClientServer* (cf. Figure 8.7) englobe une règle « *PortType1*». Cette règle stipule qu'un composant Ugatze de type *ComponentFilterClientServer* est défini par au moins un point d'interaction de type (*OIP* ou *IIP*). En outre, il englobe une règle «*PortType2*» stipulant qu'un composant  de type *ComponentFilterClientServer* possède au moins un point d'interaction de type de « *UIOP* » ou « *PIOP* ».

```
Component Type ComponentFilterClientServer extends ComponentUgatze with
{    rule  PortType1  =  invariant  exists  p  :  Port  in  self.PORTS
|declaresType(p,IIP) or declaresType(p, OIP);
    rule  PortType2  =  invariant  exists  p  :  Port  in  self.PORTS
|declaresType(p,UIOP) or declaresType(p, PIOP);
 }
```

**Figure 8.7 :** Formalisation d'un ComponentFilterClientServer en Acme/Armani

### 8.2.4 Formalisation des interactions Ugatze

Une interaction Ugatze est formalisée par un connecteur Acme/Armani  (cf. Figure 8.8). Celle-ci illustre la formalisation du type  « *Pipe* ». Ce type modélise une interaction directe de données. Des règles de cohérence relatives à une interaction directe de données sont définies telles que :

- « *exactlyTwoRoles* » : cette règle vérifie que chaque interaction « *Pipe* » est définie par deux rôles,
- « *bufferpositive*» : cette règle vérifie que la taille du buffer ne peut être négative,
- « *precondition1*» : cette règle vérifie que chaque rôle de type « *sourcePipe* » est rattaché à un port de type « *IIP* » et chaque rôle de type « *sinkPipe* » est rattaché à un port de type « *OIP* ».

La Figure 8.9 illustre la formalisation du type « *InteractionOperation* ». Ce type modélise une interaction d'opération. Des règles de cohérence relatives à ce type d'interaction sont définies telles que :

- « *exactlyTwoRoles* » : cette règle vérifie que chaque interaction d'opération est définie par deux rôles,
- « *noDanglingRoles* » : cette règle vérifie qu'un rôle ne peut être non rattaché à un point d'interaction,
- «*connectorIndependence* » : cette règle vérifie que toutes les interactions d'opération sont indépendantes. Elles ne partagent pas des points d'interaction.

```
Connector Type Pipe = {
        Property bufferSize : int <<  bufferSize : int = 0; >> ;
        Role source : sourcePipe = new sourcePipe extended with {}
        Role sink : sinkPipe = new sinkPipe extended with { }
        rule exactlyTwoRoles = invariant size(self.ROLES) == 2;
        rule bufferpositive = invariant self.bufferSize >= 0;
rule precondition1 = invariant forall r1 : Role in self.ROLES |
declaresType(r1, sinkPipe) -> forall p1 : Port in r1.ATTACHEDPORTS |
declaresType(p1, IIP) -> forall r2 : Role in self.ROLES |
declaresType(r2, sourcePipe) -> forall p2 : Port in
r2.ATTACHEDPORTS|declaresType(p2,OIP);
rule   noDanglingRoles   =   invariant   forall   r   :   Role   in   self.ROLES
|attachedOrBound(r);}
```

**Figure 8.8:** Formalisation d'un Pipe (interaction directe de donnée) en Acme/Armani

```
Connector Type InteractionOperation = {
Role source : serverT = new serverT extended with {    }
Role sink : clientT = new clientT extended with {}
rule exactlyTwoRoles = invariant size(self.ROLES) == 2;
rule   noDanglingRoles   =   invariant   forall   r   :   Role   in   self.ROLES|
attachedOrBound(r);
rule     connectorIndependence=invariant     forall     r     :     Role     in
self.ROLES|(r.ATTACHEDPORTS) == 1;}
```

**Figure 8.9 :** Formalisation d'une interaction d'opération Acme/Armani

La Figure 8.10 illustre la formalisation du type «*DataAccess*». Ce type modélise l'interaction prédéfinie d'une ressource partagée. Des règles de cohérence relatives à ce type d'interaction sont définies telles que :

- « *numberRoles* » : cette règle vérifie que chaque interaction d'opération est définie par deux rôles,
- « *typerole* » : cette règle vérifie que toutes les interactions «*DataAccess*» sont définies par un rôle de type «*sourceShared*» et un rôle de type «*sinkShared*».

```
Connector Type DataAccess = {
Role source : sourceShared = new sourceShared extended with {}
Role sink : sinkShared = new sinkShared extended with {}
rule numberRoles=invariant size(self.ROLES)>=2;
rule typerole=invariant forall r1:role in self.ROLES|
 forall r2:role in self.ROLES| r1!=r2<->!exists r in
{sourceShared,sinkShared}|declaresType(r1,r)and declaresType(r2,r);}
```

**Figure 8.10 :** Formalisation d'une interaction prédéfinie (Ressource partagée) en Acme/Armani

### 8.2.5 Formalisation des règles de configuration

Hormis les règles de cohérence relatives aux concepts architecturaux (composant, connecteur et port), d'autres règles sont définies au niveau du style *UGATZE*. Ces règles permettent de vérifier la cohérence des attachements des différents concepts architecturaux. La Figure 8.11 illustre la formalisation en Acme/Armani de ces règles. Cette formalisation propose les règles :

- « *componentType* » : vérifie que le graphe d'interaction est défini par des composants de type « UgatzeComponent »,
- « *failedAttachement*» : vérifie qu'un composant Ugatze ne peut être rattaché à lui-même,
- « *numberComponent* » : vérifie qu'un graphe d'interaction est composé par au moins deux composants.

```
rule componentType = invariant forall comp : Component in
self.COMPONENTS |declaresType(comp, ComponentUgatze)and
satisfiesType(comp,ComponentUgatze );
rule failedAttachement=invariant forall c1:Component in
self.COMPONENTS|connected(c1, c1);
rule numberComponent = invariant size(self.COMPONENTS) >= 2;
```

**Figure 8.11 :** Formalisation des règles de configuration en Acme/Armani

## 8.3 Etude de cas : diagnostic médical distribué

### 8.3.1 Cahier des charges

L'application « diagnostic médical distribué » inspirée de [Van Eenoo, 2005] a pour objectif d'établir des prescriptions médicales relatives à des patients. Les trois acteurs médecin généraliste, médecin spécialiste et pharmacien coopèrent en vue d'établir des prescriptions médicales. Le médecin généraliste fournit des données relatives à un patient au médecin spécialiste. Ce dernier établit un diagnostic basé sur les données transmises par le médecin généraliste. Le rapport du diagnostic est transmis par le médecin spécialiste au pharmacien afin que ce dernier produise la prescription correspondante. Enfin, le pharmacien fournit la prescription au médecin généraliste.

### 8.3.2 Modélisation en Ugatze

La Figure 8.12 représente l'architecture simplifiée de l'application « diagnostic médical distribué » en utilisant la syntaxe graphique d'Ugatze. Cette architecture est abstraite en ce sens qu'elle est indépendante de tout mécanisme d'implantation.

Cette architecture intègre trois styles différents (pipe-and-filter, shared-variable et client-serveur). En effet la Figure 8.12 illustre cette architecture simplifiée multi-style, on y retrouve :

- Un composant (*GP*) de type « *FilterClientServer* » doté d'un point d'information de sortie (*OIP*) et d'un point d'opération requis (*UIOP*),
- Un composant (*SP*) de type « *Filter* » doté d'un point d'information d'entrée (*IIP*) et d'un point d'information de sortie (*OIP*),
- Un composant (*PH*) de type « *FilterClientServer* » doté d'un point d'information d'entrée (*IIP*) et d'un point d'information de sortie (*PIOP*),
- Une interaction directe de données de type « *Pipe-Filter* » reliant un *IIP* (*diagnosis*) à un *OIP* (*compute_diagnosis*),
- Une interaction de type « *DataAccess* » illustrant l'accès partagé sur une variable (*shared_variable*) entre des *IIP* (patient_test_data) à des *OIP* (*write*),
- Une interaction d'opération « *Client-Server* » reliant un *UIOP* (*authenticate*) à un *PIOP* (*prescription*).

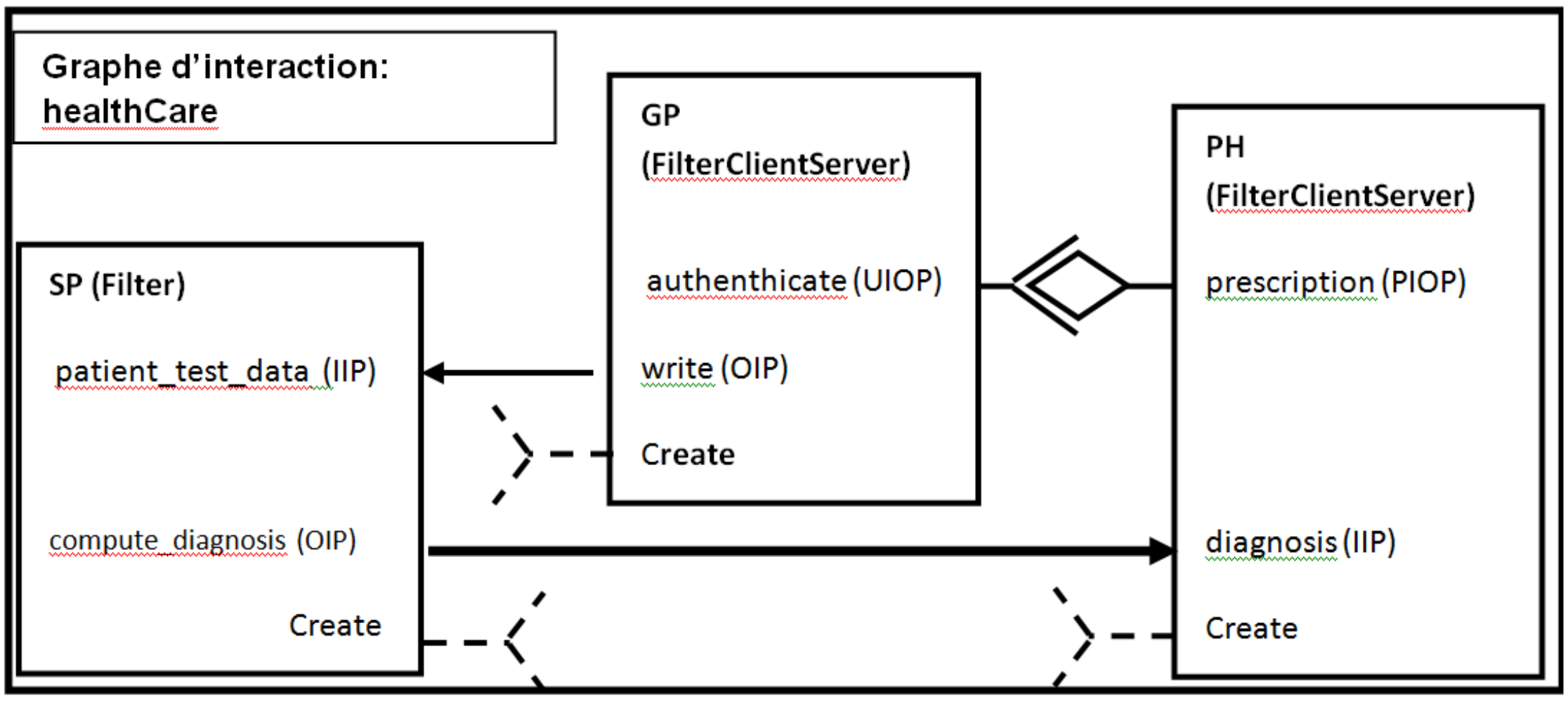

**Figure 8.12 :** Modélisation en Ugatze de l'application « diagnostic médical distribué »

### 8.3.3 Formalisation en Acme/Armani

Un assemblage de composants Ugatze est modélisé par une configuration Acme/Armani (cf. Figure 8.13) qui dérive du style *UGATZE* (cf. Figure 8.1) [Kmimech, 2009b], [Kmimech, 2009c]. La Figure 8.13 donne la traduction de la Figure 8.12 sous forme d'une configuration Acme/Armani en passant par le niveau méta fourni par le style *UGATZE*.

Les contrats propres à l'application « diagnostic médical distribué » sont au nombre de trois définis par des invariants Armani :

- « verifyProperty_GP_SP » : vérifie la compatibilité de la propriété « protocol » définie au sein du port « write » (correspondant à un OIP) et au sein du port «patient_test_data_protocol» (correspondant à un IIP). L'attachement de ces deux points d'interaction forment ainsi une interaction de type Pipe correspondant à une interaction directe d'information dans le modèle Ugatze. Ces deux ports doivent obéir au même protocole,
- «verifyProperty_PH_SP» : vérifie la compatibilité de la propriété « protocol » définie au sein du port « diagnosis » (correspondant à un IIP) et au sein du port «compute_diagnosis» (correspondant à un OIP),
- « verifyProperty_PH_GP» : vérifie la compatibilité de la propriété « operation » définie au sein du port « authenticate » (correspondant à un UIOP) et au sein du port «prescription» (correspondant à un PIOP). L'attachement de ces deux points d'interaction (UIOP et PIOP) forme ainsi une interaction de type interaction d'opération d'information dans le modèle Ugatze. Cette règle permet de vérifier la compatibilité entre la signature de l'opération requise et celle de l'opération fournie.

La vérification des contrats (ou règles) hérités du style UGATZE et spécifiques à l'application définis au sein de la configuration healthCare est confiée à la plate-forme AcmeStudio. Une règle de cohérence violée (invariant évalué à Faux) traduit forcément une incohérence dans l'assemblage de composants traité.

```
import families/UGATZE.acme;
 System healthCare : UGATZE = new UGATZE extended with {

    Component PH : ComponentFiltreClientServer = new
ComponentClientServer extended with {
        Port diagnosis : IIP = {
            Property protocol : string = "char";
        }
        Port prescription : PIOP = {
              Property operation : signature = [p = <[parameter_type =
integer_ugatze;mode = in_ugatze;]>;resultat = void_ugatze;];
        }
    }

    Component SP : ComponentFiltre = new ComponentFiltre extended with
{
        Port compute_diagnosis : OIP = {
            Property protocol : string = "char";
        }
        Port patient_test_data : IIP, p_use = {
            Property protocol : string = "char";
        }
    }
Component GP : ComponentFiltreClientServer = new ComponentClientServer
extended with {
        Port write : OIP = {
            Property protocol : string = "char";
        }
        Port authenticate : UIOP = {
              Property operation : signature = [p = <[parameter_type =
integer_ugatze;mode = in_ugatze;]>;resultat = void_ugatze;];
        }
    }

  Connector pipe : Pipe = new Pipe extended with {
       Property bufferSize = 10;
    }

    Connector sharedData : DataAccess = new DataAccess extended with {
    }
    Connector interactionOperation : InteractionOperation = new
InteractionOperation extended with {
```

```
    Attachment PH.diagnosis to pipe.sink;
    Attachment SP.compute_diagnosis to pipe.source;
    Attachment GP.write to sharedData.source;
    Attachment SP.patient_test_data to sharedData.sink;
    Attachment GP.authenticate to interactionOperation.sink;
    Attachment PH.prescription to interactionOperation.source;
//Vérification de la compatibilité des propriétés lors des attachements
rule verifyProperty_GP_SP = invariant self.GP.write.protocol ==
self.SP.patient_test_data.protocol;
rule verifyProperty_PH_SP = invariant self.PH.diagnosis.protocol ==
self.SP.compute_diagnosis.protocol;
rule verifyProperty_PH_GP = invariant self.GP.authenticate.operation ==
self.PH.prescription.operation;
  }
```

**Figure 8.13 :** Formalisation en Acme/Armani de l'application décrite en Ugatze (niveau M1)

## 8.4 Conclusion

Dans ce chapitre, nous avons formalisé en Acme/Armani les principaux concepts issus du modèle de composants semi-formel Ugatze tels que : type de données, composant, point d'interaction. En outre, nous avons établi des contrats syntaxiques et structurels afin de vérifier la cohérence d'un assemblage de composants Ugatze. Ces contrats sont formalisés en Acme/Armani sous forme des propriétés invariantes attachées aux éléments architecturaux concernés. Notre formalisation du modèle de composants Ugatze est regroupée au sein d'un style Acme/Armani appelé UGATZE. Enfin, nous avons modélisé en Ugatze l'application « diagnostic médical distribué ». La modélisation Ugatze obtenue est traduite en Acme/Armani au sein d'un système (*system*) en utilisant le style UGATZE. Les règles de cohérence (ou contrats) venant du style UGATZE et celles propres à l'application sont vérifiées par AcmeStudio.

# Conclusion générale

## 1. Bilan

Dans le cadre de cette thèse, nous avons abordé la problématique de la vérification de la cohérence d'assemblages de composants logiciels décrits par des modèles semi-formels UML2.0 et Ugatze. Pour y parvenir, nous avons adopté une approche contractuelle basée sur des contrats applicatifs à quatre niveaux : contrats syntaxiques et structurels, contrats sémantiques, contrats de synchronisation et contrats de qualité de services. Afin de vérifier ces contrats, nous avons opté pour la traduction des modèles de composants semi-formels (UML2.0 et Ugatze) vers des modèles de composants formels (Acme/Armani et Wright). Ceci favorise la continuité entre le semi-formel et le formel en choisissant le concept pivot de composant. Une telle continuité facilite la traçabilité entre les deux modèles semi-formel et formel. En effet, dans les deux modèles semi-formel et formel, l'architecte manipule des concepts issus de l'approche par composants : composant, connecteur, interface et configuration.

Afin de décrire un assemblage de composants UML2.0, nous avons préconisé les moyens de spécification suivants :

- les aspects structuraux sont spécifiés en utilisant les concepts composant, interface offerte, interface requise et connecteur d'assemblage issus d'UML2.0,
- les aspects non fonctionnels sont spécifiés en utilisant le langage de modélisation des propriétés non fonctionnelles CQML,
- les aspects comportementaux sont spécifiés en utilisant une extension au Protocol State Machine appelée Port State Machine.

Nous avons proposé une démarche *VerifComponentUML2.0* permettant de vérifier la cohérence d'assemblages de composants UML2.0 vis-à-vis des contrats syntaxiques et structurels, des contrats de qualité de services et des contrats de synchronisation. La démarche *VerifComponentUML2.0* propose des contrats syntaxiques et structurels liés à la compatibilité des types et aux règles de composition des composants UML2.0. Egalement, elle propose des contrats de qualité de services qui stipulent que chaque propriété non fonctionnelle requise par un composant doit être offerte par son environnement (les composants connectés à celui-ci). La vérification de ces contrats syntaxiques, structurels et de qualité de services est confiée à l'évaluateur de prédicats Armani moyennant la traduction d'UML2.0/CQML vers Acme/Armani. En outre, la démarche *VerifComponentUML2.0* récupère les contrats de synchronisation liés à la cohérence de composants, la cohérence de connecteurs et la compatibilité port/rôle proposés par Wright. La vérification de ces contrats est confiée au model-checker FDR moyennant notre traducteur de Wright vers CSP : Wr2fdr. Enfin, la démarche *VerifComponentUML2.0* propose un outil IDM permettant de traduire de Wright vers Ada : Wright2Ada. Celui-ci autorise l'utilisation des outils d'analyse statique et dynamique associés à Ada.

Nous avons également proposé une démarche *VerifComponentUgatze* permettant de vérifier la cohérence d'un assemblage de composants Ugatze vis-à-vis des contrats syntaxiques et structurels moyennant la traduction d'Ugatze vers Acme/Armani. Les contrats syntaxiques et structurels sont vérifiés par l'évaluateur de prédicats Armani supporté par la plate-forme AcmeStudio.

## 2. Perspectives

### 2.1 L'outil Wright2Ada

Nous pourrions envisager les prolongements relatifs à l'outil IDM Wright2Ada suivants :

- Intégrer les facilités syntaxiques offertes par CSP telles que : *where, when,* processus avec état et opération de quantification sur les ensembles,
- Traiter les composants et les connecteurs composites offerts par Wright,
- Proposer l'opération de transformation inverse d'Ada vers Wright. Ceci favorise l'extraction d'une architecture logicielle abstraite en Wright à partir d'un programme concurrent Ada,
- Améliorer éventuellement l'efficacité de notre outil Wright2Ada en étudiant l'apport des patterns d'OCL pour la transformation de modèles [Cuadrado, 2009],
- Vérifier davantage l'outil Wright2Ada en utilisant des techniques de vérification applicables sur des programmes de transformation de modèles : test structurel, analyse de mutation, problème d'oracle et analyse statique [Küster, 2006], [Mottu, 2005], [Baudry, 2009].

### 2.2 Automatisation d'UML2.0/CQML vers Acme/Armani

Les règles de traduction UML2.0/CQML vers Acme/Armani établies dans ce travail pourraient être automatisées en utilisant une approche IDM : élaboration d'un méta-modèle partiel UML2.0, méta-modèle partiel CQML, méta-modèle Acme/Armani et expression des règles de transformation UML2.0/CQML vers Acme/Armani en utilisant un langage de transformation de modèles tels que ATL.

### 2.3 D'UML2.0/PoSM vers Wright

La démarche *VerifComponentUML2.0* a identifié un module permettant de traduire un assemblage de composants UML/PoSM -dont les aspects structuraux sont décrits en UML2.0 et les aspects comportementaux sont spécifiés en PoSM- en Wright. Il est souhaitable de traiter ce module en élaborant des règles systématiques de transformation notamment de PoSM vers CSP de Wrifght.

### 2.4 Contrats sémantiques

Les contrats sémantiques permettant de spécifier la sémantique des opérations offertes et requises en utilisant une spécification pré/post ne sont pas traités dans cette thèse. Nous comptons les traiter ultérieurement. Pour le modèle de composants Ugatze, nous avons lancé des actions dans ce sens [Kmimech, 2006], [Kmimech, 2007a], [Kmimech, 2007b], [Belmabrouk, 2010].

### 2.5 Architecture orientée service

Récemment, nous avons lancé des actions [Maraoui, 2010], [Graiet, 2010] permettant d'apprécier les aptitudes d'un modèle de composants formel comme Acme/Armani vis-à-vis de la formalisation des services Web.

### 2.6 L'outil Wr2fdr

Nous comptons implémenter les propriétés 4, 6 et 7 à savoir Initialiseur unique, Substitution de paramètres et Bornes d'un intervalle (cf. 2.5.3.1) en utilisant notre analyseur sémantique de Wr2fdr.

# Bibliographie

# Liste des acronymes

| | |
|---|---|
| **AADL**<br>Architecture Analysis Design Language | **LOTOS**<br>Language Of Temporal Order Specification |
| **ADL**<br>Architecture Description Language | **LTSA**<br>Labelled Transition System Analyser |
| **AL**<br>Architecture Logicielle | **MDE**<br>Model-Driven Engineering |
| **AMMA**<br>ATLAS Model Management Architecture | **MDA**<br>Model Driven Architecture |
| **ATL**<br>ATLAS Transformation Language | **MOF**<br>Meta-Object Facility |
| **BNF**<br>Backus Naur Form | **oAW**<br>open Architecture Ware |
| **CADP**<br>Construction and Analysis of Distributed Processes | **OCL**<br>Object Constraint Language |
| **CDL**<br>Component Definition Language | **ODL**<br>Object Description Language |
| **CIM**<br>Computational Independent Model | **OMG**<br>Object Management Group |
| **CQML**<br>Component Quality Modeling Language | **PDM**<br>Platform Description Model |
| **CSP**<br>Communicating Sequential Processes | **PIM**<br>Platform Independent Model |
| **DSL**<br>Domain Specific Languages | **PNF**<br>Propriété Non-Fonctionnelle |
| **Elts**<br>extended Labelled Transition Systems | **PSM**<br>Protocol State Machine |
| **FDR**<br>Failure-Divergence Refinement | **PoSM**<br>Port State Machine |
| **FSP**<br>Finite State Processes | **QML**<br>Quality Modelling Language |
| **IDM**<br>Ingénierie Dirigée par les Modèles | **QoS**<br>Quality of Service |
| **IDL**<br>Interface Description Language | **QTL**<br>Quality of service Temporal Logic |
| **IOLTS**<br>Input Output Labeled Transition Systems | **QVT**<br>Query View Transformation |
| **CDL**<br>Component Definition Language | **SDL**<br>Specification and Description Language |
| **JML**<br>Java Modeling Language | **TINA ODL**<br>TINA Object Definition Language |
| | **XMI**<br>XML Metadata Interchange |

# Annexe A : Sémantique statique de la partie structurelle d'Ada

- Propriété 1 :

Au sein de la partie déclarative d'un sous-programme, les noms des tâches (partie spécification et implémentation) et des sous-programmes (partie spécification et implémentation) doivent être deux à deux différents.

```
context SubprogramBody
  def: col1:Sequence(String) = self.declarations ->
  select(e:Declaration|e.oclIsKindOf(SubprogramSpecification)) ->
  collect(e:SubprogramSpecification|e.designator)
  def: col2:Sequence(String) = self.declarations ->
  select(e:Declaration|e.oclIsTypeOf(SingleTaskDeclaration)) ->
  collect(e:SingleTaskDeclaration|e.identifier)
  def: col3:Sequence(String) = self.declarations ->
  select(e:Declaration|e.oclIsTypeOf(TaskBody)) ->
  collect(e:TaskBody|e.identifier)
  def: col4:Sequence(String) = self.declarations ->
  select(e:Declaration|e.oclIsTypeOf(SubprogramBody)) ->
  collect(e:SubprogramBody|e.specif.designator)
  inv: col1 -> excludesAll(col2)
  inv: col1 -> excludesAll(col3)
  inv: col2 -> excludesAll(col4)
  inv: col3 -> excludesAll(col4)
  inv: col2->includesAll(col3) and col2->size()=col3->size()
```

- Propriété 2 :

Au sein de la partie déclarative d'un sous-programme, les identificateurs des sous-programmes doivent être différents.

```
context SubprogramBody
  inv: self.declarations ->
  select(e:Declaration|e.oclIsKindOf(SubprogramSpecification)) ->
  forAll(e1:SubprogramSpecification, e2:SubprogramSpecification| e1<>e2
  implies e1.designator<>e2.designator)
  inv: self.declarations ->
  select(e:Declaration|e.oclIsTypeOf(SubprogramBody)) ->
  forAll(e1:SubprogramBody, e2:SubprogramBody| e1<>e2 implies
  e1.specif.designator<>e2.specif.designator)
```

- Propriété 3 :

Au sein de la partie déclarative d'un sous-programme les identificateurs des tâches doivent être différents.

```
context SubprogramBody
    inv: self.declarations ->
  select(e:Declaration|e.oclIsTypeOf(SingleTaskDeclaration)) ->
  forAll(e1:SingleTaskDeclaration, e2:SingleTaskDeclaration| e1<>e2
  implies e1.identifier<>e2.identifier)
    inv: self.declarations ->
  select(e:Declaration|e.oclIsTypeOf(TaskBody)) -> forAll(e1:TaskBody,
  e2:TaskBody| e1<>e2 implies e1.identifier<>e2.identifier)
```

- Propriété 4 :

Au sein de la partie déclarative d'une tâche, les identificateurs des tâches (partie spécification implémentation) et des sous-programmes (partie spécification/ implémentation) doivent être deux à deux différents.

```
context TaskBody
def: col1:Sequence(String) = self.declarations -> select(e:Declaration|e.oclIsKindOf(SubprogramSpecification)) -> collect(e:SubprogramSpecification|e.designator)
def: col2:Sequence(String) = self.declarations -> select(e:Declaration|e.oclIsTypeOf(SingleTaskDeclaration)) -> collect(e:SingleTaskDeclaration|e.identifier)
def: col3:Sequence(String) = self.declarations -> select(e:Declaration|e.oclIsTypeOf(TaskBody)) -> collect(e:TaskBody|e.identifier)
def: col4:Sequence(String) = self.declarations -> select(e:Declaration|e.oclIsTypeOf(SubprogramBody)) -> collect(e:SubprogramBody|e.specif.designator)
inv: col1 -> excludesAll(col2)
inv: col1 -> excludesAll(col3)
inv: col2 -> excludesAll(col4)
inv: col3 -> excludesAll(col4)
inv: col2 -> includesAll(col3) and col2->size()=col3->size()
```

- Propriété 5 :

Au sein de la partie déclarative d'une tâche, les identificateurs des sous-programmes doivent être différents.

```
context TaskBody
inv: self.declarations -> select(e:Declaration|e.oclIsKindOf(SubprogramSpecification)) -> forAll(e1:SubprogramSpecification, e2:SubprogramSpecification| e1<>e2 implies e1.designator<>e2.designator)
inv: self.declarations -> select(e:Declaration|e.oclIsTypeOf(SubprogramBody)) -> forAll(e1:SubprogramBody, e2:SubprogramBody| e1<>e2 implies e1.specif.designator<>e2.specif.designator)
```

- Propriété 6 :

  Au sein de la partie déclarative d'une tâche, les identificateurs des tâches doivent être différents.

```
context TaskBody
inv: self.declarations -> select(e:Declaration|e.oclIsTypeOf(SingleTaskDeclaration)) -> forAll(e1:SingleTaskDeclaration, e2:SingleTaskDeclaration| e1<>e2 implies e1.identifier<>e2.identifier)
inv: self.declarations -> select(e:Declaration|e.oclIsTypeOf(TaskBody)) -> forAll(e1:TaskBody, e2:TaskBody| e1<>e2 implies e1.identifier<>e2.identifier)
```

# Annexe B : Sémantique statique de la partie comportementale d'Ada

- Propriété 7 :

Une fonction contient au moins une instruction return.

```
context SubprogramBody
  inv: specif.oclIsTypeOf(FunctionSpecification) implies statements ->
  collect(s:Statement|s.oclIsTypeOf(ReturnStatement)) -> size()>=1
```

- Propriété 8 :

Un sous-programme ne contient pas d'instruction accept.

```
context SubprogramBody
      inv: statements -> forAll(s:Statement | not
      s.oclIsTypeOf(SimpleAcceptStatement))
```

- Propriété 9 :

Un sous-programme ne contient pas d'instruction select.

```
context SubprogramBody
      inv: statements -> forAll(s:Statement | not
      s.oclIsTypeOf(SelectOr))
```

- Propriété 10 :

Une tâche ne contient pas d'instruction return.

```
context TaskBody
      inv: statements -> forAll(s:Statement | not
      s.oclIsTypeOf(ReturnStatement))
```

- Propriété 11 :

Une tâche ne peut accepter des rendez-vous que sur ses propres entrées (entry).

```
context TaskBody
    def: c1:Sequence(String) = self.statements ->
    select(e:Statement|e.oclIsTypeOf(SimpleAcceptStatement)) ->
    collect(e:SimpleAcceptStatement|e.direct_name)
    def: c2:Sequence(String) = self.declarations ->
    collect(e:SingleTaskDeclaration|e.entryDec) ->
    collect(e:EntryDeclaration|e.identifier)
    inv: c2 -> includesAll(c1)
```

# Annexe C : Traduction des aspects comportementaux de Wright

## - Traduction de l'opérateur préfixe

L'opérateur préfixe a la forme suivante *EventExpression* -> *ProcessExpression*.
La traduction de l'opérateur préfixe consiste à traduire *EventExpression*, puis à traduire *ProcessExpression*. Le helper redéfini *transformation* permet de le faire.

```
helper context Wright!Prefix def: transformation(instance :
String):Sequence(Ada!statement)=
if self.target.oclIsTypeOf(Wright!Prefix) then
Sequence{self.event.event_transform(instance)}-
>union(self.target.transformation(instance))
else
if self.target.oclIsTypeOf(Wright!InternalChoice) or
self.target.oclIsTypeOf(Wright!ExternalChoice) then
Sequence{self.event.event_transform(instance)}->union(Sequence
{self.target.transformation(instance)})
else
Sequence{self.event.event_transform(instance)}
endif
endif;
```

Ce helper fait appel à un autre helper redéfini *event_transform* qui permet de transformer un événement selon son contexte. De plus, il fait appel au helper redéfini transformation pour transformer la cible du préfixe.

## - Traduction des événements

Dans la suite, nous présentons la traduction des événements observés, initialisés, traitements internes et l'événement succès.

### o Traduction des événements observés et initialisés

Nous rappelons que les attachements sont de la forme « nomInstanceComposant . nomPort As nomInstanceConnecteur . nomRôle »

- Traduction des événements observés et initialisés attachés à une instance de composant :

Si événement est un événement observé de la forme «nomPort.événement».
Alors
Accepter le rendez vous portant le nom nomPort_événement ; soit *accept nomPort_ événement ;*
Fin Si
Si événement est un événement initialisé de la forme « _nomPort.événement ».
Alors
Voir le nomRôle qui est attaché à nomPort de l'instance courante, demander un rendez-vous par : *Connector_nomInstanceConnecteur.nomRôle_événement ;*
Fin Si

- Traduction des événements observés et initialisés attachés à une instance de connecteur :
Si événement est un événement observé de la forme «nomRôle.événement».
Alors
Accepter le rendez-vous portant le nom nomRôle_événement; soit *accept nomRôle_ événement ;*
Fin Si

Si événement est un événement initialisé de la forme « _nomRôle.événement ».
alors
Voir le nomPort qui est attaché à nomRôle de l'instance courante, demander un rendez-vous par: *Component_nomInstanceComposant.nomPort_événement ;*
Fin Si

- Traduction en ATL :

La traduction des événements observés et initialisés se fait par l'intermédiaire des deux helpers redéfinis *event_transform* suivants :

```
helper context Wright!EventObserved def: event_transform(instance : String):Ada!simple_accept_statement=
thisModule.EventObserved2simple_accept_statement(self);
```

```
helper context Wright!EventSignalled def: event_transform(instance : String):Ada!entry_call_statement=
thisModule.EventSignalled2entry_call_statement(self,instance);
```

Ces helpers déclenchent respectivement les règles paresseuses permettant de traduire un événement observé et un événement initialisé. Ces deux règles paresseuses sont présentées par :

```
lazy rule EventObserved2simple_accept_statement{
from e:Wright!EventObserved
to s:Ada!simple_accept_statement(
direct_name<- e.name.replaceAll('.','_') )
}
```

La règle paresseuse *EventObserved2simple_accept_statement* transforme un événement observé en une instruction *accept* portant le nom de l'événement en remplaçant le point par un tiret bas.

```
lazy rule EventSignalled2entry_call_statement{
from e:Wright!EventSignalled,
  instance : String
  to ec:Ada!entry_call_statement(
  entry_name<-if(Wright!Attachment.allInstances()-
  >select(a|a.originPort.name=e.name.substring(1,e.name.indexOf('.')))-
  >select(a|a.originInstance.name=instance)->notEmpty())then
  'Connector_'
  +Wright!Attachment.allInstances()-
  >select(a|a.originPort.name=e.name.substring(1,e.name.indexOf('.')))-
  >select(a|a.originInstance.name=instance).at(1).targetInstance.name
  +'.'+Wright!Attachment.allInstances()-
  >select(a|a.originPort.name=e.name.substring(1,e.name.indexOf('.')))-
  >select(a|a.originInstance.name=instance).at(1).targetRole.name
  +'_'+e.name.substring(e.name.indexOf('.')+2,e.name.size())
  else
  'Component_'
  +Wright!Attachment.allInstances()-
  >select(a|a.targetRole.name=e.name.substring(1,e.name.indexOf('.')))-
  >select(a|a.targetInstance.name=instance).at(1).originInstance.name
  +'.'+Wright!Attachment.allInstances()-
  >select(a|a.targetRole.name=e.name.substring(1,e.name.indexOf('.')))-
  >select(a|a.targetInstance.name=instance).at(1).originPort.name
  +'_'+e.name.substring(e.name.indexOf('.')+2,e.name.size())
endif
)
}
```

La règle paresseuse *EventSignalled2entry_call_statement* transforme un événement initialisé en une instruction *entry* dont le nom dépend de l'instance de composant ou de connecteur à laquelle cet événement appartient. C'est la raison pour laquelle le passage du paramètre instance a eu lieu tout au long de la traduction du processus CSP Wright. De plus, ce nom dépend de l'attachement dans lequel le port ou le rôle est impliqué.

- **Traduction des événements succès**

L'événement succès « V » qui est toujours suivi du processus STOP représente le processus SKIP ou encore §. Ceci correspond à la terminaison avec succès. Ce cas est traduit dans Ada par l'instruction exit.
Le helper redéfini *event_transform* suivant permet le déclenchement de la règle paresseuse traduisant l'événement succès.

```
helper context Wright!SuccesEvent def: event_transform(instance :
String):Ada!exit_statement =
thisModule.SuccesEvent2exit_statement(self);
```

Traduction de l'événement par la règle paresseuse succès :

```
lazy rule SuccesEvent2exit_statement{
from p:Wright!SuccesEvent
      to e:Ada!exit_statement
}
```

**Traduction d'un événement interne**

Comme déjà cité, un événement interne est traduit en Ada par une procédure dont le corps est à raffiner. L'appel de cette procédure se fait par la règle paresseuse suivante :

```
lazy rule InternalTraitement2procedure_call_statement{
from e:Wright!InternalTraitement
      to p:Ada!procedure_call_statement(
      name<-e.name
)
}
```

Le helper redéfini *event_transform* suivant permet le déclenchement de la règle paresseuse traduisant l'appel de la procédure du traitement interne.

```
helper context Wright!InternalTraitement def: event_transform(instance :
String):Ada!procedure_call_statement=
      thisModule.InternalTraitement2procedure_call_statement(self);
```

# Annexe D: Module Wright2Ada en ATL

```
-- @path Wright=/Wright/model/Wright.ecore
-- @path Ada=/my.generator.ada/src/metamodel/Ada.ecore
module WrightToAda;
create exampleAda : Ada from exampleWright : Wright;
helper context Wright!ProcessExpression def: getEventObserved():
Set(Wright!EventObserved) =
if self.oclIsTypeOf(Wright!Prefix)then
if self.event.oclIsTypeOf(Wright!EventObserved) then
Set{self.event}->union(self.target.getEventObserved())
else
self.target.getEventObserved()
endif
else
if self.oclIsTypeOf(Wright!InternalChoice) or
self.oclIsTypeOf(Wright!ExternalChoice) then
self.elements->iterate( child1 ; elements1 : Set(Wright!EventObserved) =
Set{} | elements1->union(child1.getEventObserved()))
else
Set{}
endif
endif;
helper context Wright!ProcessExpression def: getInternalTrait():
Set(Wright!InternalTraitement) =
if self.oclIsTypeOf(Wright!Prefix)then
if self.event.oclIsTypeOf(Wright!InternalTraitement) then
Set{self.event}->union(self.target.getInternalTrait())
else
self.target.getInternalTrait()
endif
else
if self.oclIsTypeOf(Wright!InternalChoice) or
self.oclIsTypeOf(Wright!ExternalChoice) then
self.elements->iterate( child1 ; elements1 :
Set(Wright!InternalTraitement) = Set{} | elements1-
>union(child1.getInternalTrait()))
else
Set{}
endif
endif;
helper context Wright!Configuration def: getInternalTraitement:
Set(Wright!InternalTraitement)=
self.conn->iterate( child1 ; elements1 : Set(Wright!InternalTraitement) =
Set{} | elements1->union(child1.glue.getInternalTrait()))
->union(self.comp->iterate( child2 ; elements2 :
Set(Wright!InternalTraitement) = Set{} | elements2-
>union(child2.computation.getInternalTrait())));
helper context Wright!ExternalChoice def:
getPrefixInOrder():OrderedSet(Wright!Prefix) =
self.elements->select(c | c.event.oclIsTypeOf(Wright!EventObserved))
->union(self.elements->select(c |
c.event.oclIsTypeOf(Wright!SuccesEvent)));
helper context Wright!ExternalChoice def: transformation(instance :
String):Ada!select_or=
thisModule.ExternalChoice2select_or(self,instance);
helper context Wright!InternalChoice def: transformation(instance :
String):Ada!statement=
if self.elements->size()=2 then
```

```
thisModule.InternalChoice2if_else(self,instance)

else
thisModule.InternalChoice2case_statement(self,instance)
endif;
helper context Wright!Prefix def: transformation(instance :
String):Sequence(Ada!statement)=
if self.target.oclIsTypeOf(Wright!Prefix) then
Sequence{self.event.event_transform(instance)}-
>union(self.target.transformation(instance))
else
if self.target.oclIsTypeOf(Wright!InternalChoice) or
self.target.oclIsTypeOf(Wright!ExternalChoice) then
Sequence{self.event.event_transform(instance)}->union(Sequence
{self.target.transformation(instance)})
else
Sequence{self.event.event_transform(instance)}
endif
endif;
helper context Wright!EventObserved def: event_transform(instance :
String):Ada!simple_accept_statement=
thisModule.EventObserved2simple_accept_statement(self);
helper context Wright!EventSignalled def: event_transform(instance :
String):Ada!entry_call_statement=
thisModule.EventSignalled2entry_call_statement(self,instance);
helper context Wright!InternalTraitement def: event_transform(instance :
String):Ada!procedure_call_statement=
thisModule.InternalTraitement2procedure_call_statement(self);
helper context Wright!SuccesEvent def: event_transform(instance :
String):Ada!exit_statement=
thisModule.SuccesEvent2exit_statement(self);
helper context Wright!Configuration def: ICBin:Wright!InternalChoice=
Wright!InternalChoice.allInstances() ->select(e |e.elements->size()=2)-
>at(1);
helper context Wright!Configuration def: ICGen:Wright!InternalChoice=
Wright!InternalChoice.allInstances() ->select(e |e.elements->size()>2)-
>at(1);
helper context Wright!Configuration def: existICGen:Boolean=
if(Wright!InternalChoice.allInstances() ->select(e |e.elements-
>size()>2)->isEmpty() )then false else true endif;
helper context Wright!Configuration def: existICBin:Boolean=
if(Wright!InternalChoice.allInstances() ->select(e |e.elements-
>size()=2)->isEmpty() )then false else true endif;
rule Configuration2subprogram{
from c: Wright!Configuration
to sb: Ada!subprogram_body (
specif <- sp ,
statements <- st ,
declarations <- c.getInternalTraitement ->
collect(e|thisModule.InternalTraitement2subprogram(e))
->union(if (c.existICGen)then
OrderedSet{thisModule.InternalChoiceG2function(c.ICGen)} else
OrderedSet{}endif)
->union(if (c.existICBin)then
OrderedSet{thisModule.InternalChoiceB2function(c.ICBin)} else
OrderedSet{}endif)
->union(c.compInst ->
collect(e|thisModule.ComponentInstance2single_task_declaration(e)))
```

```
->union(c.connInst ->
collect(e|thisModule.ConnectorInstance2single_task_declaration(e)))
->union(c.compInst ->
collect(e|thisModule.ComponentInstance2task_body(e)))
->union(c.connInst ->
collect(e|thisModule.ConnectorInstance2task_body(e)))),
sp: Ada!procedure_specification( designator <- c.name),
st: Ada!null_statement
}
lazy rule InternalChoiceB2function{
from i:Wright!InternalChoice(not i.OclUndefined())
to s:Ada!subprogram_body(specif <- fs, statements <- r),
fs: Ada!function_specification( designator <- 'condition_interne',
returnType<-'Boolean'),
r:Ada!return_statement(exp<-e1),
e1:Ada!expression(e<-'true')
}
lazy rule InternalChoiceG2function{
from i:Wright!InternalChoice(not i.OclUndefined())
to s:Ada!subprogram_body(specif <- fs, statements <- r),
fs: Ada!function_specification( designator <- 'condition_interne1',
returnType<-'Integer'),
r:Ada!return_statement(exp<-e1),
e1:Ada!expression(e<-'1')
}
lazy rule InternalTraitement2subprogram{
from i:Wright!InternalTraitment
to sb: Ada!subprogram_body( specif <- ps,
statements <-ns),
ns:Ada!null_statement,
ps: Ada!procedure_specification( designator <- i.name)
}
lazy rule EventObserved2entry_declaration{
from eo:Wright!EventObserved
to ed:Ada!entry_declaration(
identifier<- eo.name.replaceAll('.','_')
)
}
lazy rule ComponentInstance2single_task_declaration{
from ci:Wright!ComponentInstance
to std:Ada!single_task_declaration(
identifier <- 'Component_'+ci.name,
entryDec <-ci.type.computation.getEventObserved()-
>collect(e|thisModule.EventObserved2entry_declaration(e))
)
}
lazy rule ComponentInstance2task_body{
from ci:Wright!ComponentInstance
to tb:Ada!task_body(
identifier <-'Component_'+ ci.name,
statements <- ls
),
ls : Ada!simple_loop_statement(
s<- ci.type.computation.transformation(ci.name)
)
}
lazy rule ConnectorInstance2single_task_declaration{
from ci:Wright!ConnectorInstance
```

```
to std:Ada!single_task_declaration(
identifier <- 'Connector_'+ci.name,
entryDec <-ci.type.glue.getEventObserved()-
>collect(e|thisModule.EventObserved2entry_declaration(e))
)
}
lazy rule ConnectorInstance2task_body{
from ci:Wright!ConnectorInstance
to tb:Ada!task_body(
identifier <-'Connector_'+ ci.name,
statements <- ls
),
ls : Ada!simple_loop_statement(
s<- ci.type.glue.transformation(ci.name))
}
lazy rule ExternalChoice2select_or{
from p:Wright!ExternalChoice,
instance : String
to s:Ada!select_or(
ref <- p.getPrefixInOrder()->collect(e|
if e.event.oclIsTypeOf(Wright!EventObserved)then if
e.target.oclIsTypeOf(Wright!ProcessName) then
thisModule.Prefix2accept_alternative1(e,instance)
else
thisModule.Prefix2accept_alternative2(e,instance)
endif
else
thisModule.SuccesEvent2terminate_alternative(e)
endif ))
}
lazy rule Prefix2accept_alternative1{
from p:Wright!Prefix,
instance : String
to a:Ada!accept_alternative(
as <- thisModule.EventObserved2simple_accept_statement(p.event)
)
}
lazy rule Prefix2accept_alternative2{
from p:Wright!Prefix,
instance : String
to a:Ada!accept_alternative(
as <- thisModule.EventObserved2simple_accept_statement(p.event),
s<- p.target.transformation(instance)
)
}
lazy rule SuccesEvent2terminate_alternative{
from p:Wright!SuccesEvent
to a:Ada!terminate_alternative
}
lazy rule InternalChoice2if_else{
from p:Wright!InternalChoice,
instance : String
to ls:Ada!if_else(
s1 <- p.elements->at(1).transformation(instance),
s2 <- p.elements->at(2).transformation(instance),
cond<-c
),
c:Ada!condition(c<-'condition_interne')
```

```
}
lazy rule InternalChoice2case_statement{
from p:Wright!InternalChoice,
instance : String
to ls:Ada!case_statement(
ref <- p.elements-
>collect(e|thisModule.Prefix2case_statement_alternative(e,p.elements.inde
xOf(e),instance)),
exp<-c
),
c:Ada!expression(e<-'condition_interne1')
}
lazy rule Prefix2case_statement_alternative{
from p:Wright!Prefix,
index:Integer,
instance: String
to cs: Ada!case_statement_alternative(
choice<-index.toString(),
s<- p.transformation(instance)
)
}
lazy rule SuccesEvent2exit_statement{
from p:Wright!SuccesEvent
to e:Ada!exit_statement
}
lazy rule EventObserved2simple_accept_statement{
from e:Wright!EventObserved
to s:Ada!simple_accept_statement(
direct_name<- e.name.replaceAll('.','_')
)
}
lazy rule EventSignalled2entry_call_statement{
from e:Wright!EventSignalled,
instance : String
to ec:Ada!entry_call_statement(
entry_name<-if(Wright!Attachment.allInstances()-
>select(a|a.originPort.name=e.name.substring(1,e.name.indexOf('.')))-
>select(a|a.originInstance.name=instance)->notEmpty())then
'Connector_'
+Wright!Attachment.allInstances()-
>select(a|a.originPort.name=e.name.substring(1,e.name.indexOf('.')))-
>select(a|a.originInstance.name=instance).at(1).targetInstance.name
+'.'+Wright!Attachment.allInstances()-
>select(a|a.originPort.name=e.name.substring(1,e.name.indexOf('.')))-
>select(a|a.originInstance.name=instance).at(1).targetRole.name
+'_'+e.name.substring(e.name.indexOf('.')+2,e.name.size())
else
'Component_'
+Wright!Attachment.allInstances()-
>select(a|a.targetRole.name=e.name.substring(1,e.name.indexOf('.')))-
>select(a|a.targetInstance.name=instance).at(1).originInstance.name
+'.'+Wright!Attachment.allInstances()-
>select(a|a.targetRole.name=e.name.substring(1,e.name.indexOf('.')))-
>select(a|a.targetInstance.name=instance).at(1).originPort.name
+'_'+e.name.substring(e.name.indexOf('.')+2,e.name.size())
endif
)
}
```

```
lazy rule InternalTraitement2procedure_call_statement{
from e:Wright!InternalTraitement
to p:Ada!procedure_call_statement(
name<-e.name
)
}
```

# Annexe E: Grammaire de Wright en Xtext

```
grammar org.xtext.example.Wright1 with org.eclipse.xtext.common.Terminals
generate wright1 "http://www.xtext.org/example/Wright1"
Configuration : "Configuration" name=ID
( TypeList+=Type )*
"Instances"
( InstanceList+=Instance)*
"Attachments"
( att+=Attachment )*
"End Configuration";
Instance: ComponentInstance | ConnectorInstance ;
Type: Component| Connector;
Component : "Component" name=ID
( port+=Port )+
"Computation" '=' computation=ProcessExpression ;
Port : "Port" name=ID '=' behavior=ProcessExpression;
Connector : "Connector" name=ID
( role+=Role )+
"Glue" '=' glue=ProcessExpression ;
Role : "Role" name=ID '=' behavior=ProcessExpression;
ComponentInstance : name=ID ':' "Component" type=[Component];
ConnectorInstance : name=ID ':' "Connector" type=[Connector];
Attachment : originInstance=[ComponentInstance] '-' originPort=[Port]
"As" targetInstance=[ConnectorInstance] '-' targetRole=[Role] ;
EventExpression : EventSignalled | EventObserved | InternalTraitement |
SuccesEvent;
EventSignalled: '_' name=ID (data+=Data)*;
EventObserved: name=ID (data+=Data)*;
InternalTraitement: '-' name=ID;
SuccesEvent: name='V';
Data : ('?' | '!') name=ID;
Prefix: event=EventExpression '->' target=TargetPrefix | name='§'|
name='SKIP';
TargetPrefix: Parentheses | Prefix | ProcessName ;
ProcessName: name=ID ;
Parentheses: '(' p=ProcessExpression ')';
ProcessExpression : right=Prefix (('[]' ECLeft+=Prefix)+|('|~|'
ICLeft+=Prefix)+)?;
terminal ID: ('a'..'z'|'A'..'Z') ('a'..'z'|'A'..'Z'|'_'|'.'|'0'..'9')*;
```

# Annexe F: Spécifications en Xpand des instructions de la partie exécutive d'Ada.

✓ **L'instruction if :**

```
«DEFINE statement FOR if_else»
if «this.cond.c» then
«EXPAND statement FOREACH this.s1»
else
«EXPAND statement FOREACH this.s2»
end if;
«ENDDEFINE»
```

✓ **L'instruction case :**

```
«DEFINE statement FOR case_statement»
case «this.exp.e» is
«IF this.ref.notExists(e|e.choice=="others")»
«EXPAND Case FOREACH this.ref.reject(e|e.choice=="others")»
others => null;
«ELSE»
«EXPAND Case FOREACH this.ref.reject(e|e.choice=="others")»
«EXPAND Case FOREACH this.ref.select(e|e.choice=="others")»
«ENDIF»
end case;
«ENDDEFINE»
«DEFINE Case FOR case_statement_alternative»
when «this.choice» => «EXPAND statement FOREACH this.s»
«ENDDEFINE»
```

✓ **L'instruction select_or :**

```
«DEFINE statement FOR select_or»
select
«EXPAND Alternative FOREACH
this.ref.reject(e|e.metaType==ada::terminate_alternative) SEPARATOR 'or'-
»
«IF !this.ref.select(e|e.metaType==ada::terminate_alternative).isEmpty»
or
«ENDIF»
«EXPAND Alternative FOREACH
this.ref.select(e|e.metaType==ada::terminate_alternative) SEPARATOR 'or'-
»
end select;
«ENDDEFINE»
«DEFINE Alternative FOR select_alternative»
«ENDDEFINE»
```

✓ L'instruction terminate :

```
«DEFINE Alternative FOR terminate_alternative»
terminate;
«ENDDEFINE»
```

✓ L'instruction accept :

```
«DEFINE Alternative FOR accept_alternative»
accept «this.as.direct_name»;
«EXPAND statement FOREACH this.s»
```

```
«ENDDEFINE»
```

✓ **L'instruction loop :**

```
«DEFINE statement FOR simple_loop_statement»
loop
«EXPAND statement FOREACH this.s»
end loop;
«ENDDEFINE»
```

✓ L'instruction nulle :

```
«DEFINE statement FOR null_statement»
null;
«ENDDEFINE»
```

✓ L'instruction exit :

```
«DEFINE statement FOR exit_statement»
exit;
«ENDDEFINE»
```

✓ L'instruction return :

```
«DEFINE statement FOR return_statement»
return «this.exp.e»;
«ENDDEFINE»
```

✓ L'appel d'une procédure :

```
«DEFINE statement FOR procedure_call_statement»
«this.name»;
«ENDDEFINE»
```

✓ L'appel des entrées :

```
«DEFINE statement FOR entry_call_statement»
«this.entry_name»;
«ENDDEFINE»
```

# Annexe G : Template de génération de code Ada en Xpand

```
«IMPORT ada»
«DEFINE main FOR subprogram_body»
«FILE "adaCode.adb"»
«EXPAND specification FOR this.specif-»
«EXPAND declaration FOREACH this.declarations-»
begin
«EXPAND statement FOREACH this.statements-»
end «this.specif.designator»;
«ENDFILE»
«ENDDEFINE»
«DEFINE specification FOR subprogram_specification»
«ENDDEFINE»
«DEFINE specification FOR procedure_specification»
procedure «this.designator» is
«ENDDEFINE»
«DEFINE specification FOR function_specification»
function «this.designator» return «this.returnType» is
«ENDDEFINE»
«DEFINE declaration FOR declaration»
«ENDDEFINE»
«DEFINE declaration FOR procedure_specification»
procedure «this.designator» ;
«ENDDEFINE»
«DEFINE declaration FOR function_specification»
function «this.designator» return «this.returnType» ;
«ENDDEFINE»
«DEFINE declaration FOR subprogram_body»
«EXPAND specification FOR this.specif-»
«EXPAND declaration FOREACH this.declarations-»
begin
«EXPAND statement FOREACH this.statements-»
end «this.specif.designator»;
«ENDDEFINE»
«DEFINE declaration FOR single_task_declaration»
task «this.identifier» «IF this.entryDec.isEmpty» ; «ELSE» is
«EXPAND Entry FOREACH this.entryDec-»
end «this.identifier»;
«ENDIF»
«ENDDEFINE»
«DEFINE Entry FOR entry_declaration»
entry «this.identifier» ;
«ENDDEFINE»
«DEFINE declaration FOR task_body»
task body «this.identifier» is
«EXPAND declaration FOREACH this.declarations-»
begin
«EXPAND statement FOREACH this.statements-»
end «this.identifier»;
«ENDDEFINE»
«DEFINE statement FOR statement»
«ENDDEFINE»
«DEFINE statement FOR null_statement»
null;
«ENDDEFINE»
«DEFINE statement FOR simple_loop_statement»
loop
«EXPAND statement FOREACH this.s-»
```

```
end loop;
«ENDDEFINE»
«DEFINE statement FOR if_else»
if «this.cond.c» then
«EXPAND statement FOREACH this.s1-»
else
«EXPAND statement FOREACH this.s2-»
end if;
«ENDDEFINE»
«DEFINE statement FOR select_or»
select
«EXPAND Alternative FOREACH this.ref.reject(e|e.metaType==ada::terminate_alternative) SEPARATOR 'or'-»
«IF !this.ref.select(e|e.metaType==ada::terminate_alternative).isEmpty»
or
«ENDIF»
«EXPAND Alternative FOREACH this.ref.select(e|e.metaType==ada::terminate_alternative) SEPARATOR 'or'-»
end select;
«ENDDEFINE»
«DEFINE statement FOR case_statement»
case «this.exp.e» is
«IF this.ref.notExists(e|e.choice=="others")»
«EXPAND Case FOREACH this.ref.reject(e|e.choice=="others")-»
when others => null;
«ELSE»
«EXPAND Case FOREACH this.ref.reject(e|e.choice=="others")-»
«EXPAND Case FOREACH this.ref.select(e|e.choice=="others")-»
«ENDIF»
end case;
«ENDDEFINE»
«DEFINE statement FOR return_statement»
return «this.exp.e»;
«ENDDEFINE»
«DEFINE statement FOR exit_statement»
exit;
«ENDDEFINE»
«DEFINE statement FOR procedure_call_statement»
«this.name»;
«ENDDEFINE»
«DEFINE statement FOR entry_call_statement»
«this.entry_name»;
«ENDDEFINE»
«DEFINE statement FOR simple_accept_statement»
accept «this.direct_name»;
«ENDDEFINE»
«DEFINE Alternative FOR select_alternative»
«ENDDEFINE»
«DEFINE Alternative FOR terminate_alternative»
terminate;
«ENDDEFINE»
«DEFINE Alternative FOR accept_alternative»
accept «this.as.direct_name»;
«EXPAND statement FOREACH this.s-»
«ENDDEFINE»
«DEFINE Case FOR case_statement_alternative»
```

```
when «this.choice» => «EXPAND statement FOREACH this.s-»
«ENDDEFINE»
```

# Résumé

L'approche par composants vise la réutilisation par assemblage aisé et cohérent des composants. Mais l'obtention d'un assemblage de composants cohérent n'est pas un exercice facile. Pour y parvenir, nous préconisons une approche contractuelle distinguant divers contrats syntaxiques, structurels, sémantiques, de synchronisation et de qualité de services. Nous avons appliqué avec succès cette approche contractuelle sur deux modèles de composants semi-formels : UML2.0 et Ugatze. En effet, nous proposons deux démarches *VerifComponentUML2.0* et *VerifComponentUgatze*.

La démarche *VerifComponentUML2.0* vise la vérification des contrats syntaxiques, structurels, de synchronisation et de qualité de services sur une assemblage de composants UML2.0 en passant par les deux modèles de composants formels Acme/Armani et Wright. *VerifComponentUML2.0* est équipé de deux outils : Wr2fdr et Wright2Ada. L'outil Wr2fdr permet de traduire des expressions Wright vers CSP afin de vérifier les contrats de synchronisation en utilisant le model-checker FDR. L'outil Wright2Ada est un outil IDM permettant de transformer un code Wright en Ada afin d'ouvrir UML2.0 sur les outils d'analyse statique et dynamique associés à Ada.

La démarche *VerifComponentUgatze* offre un cadre permettant de vérifier les contrats syntaxiques et structurels d'un assemblage de composants Ugatze en passant par Acme/Armani.

**Mots-clés :** Assemblage de composants cohérent, Approche contractuelle, Vérification, Modèle de composants semi-formel, Modèle de composants formel.

# Abstract

The component approach aims for the reuse by a coherent and easy components assembly. But obtaining a coherent components assembly is not an easy exercise. To achieve this, we advocate a contractual approach distinguishing different syntactic, structural, semantic, synchronization and service quality contracts. We have successfully applied this approach on two models of semi-formal contractual components: UML2.0 and Ugatze. Indeed, we propose two approaches: *VerifComponentUML2.0* and *VerifComponentUgatze*.

The *VerifComponentUML2.0* approach aims the verification of syntactic, structural, synchronization and quality service contracts on a UML2.0 component assembly through two formal component models Acme/Armani and Wright. *VerifComponentUML2.0* has two tools: Wr2fdr and Wright2Ada. The tool Wr2fdr allows translating Wright expression to CSP contracts in order to verify synchronization using the model checker FDR. It is a IDM tool Wright2Ada which allow is transforming Wright code to Ada, in order to open UML2.0 on static analysis and dynamic tools associated with Ada.

*VerifComponentUgatze* approach provides a frame allowing to check syntactic and structural contracts of an Ugatze component assembly through Acme/Armani.